\documentclass[preprint2]{emulateapj}

\shortauthors{Oliveira et al.}
\shorttitle{Disk Structure in Serpens}

\begin{document}

\title{The Physical Structure of Protoplanetary Disks: the Serpens
  Cluster Compared with Other Regions}
\author{ Isa Oliveira\altaffilmark{1,2,6}, Bruno
  Mer\'{i}n\altaffilmark{3}, Klaus M. Pontoppidan\altaffilmark{4}
and Ewine F. van Dishoeck\altaffilmark{1,5} }
\altaffiltext{1}{Leiden Observatory, Leiden University, P.O. Box 9513,
  NL-2300 RA Leiden, The Netherlands}
\altaffiltext{2}{McDonald Observatory, The University of Texas at Austin,
  Austin, TX 78712, USA, email: oliveira@astro.as.utexas.edu}
\altaffiltext{3}{Herschel Science Center, European Space Astronomy
  Centre (ESA), P.O. Box 78, E-28691 Villanueva de la Ca\~nada (Madrid),
  Spain}
\altaffiltext{4}{Space Telescope Science Institute, Baltimore, MD
  21218, USA}
\altaffiltext{5}{Max-Planck Institut f\"ur Extraterrestrische Physik,
  Giessenbachstrasse 1, D-85748 Garching, Germany}
\altaffiltext{6}{Harlan J. Smith Postdoctoral Fellow}

\begin{abstract}

  Spectral energy distributions are presented for 94 young stars
  surrounded by disks in the Serpens Molecular Cloud, based on
  photometry and {\it Spitzer} IRS spectra. Most of the stars have
  spectroscopically determined spectral types. Taking a distance to
  the cloud of 415 pc rather than 259 pc, the distribution of ages is
  shifted to lower values, in the 1 -- 3 Myr range, with a tail up to
  10 Myr. The mass distribution spans 0.2 -- 1.2 M$_\odot$, with
  median mass of 0.7 M$_\odot$. The distribution of fractional disk
  luminosities in Serpens resembles that of the young Taurus Molecular
  Cloud, with most disks consistent with optically thick, passively
  irradiated disks in a variety of disk geometries ($L_{{\rm
      disk}}/L_{{\rm star}} \sim$ 0.1). In contrast, the distributions
  for the older Upper Scorpius and $\eta$ Chamaeleontis clusters are
  dominated by optically thin lower luminosity disks ($L_{{\rm
      disk}}/L_{{\rm star}} \sim$ 0.02). This evolution in fractional
  disk luminosities is concurrent with that of disk fractions: with
  time disks become fainter and the disk fractions decrease. The
  actively accreting and non-accreting stars (based on H$\alpha$ data)
  in Serpens show very similar distributions in fractional disk
  luminosities, differing only in the brighter tail dominated by
  strongly accreting stars. In contrast with a sample of Herbig Ae/Be
  stars, the T Tauri stars in Serpens do not have a clear separation
  in fractional disk luminosities for different disk geometries: both
  flared and flat disks present wider, overlapping distributions. This
  result is consistent with previous suggestions of a faster evolution
  for disks around Herbig Ae/Be stars. Furthermore, the results for
  the mineralogy of the dust in the disk surface (grain sizes,
  temperatures and crystallinity fractions, as derived from {\it
    Spitzer} IRS spectra) do not show any correlation to either
  stellar and disk characteristics or mean cluster age in the 1 -- 10
  Myr range probed here. A possible explanation for the lack of
  correlation is that the processes affecting the dust within disks
  have short timescales, happening repeatedly, making it difficult to
  distinguish long lasting evolutionary effects.

\end{abstract}


\section{Introduction}
\label{7sintro}

Protoplanetary disks are a natural consequence of the star formation
process. They are created as a result of the conservation of angular
momentum when a dense slowly rotating core in a molecular cloud
collapses to form a star \citep{SH93,MY00}. There is evidence that the
initial disk mass is a function of the stellar mass
\citep{AW05,GR10}. In addition, different disk lifetimes have been
suggested for stars of different masses, with disks around low-mass
stars lasting longer \citep{LA06,CA06,KK09}. If true, these relations
put strong constraints on the number of planets, and of which type,
could be formed in such disks. A great diversity in planetary systems
is observed for the more than 750 exo-planets confirmed to date
(\citealt{US07},\footnote{http://exoplanet.eu}) and it is important to
explore whether the variety of planets is a consequence of the
diversification in stars and their protoplanetary disks.

Combining theory, observations and laboratory experiments, there has
been significant progress in our understanding on initial growth from
dust into pebbles \citep{WE80,DT97,BW08}. The further growth is still
under debate, and is a very active field in simulations of planet
formation. In addition to growth, a change in dust mineralogy has been
observed.  Crystallization of the originally amorphous interstellar
grains is necessary to understand the high crystallinity fraction
found in many comets and interplanetary dust grains (see
\citealt{WO07,PB10,HE10} for recent reviews). An open question is to
what extent these dust properties are related to the stellar and disk
characteristics.

This work presents the spectral energy distributions (SEDs) of the
young stellar population of the Serpens Molecular Cloud discovered by
the {\it Spitzer Space Telescope} legacy program ``From molecular
cores to planet-forming disks'' (c2d, \citealt{EV03}), together with
{\it Spitzer} IRS spectra \citep{OL10,OL11}. Combined with photometry
(from optical to mid-IR, \citealt{HA06,HB07,HA07,SP10}) and stellar
spectral types, these data provide the necessary ingredients to
construct the SEDs and study the physical structure of disks (and its
dust) surrounding the young stars in Serpens.

For low-mass stars, the stellar and disk characteristics cannot be
easily separated as is the case for higher mass Herbig stars
(e.g. \citealt{ME01}), unless the stellar characteristics are well
known. In the last decade, a growing number of low-mass star forming
regions have been surveyed throughout the wavelength spectrum. The
original prototype, Taurus (e.g. \citealt{KH87,KH95}), is joined by
Ophiuchus, Chamaeleon, Lupus amongst others
\citep{KL04,KL08,FC08,CE08,EV09}, probing different stellar densities,
environments, sample sizes and mean cluster ages. To test the
universality of the results achieved in this field, we have engaged in
a systematic study of stars and their disks in several of the nearby
low-mass star-forming regions
\citep{JA08,SP08,SP10,BM10,AM11}. Similar procedures to those
presented here for constructing SEDs are being performed for a large
number of systems in most of the nearby star-forming regions observed
by {\it Spitzer}, considering all young stellar objects YSOs) for which the
central star has been optically characterized in the literature.  This
large database allows comparison between the disks in Serpens with
those in other star-forming regions, of different mean ages and
environments. 

The well characterized Taurus sample (2 Myr, \citealt{LH01,LU10}) is
used here in comparison with Serpens, both probing the young bin of
disk evolution. Taurus has been studied over a wide range of
wavelengths, from X-rays to radio, which allows an extensive
characterization of its members that are still surrounded by disks, as
well as the lower fraction of young stars ($\sim$40\%) around which
disks have already dissipated (e.g. \citealt{DP99,AW05,MG07}). Older
populations are probed using well-studied samples in $\eta$
Chamaeleontis ($\sim$6 Myr, \citealt{LS04,SI09}) and Upper Scorpius
(originally thought to be 5 Myr, but recently found to be 11 Myr,
\citealt{BL78,PE12,DC09}). The stellar and disk characteristics for
hundreds of objects in Taurus, Upper Scorpius and $\eta$ Chamaeleontis
with well studied stars and disks (making these samples statistically
robust) will be used in this work to place Serpens into an
evolutionary context.

This article is presented as follows: the SEDs are constructed in \S{}
\ref{7sseds}. Specifically, the data are presented in \S{}
\ref{7sdata}, the procedure to construct the SEDs is described in \S{}
\ref{7sproced}. Using the new distance estimate for the cloud ($d$ =
415 pc, \citealt{DZ10}), an updated distribution of ages and masses is
derived in \S{} \ref{7shrd}. The disk characteristics are discussed in
\S{} \ref{7sdisk}. With stars and disks well characterized, \S{}
\ref{7sconnect} investigates to what extent they affect each other and
whether the dust properties are correlated with either. Finally, the
conclusions are presented in \S{} \ref{7scon}.

\section{Spectral Energy Distributions}
\label{7sseds}

\subsection{Data}
\label{7sdata}

The Serpens Molecular Cloud has been imaged by {\it Spitzer} as part
of the c2d program. The detected sources in the IRAC and MIPS bands
were published by \citet{HA06} and \citet{HB07}, respectively. By
combining the data in all bands, \citet{HA07} could identify a red
population classified as YSO candidates, which is interpreted as being
due to emission from the disk. Confirmation of their nature as young
object members of the cloud was done through spectroscopy. The final
catalog is band-merged with the Two Micron All Sky Survey (2MASS),
providing data at $J$, $H$, $K_s$ (at 1.2, 1.6 and 2.2 $\mu$m,
respectively), IRAC bands 1 through 4 (at 3.6, 4.5, 5.8 and 8.0
$\mu$m) and MIPS bands 1 and 2 (at 24 and 70 $\mu$m), when
available. \citet{OL10} describe the complete, flux-limited sample of
YSOs in Serpens that is used in this work, for which {\it Spitzer} IRS
spectra have been taken. The 115 objects comprise the entire young
IR-excess population of Serpens that is brighter than 3 mJy at 8
$\mu$m (from the catalog of \citealt{HA07}). With this sensitivity, we
can detect YSOs close to the brown dwarf limit. Of these 115 young
objects, 21 are shown to be still embedded in a dusty envelope. The
remaining 94 objects, classified as disk sources, are the subject of
this work. \citet{OL09} derived spectral types (and therefore also
temperatures) from optical spectroscopy for 62\% of the Serpens
flux-limited disk sample (58 objects). The remaining 36 objects are
too extincted and could not be observed spectroscopically using
4-m class telescopes. These objects have spectral types derived
from photometry alone, which is less reliable than derivations from
spectroscopy. Optical $R$-band photometry is available covering
exactly the same area of Serpens as was covered by the c2d {\it
  Spitzer} observations \citep{SP10}, however the high extinction
toward a few directions in Serpens makes it impossible for optical
detection. That means that not all objects have optical photometric
data available.

The {\it Spitzer} IRS mid-IR spectroscopy (5 -- 35 $\mu$m) for this
sample cover the silicate bands at 10 and 20 $\mu$m that are emitted
by the dust in the surface layers of optically thick protoplanetary
disks \citep{OL10}. Information about the typical sizes, composition
and crystalline fractions of the emitting dust can be obtained from
fitting models to these silicate bands. Those results are presented in
\citet{OL11}.

\subsection{Building the SEDs}
\label{7sproced}

The first step to build the SED of a given object is to determine the
stellar emission. For each object, a NextGen stellar photosphere
\citep{HA99} corresponding to the spectral type of said star is
selected. This model photosphere is scaled to either the optical or
the 2MASS $J$ photometric point to account for the object's
brightness. The observed photometric data are corrected for extinction
from its visual extinction ($A_V$) using the \citet{WD01} extinction
law, with $R_V$ = 5.5. For objects without $A_V$ values derived from
the optical spectroscopy, these values are estimated by the best fit
of the optical/near-IR photometry to the NextGen photosphere, on a
close visual inspection of the final result SEDs.

\begin{figure*}[!h]
\begin{center}
\includegraphics[width=0.2\textwidth]{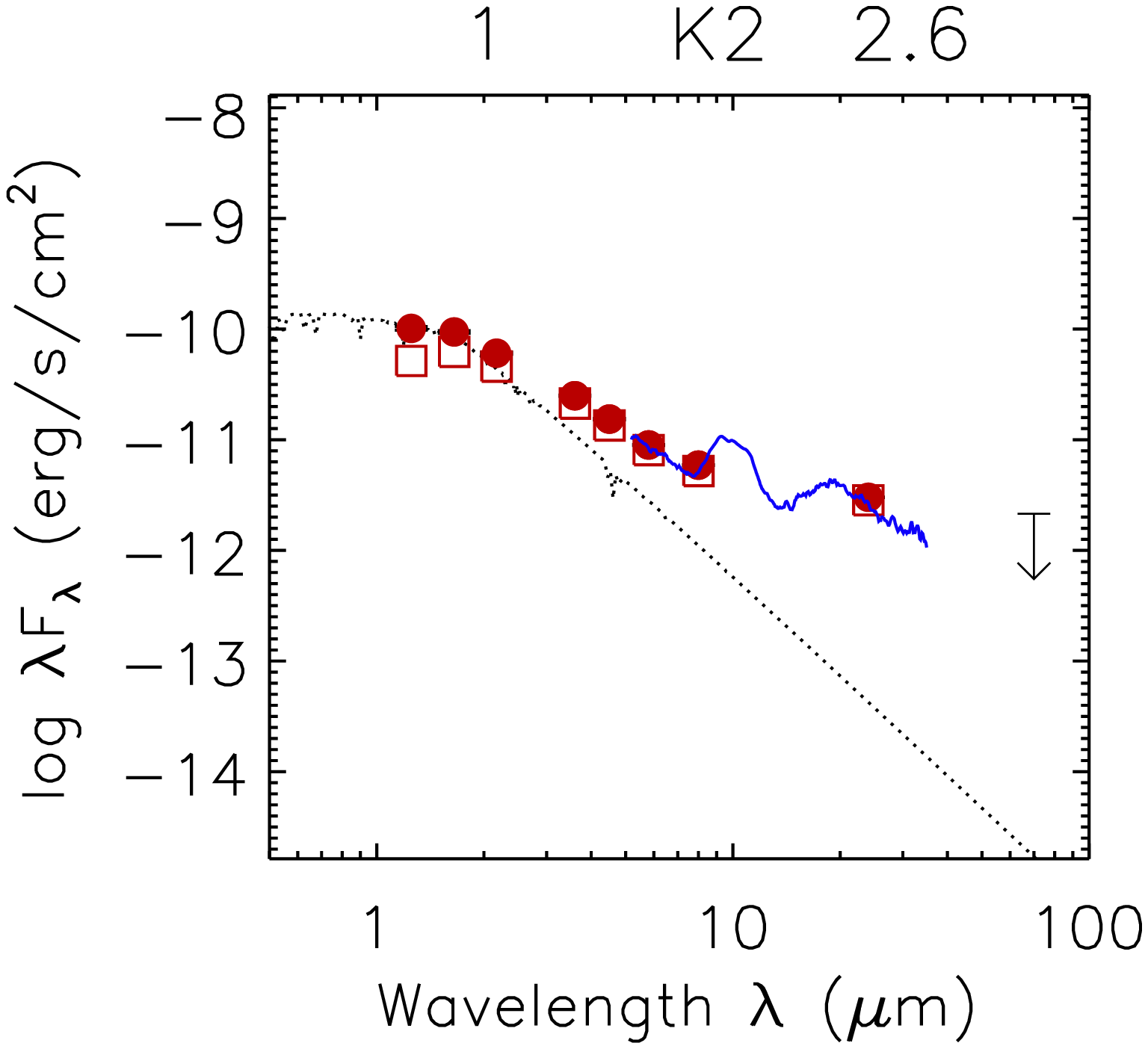}
\includegraphics[width=0.2\textwidth]{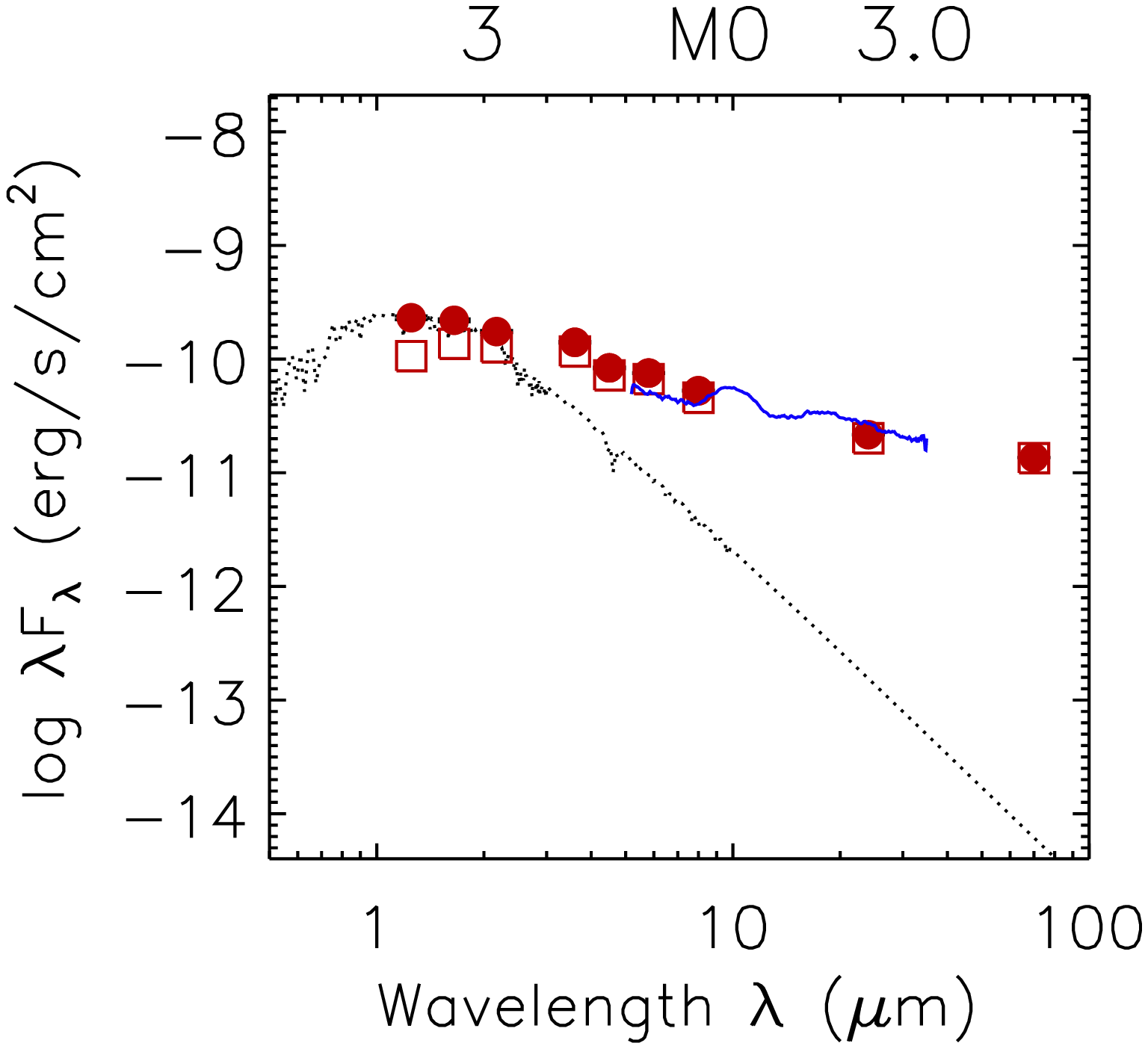}
\includegraphics[width=0.2\textwidth]{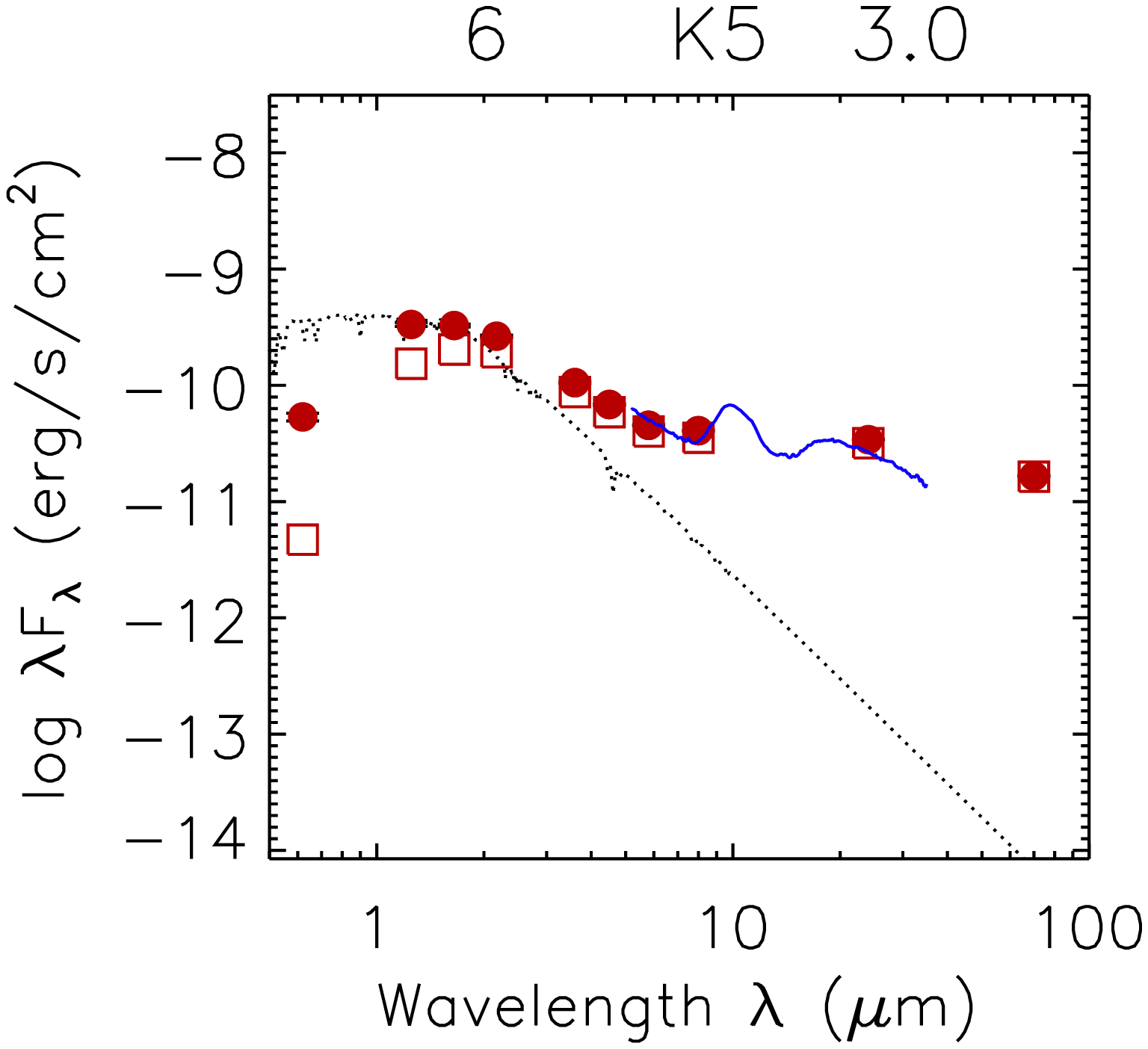}
\includegraphics[width=0.2\textwidth]{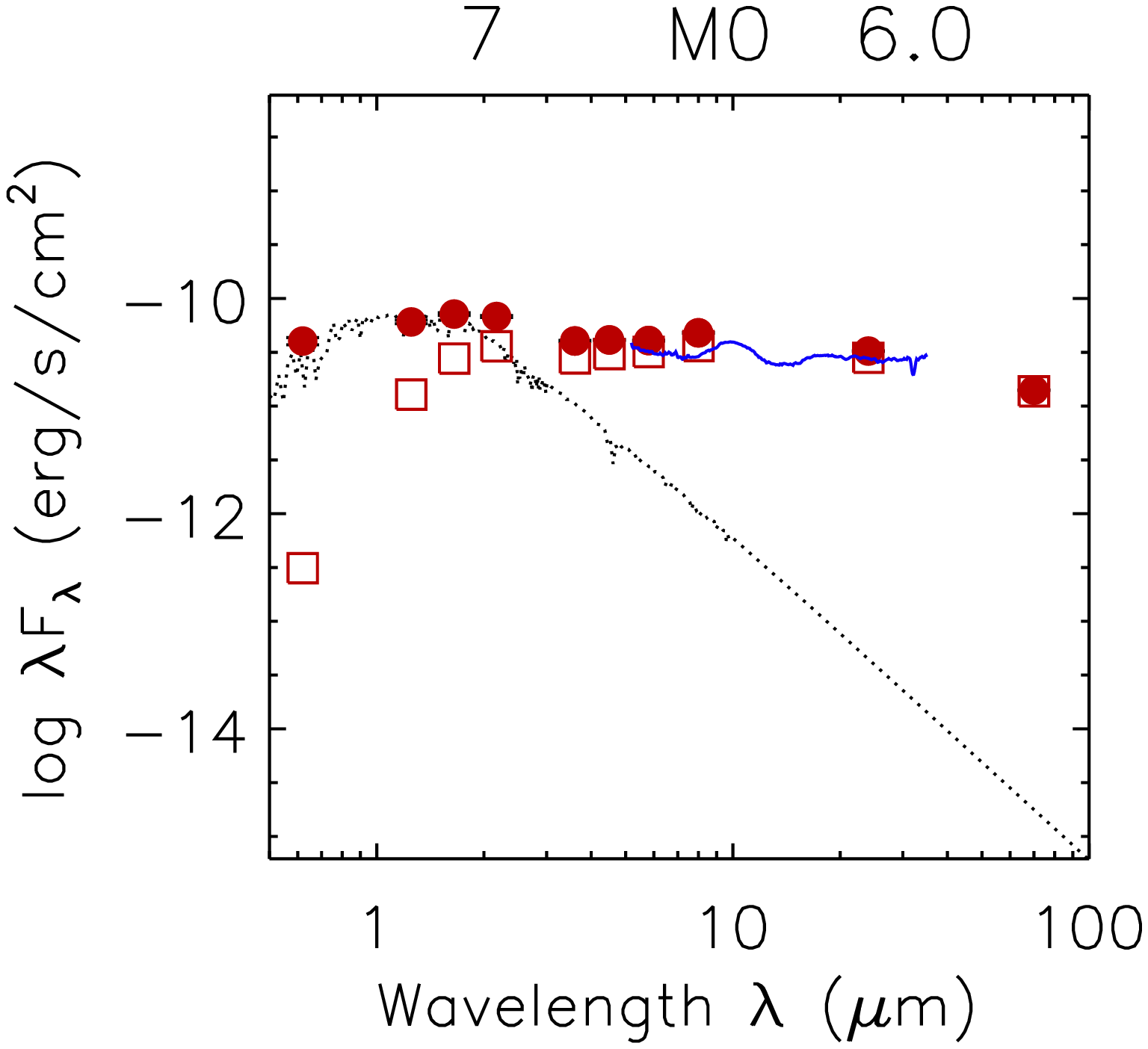}
\includegraphics[width=0.2\textwidth]{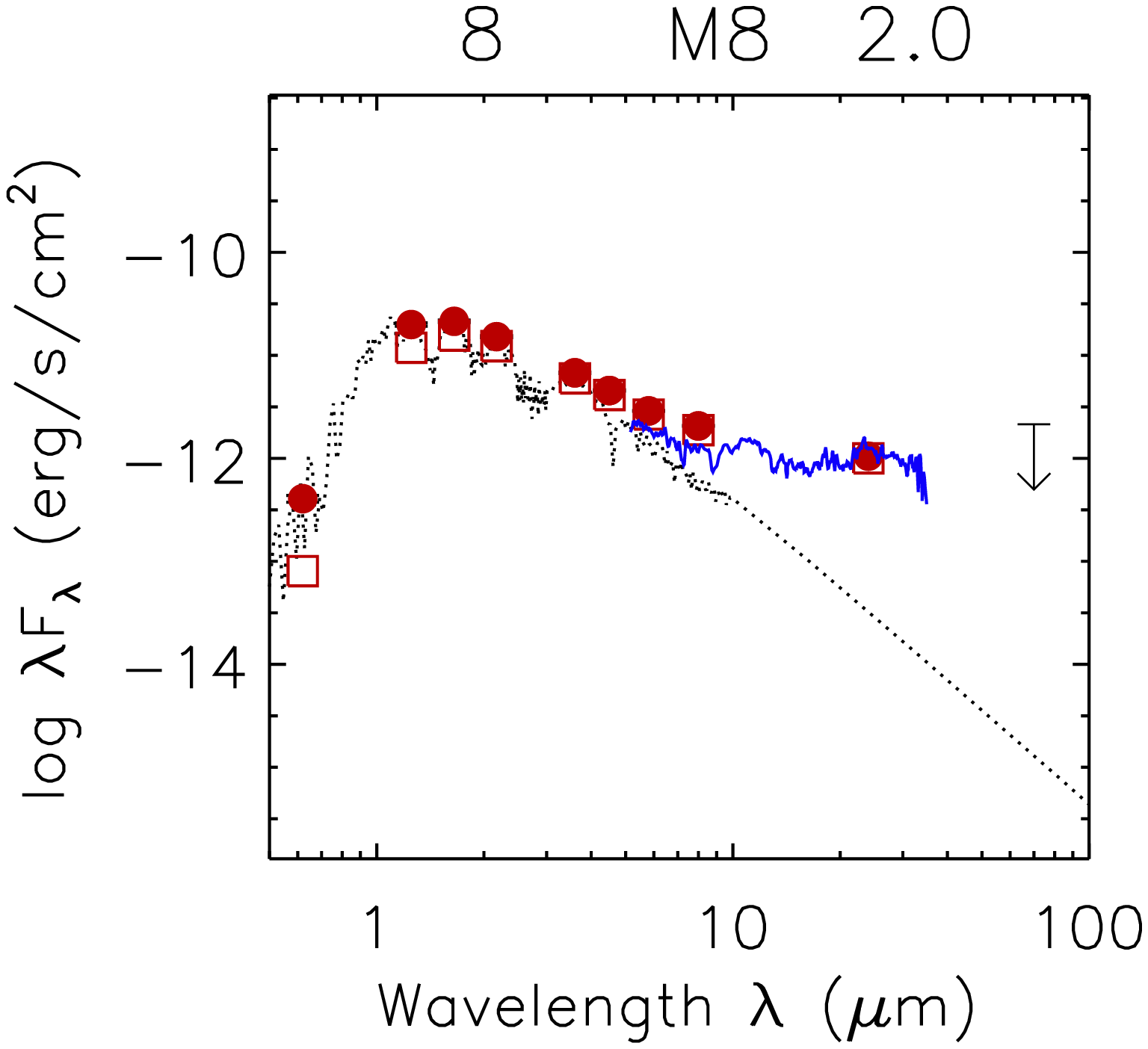}
\includegraphics[width=0.2\textwidth]{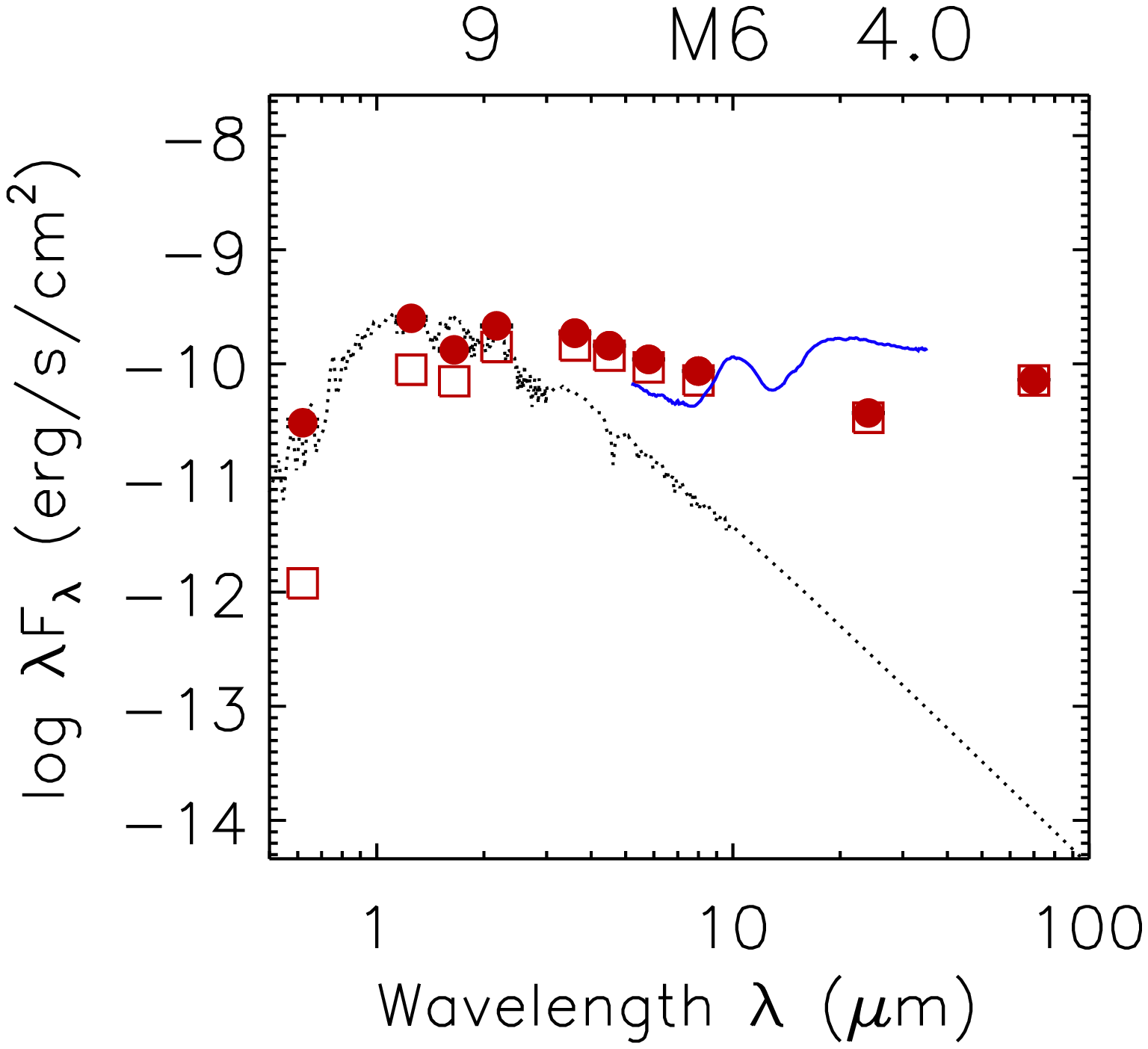}
\includegraphics[width=0.2\textwidth]{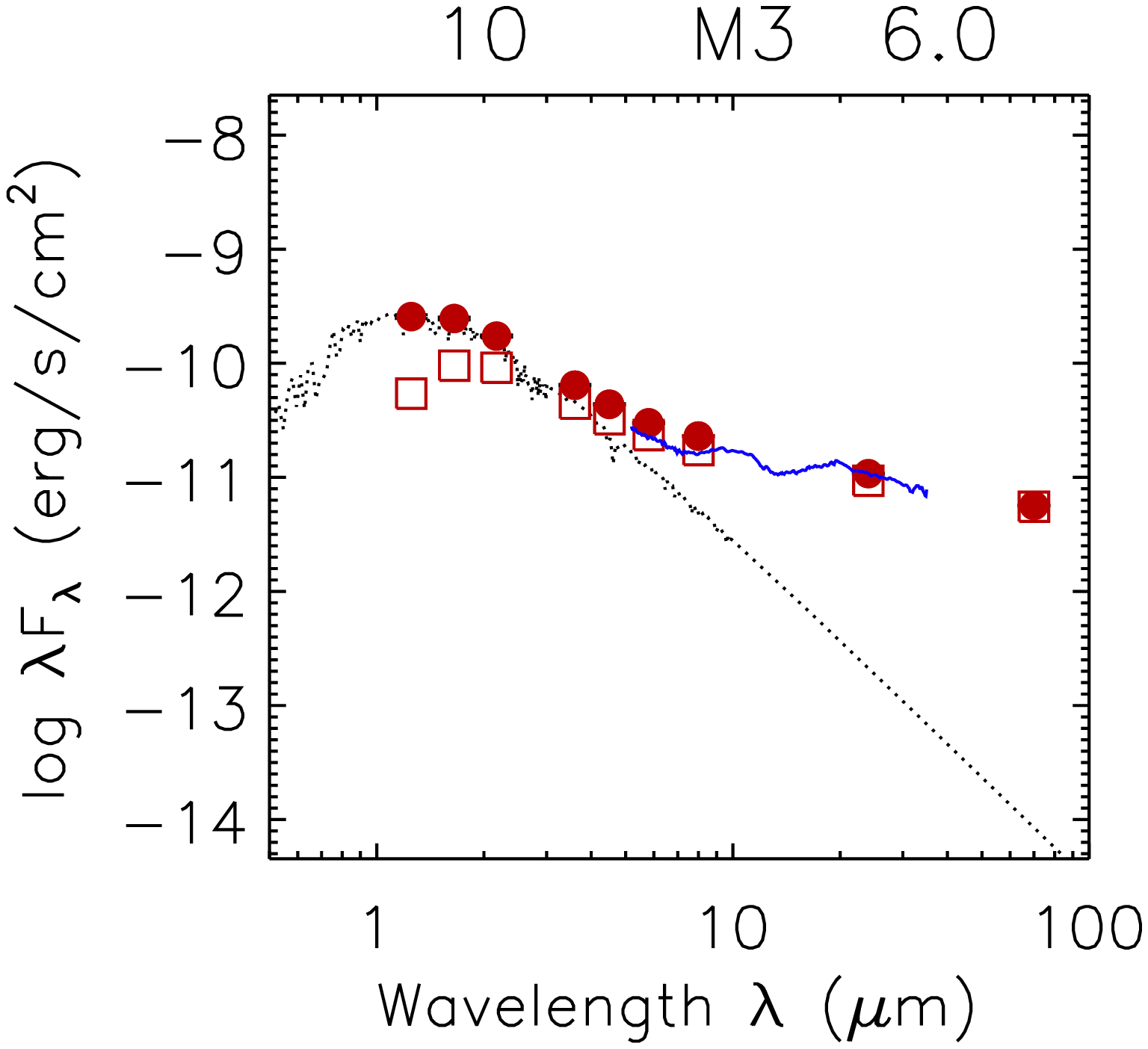}
\includegraphics[width=0.2\textwidth]{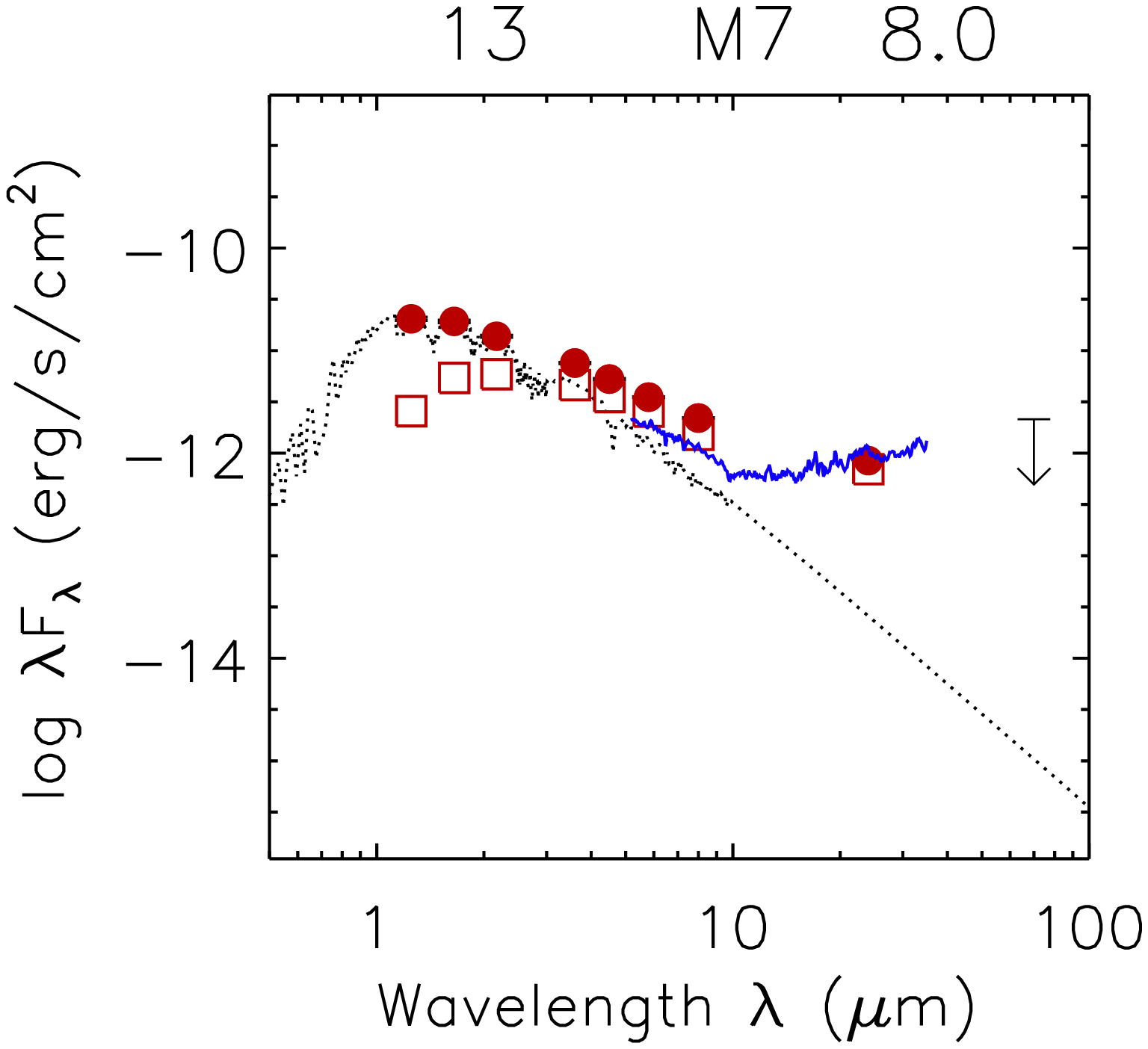}
\includegraphics[width=0.2\textwidth]{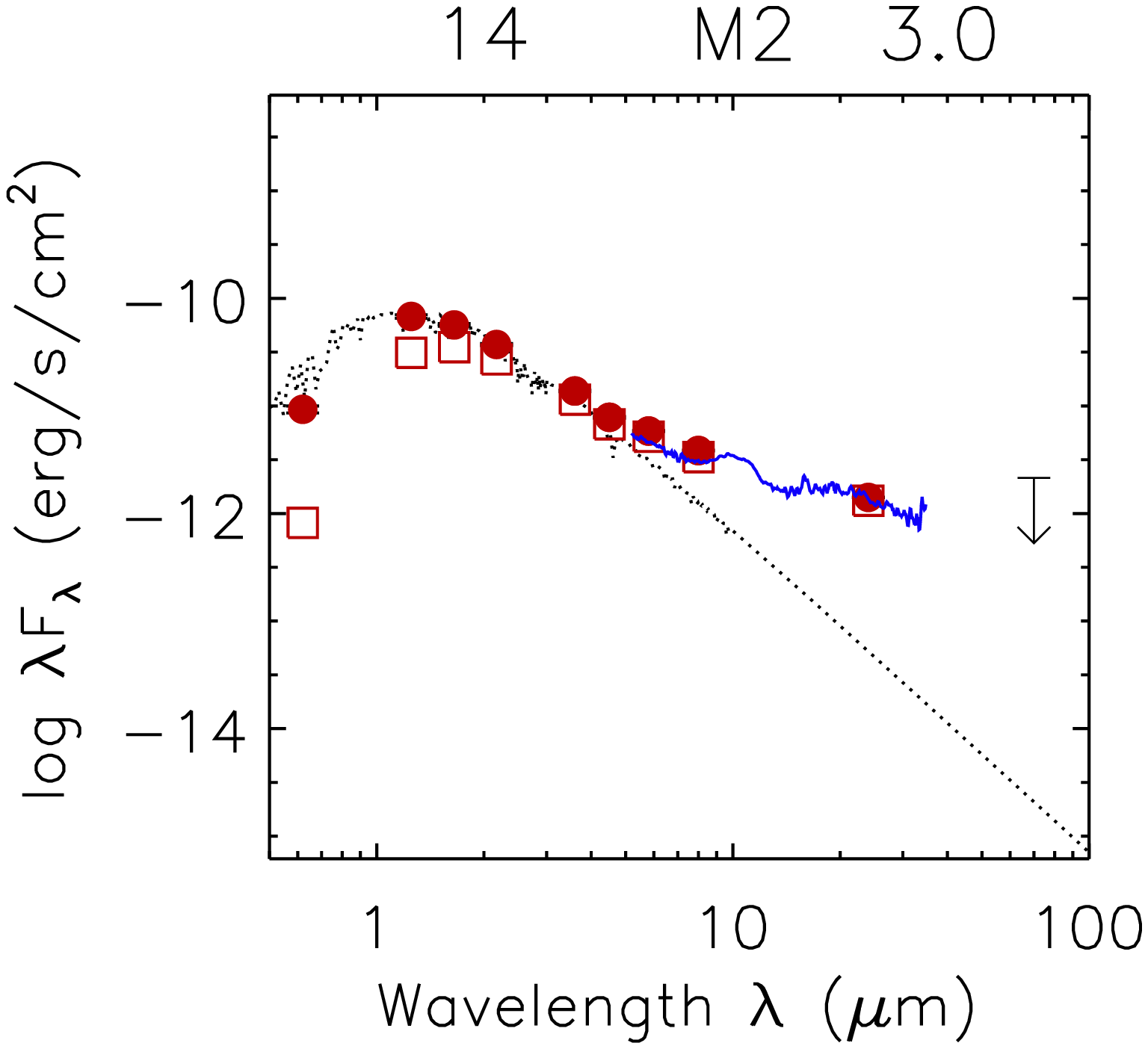}
\includegraphics[width=0.2\textwidth]{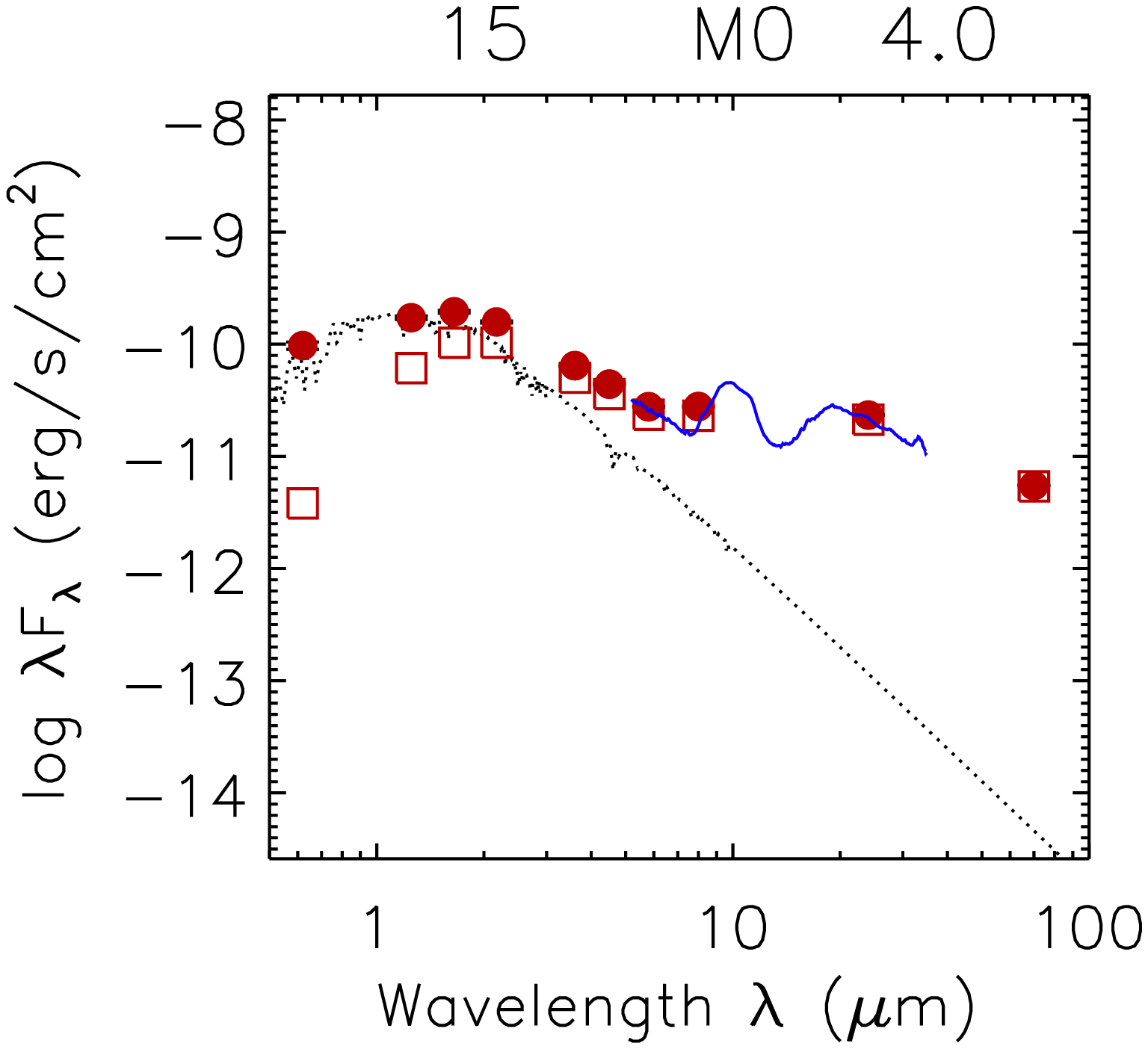}
\includegraphics[width=0.2\textwidth]{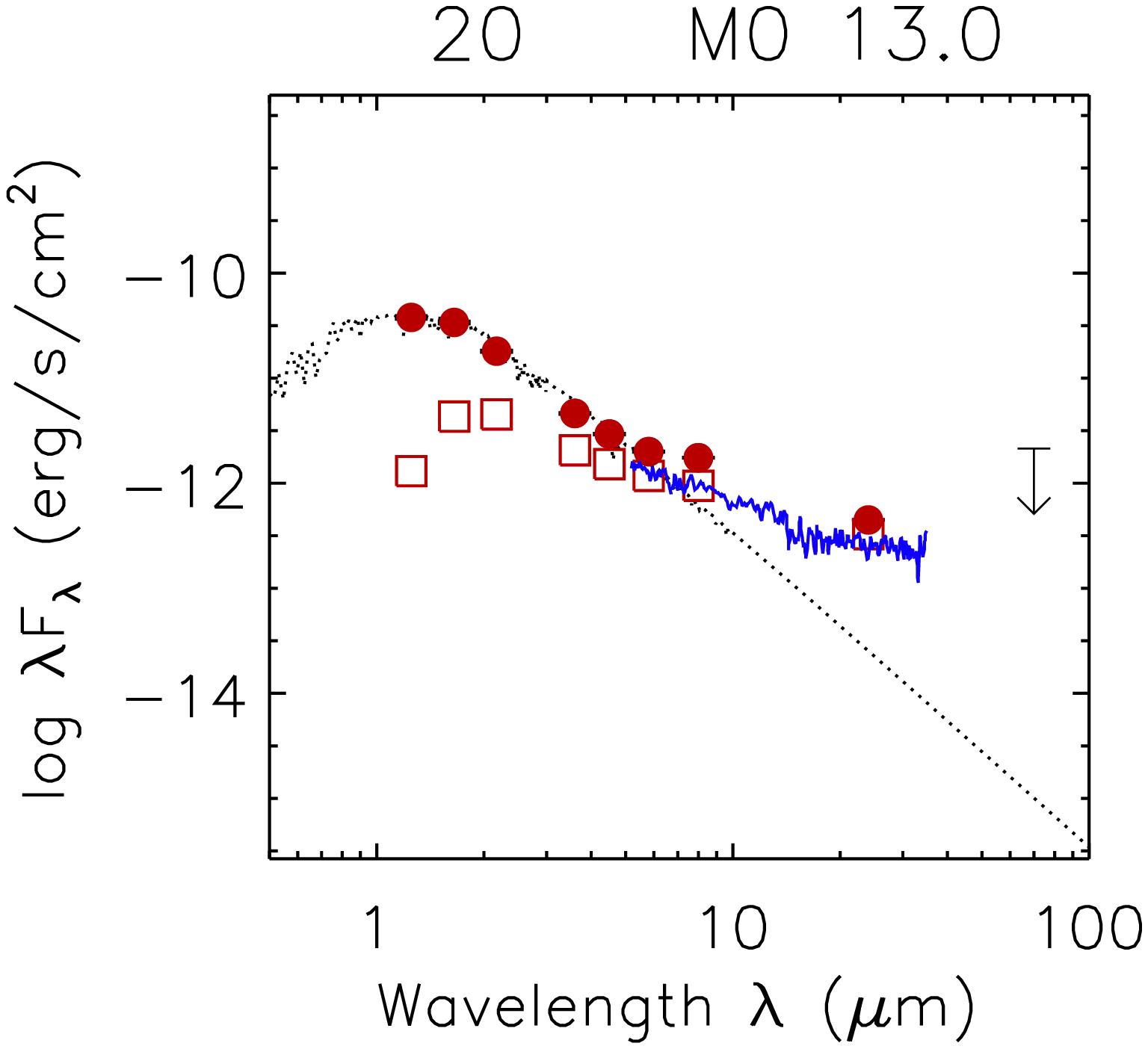}
\includegraphics[width=0.2\textwidth]{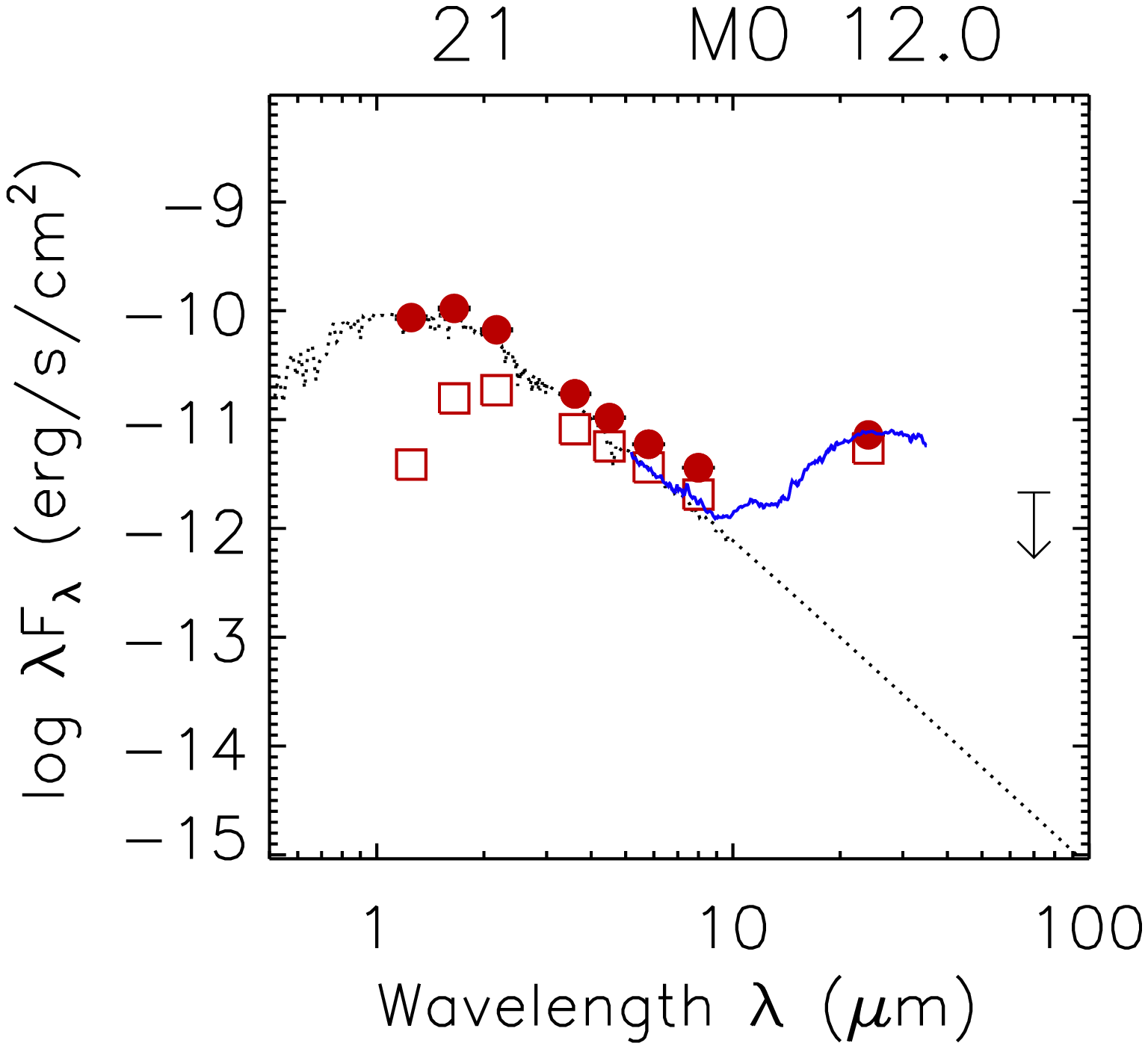}
\includegraphics[width=0.2\textwidth]{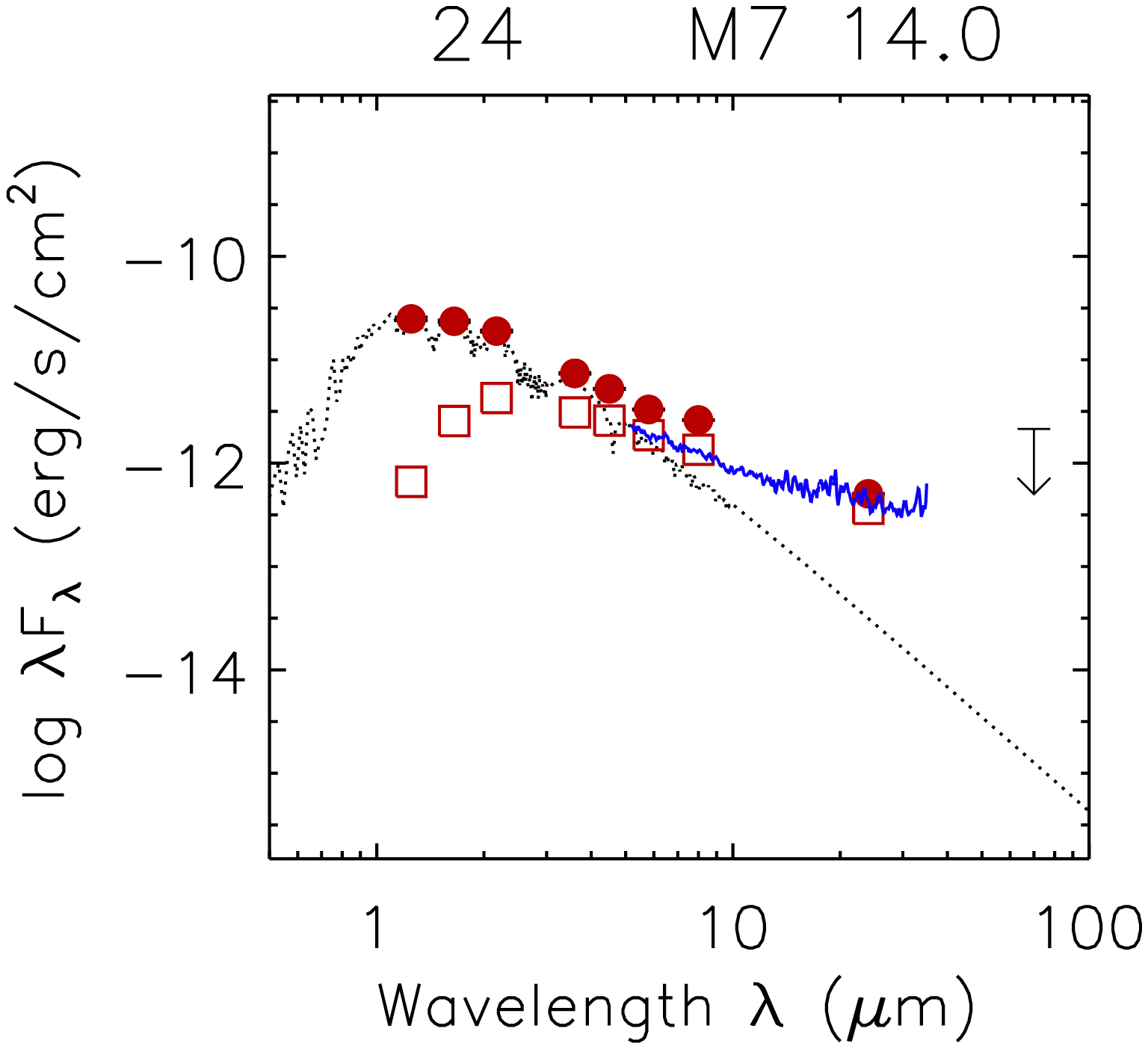}
\includegraphics[width=0.2\textwidth]{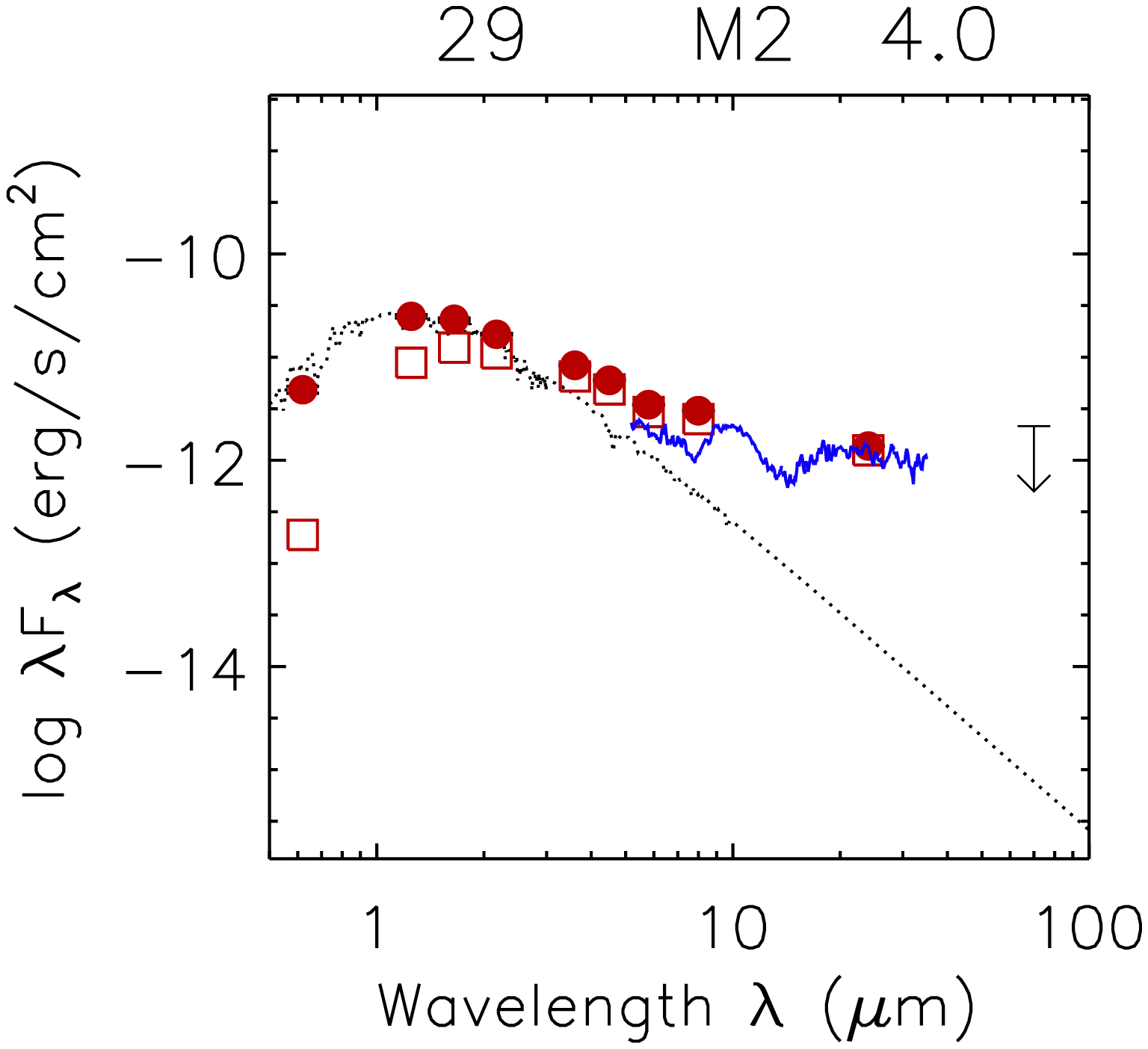}
\includegraphics[width=0.2\textwidth]{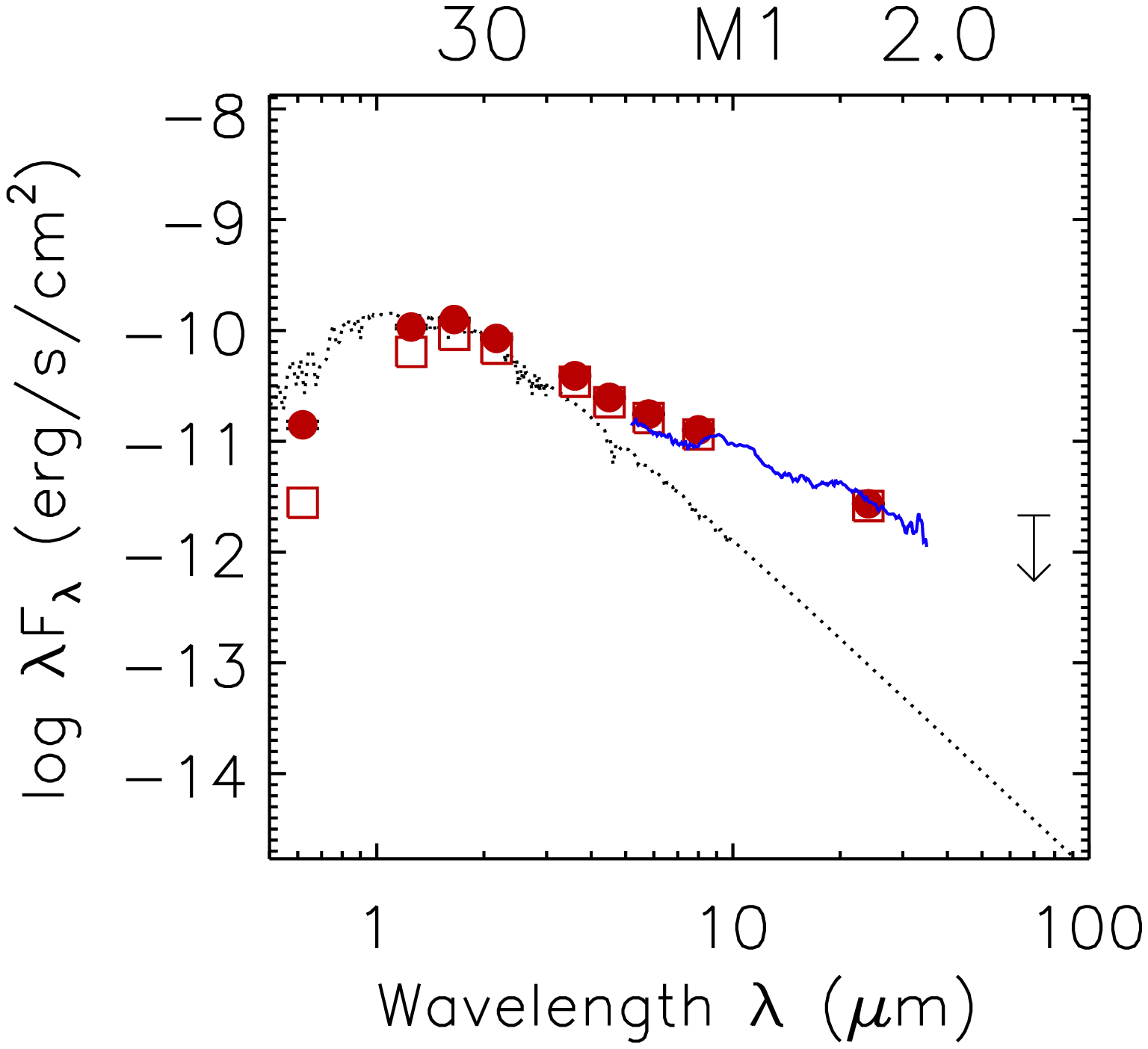}
\includegraphics[width=0.2\textwidth]{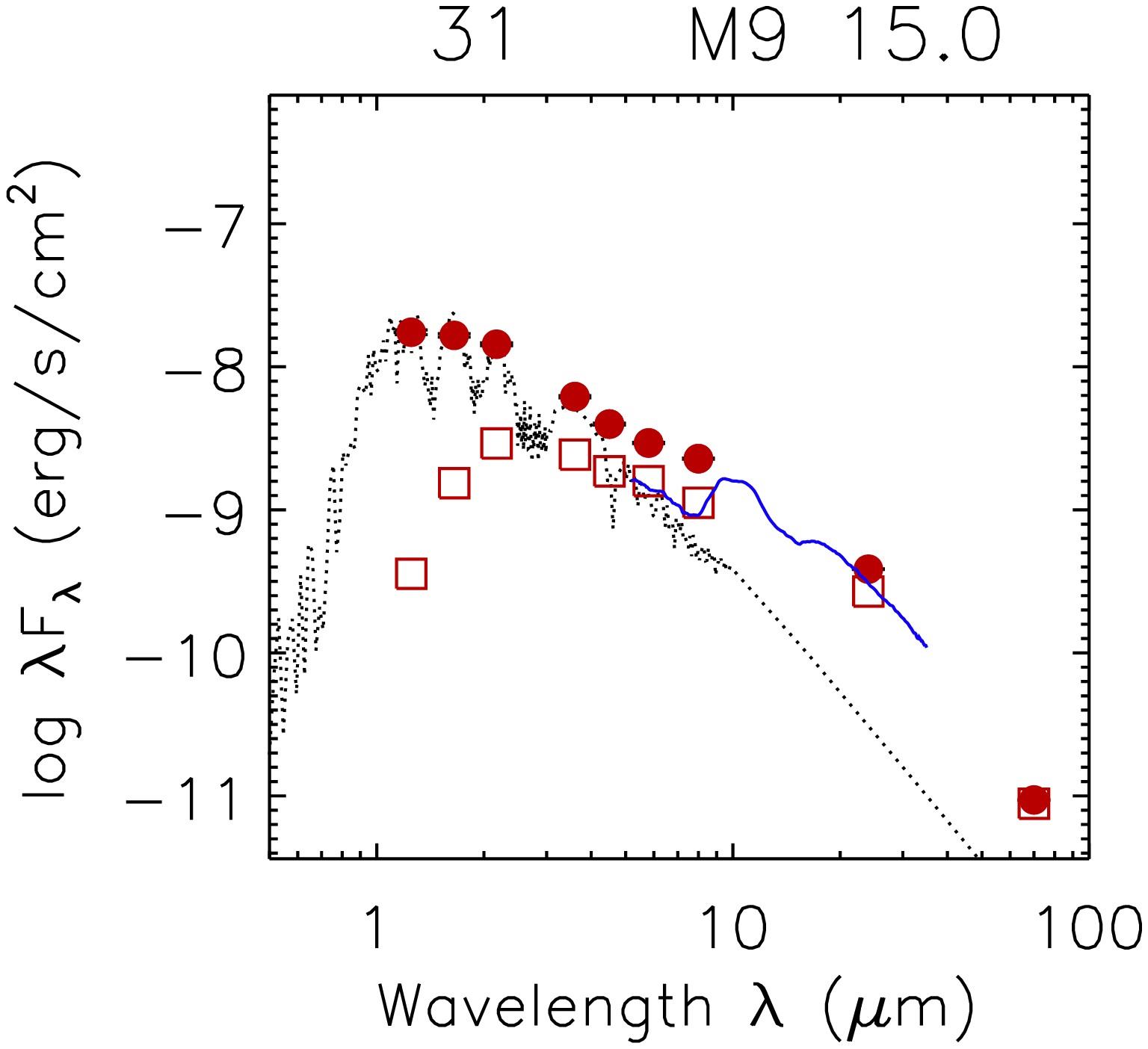}
\end{center}
\caption{\label{7f_seds} SEDs of the young stellar population with
  disks of Serpens. Each SED has the corresponding object ID in Table
  \ref{7tab1} (as in \citealt{OL10}) on the top left. The solid black
  line indicates the NextGen stellar photosphere model for the
  spectral type indicated on the top of each plot. Open squares are
  the observed photometry while the solid circles are the dereddened
  photometry. The visual extinction (mag) of each object can be seen
  on the top right. The solid blue line is the object's IRS
  spectrum. The first 16 objects are shown here, while the remaining
  76 SEDs are shown in Appendix A. }
\end{figure*}

Figure \ref{7f_seds} shows the first 16 SEDs constructed for the
objects in the sample (the remaining SEDs are shown in the
appendix). No SEDs could be constructed for objects \#42 and 94 due to
the lack of either optical or 2MASS near-IR photometric
detections. For the other sources, Figure \ref{7f_seds} shows the
NextGen model photosphere (dashed black line), observed photometry
(open squares), dereddened photometry (filled circles) and IRS
spectrum (thick blue line). When there is no detection for the MIPS2
band at 70 $\mu$m, an upper limit is indicated by a downward
arrow. Significant differences in the amounts of IR radiation in
excess of the stellar photosphere are evident. This translates into a
diversity of disk geometries, as inferred by mid-IR data
\citep{OL10}.

Once the SEDs are built, it is possible to separate the radiation that
is being emitted by the star from that re-emitted by the disk -- the
integration of the radiation emitted by the system at all wavelengths
gives the bolometric brightness of the entire system. This is a direct
integration up to 70 $\mu$m. For longer wavelengths, an extrapolation
is applied as suggested by \citet{CH81}, which assumes the hottest
black body that fits the data at the longest available
wavelength. This results in a typical contribution to the total
luminosity on the order of 10\%. By integrating the scaled NextGen
model photosphere, the stellar luminosity ($L_{{\rm star}}$) can be
obtained. Similar methods for luminosity estimates are widely used in
the literature (e.g. \citealt{KH95, JA08}). If $L_{{\rm star}}$ is
subtracted from the emission of the entire system, the disk luminosity
($L_{{\rm disk}}$) can be derived. These integrations take into
consideration the distance to the star, besides the fluxes at
different bands. The errors in the derivation of $L_{{\rm star}}$ and
$L_{{\rm disk}}$ are propagated from the errors in the distance,
extinction ($\pm$ 2 mag) and in the spectral type determination, and
can be found in Table \ref{7tab1}. The stellar and fractional disk
luminosities for the objects in Taurus, Upper Scorpius and $\eta$
Chamaeleontis were calculated in the exact same manner as for Serpens.

\subsection{Masses and Ages Revisited}
\label{7shrd}

In their derivation of stellar luminosities for the Serpens YSOs with
optical spectroscopy, \citet{OL09} adopted a distance to Serpens of
259 $\pm$ 37 pc (\citealt{ST96}, a discussion using $d$ = 193 $\pm$ 13
pc of \citealt{JK10} is included). However, since then the distance to
the cloud has been revisited. \citet{DZ10} find a distance of 415
$\pm$ 15 pc to the Serpens Core from Very Long Baseline Array parallax
observations of one star. This new distance is used in this work,
which is also compatible with the {\it Chandra} observations of the
Serpens Core by \citet{WI10}.

The new stellar luminosities, derived for the distance of 415 pc,
imply that the young stars in Serpens move up in the
Hertzsprung-Russell (HR) diagram. Following \citet{OL09}, $T_{{\rm
    eff}}$ is determined from the object's spectral type as follows:
for stars earlier than M0 the relationship of \citet{KH95} is used,
while for stars of later spectral type that of \citet{LU03} is
used. The errors in temperature are translated directly from the
errors in spectral types. With $L_{{\rm star}}$ and $T_{{\rm eff}}$ in
hand, the objects can be placed in the HR diagram. For YSOs, ages and
masses can be derived by overlaying pre-main sequence (PMS)
evolutionary tracks on the HR diagram, and comparing the position of
an object to the isochrones and mass tracks. Due to the intrinsic
physics and validation of parameters, the models of \citet{BA98} are
used for stars with masses smaller than 1.4 M$_\odot$, while more
massive stars are compared to the models of \citet{SI00}. The new
individual ages and masses are presented in Table \ref{7tab1}.

\begin{figure*}[!h]
\begin{center}
\includegraphics[width=0.7\textwidth]{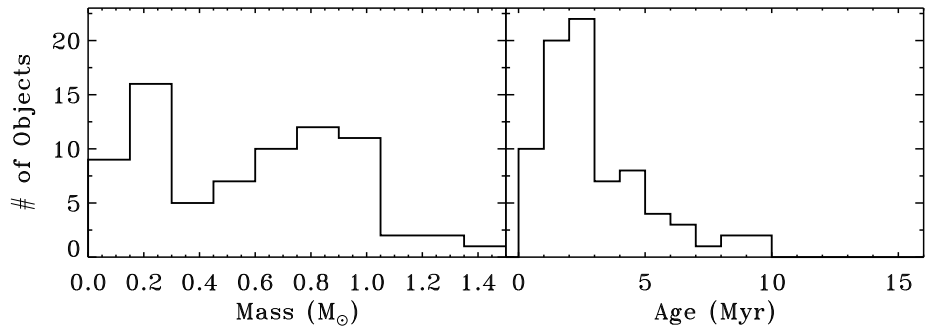}
\end{center}
\caption{\label{7f_age} Updated distribution of masses and ages of the
  young stellar objects in Serpens, assuming $d$ = 415 pc. }
\end{figure*}

Figure \ref{7f_age} shows this updated distribution of masses and ages
for the YSOs in Serpens. Compared to the results of \citet{OL09} for
$d$ = 259 pc, it is seen that the mass distribution does not change
much, while the age distribution does. This is understood by looking
at the isochrones and mass tracks of a given model (e.g. Figure 7 of
\citealt{OL09}): for the temperature range of the stars in Serpens
(mostly K- and M-type), mass tracks are almost vertical. This means
that a change in luminosity due to the new distance hardly affects the
inferred mass. From the isochrones, however, it can be noted that in
general higher stellar luminosities (for this further distance) imply
younger ages. The median mass derived here is $\sim$0.7 M$_\odot$ and
median age is $\sim$2.3 Myr, while \citet{OL09} found $\sim$0.7
M$_\odot$ and $\sim$5 Myr. As for most star-forming regions studied to
date, a spread around the median age is seen for Serpens, with a tail
up to 10 Myr. The spread, however, does not resemble a bimodal
distribution of young stars as it is seen for Orion, which has been
found to be the consequence of the projection of another potentially
unrelated foreground stellar population \citep{AB12}.

\subsection{Notes on Individual Objects}
\label{7sobj}

Since the quantity of data available for each object in this sample
varies, not all SEDs produce good results or yield physical
parameters. Objects 31, 62, 80, 81, 86 and 103 are found to be much
too luminous, which is not consistent with them being members of
Serpens. Thus, they could not be placed in the HR diagram, and
therefore no ages and masses could be determined. For objects 31, 80
and 103 the degeneracy between spectral type and extinction due to the
lack of optical spectroscopy makes it difficult to establish good
values for both parameters. Confirmation of spectral types, better
extinction determination, and the addition of optical photometry is
necessary to revisit these objects and precisely determine their
stellar parameters and whether they belong to the cloud or are
contaminants.

Furthermore, objects 7, 40, 48, 54, 56, 59, 60, 65, 74, 88, 101,
117 and 129 show flat SEDs. This produces large fractional disk
luminosities that deserve attention. None of these objects show signs
of being (close to) edge-on. Edge-on systems will indeed produce high
relative disk luminosities, but will also produce other signatures
(e.g. inability of fitting optical/near-IR photometry in its SED;
\citealt{BM10}), which is not the case of any for the high luminosity
disks shown here. Most likely, those objects are in transition from
stage I (embedded) to stage II (disks) or surrounded by a nebulosity,
leading to their classification as flat sources.

In the SED of object 64 it can be seen that its photometry and IRS
spectra do not match. This could be due to IR variability
(e.g. \citealt{MU09}). The photometry was used for the luminosity
derivation. Lastly, object 41 seems to have a mismatch in the 2MASS
(photometry) making the results unreliable. For all these objects, the
addition of more data, especially at longer wavelength, will help in
understanding their nature and the derivation of accurate parameters.

\section{Disk Properties}
\label{7sdisk}

\subsection{Completeness of the Sample}
\label{comp}

The Serpens sample presented here is flux-limited and selected based
on IR excess. That means that, by definition of the selection
criteria, stars without disks and with disks fainter than 3 mJy at 8
$\mu$m are not part of the sample. A conservative calculation of the
fractional disk luminosity of the missed sources (considering a flux
lower than 3 mJy at 8 $\mu$m) was performed as described below.

Due to the selection criteria, the disk population missed in Serpens
should be fainter than that presented here. \citet{HA07} identified a
population of 235 IR-excess sources in Serpens, called YSO
candidates. 147 of the original sample were further studied with the
IRS spectrograph onboard {\it Spitzer}, comprising the sample
presented here. This means that about 88 potential young stars with
disks are missing. Considering the $\sim$20 \% contamination fraction
of background sources in the direction of Serpens (due to its low
galactic latitude, \citealt{OL09}), conservatively about 70 of these
88 objects could be young stars that were missed, which should
populate the faint end of the $L_{{\rm disk}}/L_{{\rm star}}$ distribution. 

\subsection{Disk Luminosities}

\begin{figure}[!h]
\begin{center}
\includegraphics[width=0.5\textwidth]{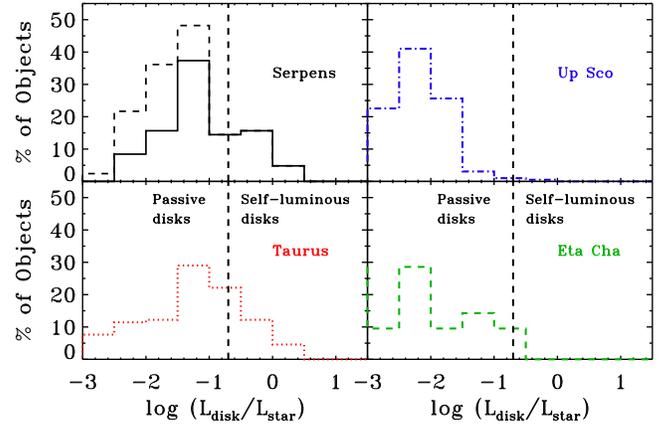}
\end{center}
\caption{\label{7f_disk1} Fractional disk luminosity ($L_{{\rm
      disk}}/L_{{\rm star}}$) derived for the objects in Serpens (top
  left), compared to those in Taurus (bottom left), Upper Scorpius
  (top right) and $\eta$ Chamaeleontis (bottom right). The dashed line
  in the Serpens distribution accounts for completeness (see text for
  details). An indicative boundary for self-luminous vs. passive disks is put at
  $L_{{\rm disk}}/L_{{\rm star}} \sim$ 0.2 (see text for details). }
\end{figure}

The construction of the SEDs is one way to study the diversity of
disks in the same region, most of which have ages with a narrow span
around a few Myr (Figure \ref{7f_age}). It is clear from the SEDs that
different types of disks are present in Serpens, some with substantial
IR excess, others almost entirely dissipated. This is even more clear
by looking at the distribution of fractional disk luminosities
($L_{{\rm disk}}/L_{{\rm star}}$) for this sample, which is presented
in Figure \ref{7f_disk1}. Here, Serpens (solid black line) is compared
to Taurus, equally young yet very different in terms of cloud
structure and environment (dotted red line). The peak and distribution
of these two samples are very similar, with the bulk of each
population showing fairly bright disks (peak $L_{{\rm disk}}/L_{{\rm
    star}} \sim$ 0.1, median $\sim$ 0.2), the majority of which are
consistent with passively irradiated disks ($L_{{\rm disk}}/L_{{\rm
    star}} \le$ 0.2, \citealt{KH87}). This is in agreement with
studies of disk geometry as inferred from IR colors, which show a
large fraction of disks in young clusters to be flared
(e.g. \citealt{ME05,FU06,SI06,GU08,MU10,OL10}).  Figure \ref{7f_disk1}
includes a correction for the possible missed sources discussed in
\S{} \ref{comp} (dashed black line), distributed in fractional disk
luminosity bins according to their IR fluxes (from
\citealt{HA07}). Those could account for the difference between
Serpens and Taurus in the faintest bin of $L_{{\rm disk}}/L_{{\rm
    star}}$ in Figure \ref{7f_disk1}, but should not be able to shift
the peak of the $L_{{\rm disk}}/L_{{\rm star}}$ distribution for
Serpens. These findings support the idea that these two star-forming
regions are similar in spite of their different environments and star
formation rates, and that together they provide a good probe of the
young bin of disk evolution.

Furthermore, Figure \ref{7f_disk1} shows the distribution of $L_{{\rm
    disk}}/L_{{\rm star}}$ for samples in the older Upper Scorpius and
$\eta$ Chamaeleontis clusters with optical and IR data
\citep{DZ99,PZ99,PR02,MA02,HA05,ME05,DC09,SI09}. These older regions
are known to have lower disk fractions (40\% for $\eta$ Cha and 17\%
for Up Sco; \citealt{ME05,CA06}), meaning that most of the member
stars have already fully dissipated their disks. Figure \ref{7f_disk1}
clearly shows this difference in relation to the young Serpens and
Taurus clouds, with distributions that peak (and spread) at
considerably lower disk luminosities. The vertical dotted lines
roughly separate luminosity ratios that can be explained by different
mechanisms: self-luminous disks ($L_{{\rm disk}}/L_{{\rm star}} >
0.2$, \citealt{KH87}) and passive disks. This illustrative boundary
was calculated by taking the maximum amount of light that a flared
disk would be able to re-radiate by only reprocessing the stellar
radiation. ``Debris''-like disks are considerably fainter ($L_{{\rm
    disk}}/L_{{\rm star}} < 0.001$; \citealt{CC05}).

The difference in observed fractional disk luminosities between the
young Serpens and Taurus and the old Upper Sco and $\eta$ Cha
populations has implications on our understanding of disk
evolution. Figure \ref{7f_disk1} shows an evolution in disk brightness
that is concurrent with that of disk fraction
\citep{HA01,HE08,MA09}. With time, not only the fraction of stars that
have disks diminishes, but the remaining disks tend to be fainter (see
also \citealt{SI06,HE07,HE08,CU09}). This conclusion is in agreement
with models of disk evolution that include long-term dust growth and
settling and predict disks to become flatter and fainter within a few
million years (e.g. \citealt{CG97,DD04b}). Moreover, Figure
\ref{7f_disk1} is consistent with the new younger age of Serpens
derived in \S{} \ref{7shrd}, since the distribution in Serpens is so
similar to that in Taurus and very different than those in Upper Sco
and $\eta$ Cha.

\subsection{Accretion Properties}

\begin{figure}[!h]
\begin{center}
\includegraphics[width=0.45\textwidth]{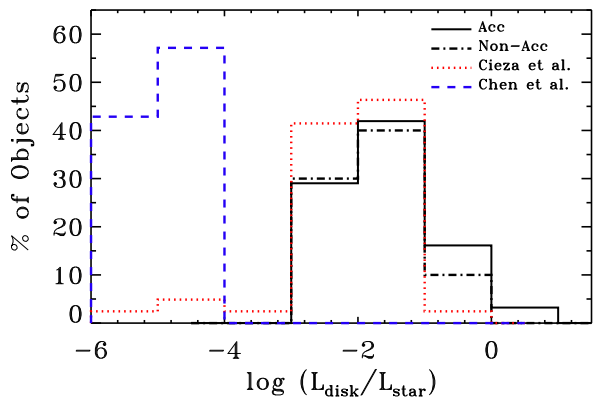}
\end{center}
\caption{\label{7f_disk2} Fractional disk luminosity ($L_{{\rm
      disk}}/L_{{\rm star}}$) derived for the accreting stars (based
  on H$\alpha$ data, solid black line) and non-accreting stars
  (dot-dashed black line) in Serpens, compared to a sample of
  weak-lined T Tauri stars (dotted red line; \citealt{LC07}) and
  a sample of debris disks (dashed blue line; \citealt{CC05}). }
\end{figure}

Figure \ref{7f_disk2} shows an additional comparison of the disks in
Serpens with a sample of weak-line T Tauri stars (WTTS,
\citealt{LC07}) and a sample of debris disks around T Tauri and Herbig
Ae/Be stars \citep{CC05}. The WTTS sample consists of sources selected
based on the original definition of weak H$\alpha$ emission. This
criterion also implies low accretion rates. In our samples, objects
are classified as accreting according to two prescriptions. The first
method is based on the width at 10\% of peak intensity of the
H$\alpha$ line (from the relationship of \citealt{NA04}) where objects
are classified as accreting if the width is greater than 270 km
s$^{-1}$ \citep{WB03}. The second method is based on the equivalent
width of H$\alpha$ and its spectral type, according to
\citet{WB03}. \citet{OL09} present mass accretion rates based on the
H$\alpha$ data for the Serpens sample, with the majority of objects
fulfilling both criteria for classification as either accreting or
non-accreting. It can be seen in Figure \ref{7f_disk2} that $L_{{\rm
    disk}}/L_{{\rm star}}$ of the accreting (CTTS, solid black line)
and non-accreting (WTTS, dot-dashed line) stars in Serpens overlap
with the WTTS sample of \citet{LC07} (dotted red line). The Serpens
population and the Cieza WTTSs differ in the distribution tails. The
Cieza WTTS sample has a faint tail that overlaps with the debris disk
population (dashed blue line), while the Serpens population shows a
bright tail.

Within Serpens, the accreting and non-accreting sub-samples overlap,
differing slightly at the brighter end of the distribution, which is
dominated by accreting objects. This is more clearly seen by looking
at the median fractional disk luminosity $\langle L_{{\rm
    disk}}/L_{{\rm star}} \rangle$ which is 0.21 and 0.11 for
accreting and non-accreting objects, respectively. The median
fractional disk luminosity $\langle L_{{\rm disk}}/L_{{\rm star}}
\rangle$ for the entire population of Serpens is 0.20. The few very
bright (self-luminous) disks are actively accreting. These results
confirm the finding by several authors that WTTS may very well have
massive disks not much different from those of CTTS
(e.g. \citealt{ST89,LC07,ZW10}). At the other end, the faint tail of
the Cieza WTTS population overlaps with the debris sample and should
represent non-accreting stars surrounded by very flat optically thin
disks. Diskless WTTS in Serpens are not yet identified and therefore
not shown in this plot.

\subsection{Comparison with Herbig Ae/Be Stars}
\label{sHAeBe}

\citet{ME01} found that the disks around higher mass Herbig Ae/Be
stars can be divided into two groups, according to the disk geometry:
group I comprises sources with considerable IR excess, associated with
a flared geometry; group II consists of little IR excess, associated
with a geometrically thin midplane, shadowed by the puffed-up disk
inner rim. \citet{ME01} showed that the distributions of fractional
disk luminosities for the two groups are different, with a mean
$L_{{\rm disk}}/L_{{\rm star}}$ of 0.52 for group I and 0.17 for group
II.

\begin{figure}[!h]
\begin{center}
\includegraphics[width=0.45\textwidth]{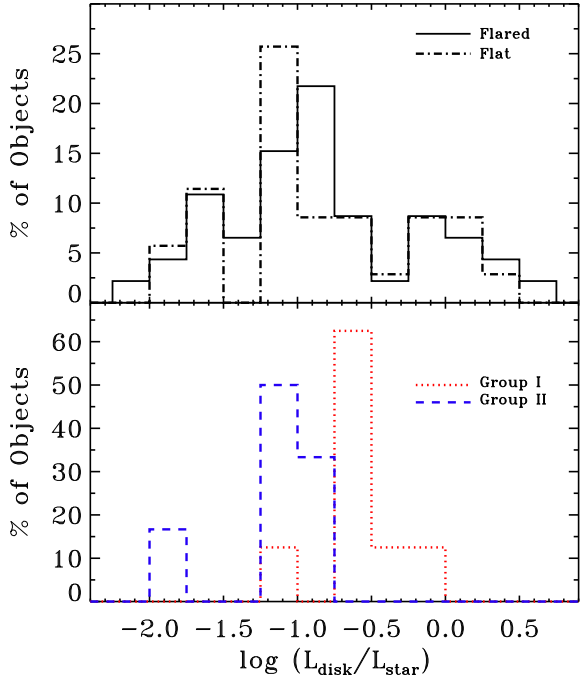}
\end{center}
\caption{\label{7f_disk3} $L_{{\rm disk}}/L_{{\rm star}}$ derived for
  the flared (solid black line) and flat (dot-dashed black line) disks
  in Serpens (top), compared to the sample of Herbig Ae/Be of
  \citet{ME01} (bottom). Objects belonging to group I (flared, dotted
  red line) and group II (self-shadowed, dashed blue line) are shown
  separately. }
\end{figure}

Figure \ref{7f_disk3} compares the two groups of Herbig Ae/Be stars
with the young stars in Serpens, separated in disk geometry according
to the ratio between the fluxes at 30 and 13 $\mu$m ($F_{30}/F_{13}$,
\citealt{OL10}). Although the mid-IR data for the cooler disks around
T Tauri stars probe a smaller portion of the disk compared to the more
massive Herbig Ae/Be counterparts \citep{KE07}, the fractional disk
luminosities calculated here account for the bulk of the disk. A
comparison between cooler (T Tauri) and hotter (Herbig Ae/Be) stars
can inform about the universality of processes taking place in these
disks, and whether they evolve in a similar manner despite the
differences in masses and incident radiation field. It is seen that
the geometry separation between flared and flat disks at
$F_{30}/F_{13}$ = 1.5 for T Tauri stars is not reflected with an
accompanying separation in $L_{{\rm disk}}/L_{{\rm star}}$, which is
the case for groups I and II of the Herbig Ae/Be stars (dotted red and
dashed blue lines, respectively). Although both the flared and flat T
Tauri disks span the same luminosity range, the peaks of the
distributions are slightly different, yielding marginally distinctive
median fractional disk luminosities: $\langle L_{{\rm disk}}/L_{{\rm
    star}} \rangle$ is 0.21 for the flared disks and 0.17 for flat
disks.

It can be noted from Figure \ref{7f_disk3} that the great majority of
disks around Herbig Ae/Be stars are concentrated in narrow ranges of
fractional disk luminosities, right at the border between
self-luminous and passively irradiated disks, showing a bimodal
distribution for groups I and II. The T Tauri stars, on the other
hand, span a much wider range of $L_{{\rm disk}}/L_{{\rm star}}$. The
most striking difference between T Tauri and Herbig Ae/Be stars are
both tails of the distribution. The lack of relatively very faint and
very bright disks around Herbig Ae/Be stars could be a bias effect due
to the considerably lower number of disks observed compared to their
lower mass counterparts.  Another possibility is that indeed disks
around higher mass stars evolve faster, as suggested by previous
studies \citep{LA06,CA06,KK09}. That would mean that the bright phase
of disk evolution happens when the disks are still embedded in a
spherical collapse envelope and consequently not visible, while the
lack of the faint end of the distribution would imply a very fast
evolution from flat disks to no disks at all, being only visible again
in the debris stage. The latter finding is consistent with models of
photoevaporation by high-energy photons
\citep{CC01,GO09a,GO09b,ER09}. In those models, photoevaporation
becomes important once the viscous transport declines below a certain
threshold, rendering a quick dispersion of the disk on a timescale
that is a small fraction of its lifetime. It is predicted that more
massive stars could lose their disks in $\sim$10$^{5}$ yr, which could
explain the difference in the faint end of the distributions seen in
Figure \ref{7f_disk3}.

\section{Connection Between Stars and Disks}
\label{7sconnect}

\subsection{Variations with Stellar Type}

While the late-type (K and M) population of Serpens spans a wide
variety in disk shapes, the early-type (A, F and G) stars catch the
attention. Two of the 9 early-type stars (\#52 and 114) are surrounded
by so-called cold or transitional disks, i.e. disks depleted of warm
dust close to the star but otherwise massive \citep{OL10,BM10}. The
majority, however, show very little IR excess (\#70, 80, 98, 120, 131,
139, and 145) consistent with a rapid transition from stage II to
III. Assuming that the stars in Serpens are nearly coeval, this result
supports the conclusion of \S{} \ref{sHAeBe} that disks around more
massive stars evolve on faster timescales, albeit with lower number
statistics for early-type stars.

\begin{figure}[!h]
\begin{center}
\includegraphics[width=0.45\textwidth]{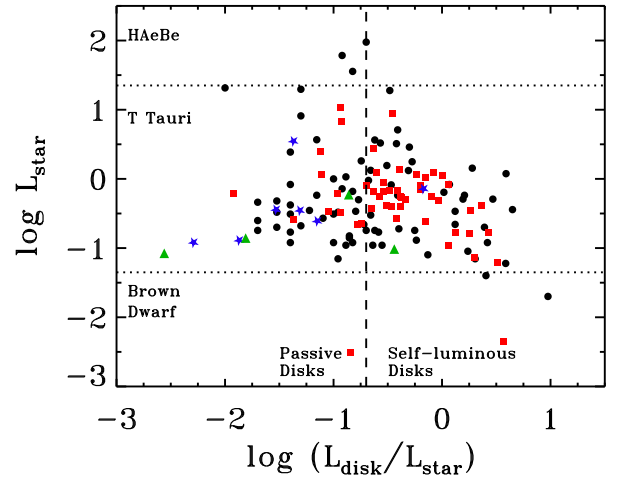}
\end{center}
\caption{\label{7f_lum} Fractional disk luminosity ($L_{{\rm
      disk}}/L_{{\rm star}}$) versus the stellar luminosity ($L_{{\rm
      star}})$ derived for the objects in Serpens (black circles),
  compared to the objects in Taurus (red squares), in Upper Sco (blue
  stars), and in $\eta$ Cha (green triangles). }
\end{figure}

Figure \ref{7f_lum} shows the stellar luminosity related to the
fractional disk luminosity for the sample in Serpens (black points),
Taurus (red points), Upper Sco (blue points), and $\eta$ Cha (green
points). Horizontal dotted lines separate stellar luminosities of
Herbig Ae/Be stars (earlier than F0), T Tauri stars (down to M7) and
brown dwarfs (below M7), while the vertical line roughly separates
self-luminous from passive disks, as in Figure \ref{7f_disk1}. It can be
seen that the few Herbig Ae stars in Serpens follow the locus of
$L_{{\rm disk}}/L_{{\rm star}}$ established by the larger sample of
Herbig Ae/Be stars (Figure \ref{7f_disk3}) by \citet{ME01}, not
occupying either tail of the distribution.

\subsection{Dust Characteristics}

Besides the stellar and disk characteristics discussed in \S{}
\ref{7sseds}, \citet{OL11} present the dust mineralogy, crystalline
fractions and mean grain sizes in the surface of disks around the
stars in Serpens, together with those disks in Taurus, Upper Sco and
$\eta$ Cha for which IRS spectra are available, obtained using the
same procedure: the B2C decomposition method \citep{OF10}. This method
reproduces the IRS spectra over the full spectral range (5 -- 35
$\mu$m), assuming two dust populations: a warm component responsible
for the 10 $\mu$m emission arising from the inner disk ($\lesssim$1
AU) and a colder component responsible for the 20 -- 30 $\mu$m
emission, arising from more distant regions ($\lesssim$10 AU). Each
component is a combination of five different dust species (three
amorphous and two crystalline) for different grain sizes (3 for
amorphous and 2 for crystalline silicates). The fitting strategy
relies on a random exploration of parameter space coupled with a
Bayesian inference method. Those results, presented in \citet{OL11},
combined with the analysis of their SEDs, allow the comparison of
different disk and dust characteristics with those of their host stars
for all 4 regions.

\begin{figure}[!h]
\begin{center}
\includegraphics[width=0.45\textwidth]{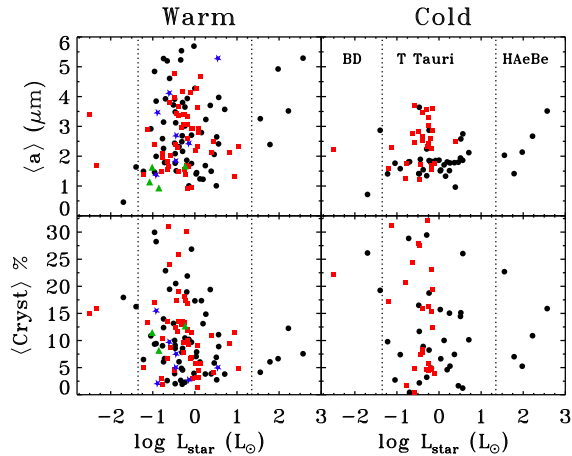}
\end{center}
\caption{\label{7f_ls_min} Mean grain size (top) and mean
  crystallinity fraction (bottom) of the dust in the disk surface
  versus the stellar luminosity ($L_{{\rm star}}$) derived for the
  objects in Serpens (black circles), compared to the objects in
  Taurus (red squares), in Upper Sco (blue stars), and in $\eta$ Cha
  (green triangles). }
\end{figure}

\begin{figure}[!h]
\begin{center}
\includegraphics[width=0.45\textwidth]{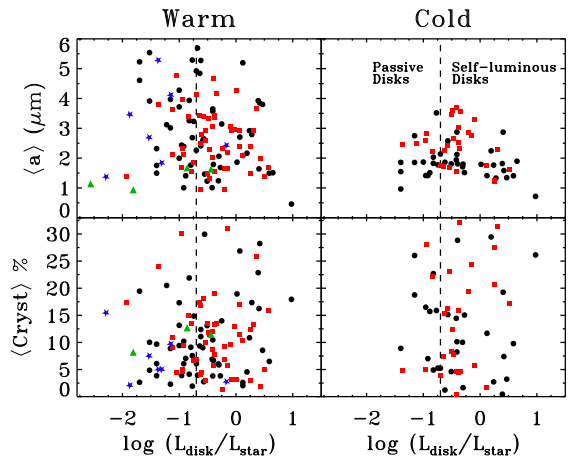}
\end{center}
\caption{\label{7f_ld_min} Mean grain size (top) and mean
  crystallinity fraction (bottom) of the dust in the disk surface
  versus the fractional disk luminosity ($L_{{\rm disk}}/L_{{\rm
      star}}$) derived for the objects in Serpens (black circles),
  compared to the objects in Taurus (red squares), in Upper Sco (blue
  stars), and in $\eta$ Cha (green triangles). }
\end{figure}

\begin{figure}[!h]
\begin{center}
\includegraphics[width=0.45\textwidth]{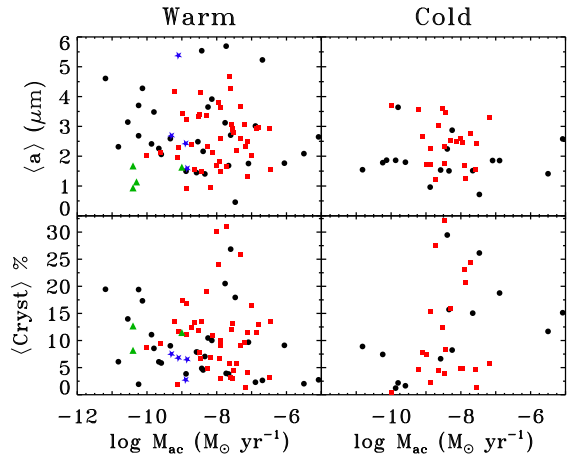}
\end{center}
\caption{\label{7f_mac_min} Mean grain size (top) and mean
  crystallinity fraction (bottom) of the dust in the disk surface
  versus the mass accretion rate as derived from H$\alpha$ of the
  objects in Serpens (black circles), compared to the objects in
  Taurus (red squares), in Upper Sco (blue stars), and in $\eta$ Cha
  (green triangles). }
\end{figure}

Figures \ref{7f_ls_min}, \ref{7f_ld_min} and \ref{7f_mac_min} relate
the stellar and disk fractional luminosities and mass accretion rate,
respectively, with the results from the B2C decomposition method of
objects in Serpens, Taurus, $\eta$ Cha and Upper Sco. The two upper
panels show the mean mass-averaged grain size and the two lower panels
show the mean crystallinity fraction of the dust in the surface of the
disks. The two left panels are the results for the warm component
close to the stars, while the right panels show the results for the
cold component, further away and deeper into the disk. The low number
of Herbig Ae/Be stars and brown dwarfs in these samples do not allow a
study across the stellar mass regime.

No strong correlations are seen in either Figures \ref{7f_ls_min},
\ref{7f_ld_min} or \ref{7f_mac_min}, pointing to no direct
cause-effect relationships between either stellar, disk fractional
luminosity or mass accretion rates and the dominant grain size or
crystallinity fraction of the surface dust in a disk. Similar results
are found for the populations in Taurus \citep{WA09}, Tr 37 and NGC
7160 \citep{SI07} and Cep OB2 \citep{SI11} for smaller samples, with
different methods of analysis. This picture is unchanged by separating
objects into 3 classes of disk geometry: flared ($1.5 \lesssim
F_{30}/F_{13} \lesssim 5$), flat ($F_{30}/F_{13} \lesssim 1.5$) and
cold or transitional disks ($5 \lesssim F_{30}/F_{13} \lesssim 15$,
\citealt{JB07,BM10,OL10}). Distinct disk types show similar scatter.

The results in Figure \ref{7f_ld_min} for T Tauri stars differ from
those for Herbig Ae/Be stars \citep{ME01}, which show a correlation
between the mean grain size in the disk surface (as derived from the
silicate features) and the geometry of the disk. Their study of a
small number of disks (14 objects) argues that as the disk becomes
flat (transitioning from group I into group II, and therefore
decreasing $L_{{\rm disk}}/L_{{\rm star}}$, as they interpret), small
dust grains are removed from the disk surface (by coagulation into
bigger grains or blown away by the stellar radiation or stellar
winds), which yields larger dominating grain sizes for flatter
disks. Their results are supported by a less steep submilimeter slope
for group II sources than for group I. \citet{BA04} studied the
milimeter slope of a sample of 26 Herbig Ae/Be stars and found a
correlation between this parameter and the geometry of the disk. It is
important to note, however, that the milimeter data probe the dust population
throughout the entire disk and do not say anything about the size of
the dust in the disk surface, as is discussed here. \citet{BA04}
suggest a geometry evolution from flared to self-shadowed with a
concurrent evolution of the size of grains in the disk.

Similar results as those of \citet{ME01} are found for T Tauri stars
by \citet{BO08}, albeit for a very small number of objects (7
systems). \citet{LO10} also find a tentative correlation between
submilimeter slope and grain size probed by the 10 $\mu$m feature for
a set of T Tauri disks. Those conclusions do not hold up for larger
samples of disks around T Tauri stars. In contrast, \citet{OL10,OL11}
show for much larger samples of T Tauri stars that the dust population
in the disk surface is not the result of grain growth alone, but that
also fragmentation of bigger grains enriches the population of small
grain. This argument explains the presence of small grains in the
surfaces of disks in all geometries (and even debris disks) and mean
cluster ages. Larger samples of Herbig stars are needed to test
whether this is also true for higher mass systems, or whether the
concurrent dust and disk evolution is indeed mass dependent.

In addition to stellar and disk fractional luminosities and mass
accretion rate, other stellar and disk parameters (such as stellar
mass, disk colors and slopes) were investigated in relation to the
mineralogical results for the T Tauri stars in our sample. Similar to
Figures \ref{7f_ls_min}, \ref{7f_ld_min} and \ref{7f_mac_min}, no
strong correlation was found for any combination of parameters. This
lack of direct correlations between the stellar and disk
characteristics presents itself as a strong argument for the
non-direct relationship of stellar and disk characteristics, in the
range of parameters (time, mass, environment) probed by the objects
presented here. That is, no direct causal relationship between stellar
and disk characteristics is seen for T Tauri stars within a few Myr
($\sim$1 -- 10 Myr). 

One possible explanation is that the physical correlations are washed
away due to short timescale events. In addition to the above mentioned
continuous balance between grain growth and destruction, episodic
accretion events my play a role. An example is the case of the
eruptive young star EX Lupi, which showed an increase in crystallinity
fraction with increase in stellar luminosity right after outburst
\citep{AB09}. \citet{JU12} showed that months after outburst the
crystallinity fraction decreased from the post-outburst
value. Processes that are efficient on short timescales such as
variability \citep{CA01,CE02,BO07,MU09}, vertical or radial mixing
(e.g. \citealt{CI07,VD10,JU12}) or dust crystallization/amorphization
\citep{GL09} reconcile the notion that evolution of the dust takes
place in disks but that no systematic evolution is detected in this
work.

On the other hand, larger samples of sources that span a wider range
in parameters could reveal relationships that are not found here. For
instance, the dependence of disk dispersal timescale on stellar mass
\citealt{LA06,CA06,KK09}) is reflected in our data. It will be
interesting in the future to have a similar analysis as presented here
(for stellar and disk characteristics, plus dust mineralogy) probing a
wider range in stellar mass and in time, reaching the debris disk
population.

\section{Conclusions}
\label{7scon}

We have studied a flux-limited population of young stars still
surrounded by disks in the Serpens Molecular Cloud. Aided by
spectroscopic characterization of the central sources of star+disk
systems combined with IR photometry, SEDs of the objects could be
constructed.

The SEDs of Serpens show a considerable spread in IR excess. This
implies the presence of disks with different geometries and in
different stages of dissipation around stars that are nearly coeval,
indicating that time is not the dominant parameter in the evolution of
protoplanetary disks. The distribution of disk to star luminosity as a
function of the stellar luminosity shows a trend in which lower mass
stars have relatively brighter disks, consistent with other evidence
in the literature that disks around lower mass stars have generally
longer lifetimes or that disks around higher mass stars evolve faster.

Adopting the new distance of 415 pc for Serpens, higher stellar
luminosities are found than previously inferred by \citet{OL09}. The
higher luminosities, in turn, combined with PMS evolutionary models,
allude to a distribution of ages that is younger than that found by
\citet{OL09}. The great majority of young stars in Serpens are in the
1 -- 3 Myr range, with a median age of $\sim$2.3 Myr. This result
supports the observational evidence that Serpens joins the Taurus
Molecular Cloud in probing the young bin of disk evolution, in spite
of the different environment and star formation rates.

The distribution of fractional disk luminosity of the objects in
Serpens also closely resembles that in Taurus, both of which are very
different from those in the older regions Upper Scorpius and $\eta$
Chamaeleontis, where most disks have already dissipated. Furthermore,
the majority of the Serpens population is consistent with passively
reprocessing disks. When comparing the actively accreting and
non-accreting stars of Serpens (based on H$\alpha$ data), the main
difference is seen at the bright tail of the fractional disk
luminosity, dominated by strongly accreting stars.

The disks around T Tauri stars in Serpens are compared to those around
Herbig Ae/Be stars \citep{ME01}. Herbig Ae/Be stars show a clear
separation in fractional disk luminosity for different disk geometries
(flared versus flat) but this difference is not apparent for T Tauri
stars. The disks around Herbig Ae/Be stars present a very narrow range
of $L_{{\rm disk}}/L_{{\rm star}}$, concentrated around the border
between self-luminous and passively irradiated disks, while the disks
around T Tauri stars span a wider range of fractional disk
luminosities. The absence of the tail distributions for Herbig Ae/Be
could be due a faster evolution of these disks, or a bias effect due
to the smaller number of disks observed around those higher mass
stars.

The stellar and disk characteristics are combined with dust mineralogy
results delivered for the same regions by \citet{OL11}. By combining
all these data, the effects of stellar and disk characteristics on the
surface dust of disks are studied. No strong correlations are found,
suggesting that the many processes taking place in disks somehow
conspire in complicated ways that make it difficult to isolate the
effect of each process/parameter individually. One possibility is that
the processes that change the structure and size distribution of dust
within disks are recurring and have short timescales, making it
difficult to discern long timescale evolutionary effects.

\acknowledgements Astrochemistry at Leiden is supported by a Spinoza
grant from the Netherlands Organization for Scientific Research (NWO)
and by the Netherlands Research School for Astronomy (NOVA)
grants. This work is based on observations made with the {\it Spitzer
  Space Telescope}, which is operated by the Jet Propulsion
Laboratory, California Institute of Technology under a contract with
NASA. Support for this work was provided by NASA through an award
issued by JPL/Caltech.

\clearpage

\appendix

\section{The Remaining SEDs}
\label{7s_app}

Seventy-six SEDs are shown here in Figure 10.

\begin{figure*}[!h]
\begin{center}
\includegraphics[width=0.2\textwidth]{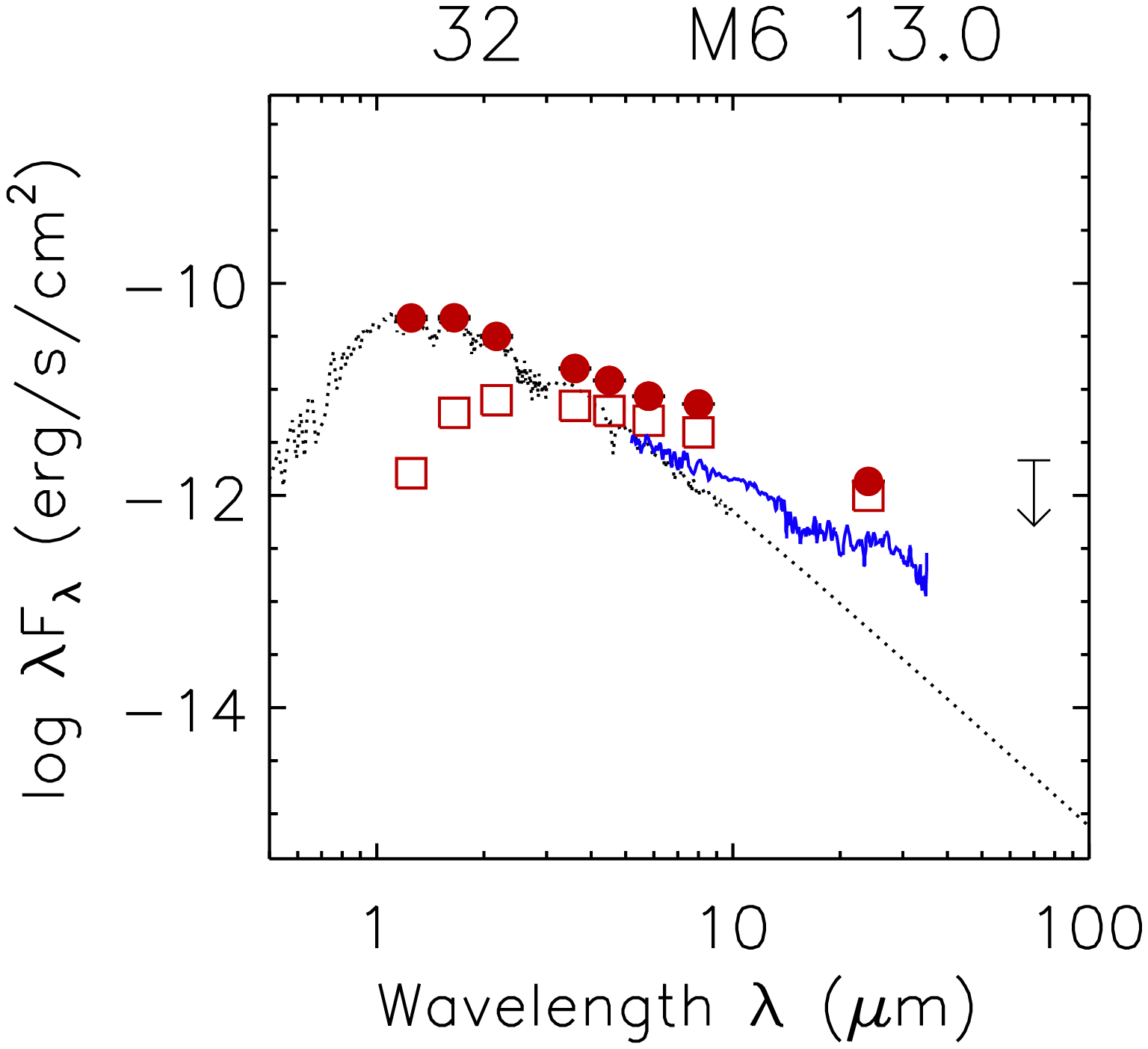}
\includegraphics[width=0.2\textwidth]{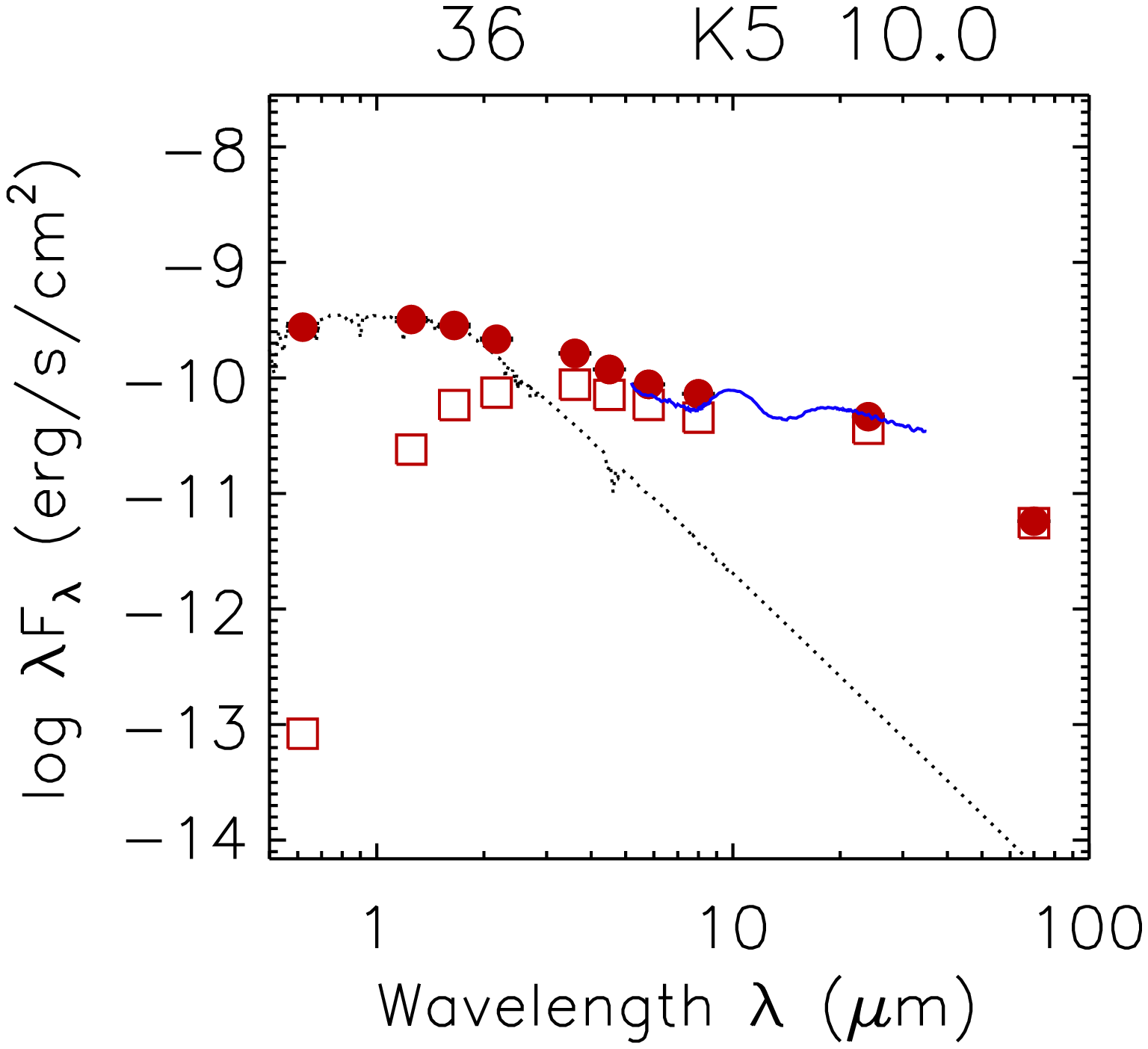}
\includegraphics[width=0.2\textwidth]{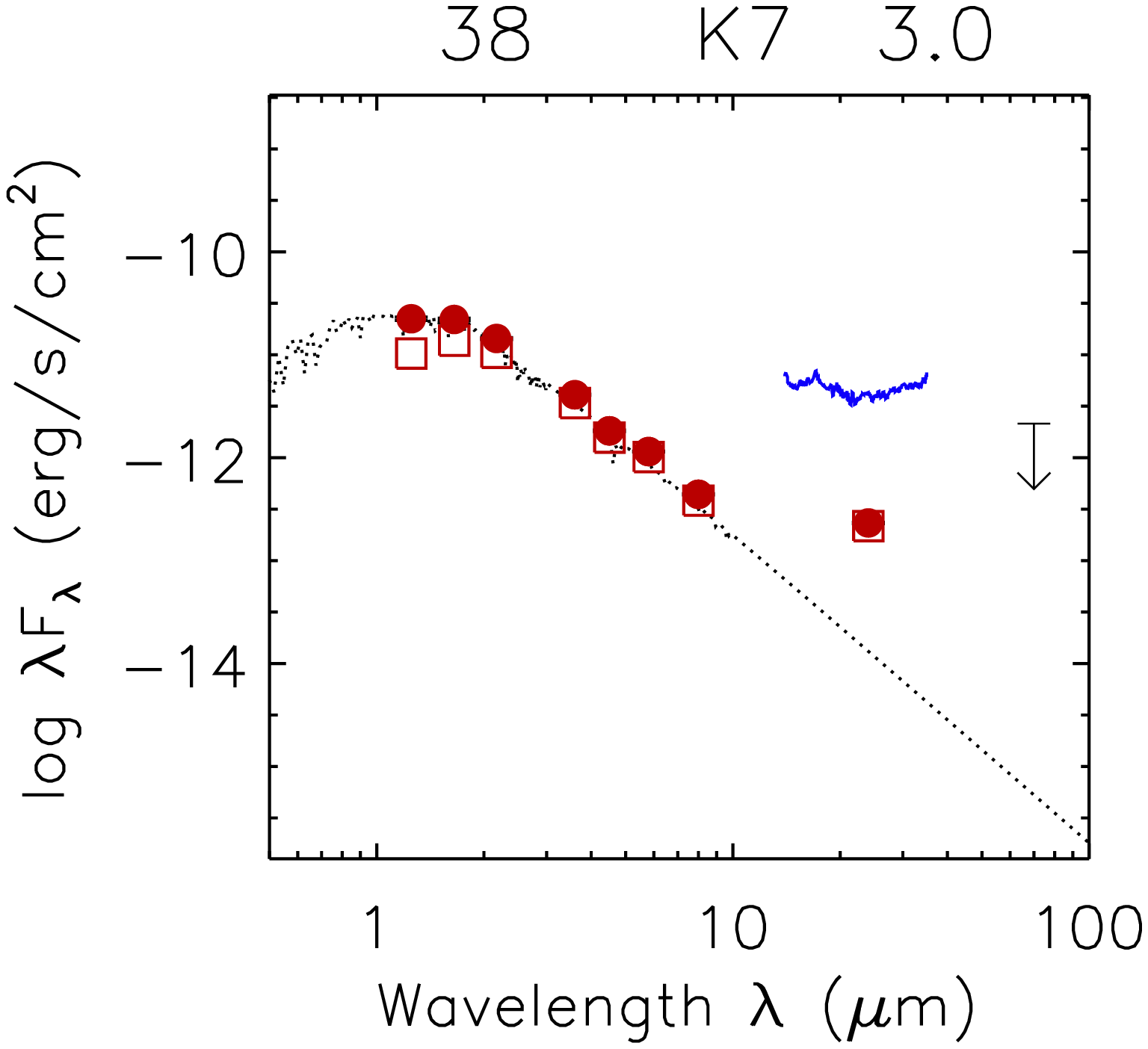}
\includegraphics[width=0.2\textwidth]{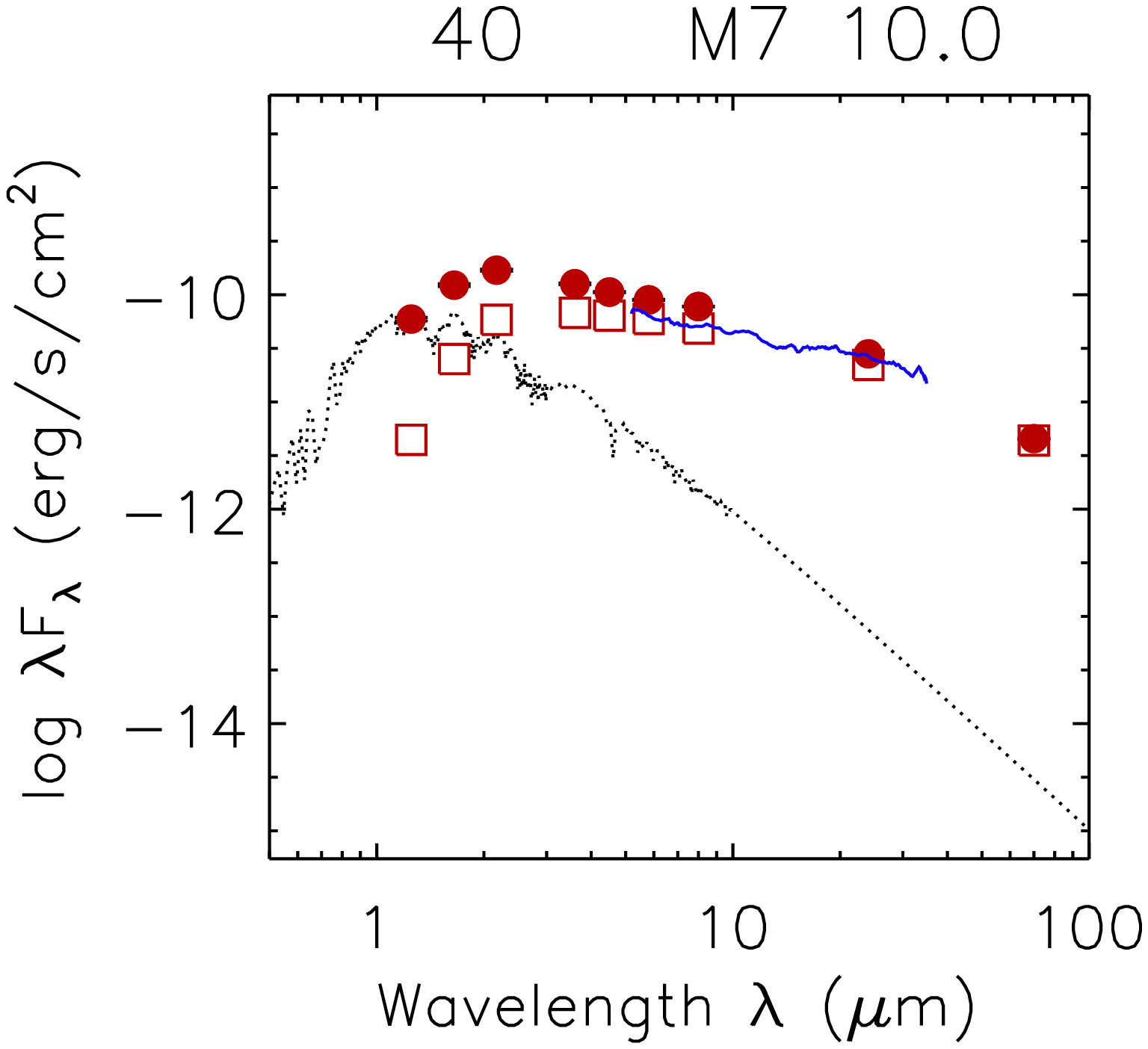}
\includegraphics[width=0.2\textwidth]{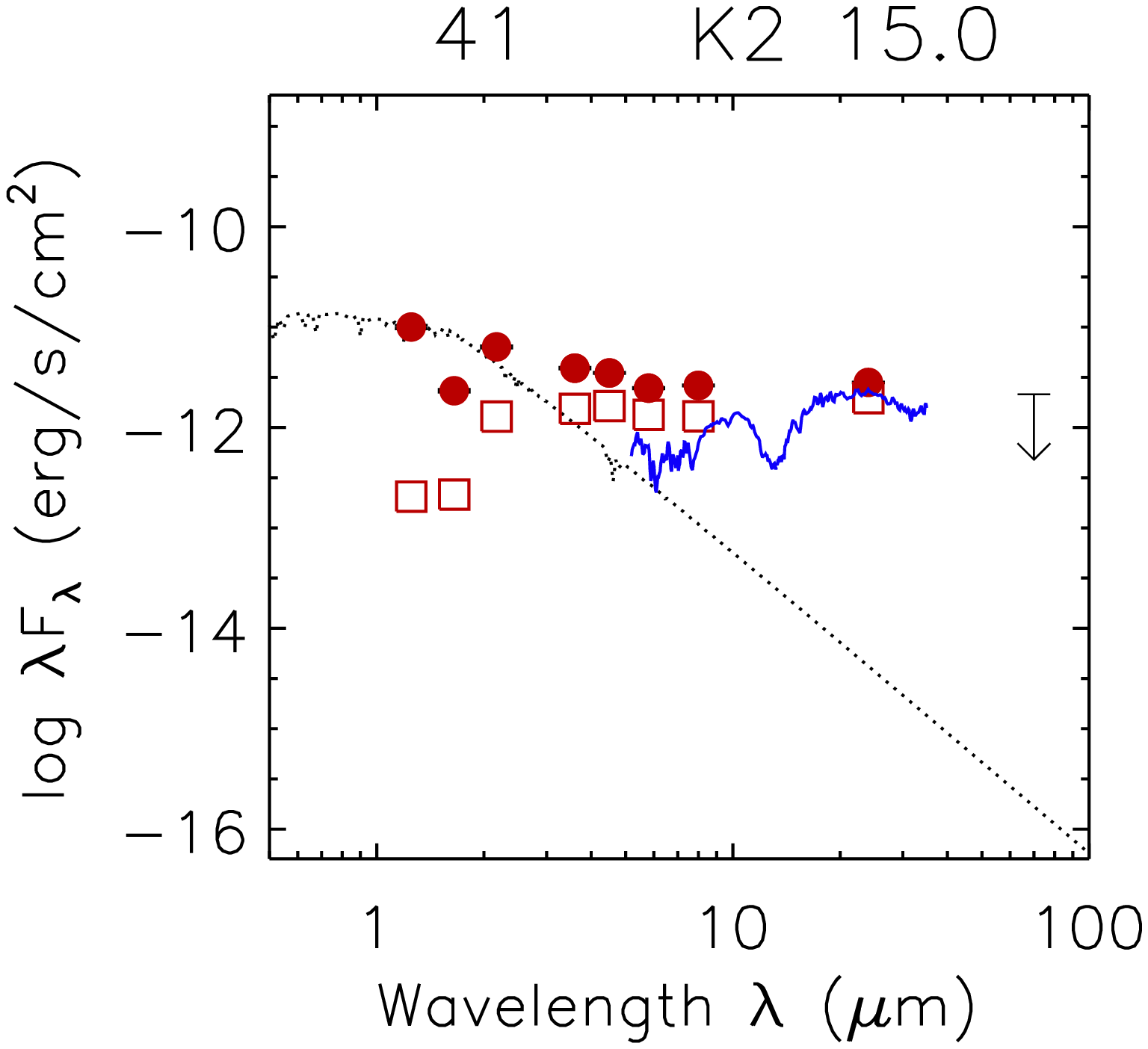}
\includegraphics[width=0.2\textwidth]{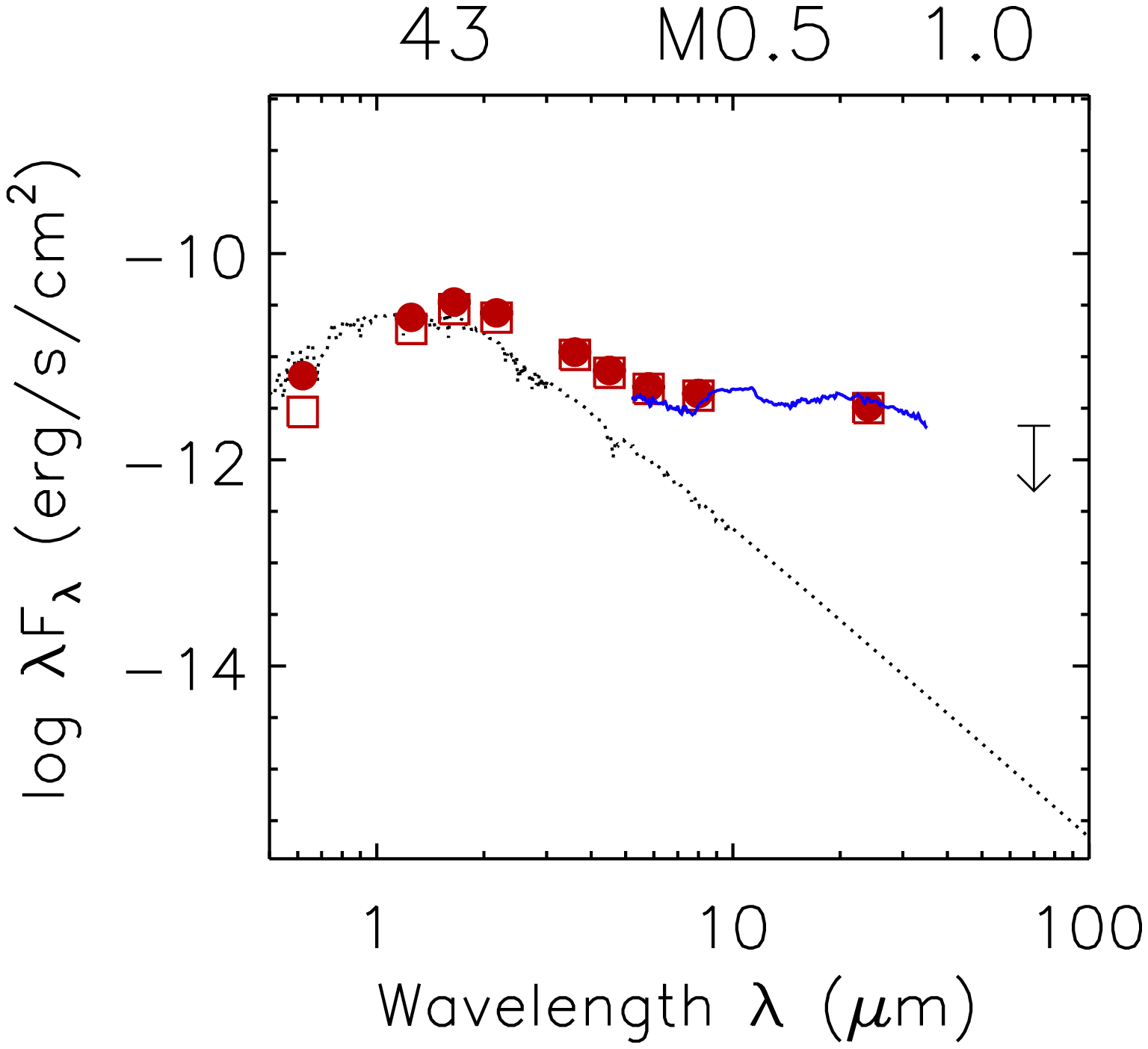}
\includegraphics[width=0.2\textwidth]{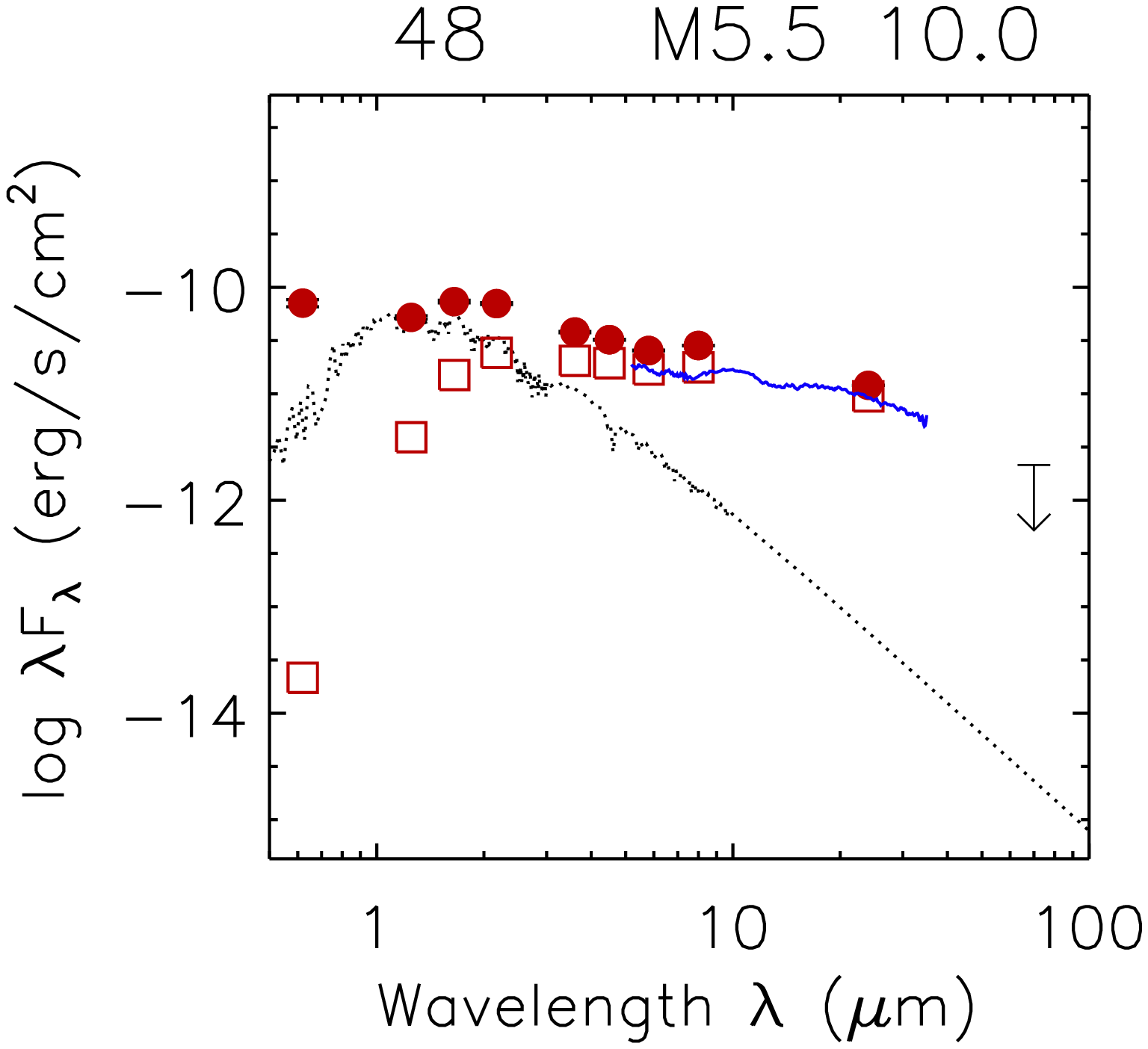}
\includegraphics[width=0.2\textwidth]{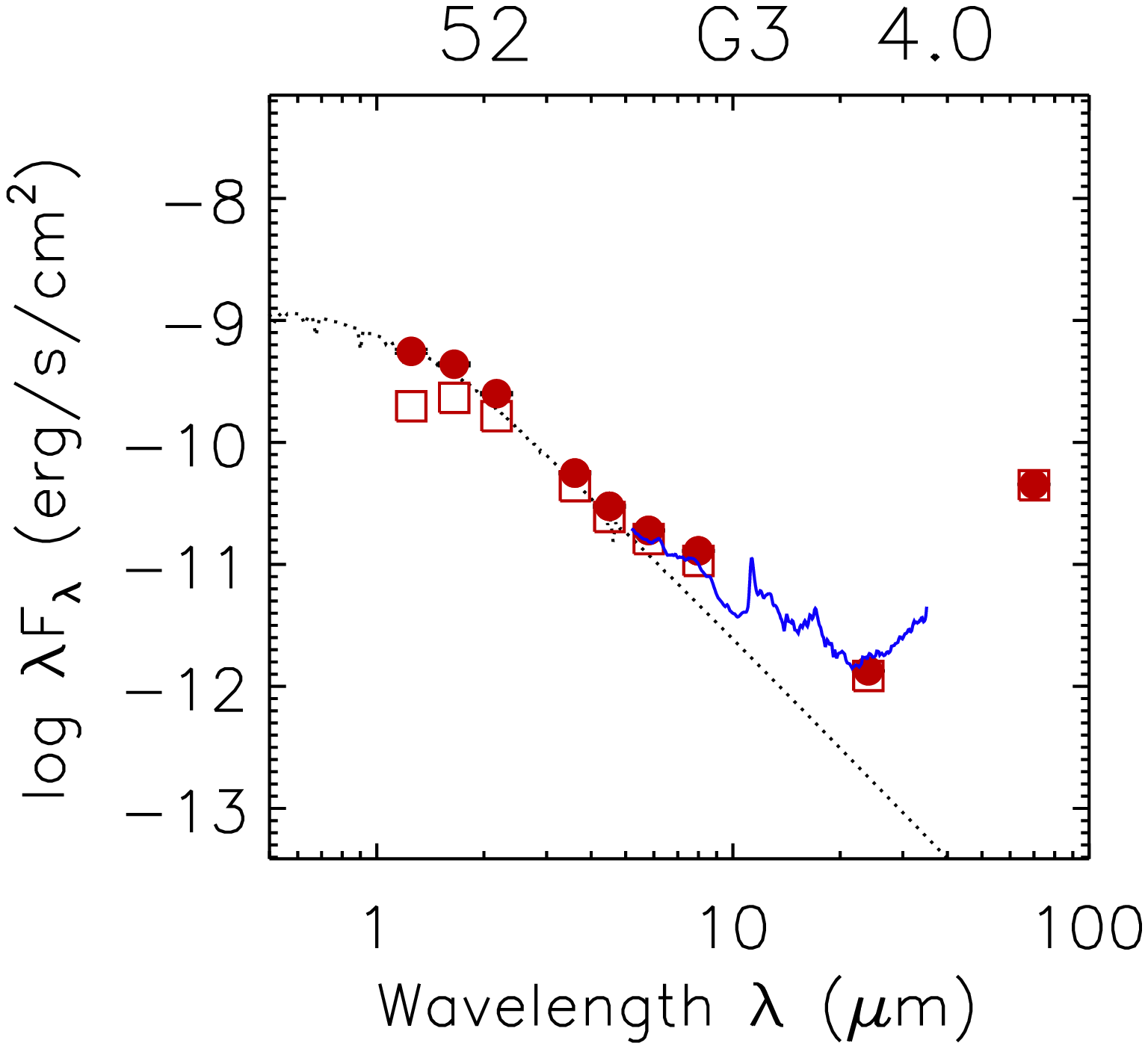}
\includegraphics[width=0.2\textwidth]{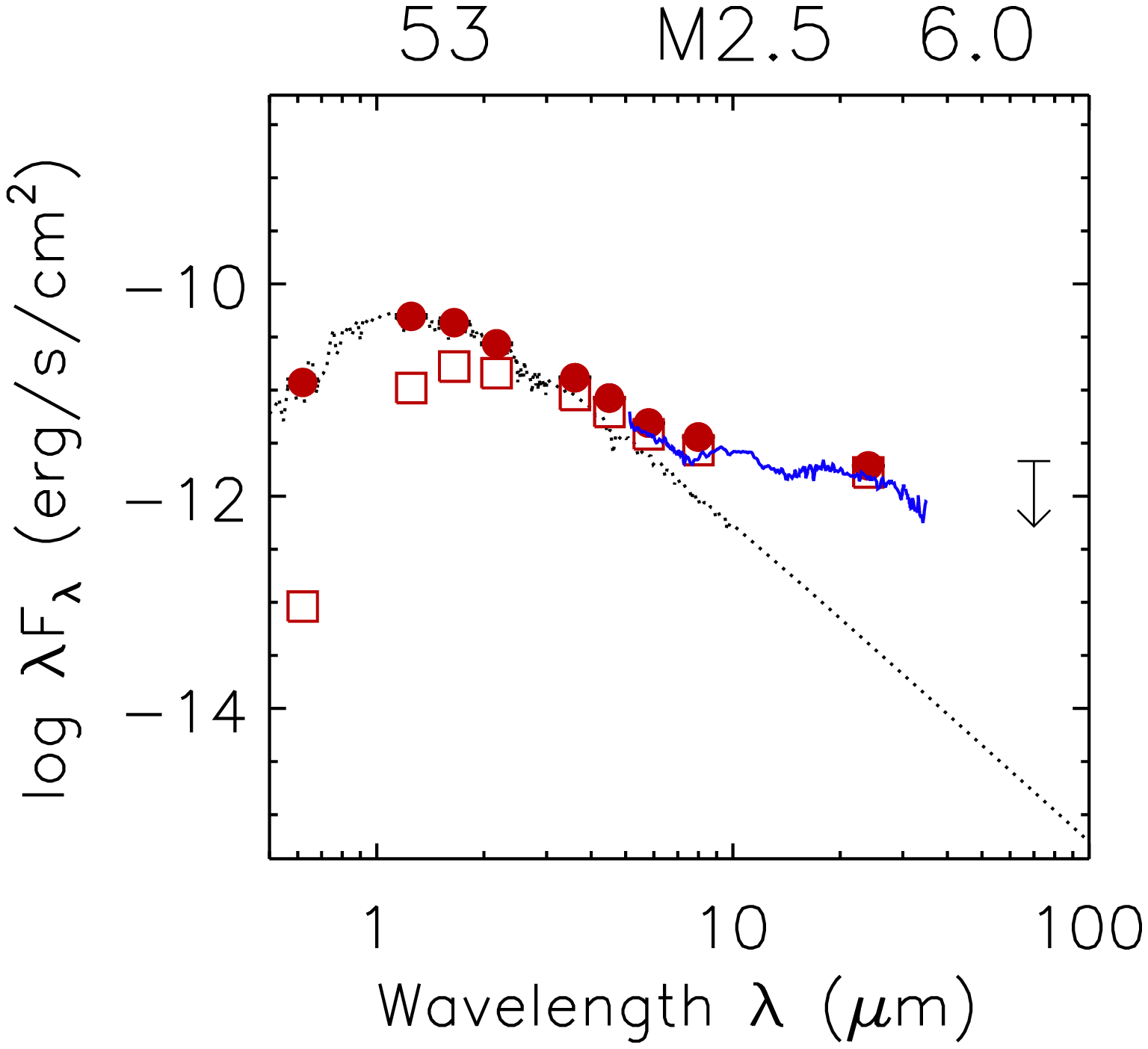}
\includegraphics[width=0.2\textwidth]{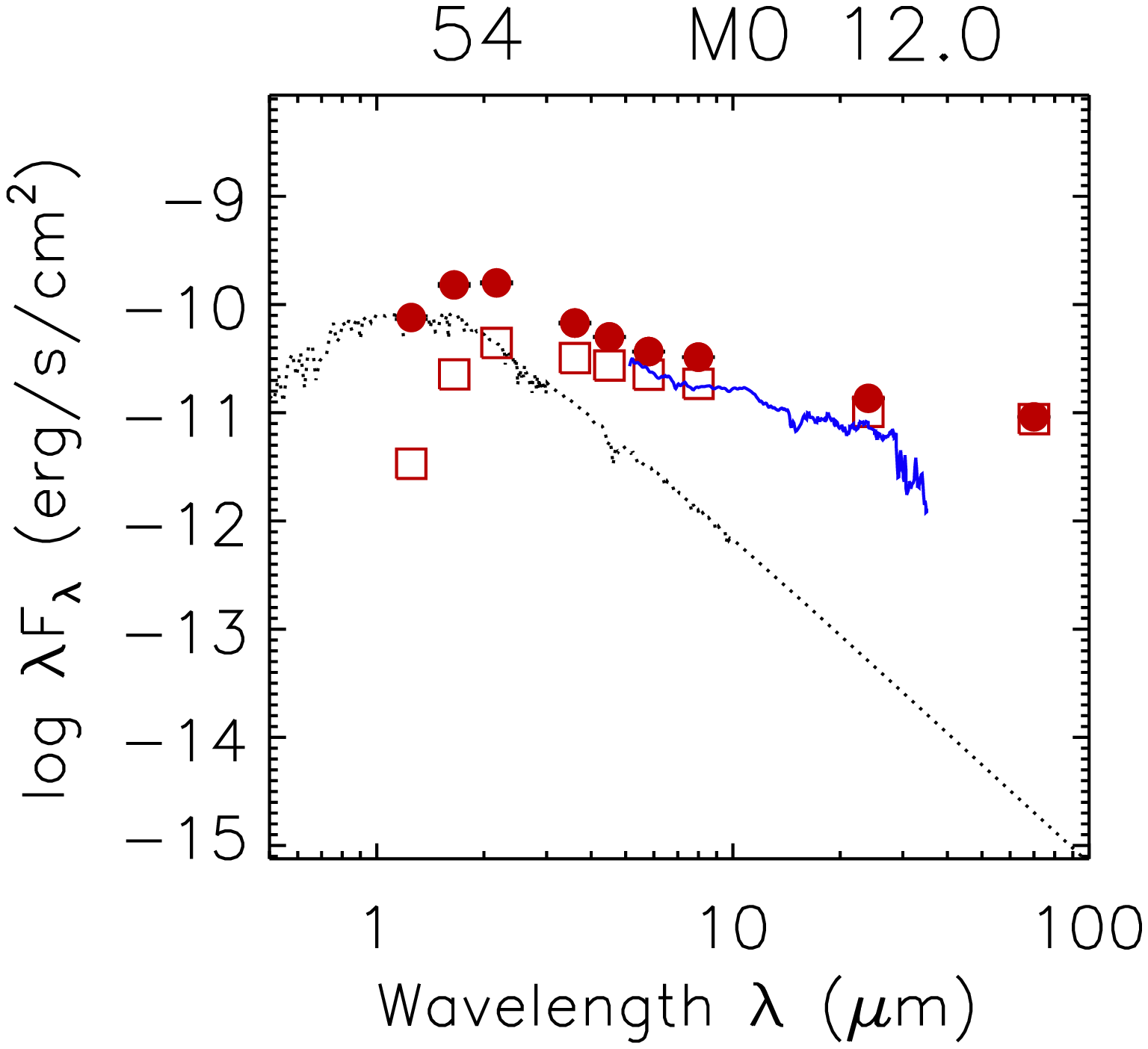}
\includegraphics[width=0.2\textwidth]{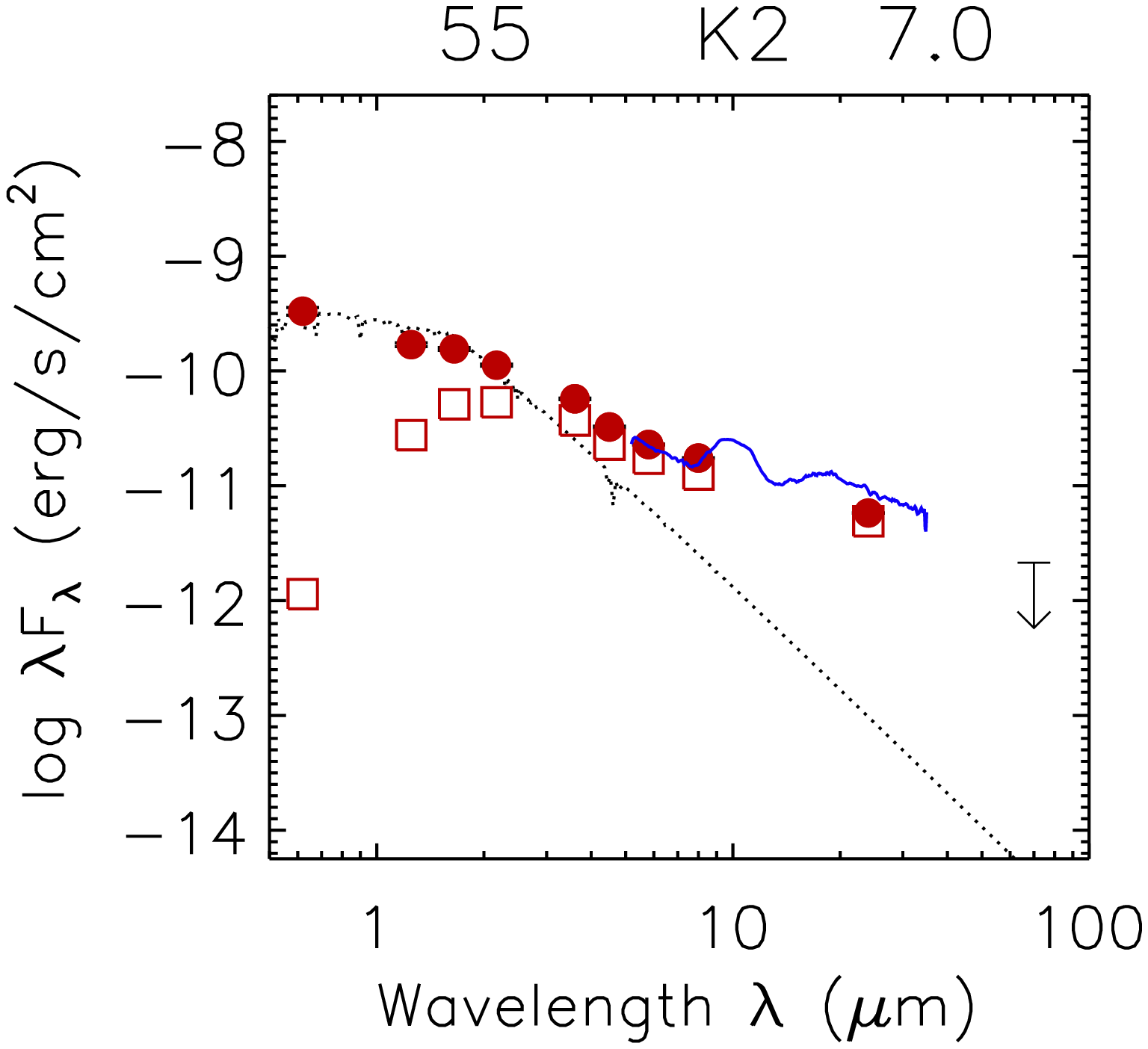}
\includegraphics[width=0.2\textwidth]{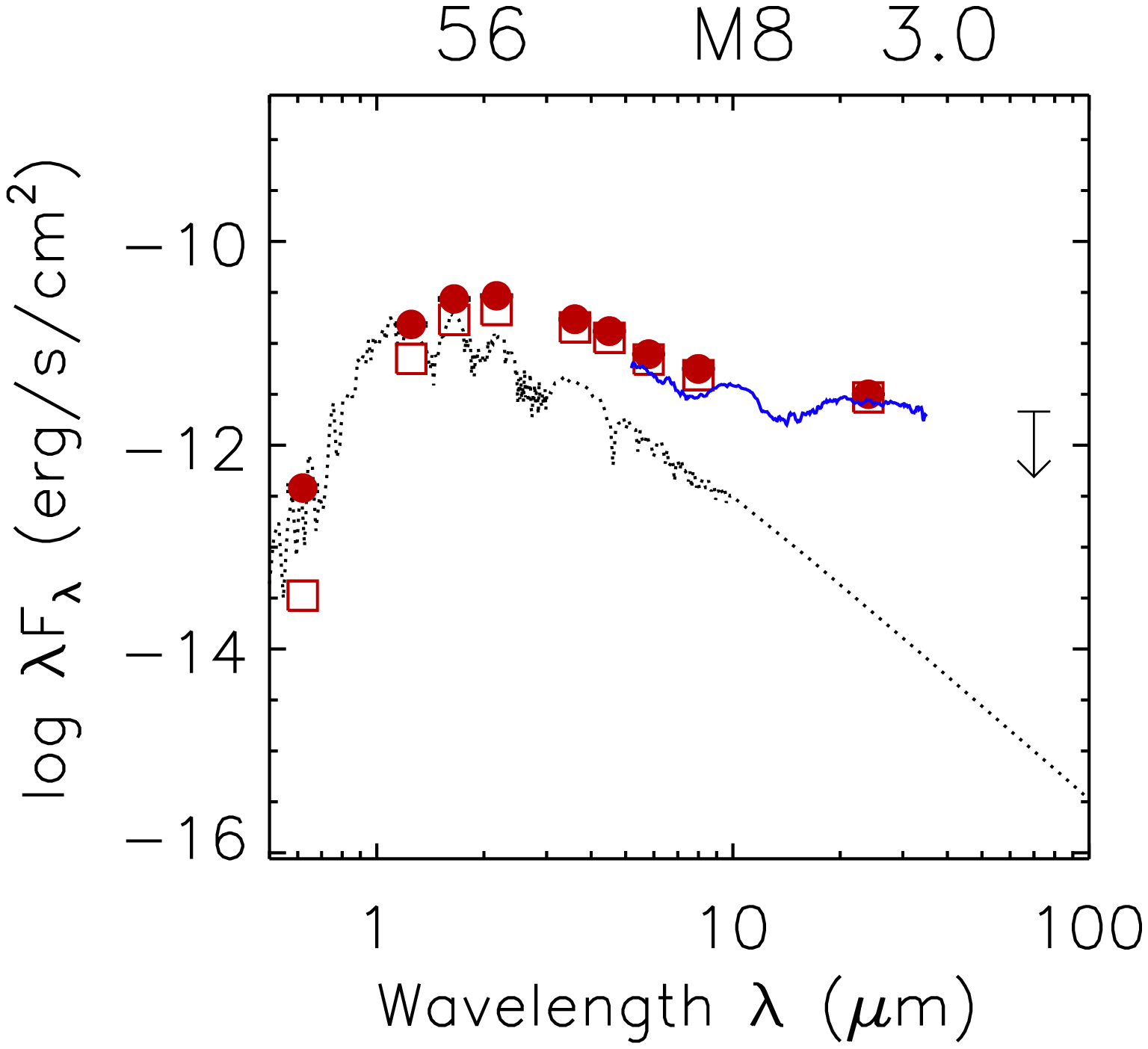}
\includegraphics[width=0.2\textwidth]{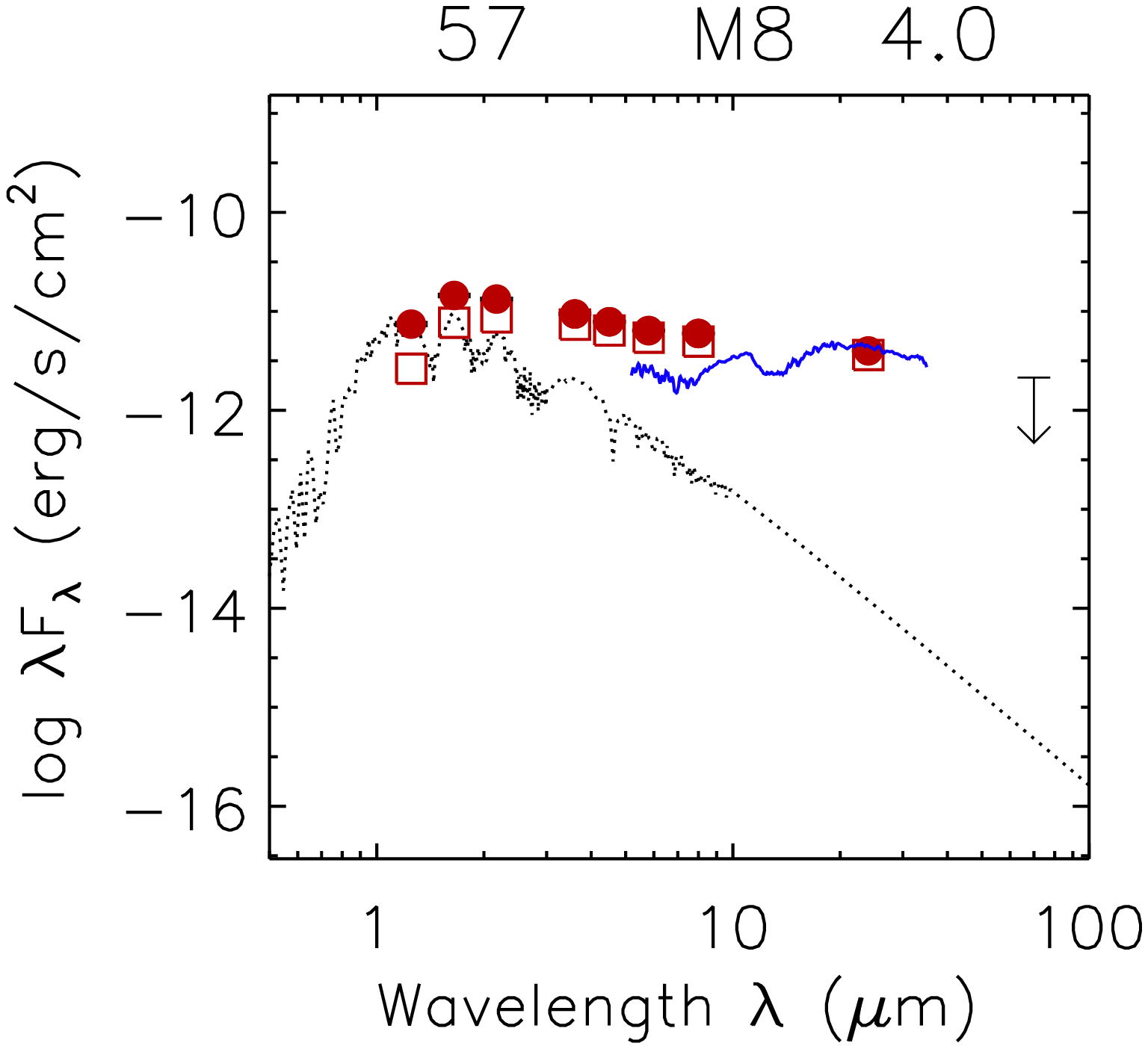}
\includegraphics[width=0.2\textwidth]{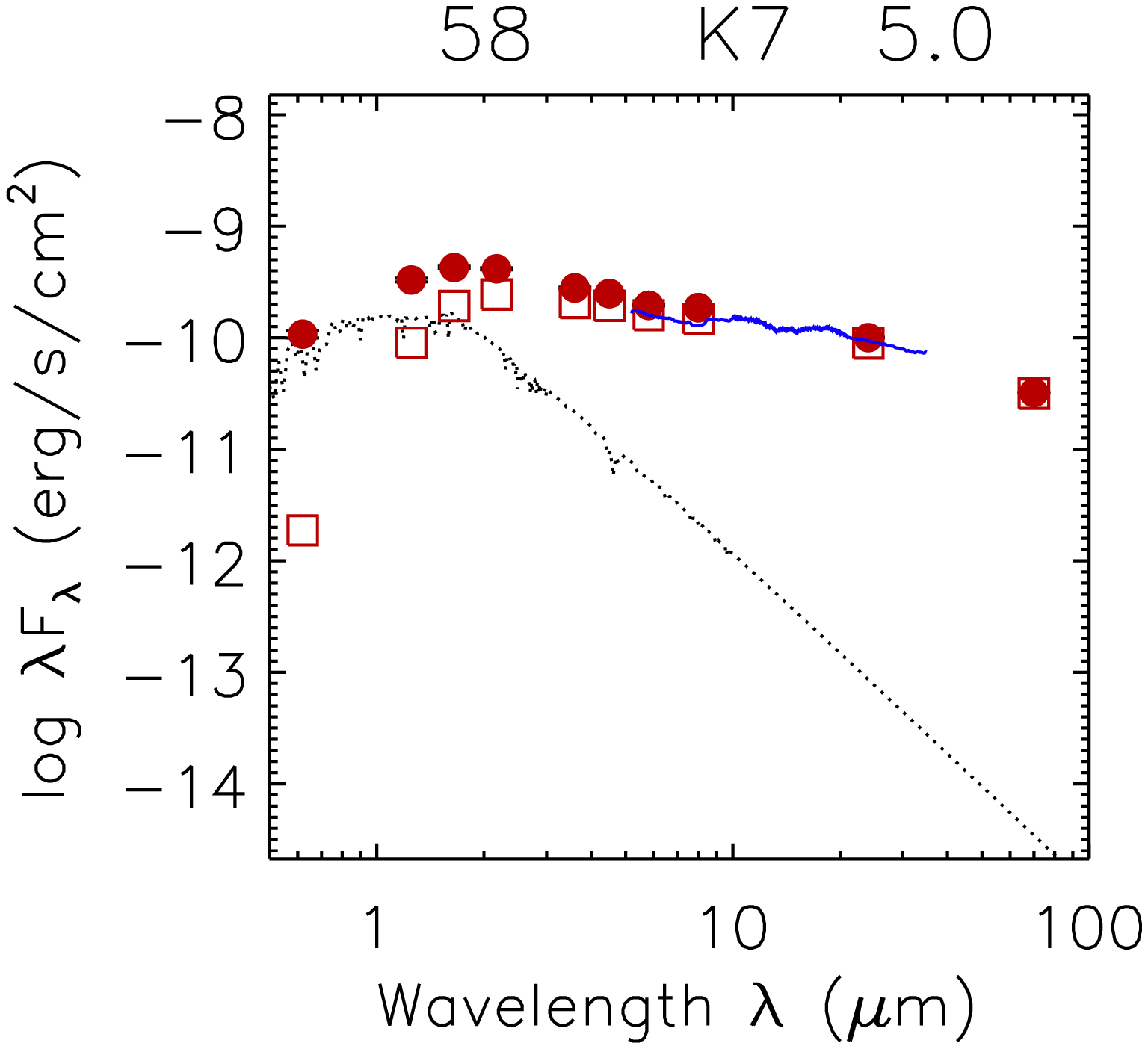}
\includegraphics[width=0.2\textwidth]{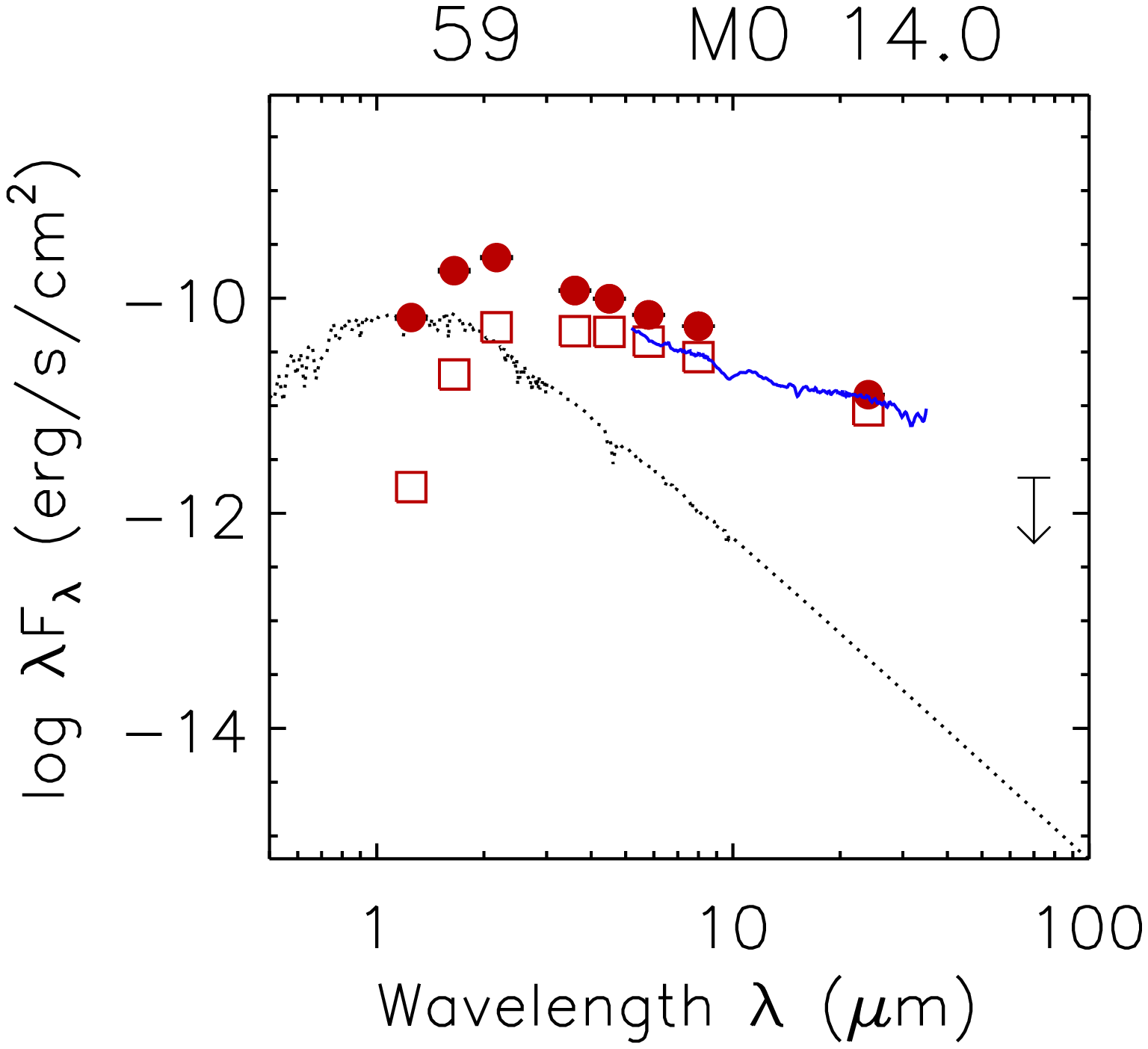}
\includegraphics[width=0.2\textwidth]{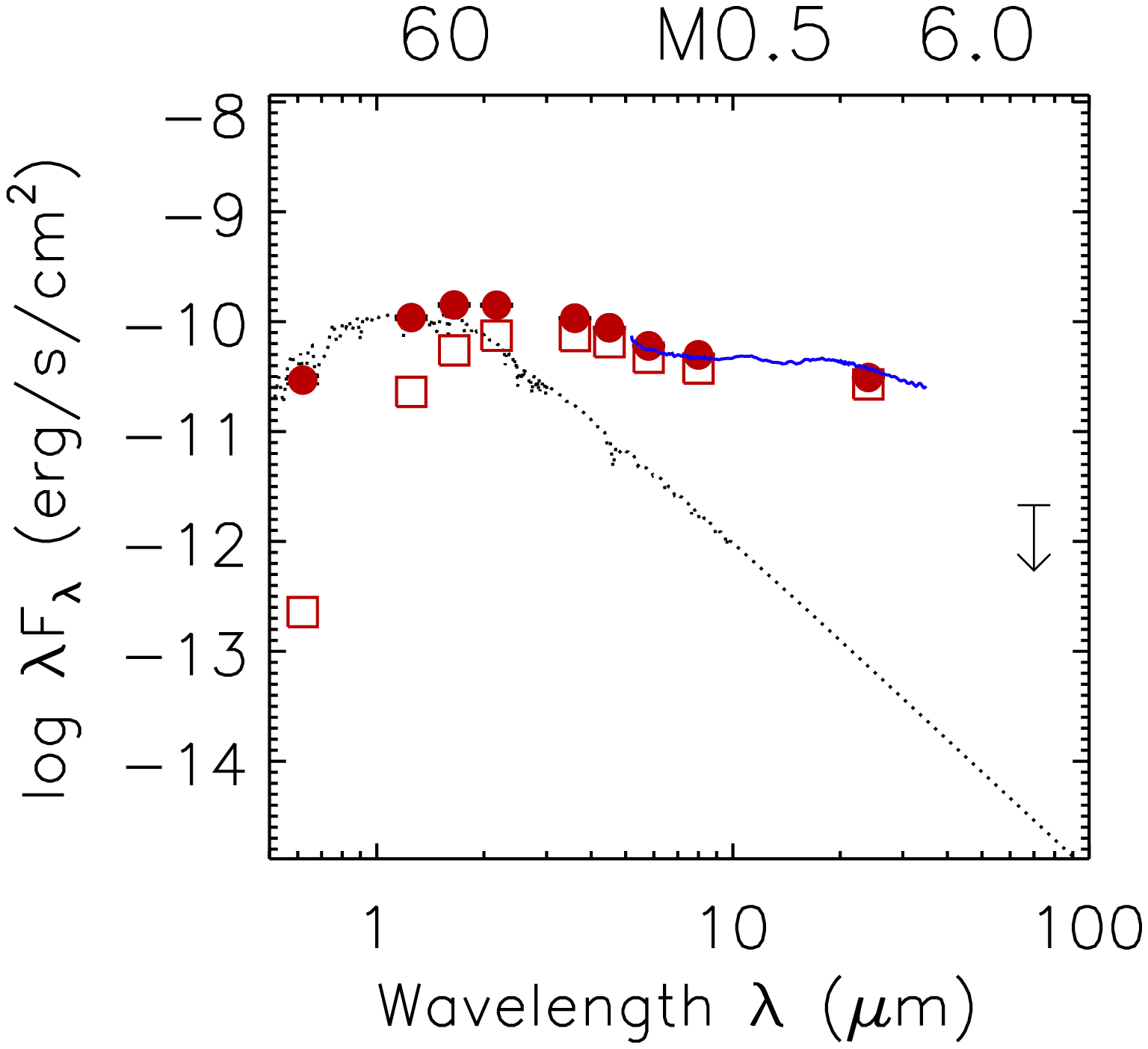}
\includegraphics[width=0.2\textwidth]{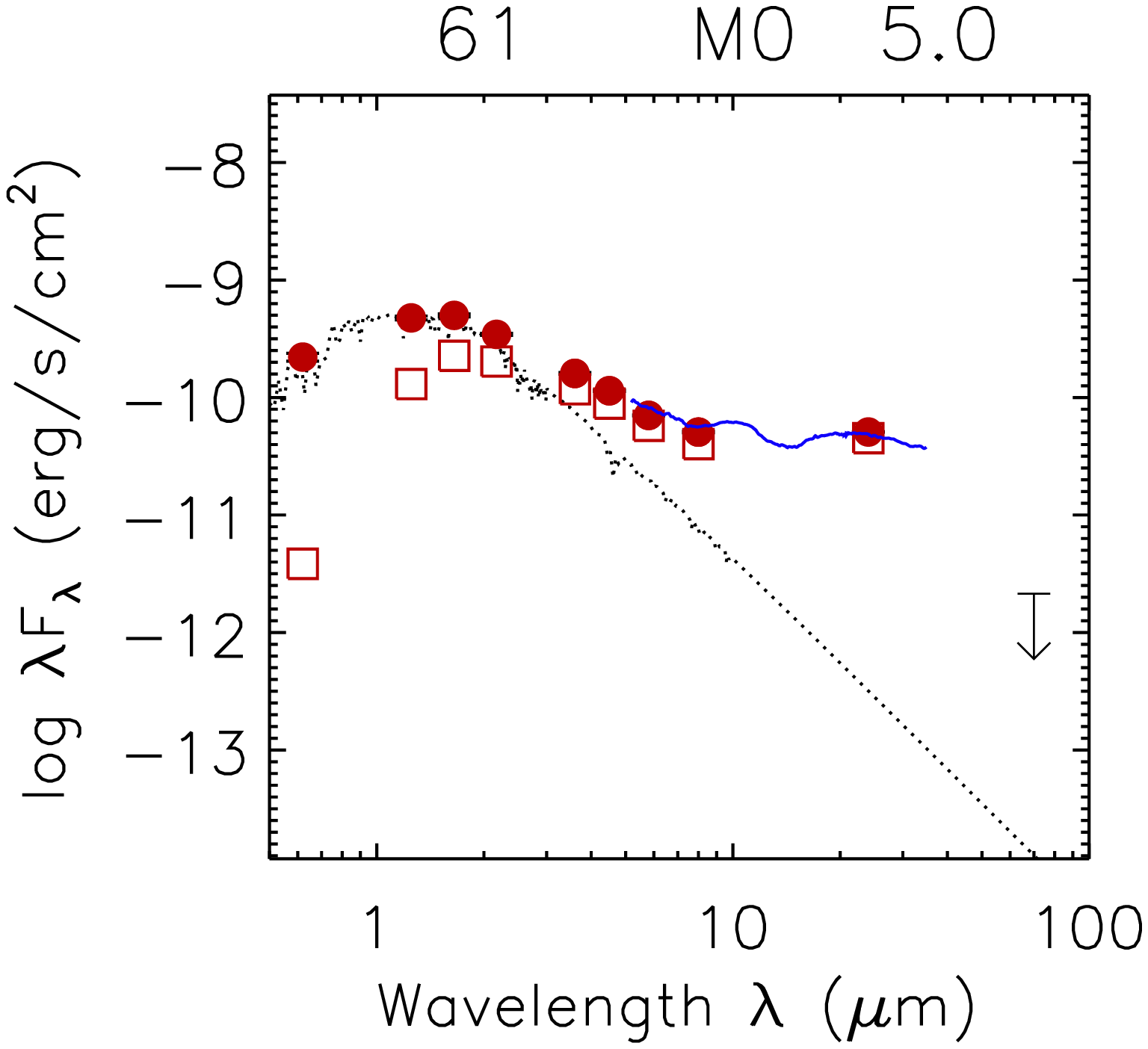}
\includegraphics[width=0.2\textwidth]{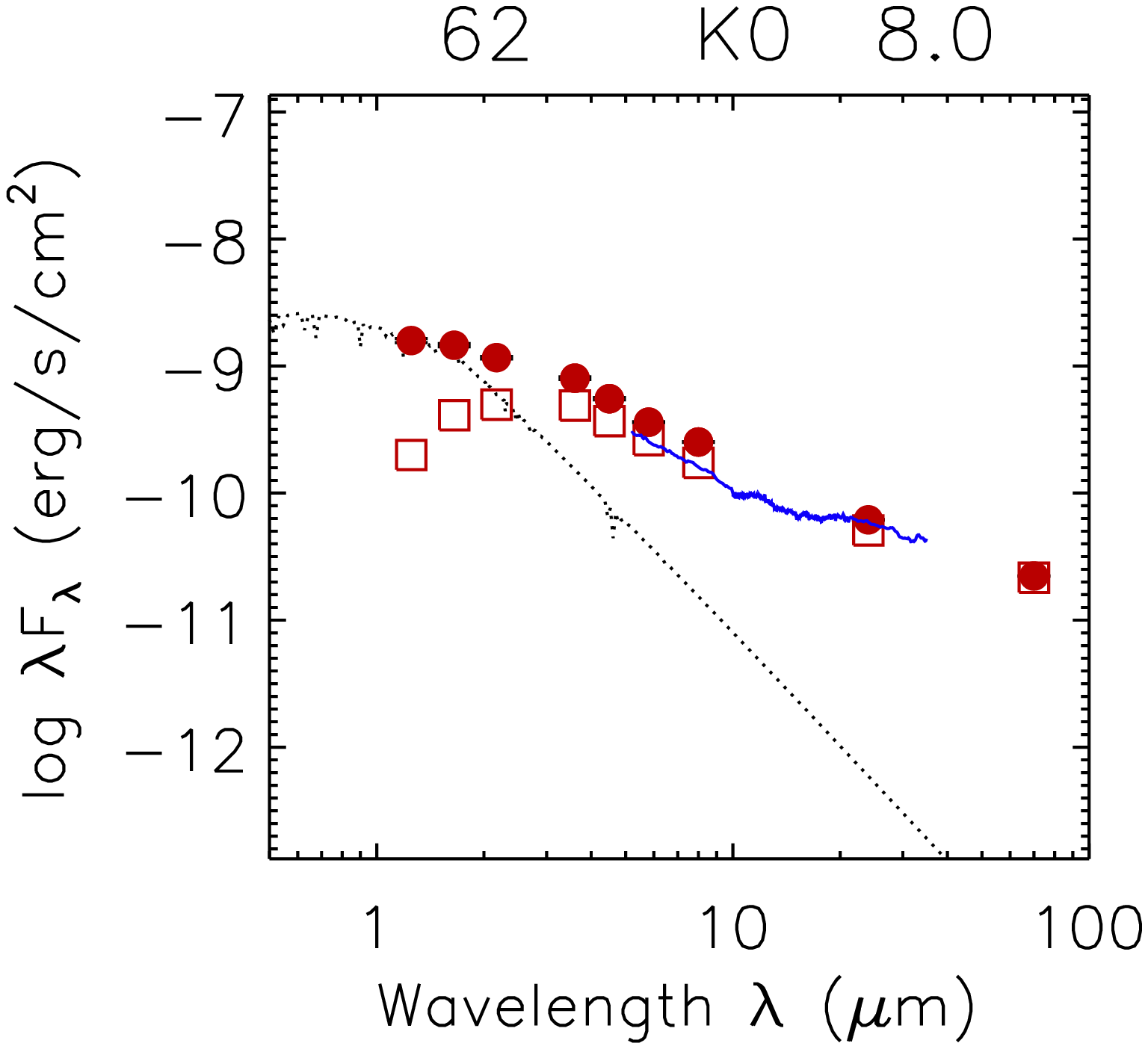}
\includegraphics[width=0.2\textwidth]{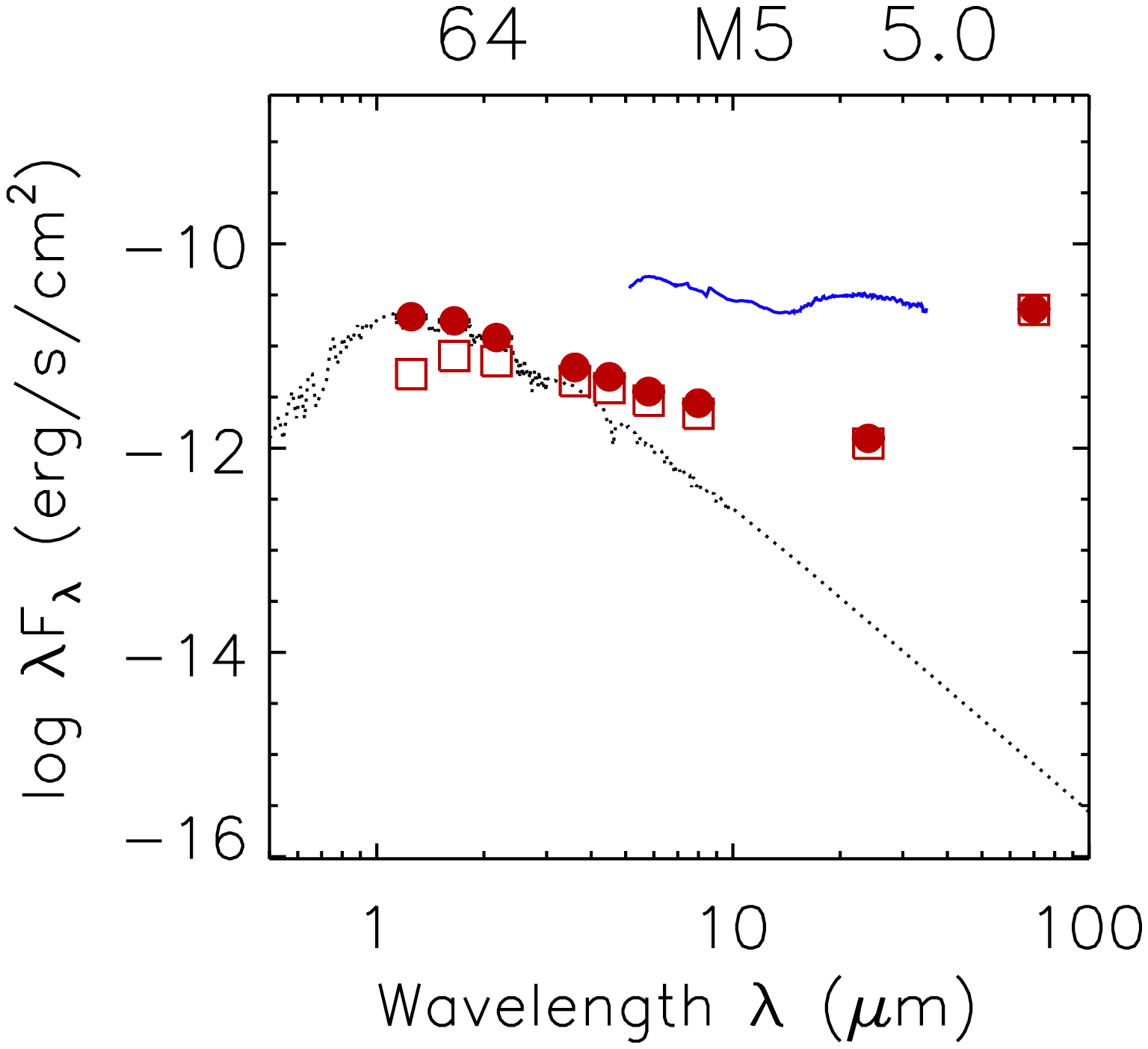}
\includegraphics[width=0.2\textwidth]{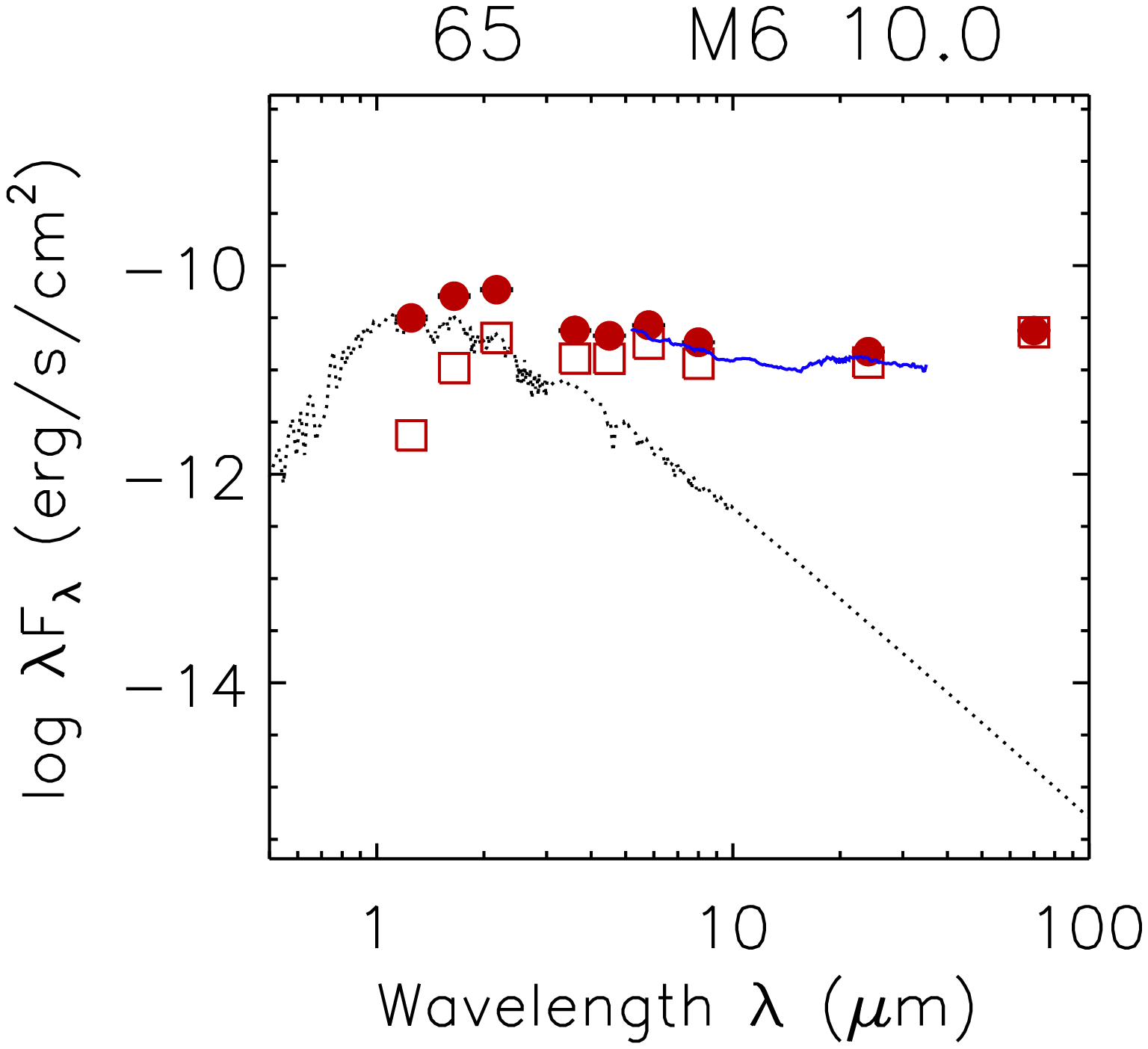}
\includegraphics[width=0.2\textwidth]{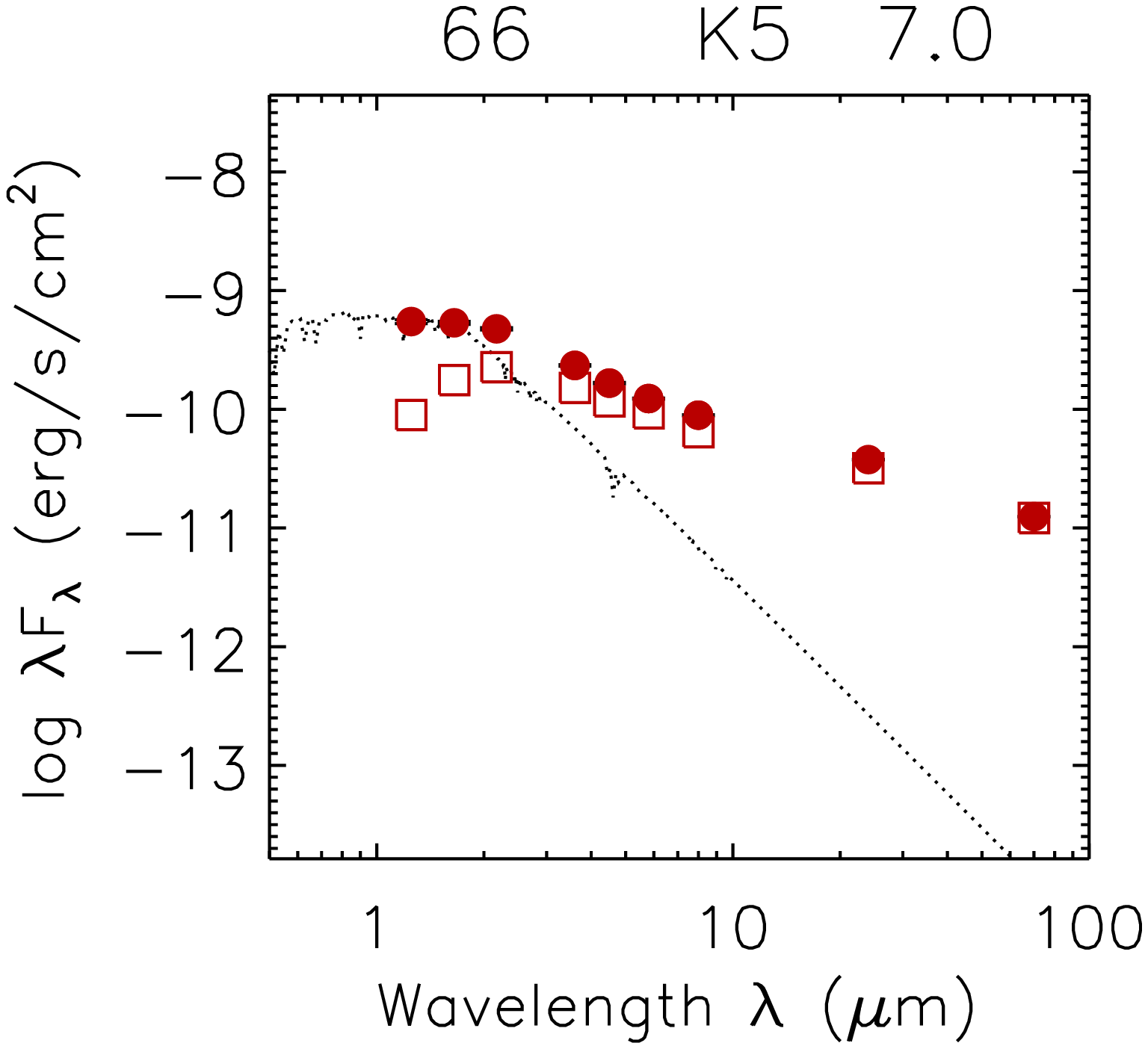}
\includegraphics[width=0.2\textwidth]{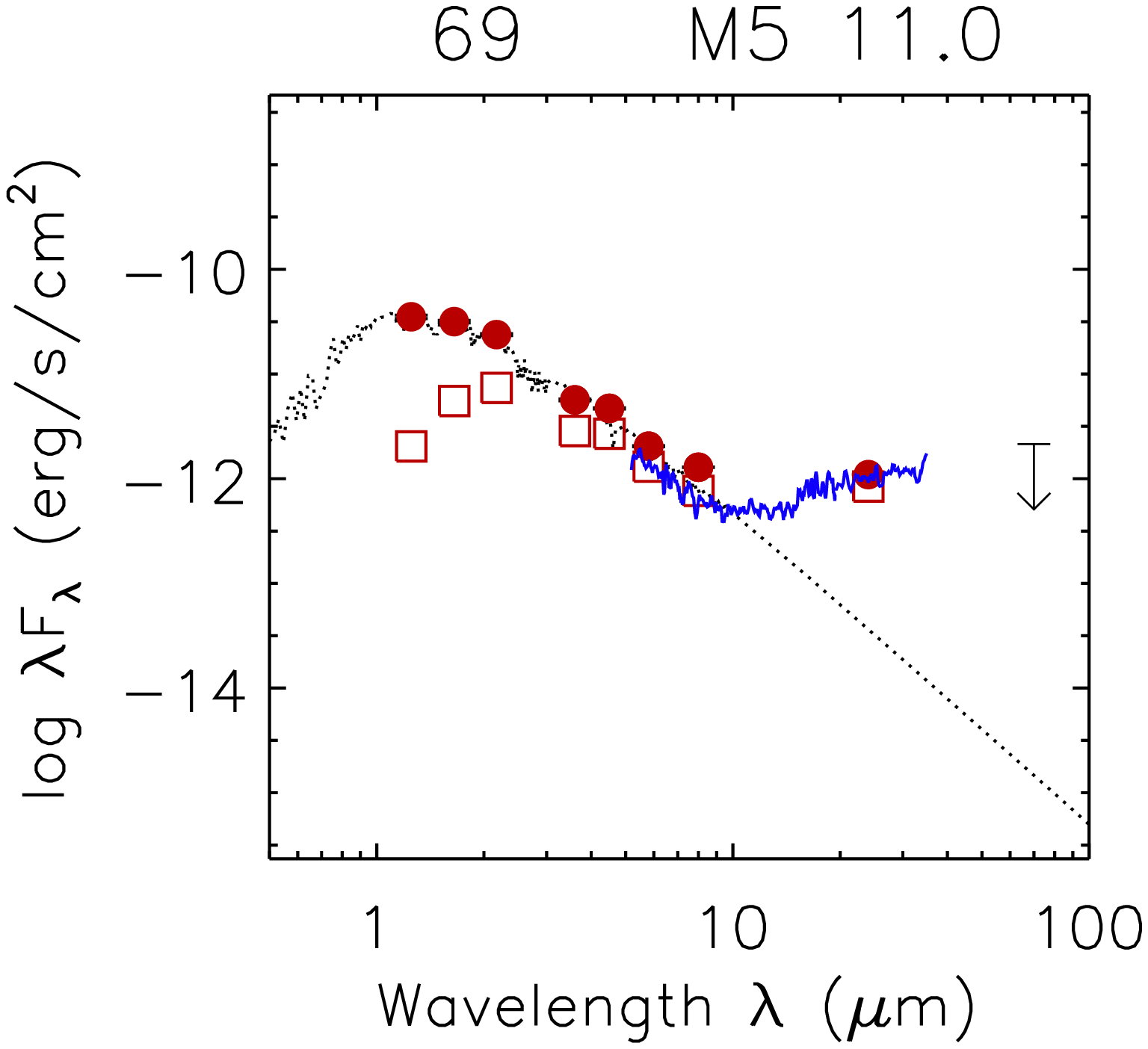}
\includegraphics[width=0.2\textwidth]{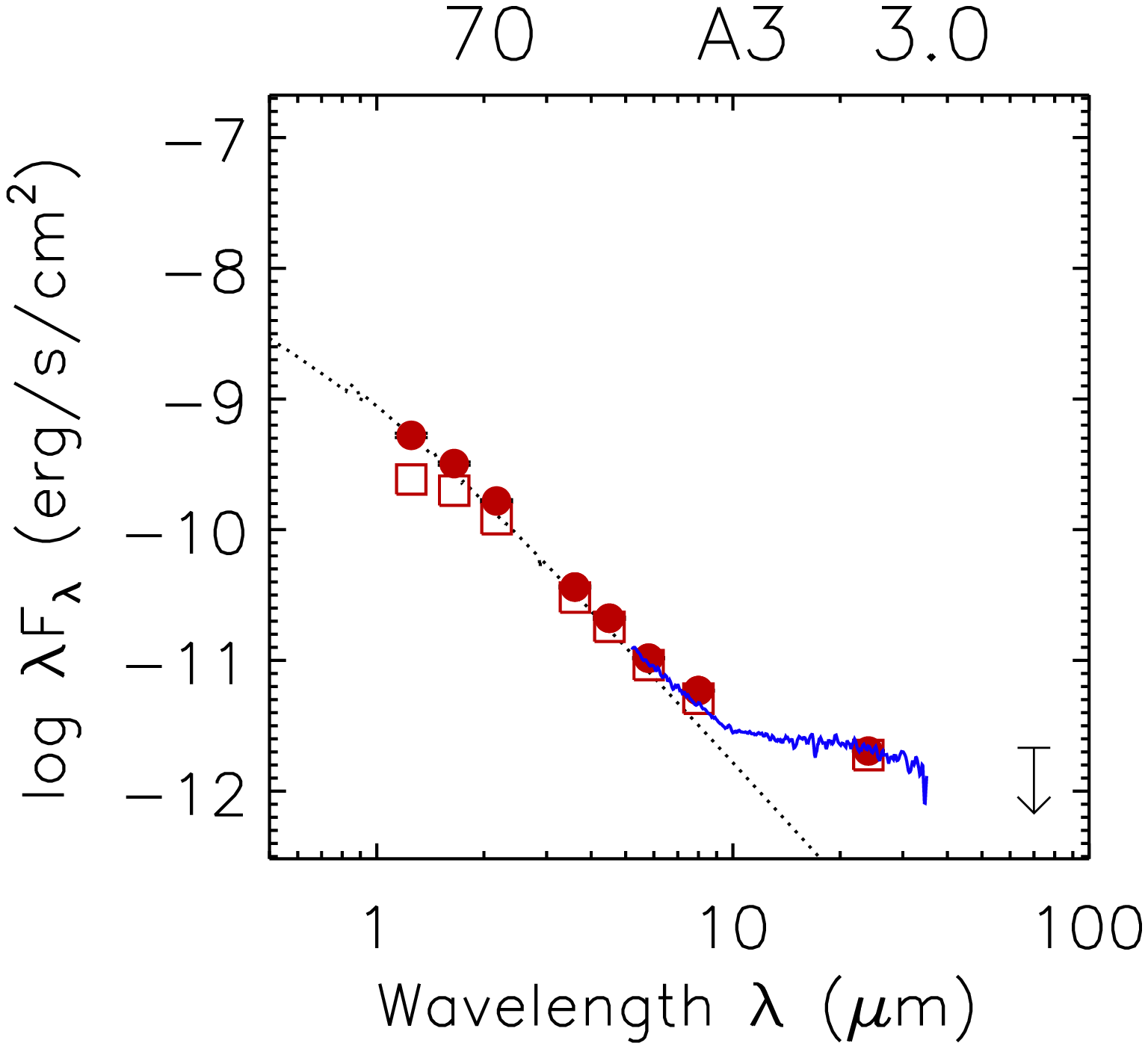}
\includegraphics[width=0.2\textwidth]{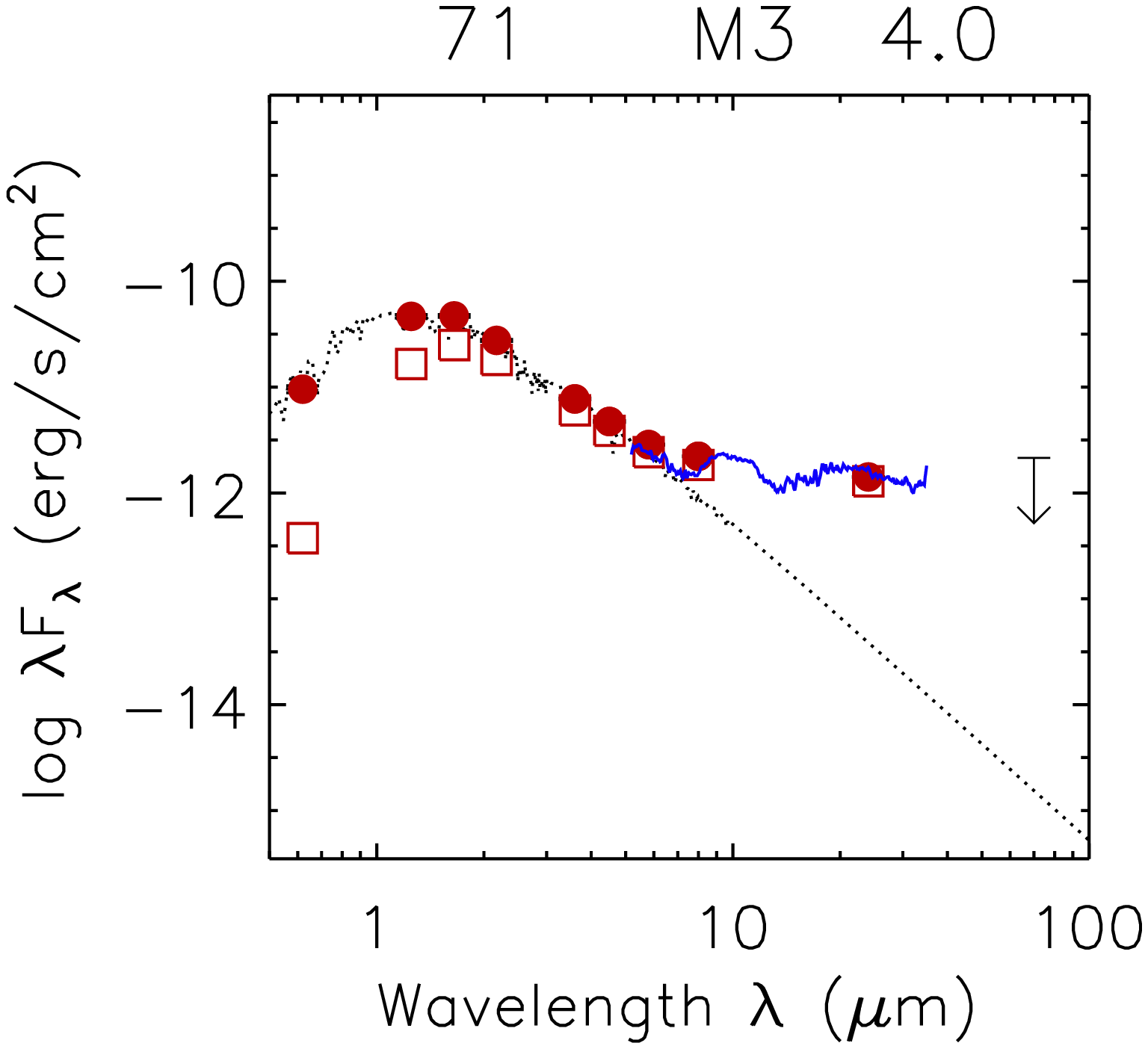}
\end{center}
\caption{\label{7f_ap_seds1} SEDs of the young stellar population with
  disks of Serpens. Each SED has the corresponding object ID (as in
  \citealt{OL10}) on the top left. The solid black line indicates the
  NextGen stellar photosphere model for the spectral type indicated on
  the top of each plot. Open squares are the observed photometry while
  the solid circles are the dereddened photometry. The visual
  extinction of each object can be seen on the top right. The solid
  gray line is the object's IRS spectrum.  }
\end{figure*}

\newpage

\begin{figure*}[!h]
\begin{center}
\includegraphics[width=0.2\textwidth]{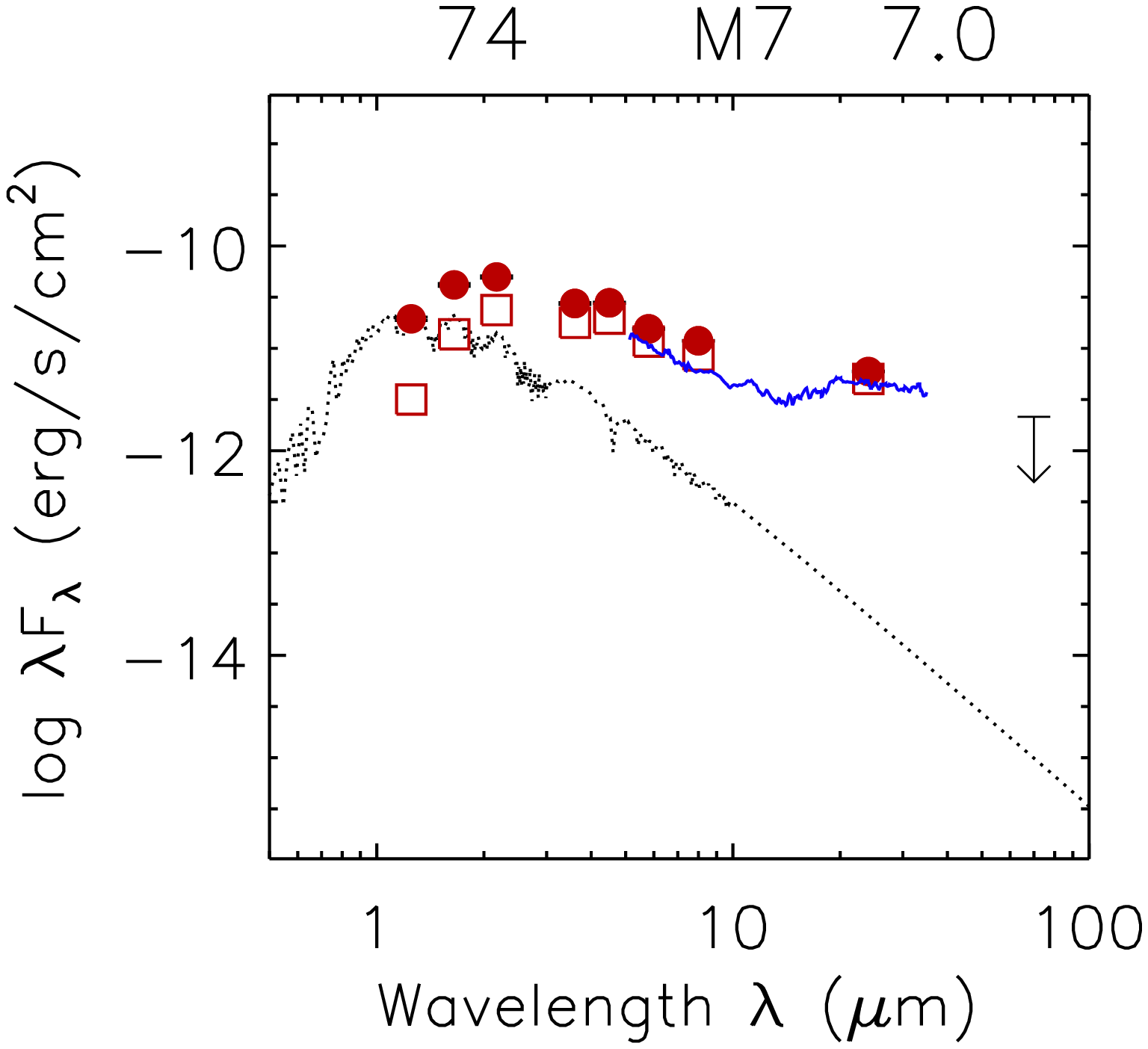}
\includegraphics[width=0.2\textwidth]{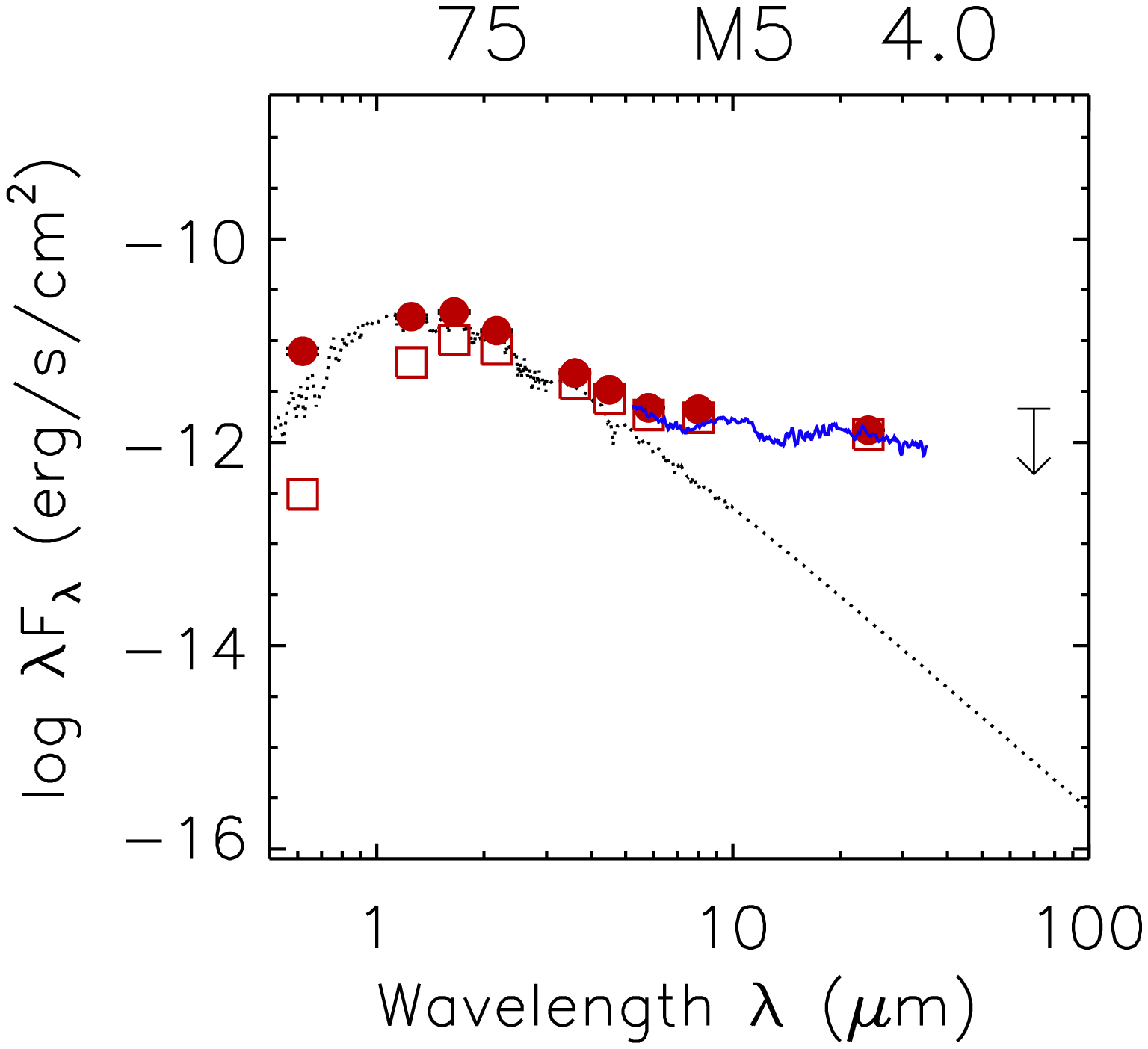}
\includegraphics[width=0.2\textwidth]{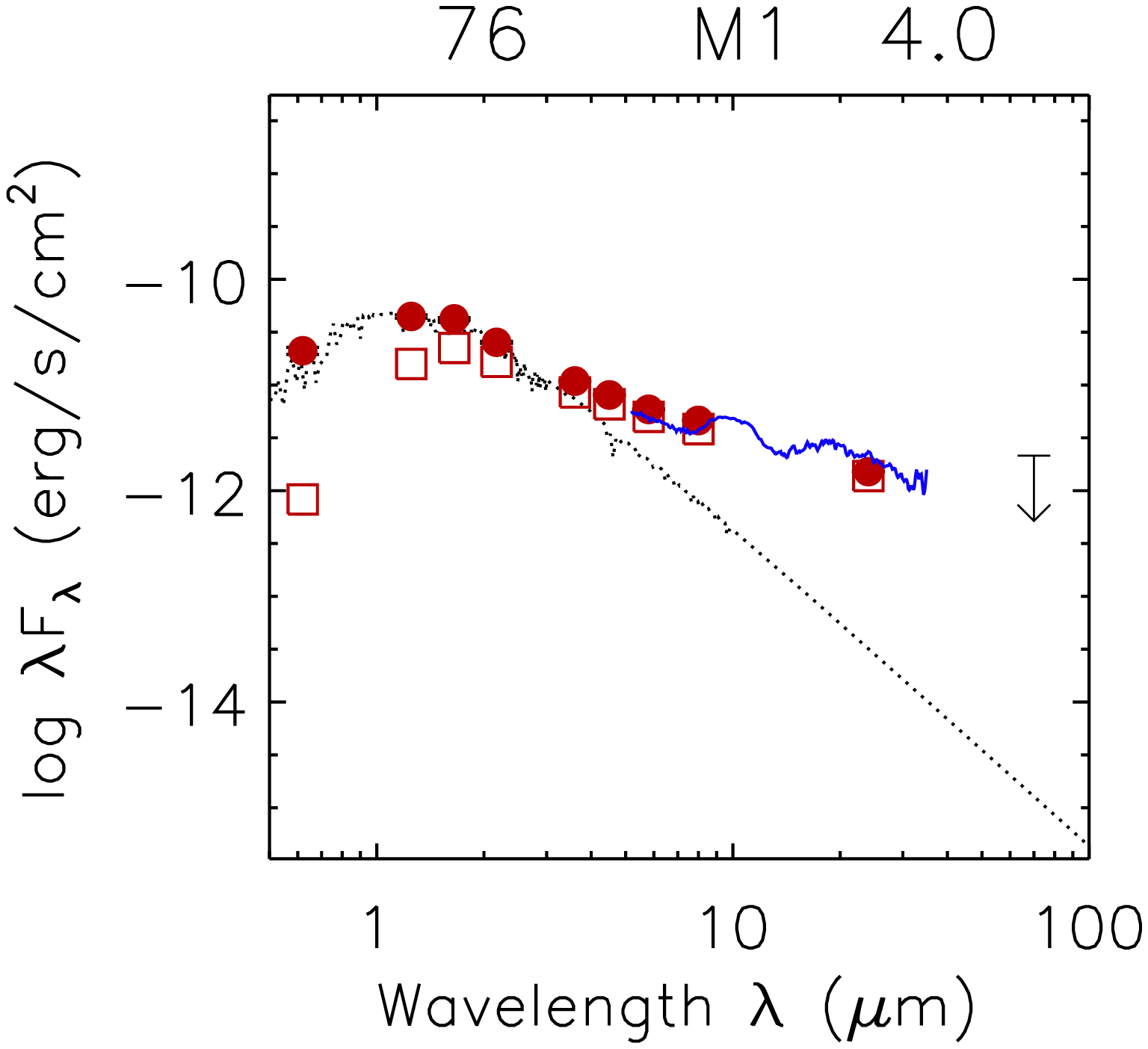}
\includegraphics[width=0.2\textwidth]{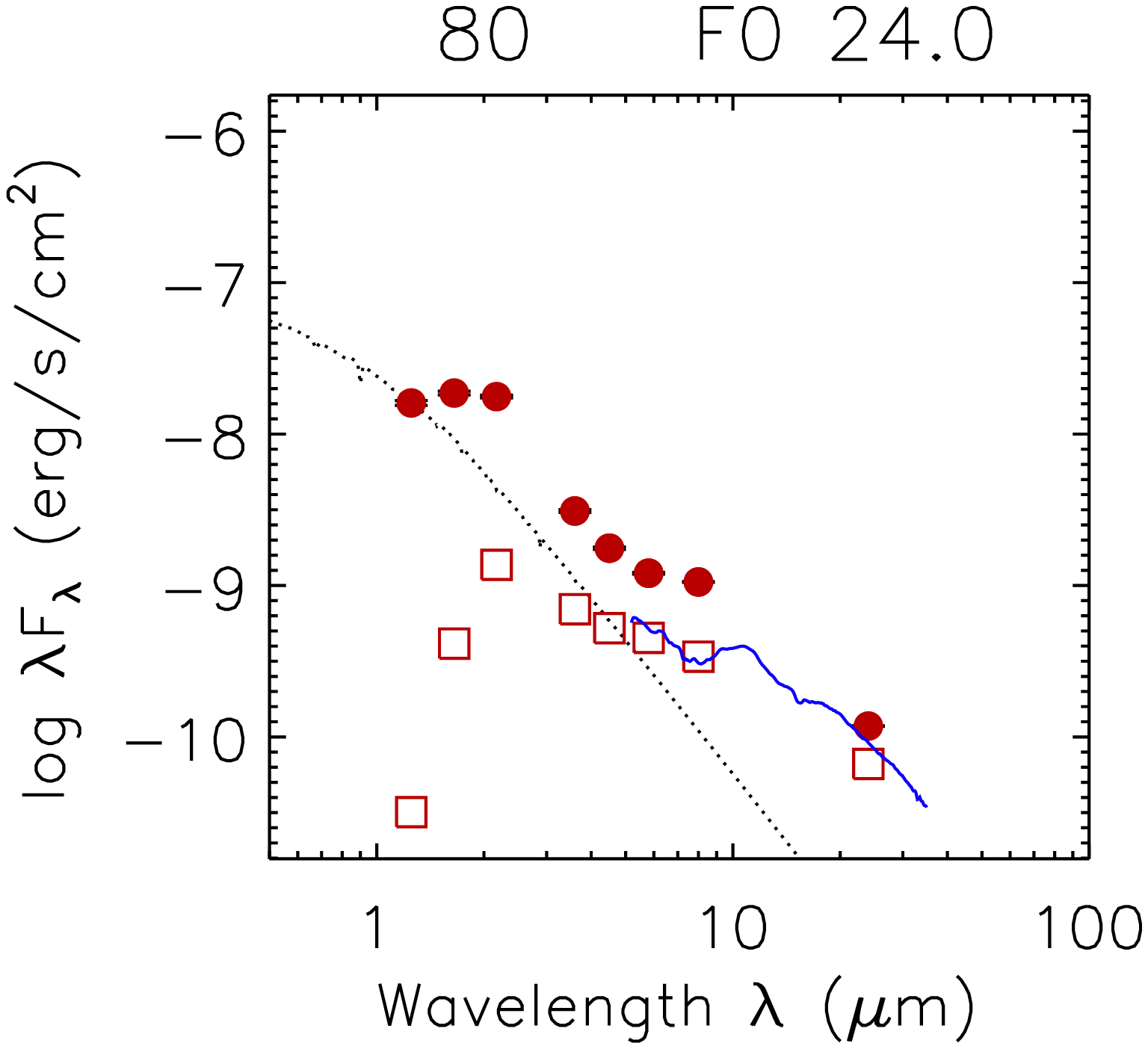}
\includegraphics[width=0.2\textwidth]{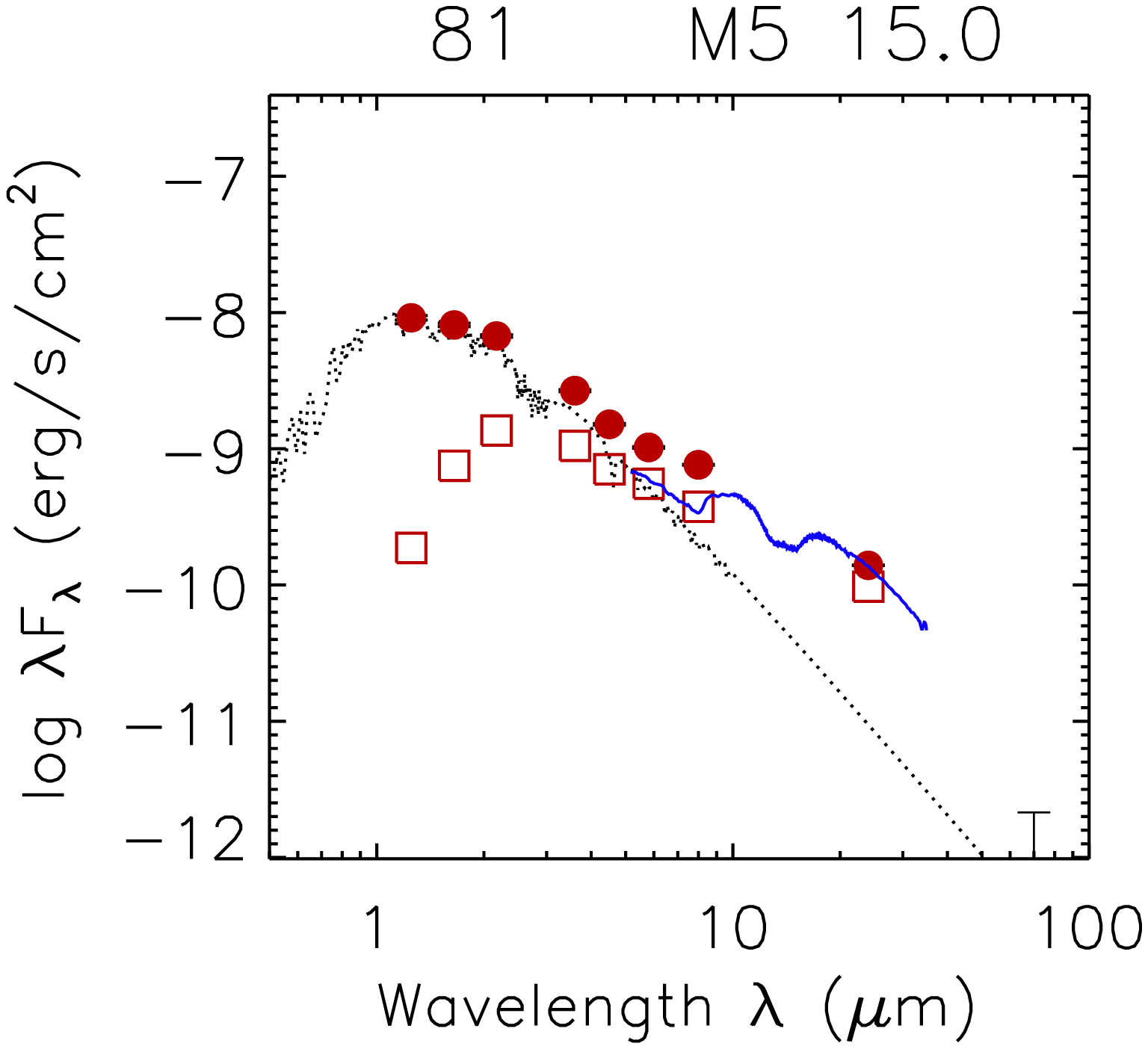}
\includegraphics[width=0.2\textwidth]{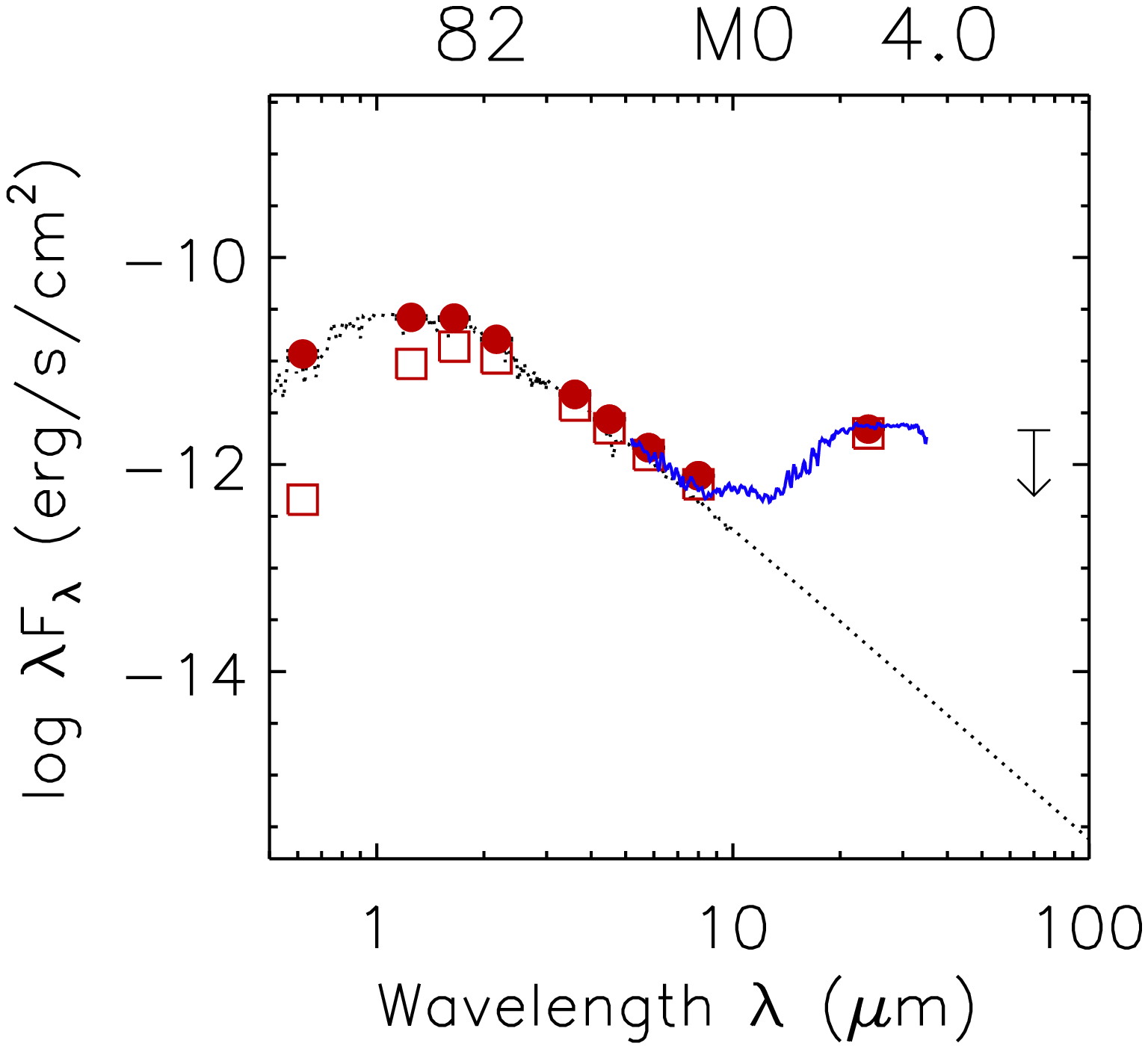}
\includegraphics[width=0.2\textwidth]{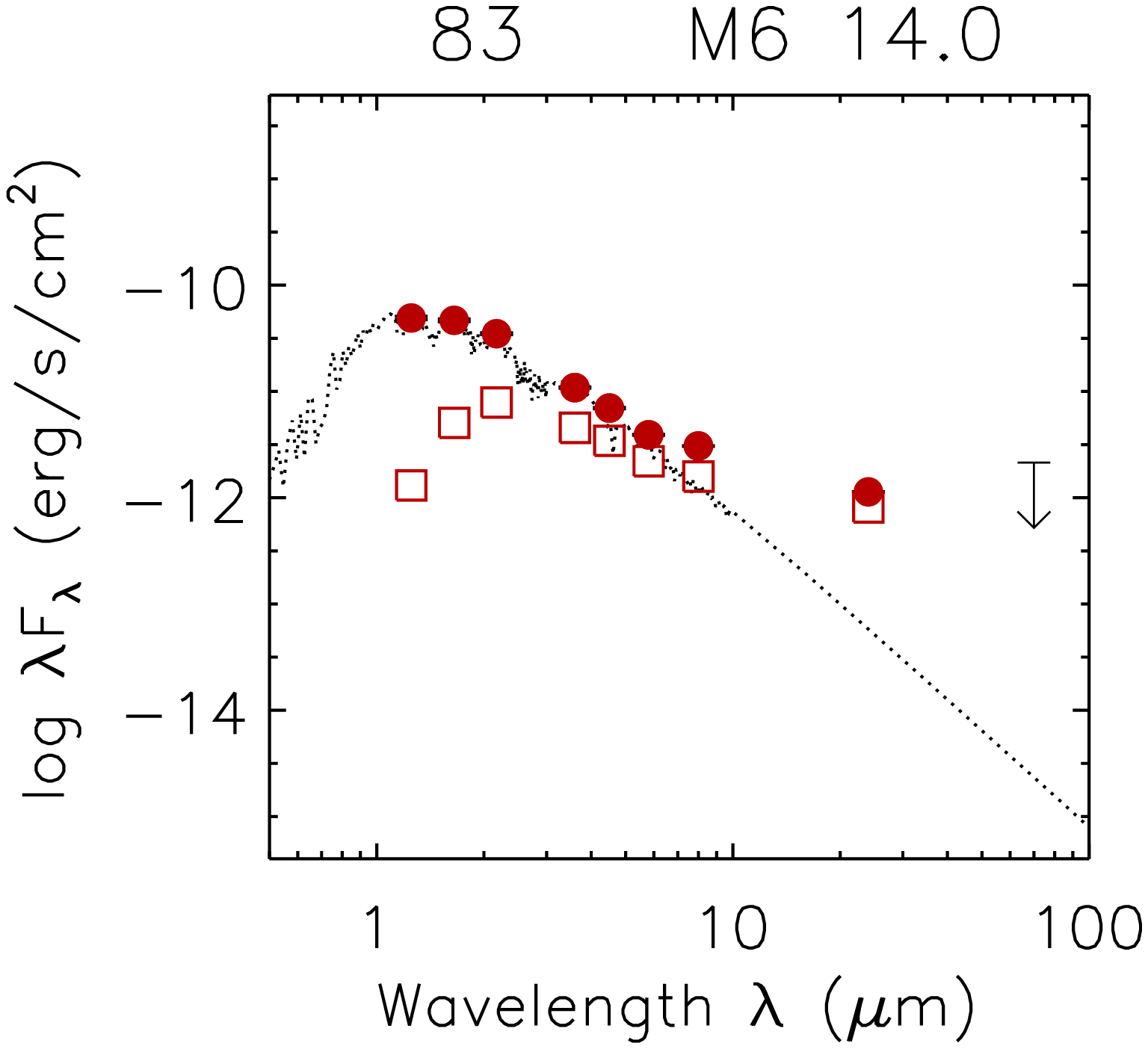}
\includegraphics[width=0.2\textwidth]{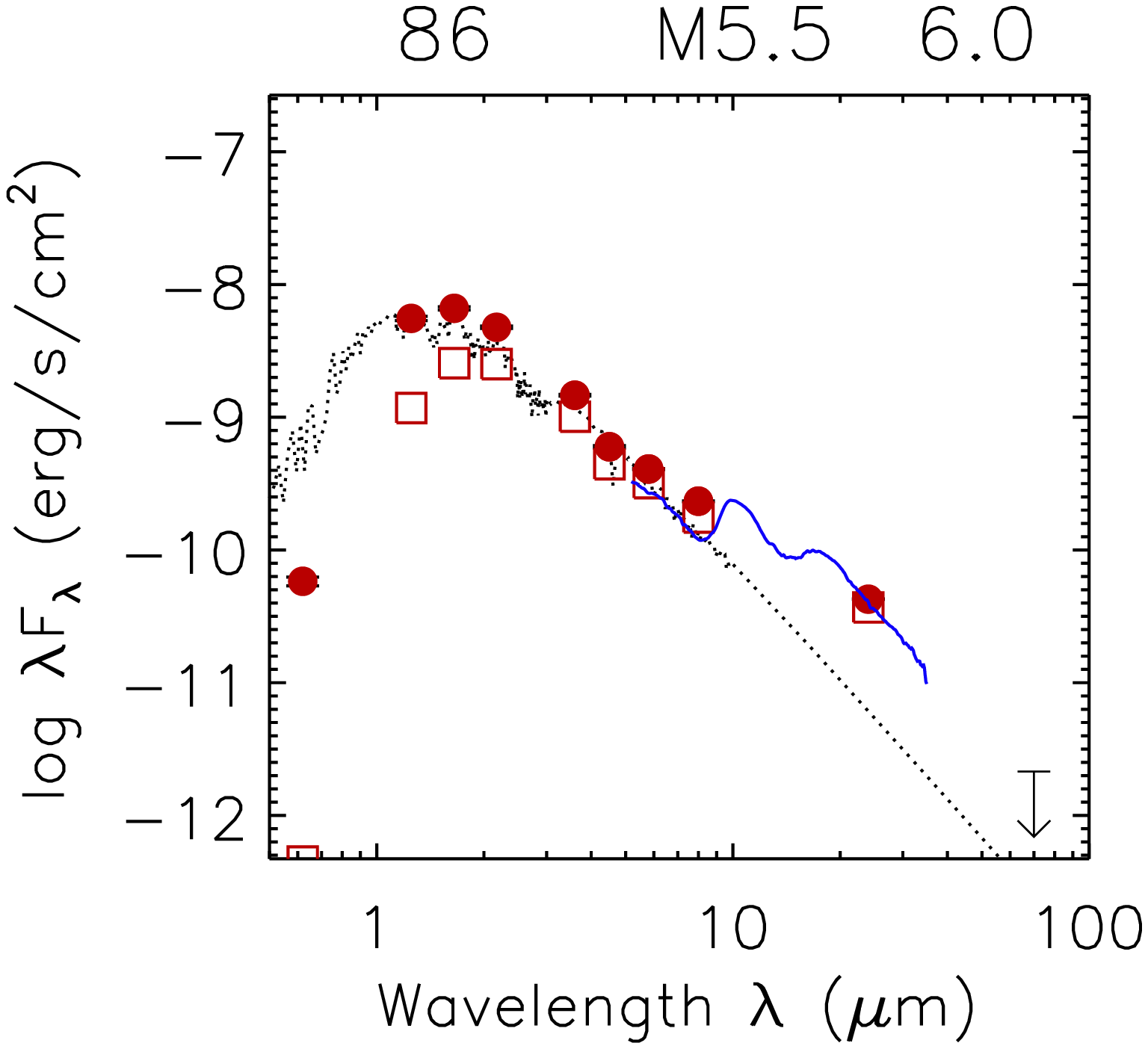}
\includegraphics[width=0.2\textwidth]{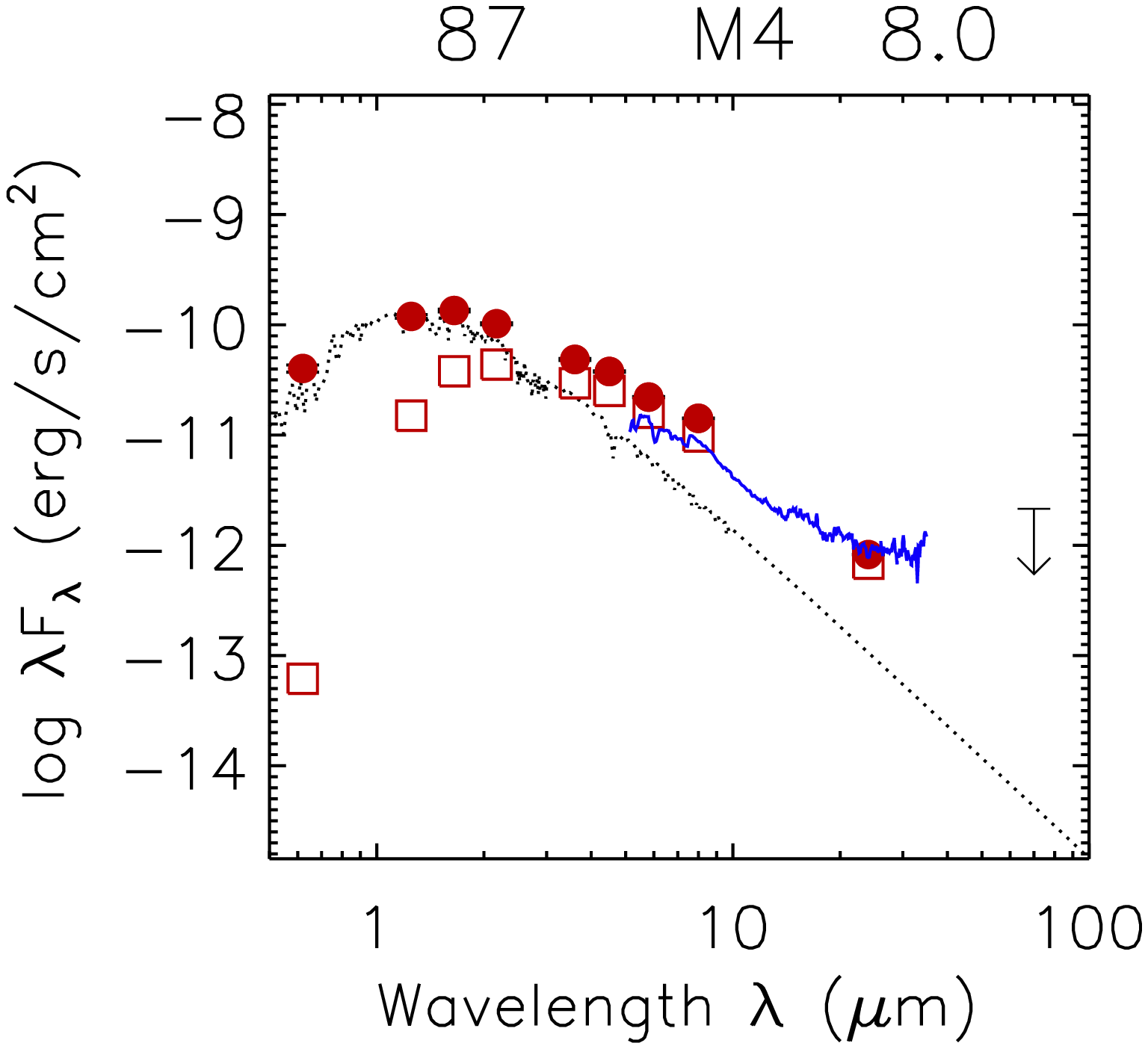}
\includegraphics[width=0.2\textwidth]{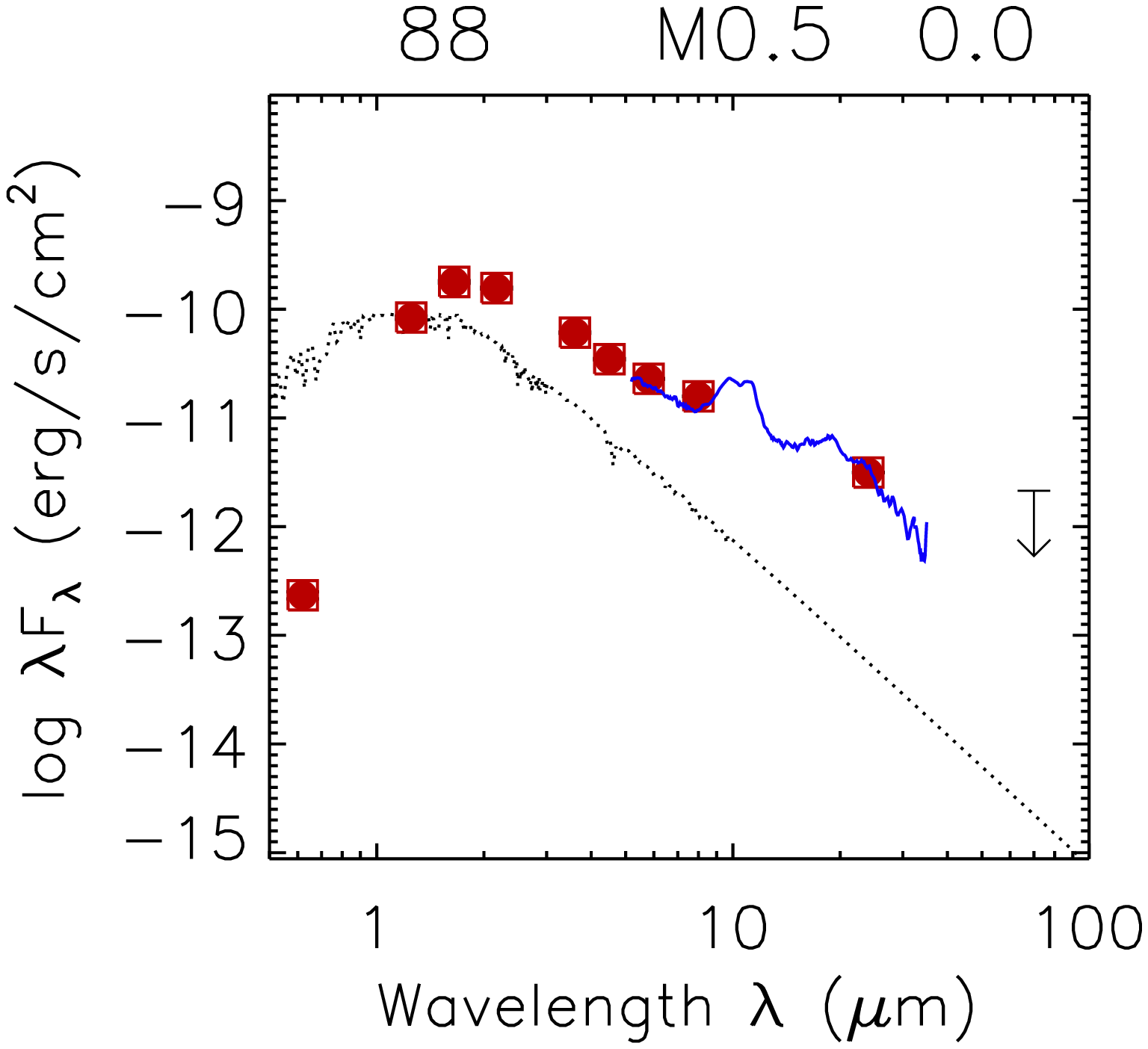}
\includegraphics[width=0.2\textwidth]{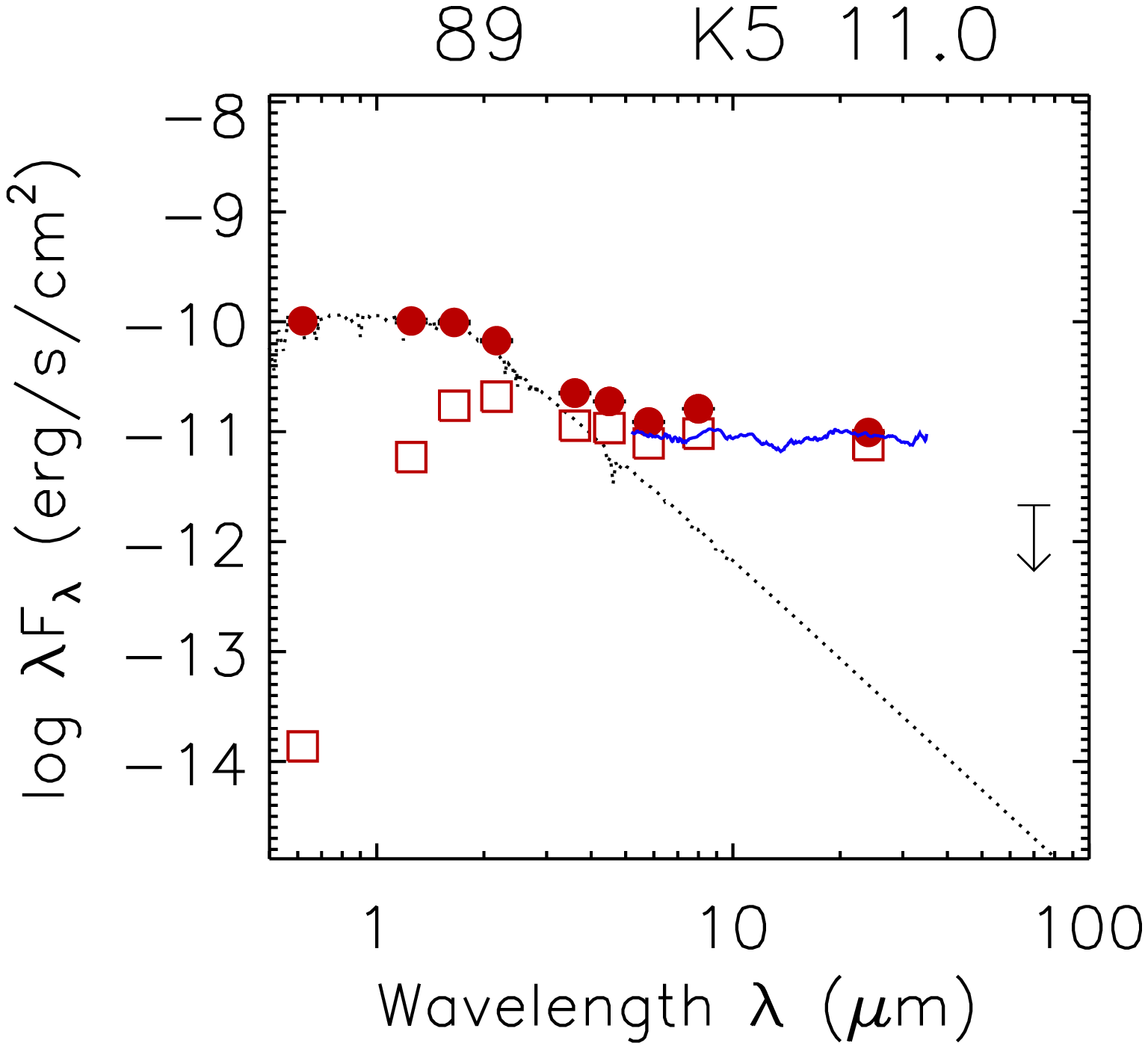}
\includegraphics[width=0.2\textwidth]{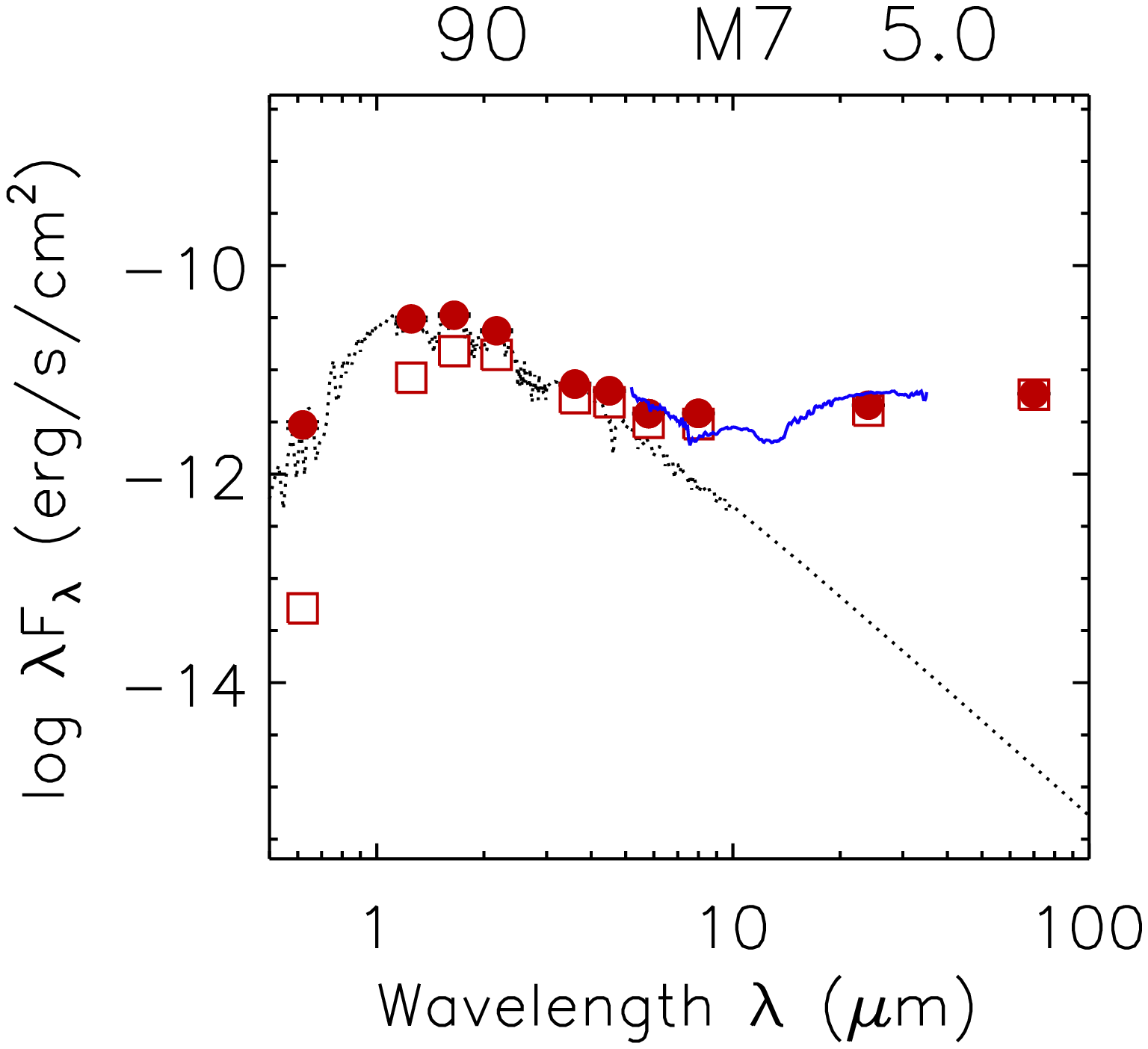}
\includegraphics[width=0.2\textwidth]{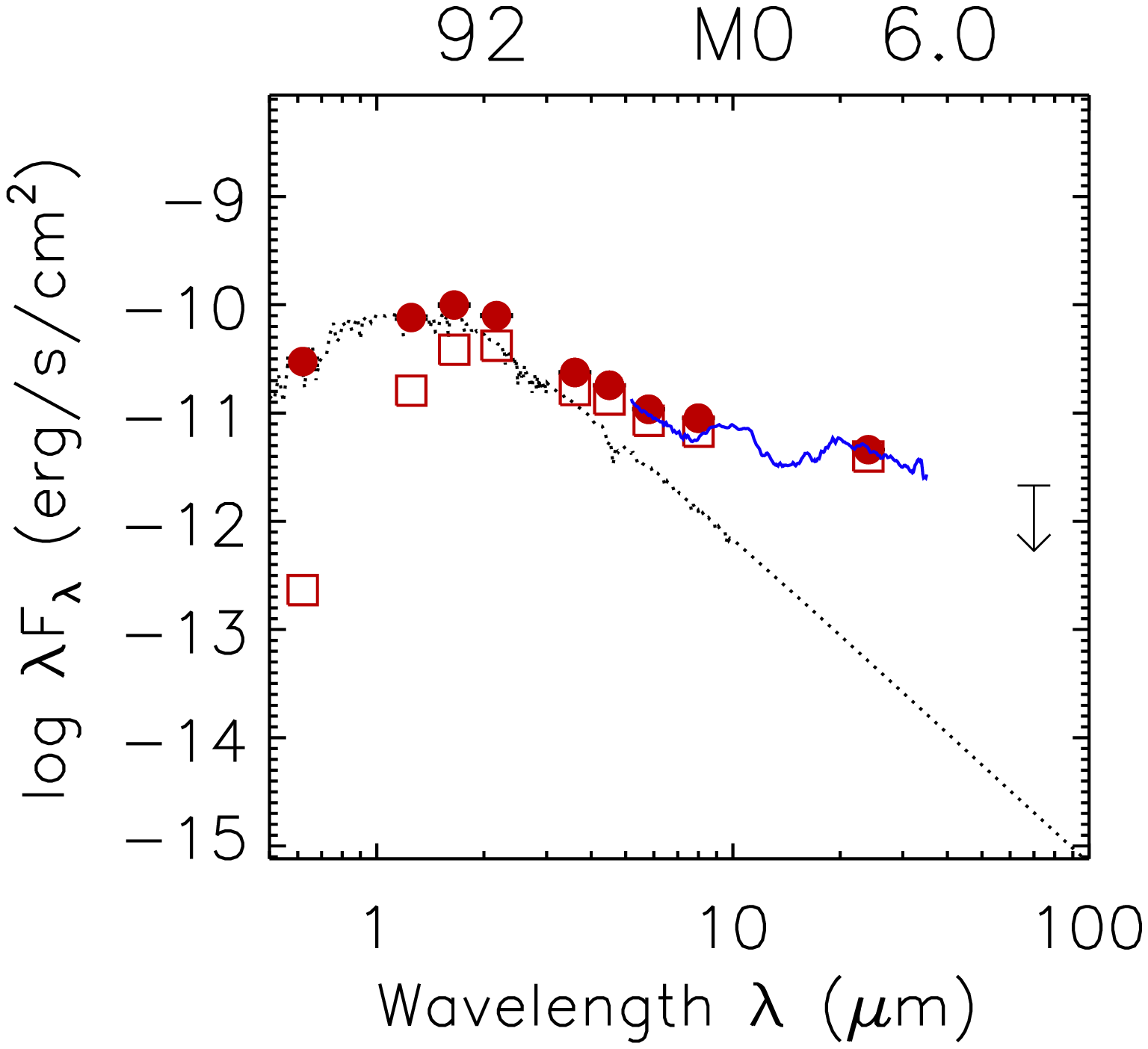}
\includegraphics[width=0.2\textwidth]{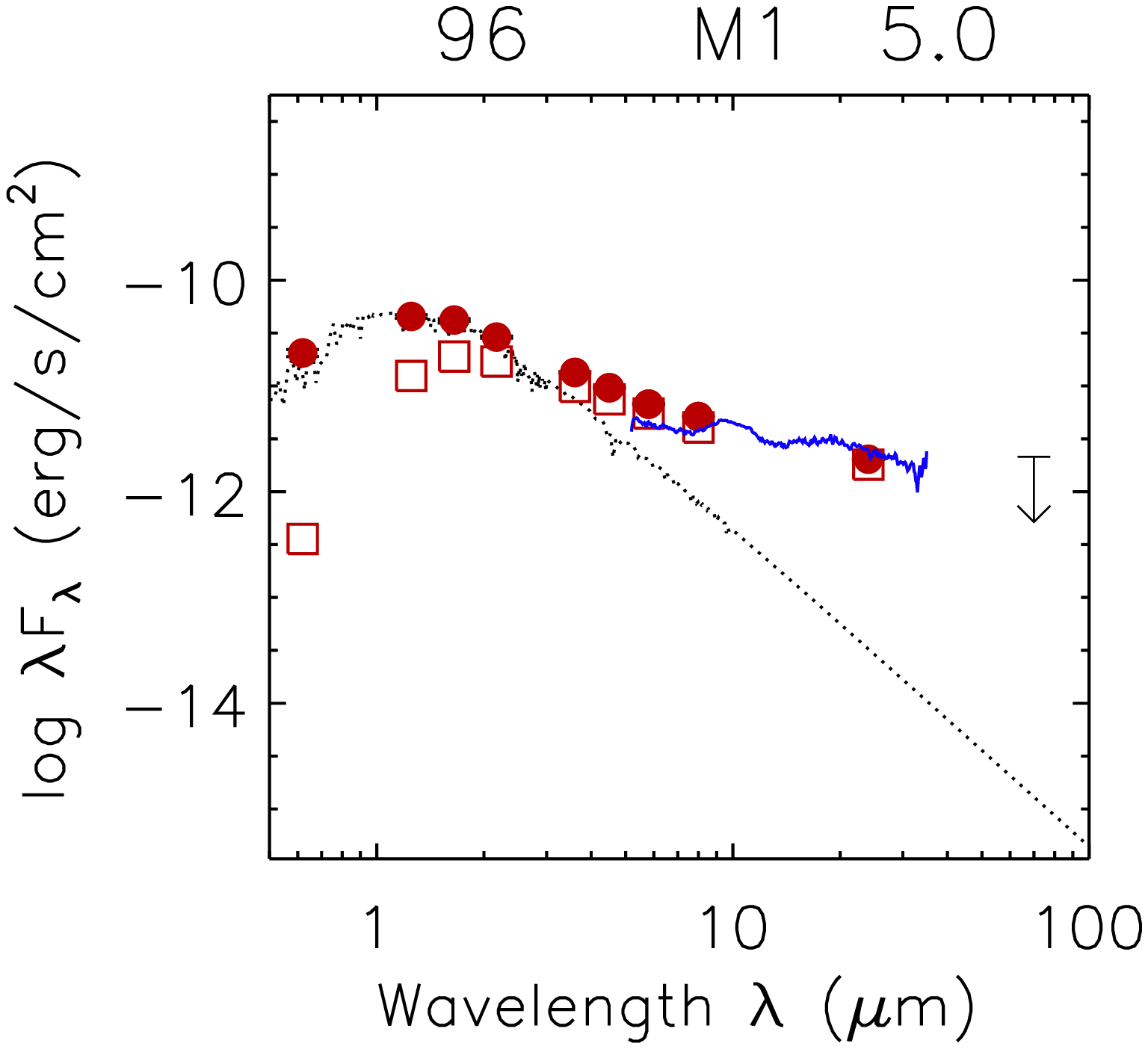}
\includegraphics[width=0.2\textwidth]{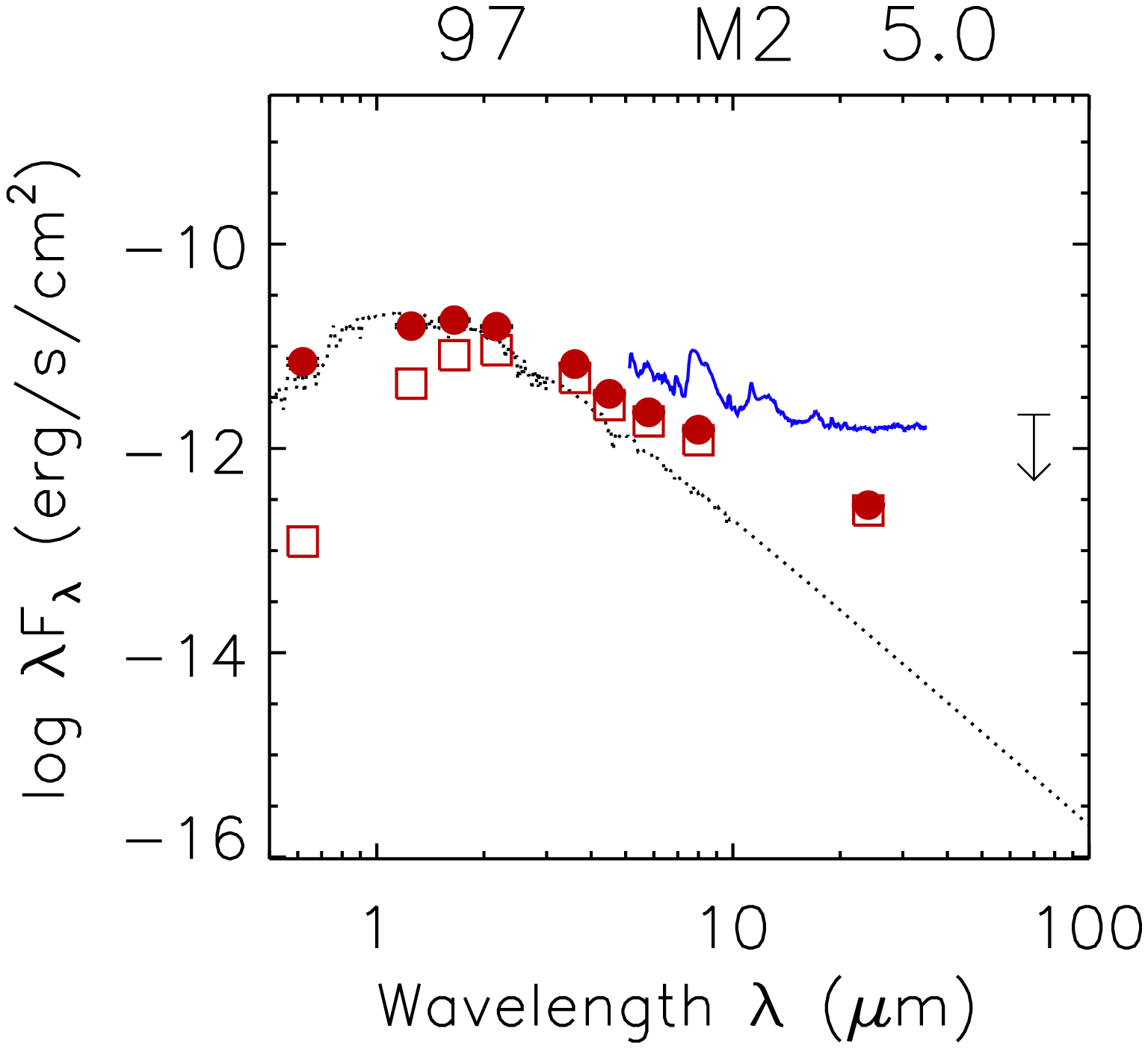}
\includegraphics[width=0.2\textwidth]{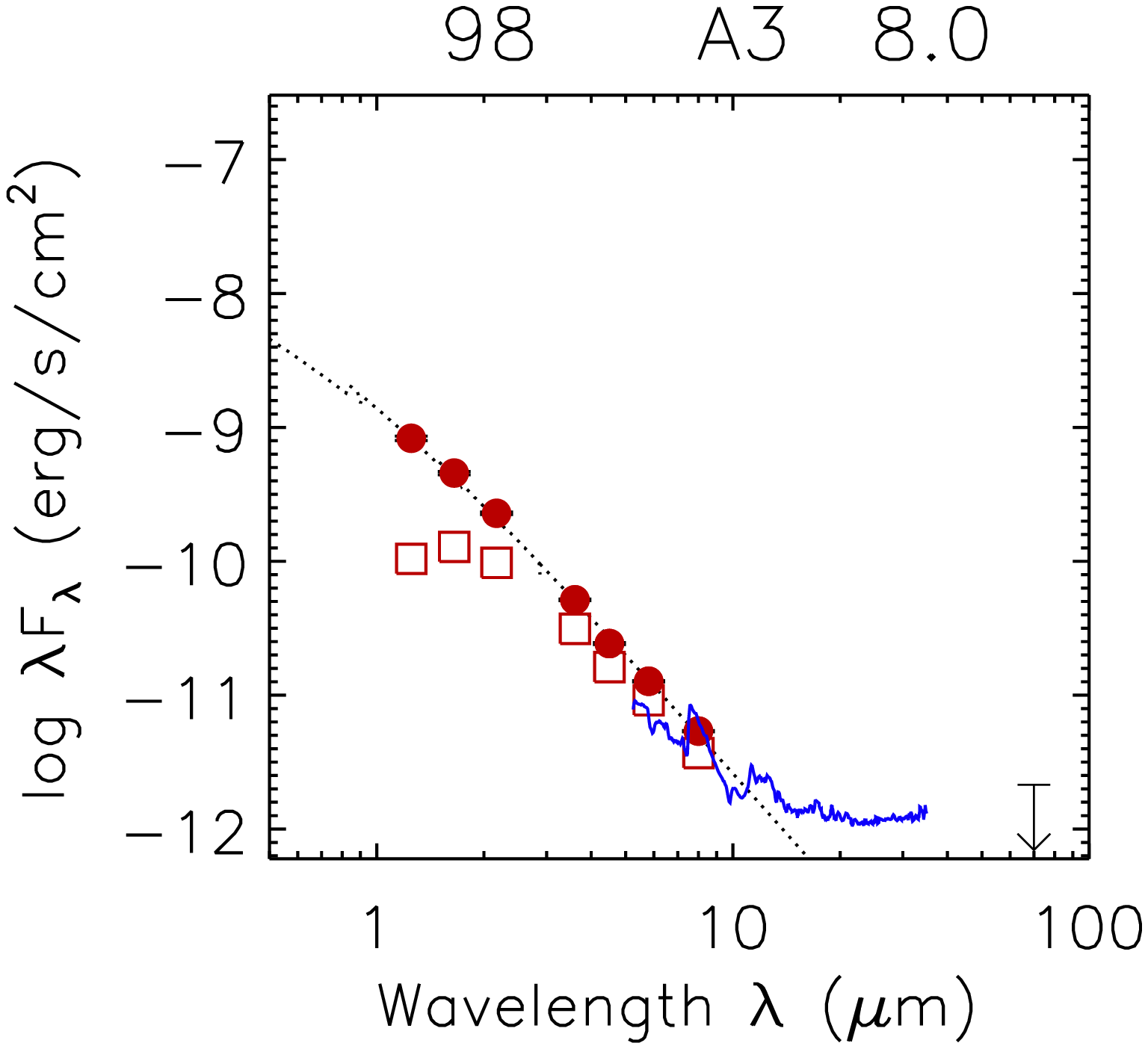}
\includegraphics[width=0.2\textwidth]{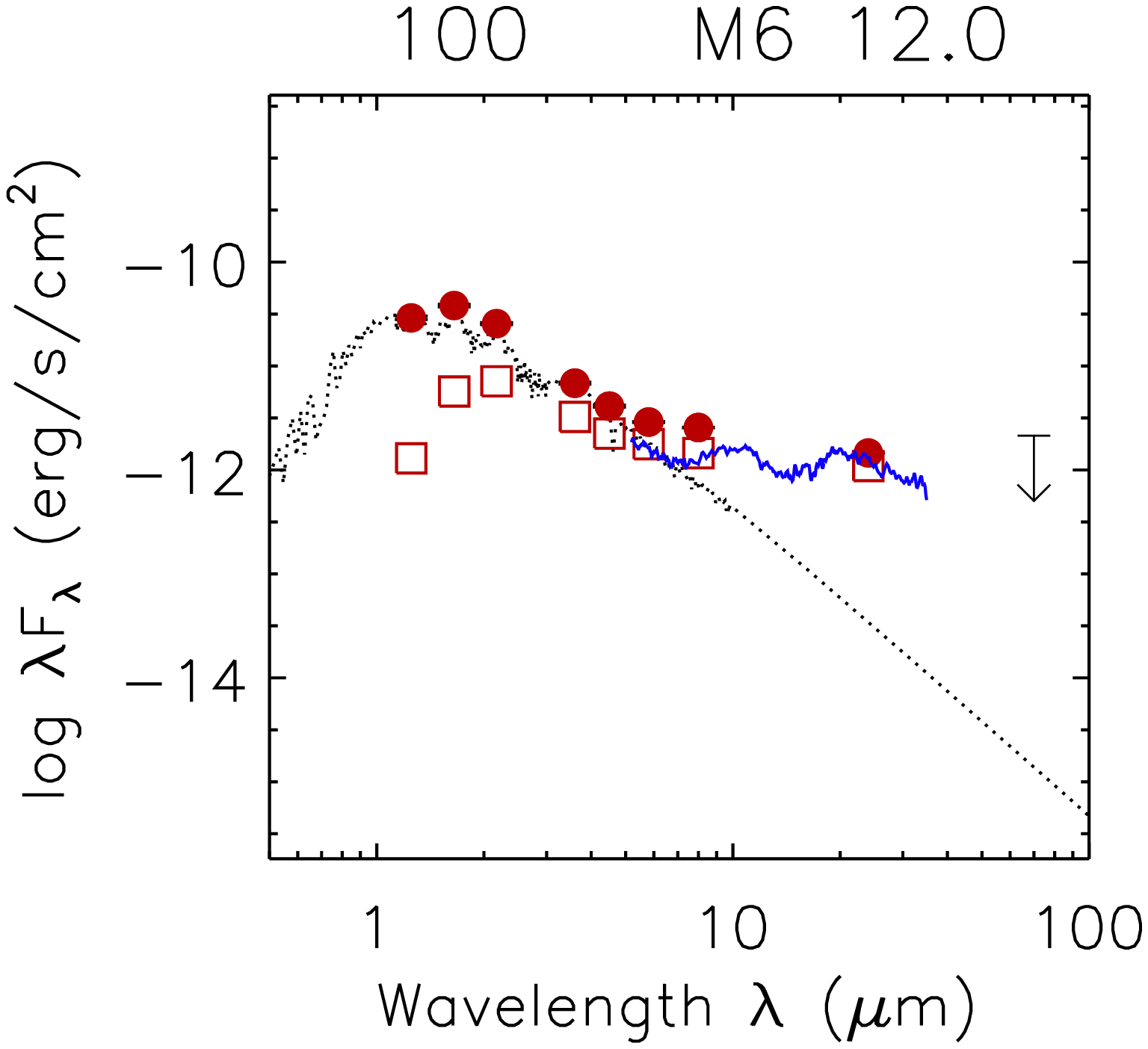}
\includegraphics[width=0.2\textwidth]{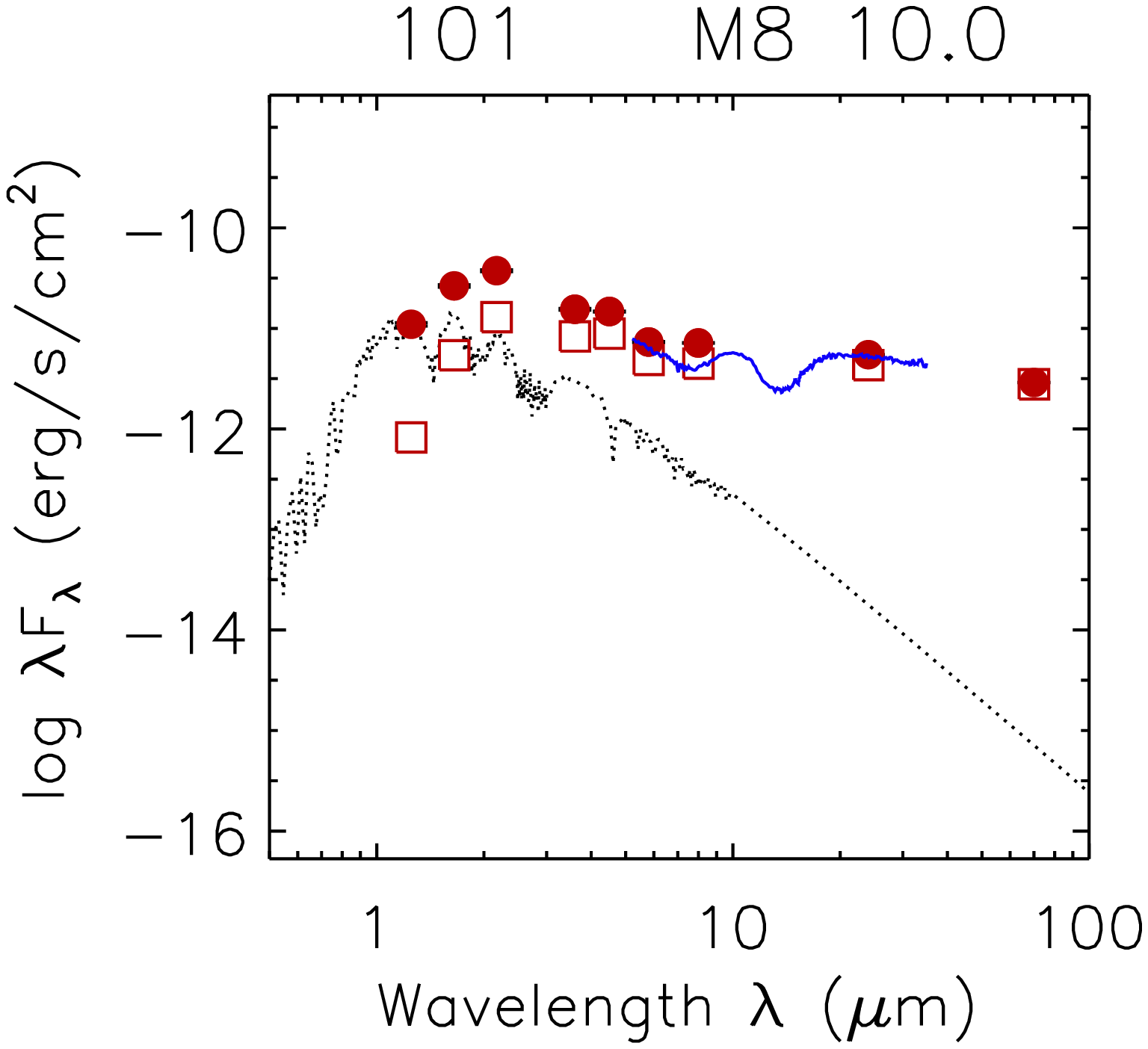}
\includegraphics[width=0.2\textwidth]{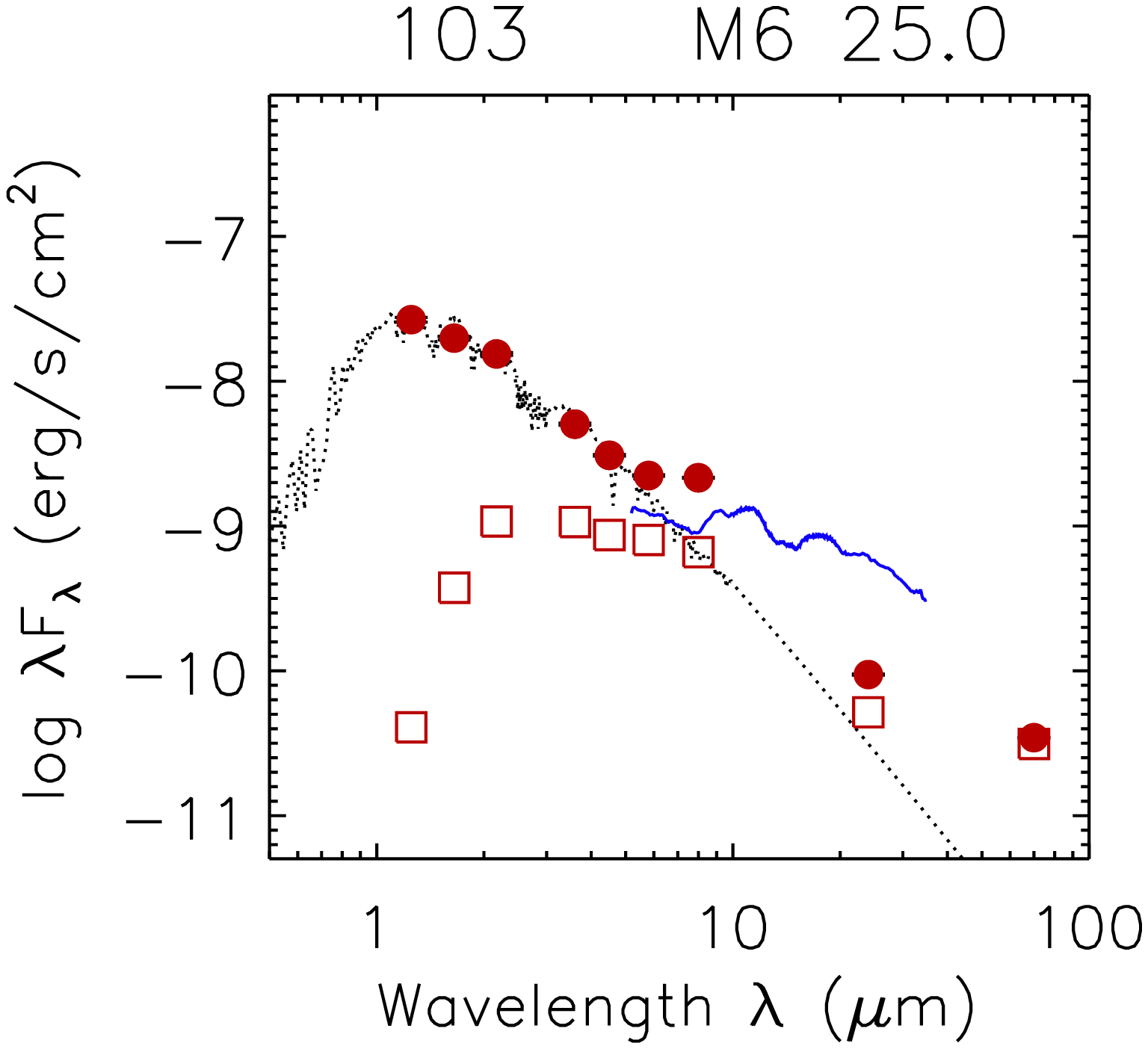}
\includegraphics[width=0.2\textwidth]{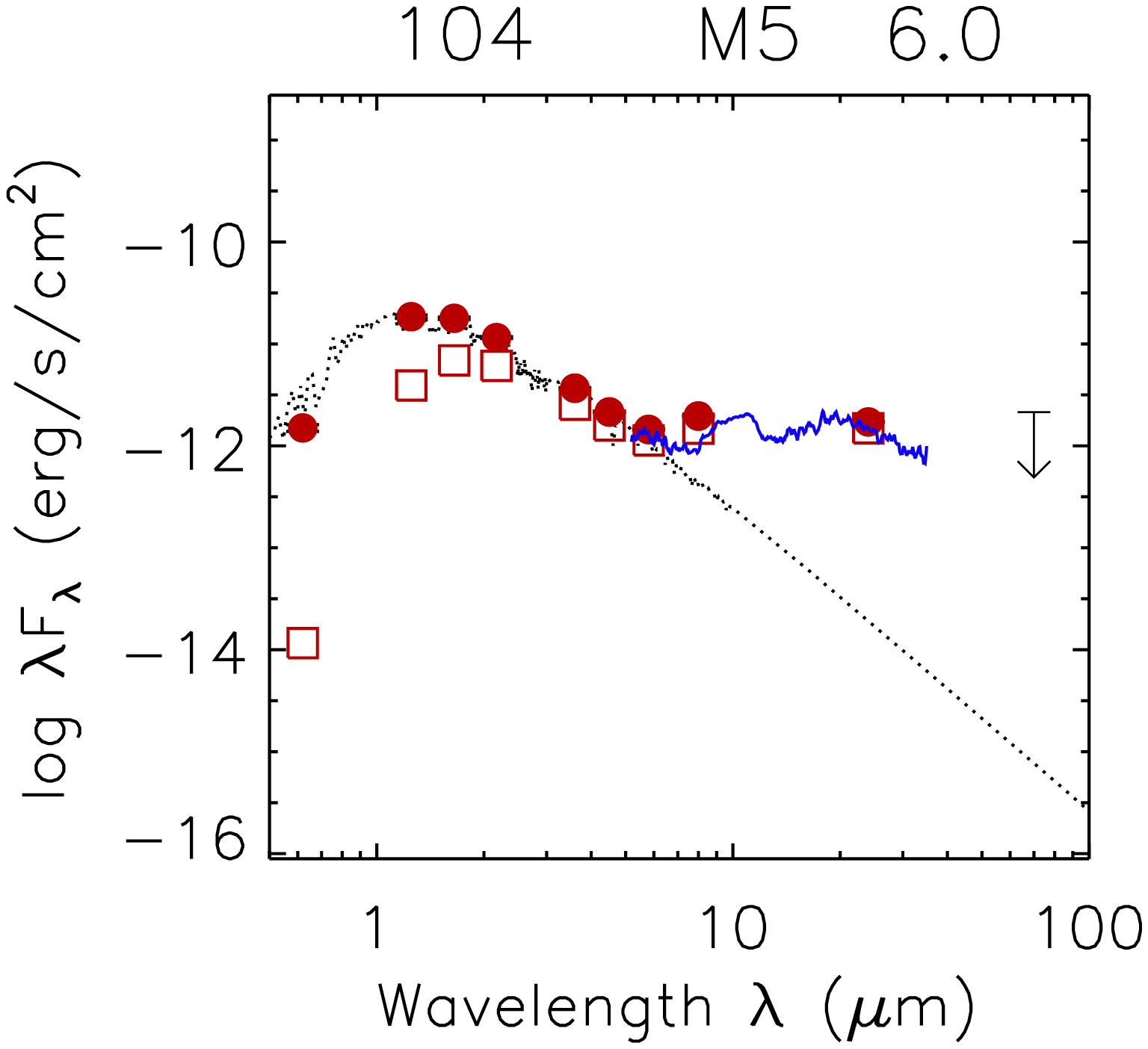}
\includegraphics[width=0.2\textwidth]{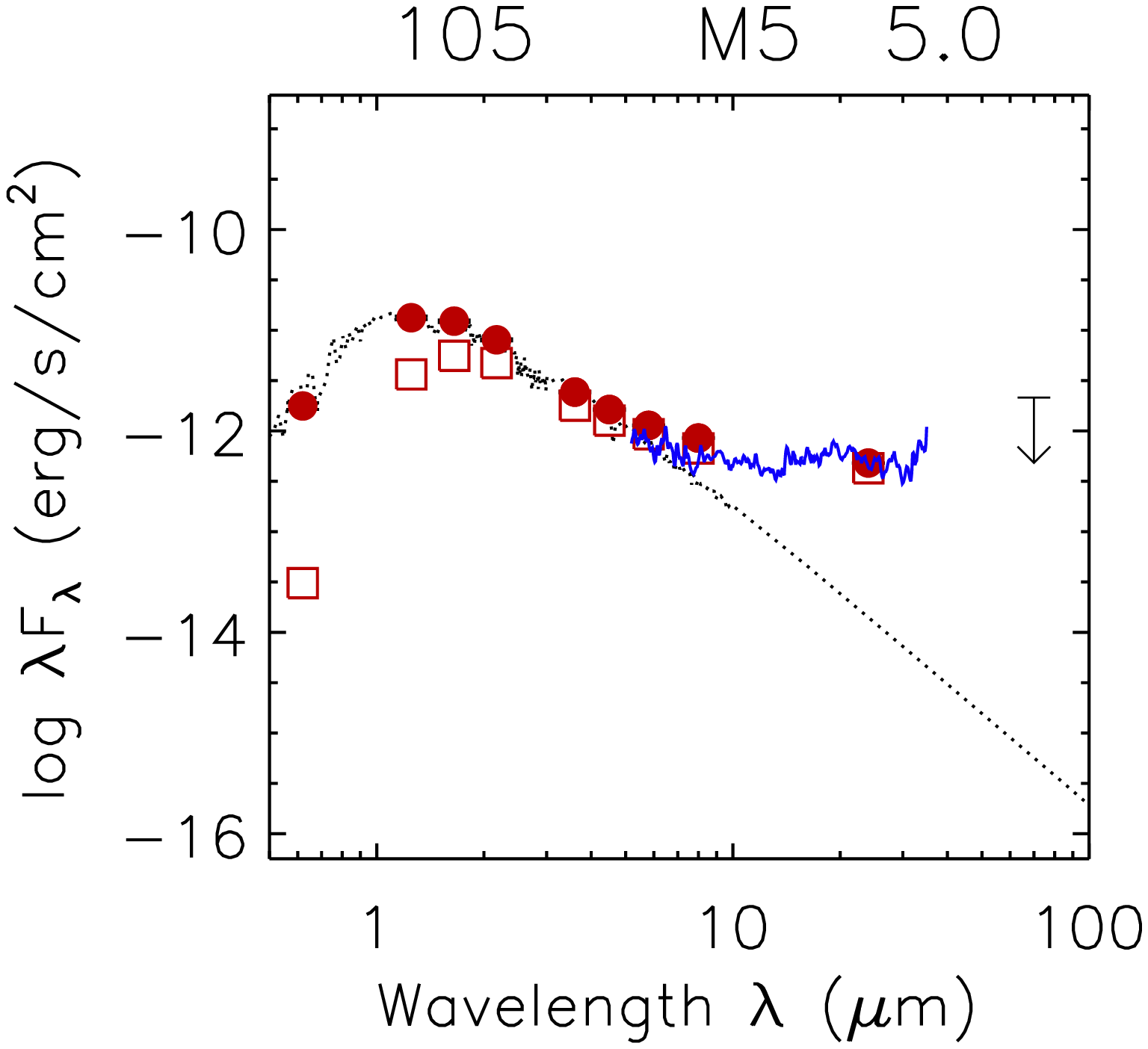}
\includegraphics[width=0.2\textwidth]{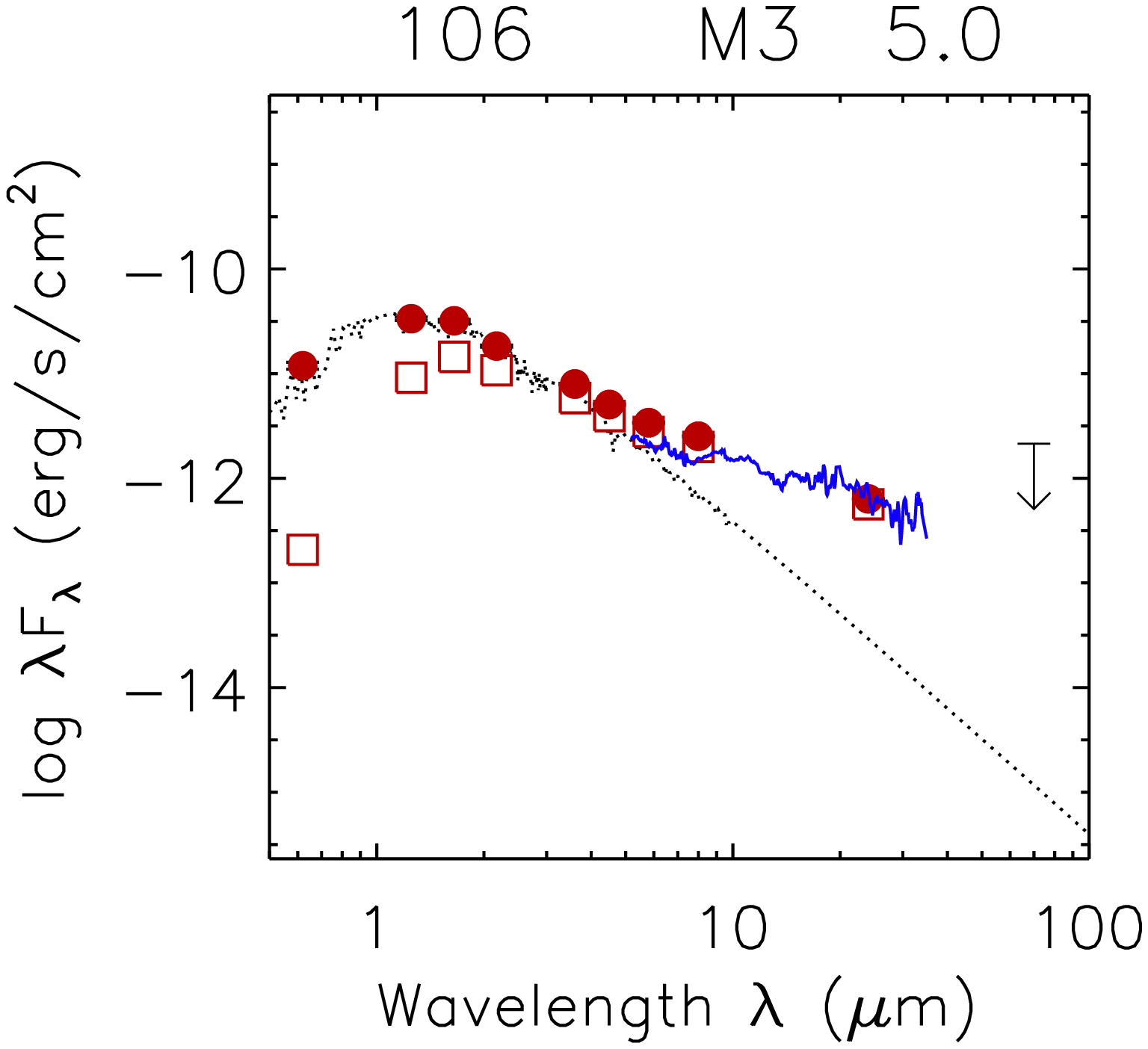}
\includegraphics[width=0.2\textwidth]{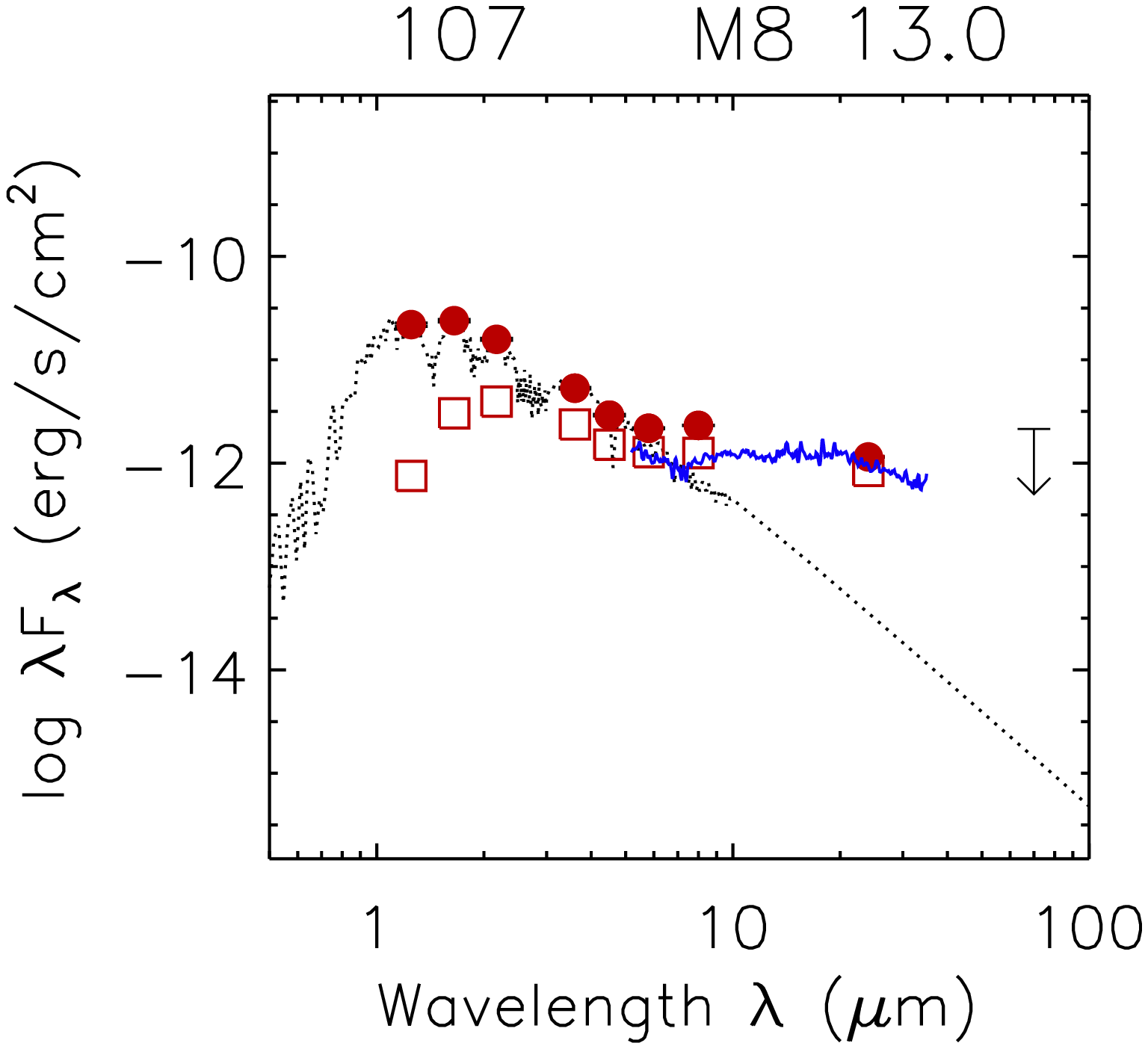}
\includegraphics[width=0.2\textwidth]{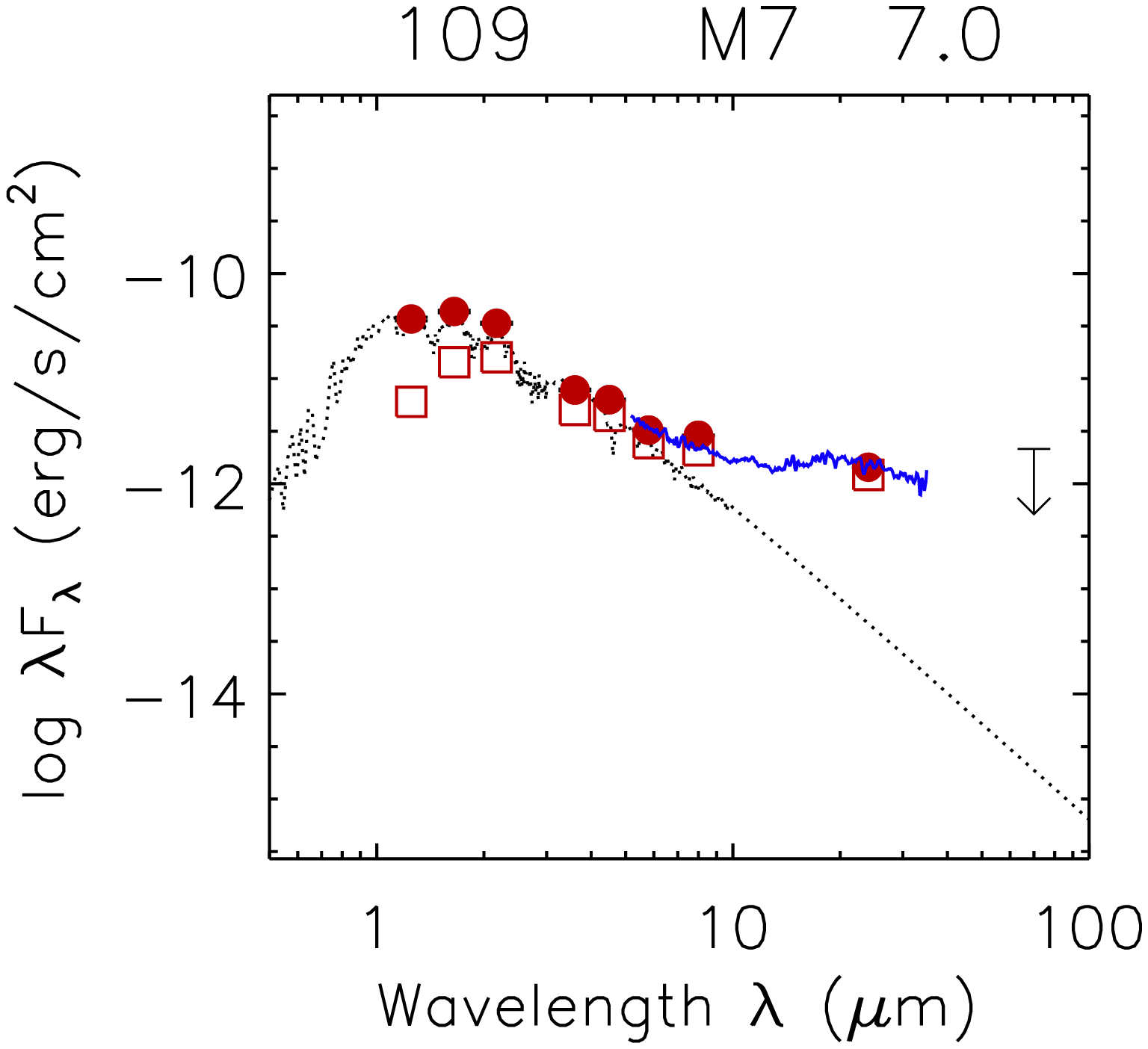}
\end{center}
\caption{\label{7f_ap_seds2} SEDs, continued.  }
\end{figure*}

\newpage

\begin{figure*}[!h]
\begin{center}
\includegraphics[width=0.2\textwidth]{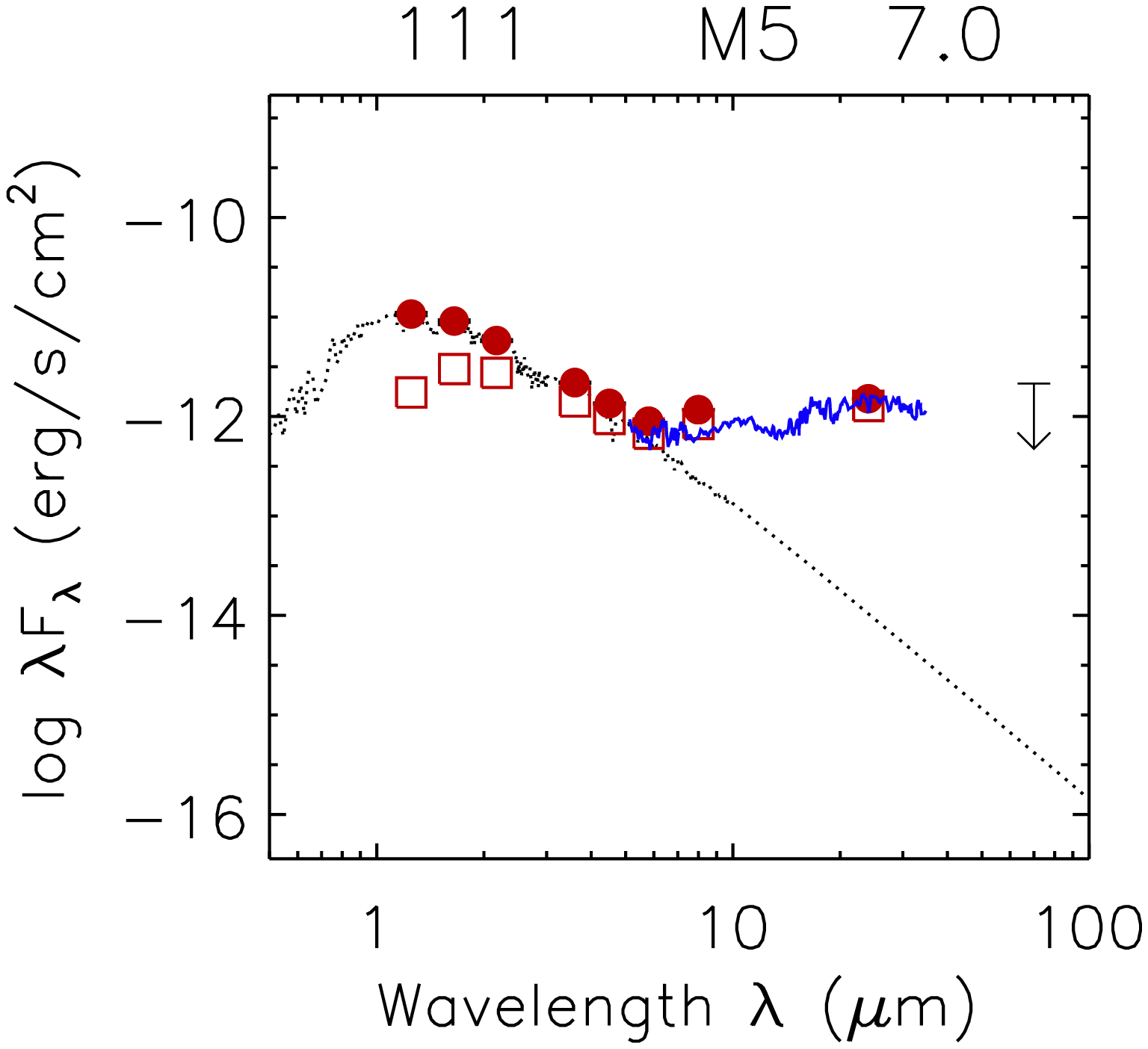}
\includegraphics[width=0.2\textwidth]{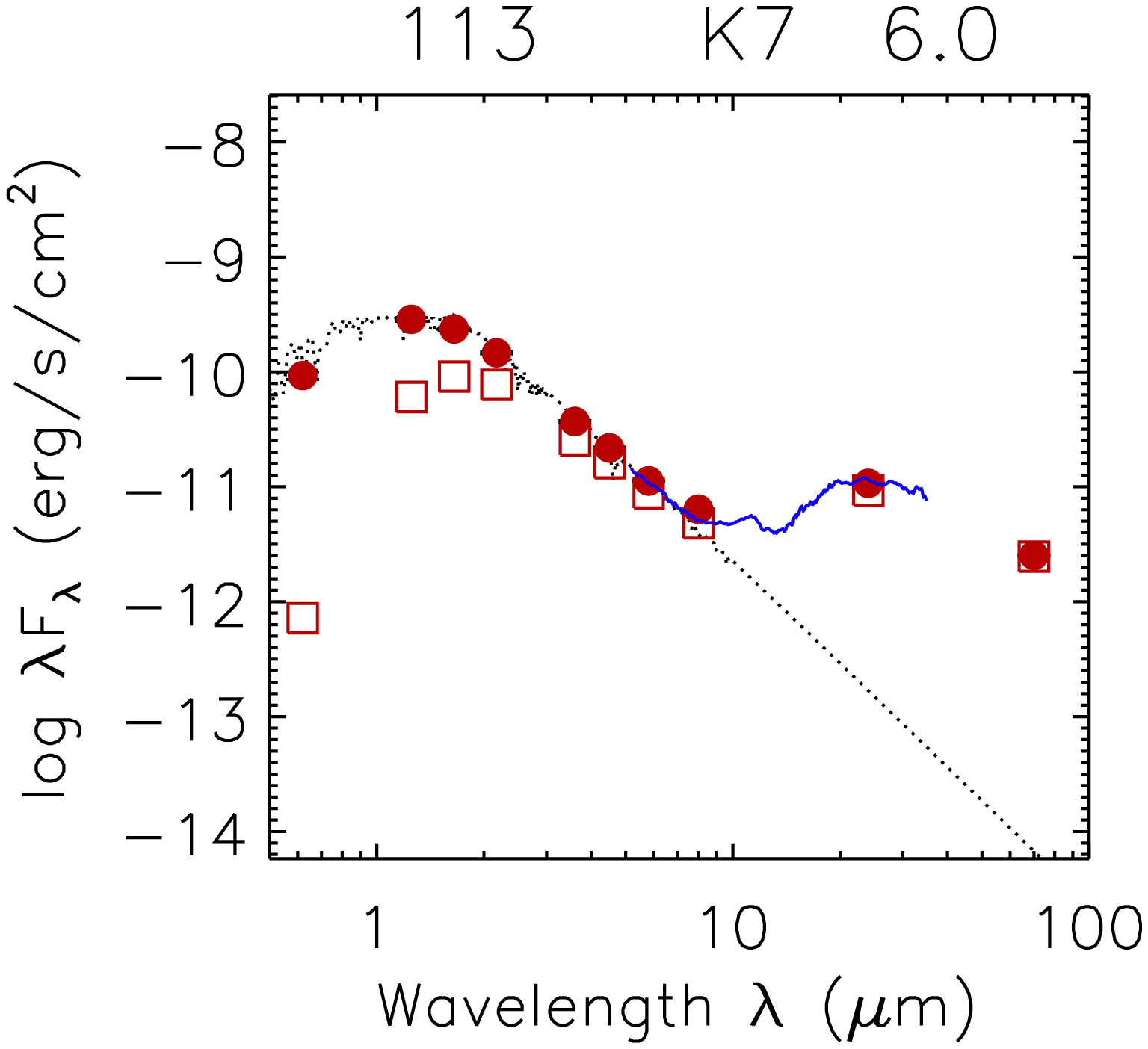}
\includegraphics[width=0.2\textwidth]{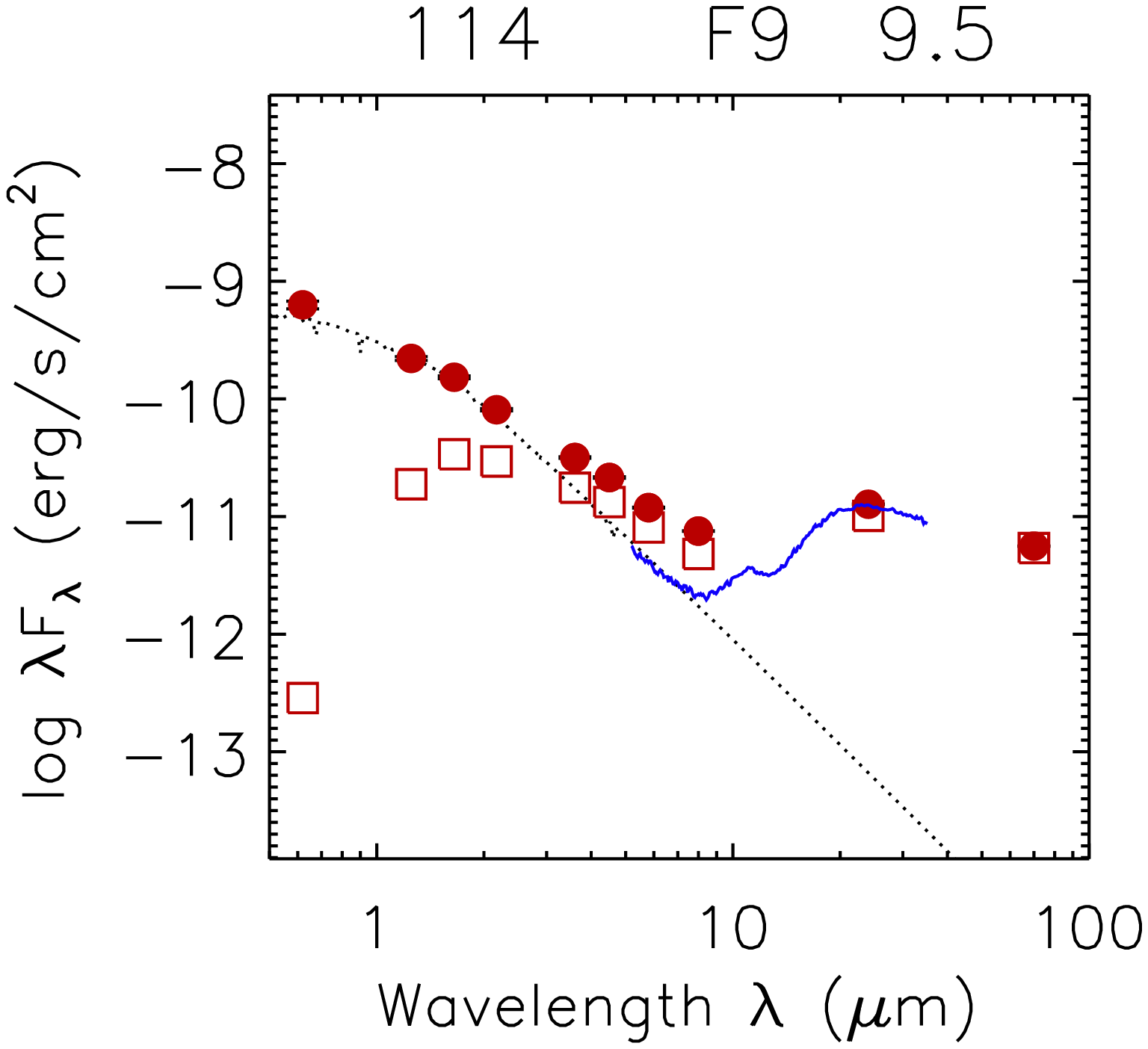}
\includegraphics[width=0.2\textwidth]{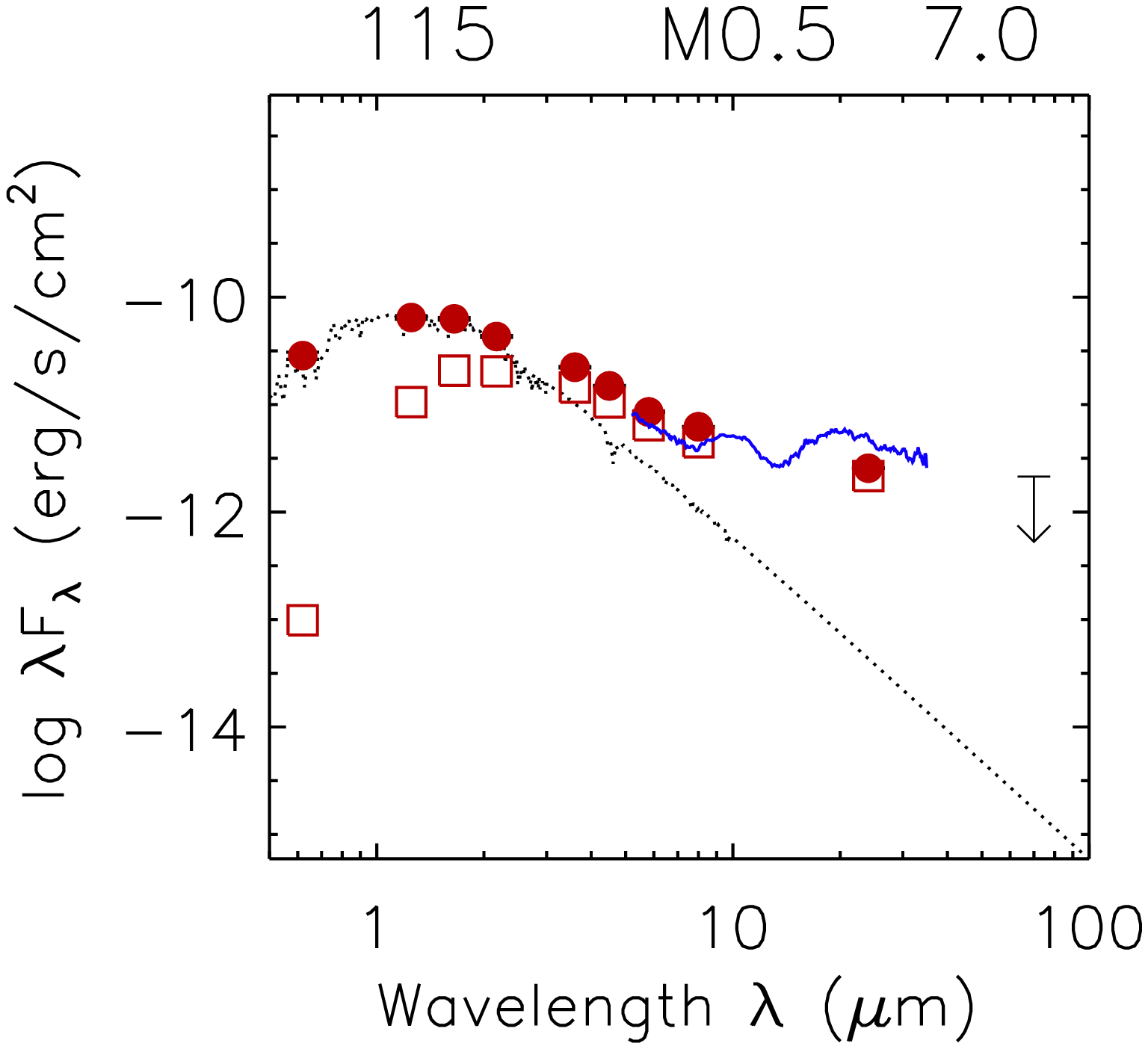}\\
\includegraphics[width=0.2\textwidth]{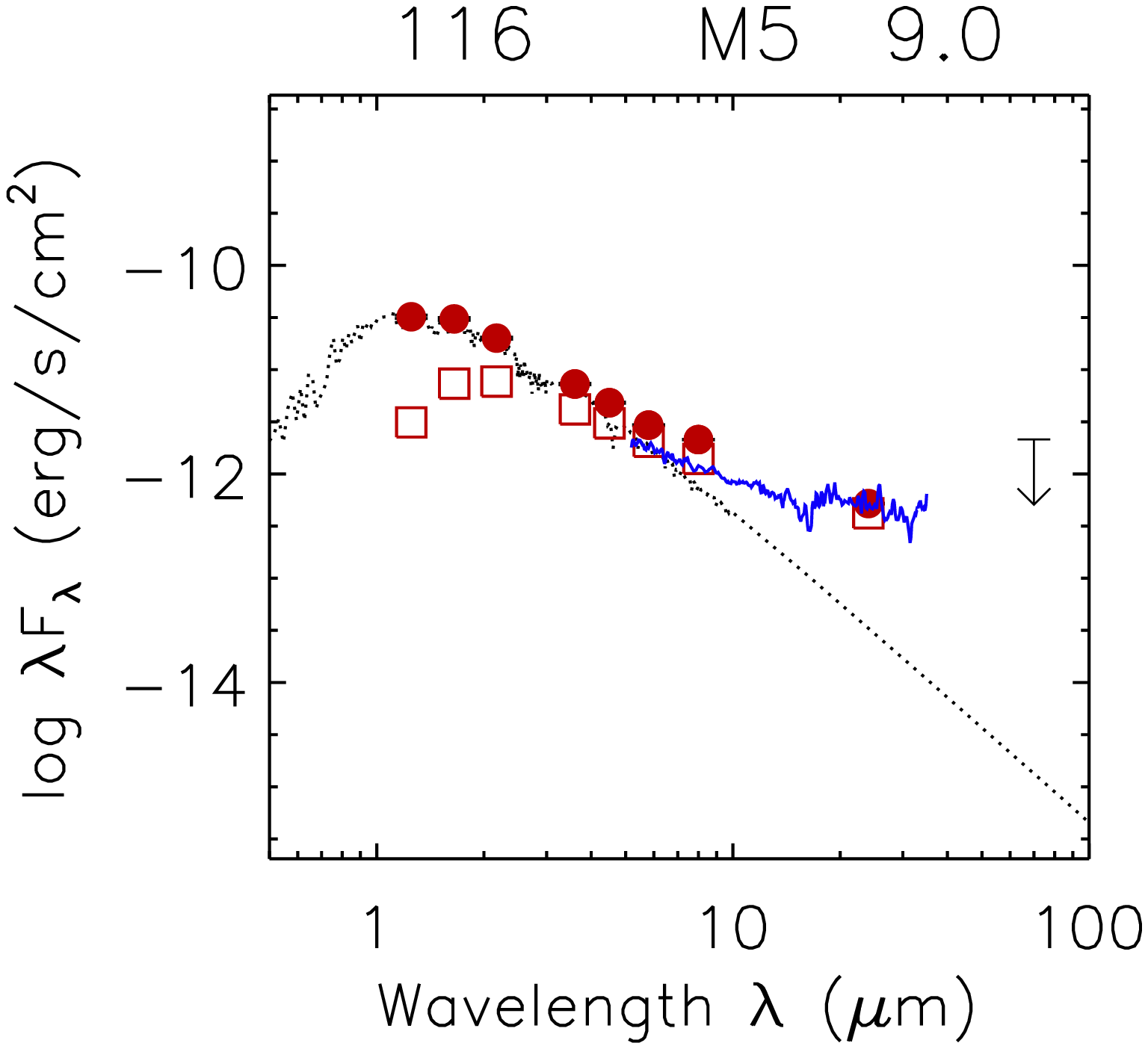}
\includegraphics[width=0.2\textwidth]{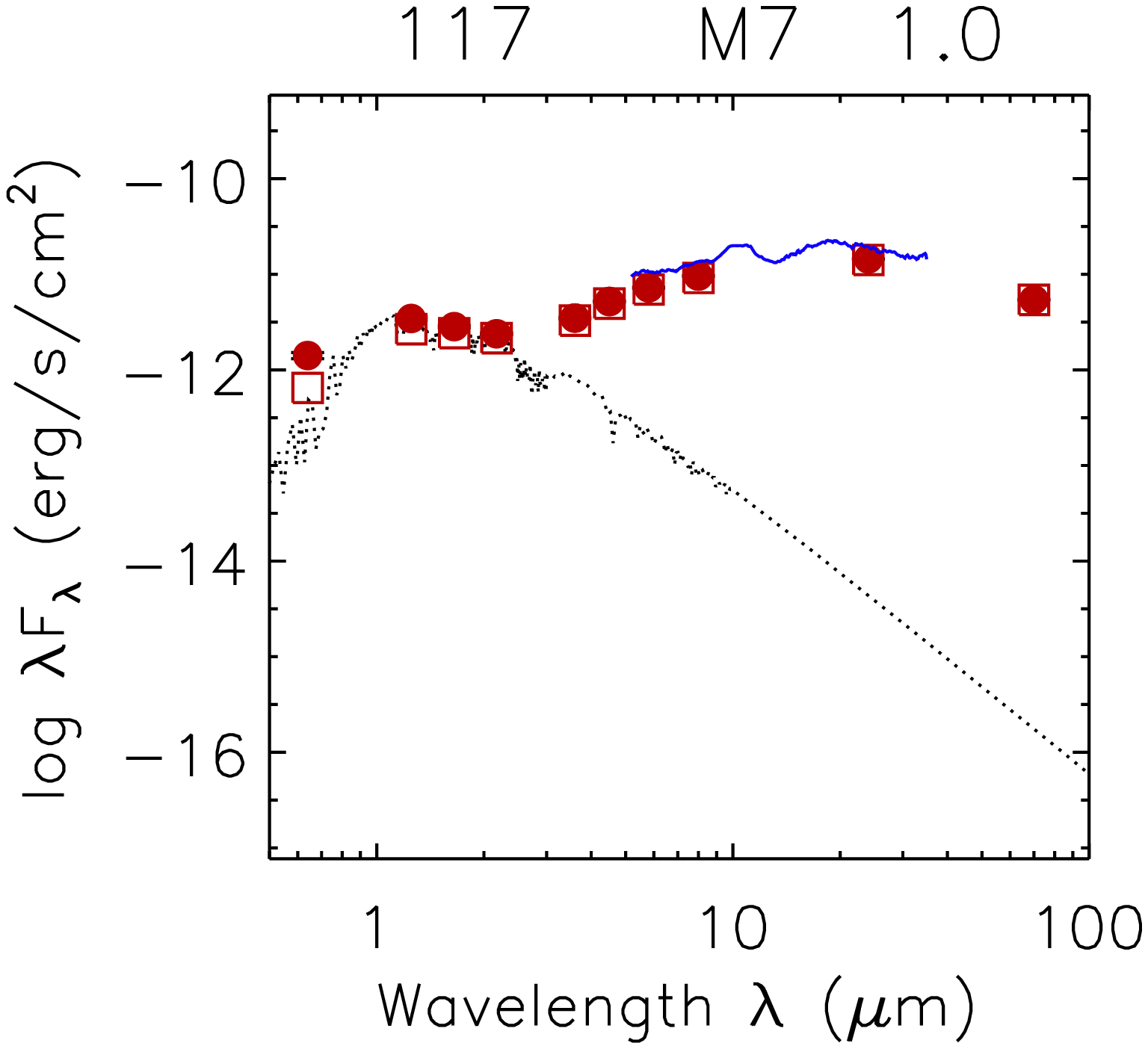}
\includegraphics[width=0.2\textwidth]{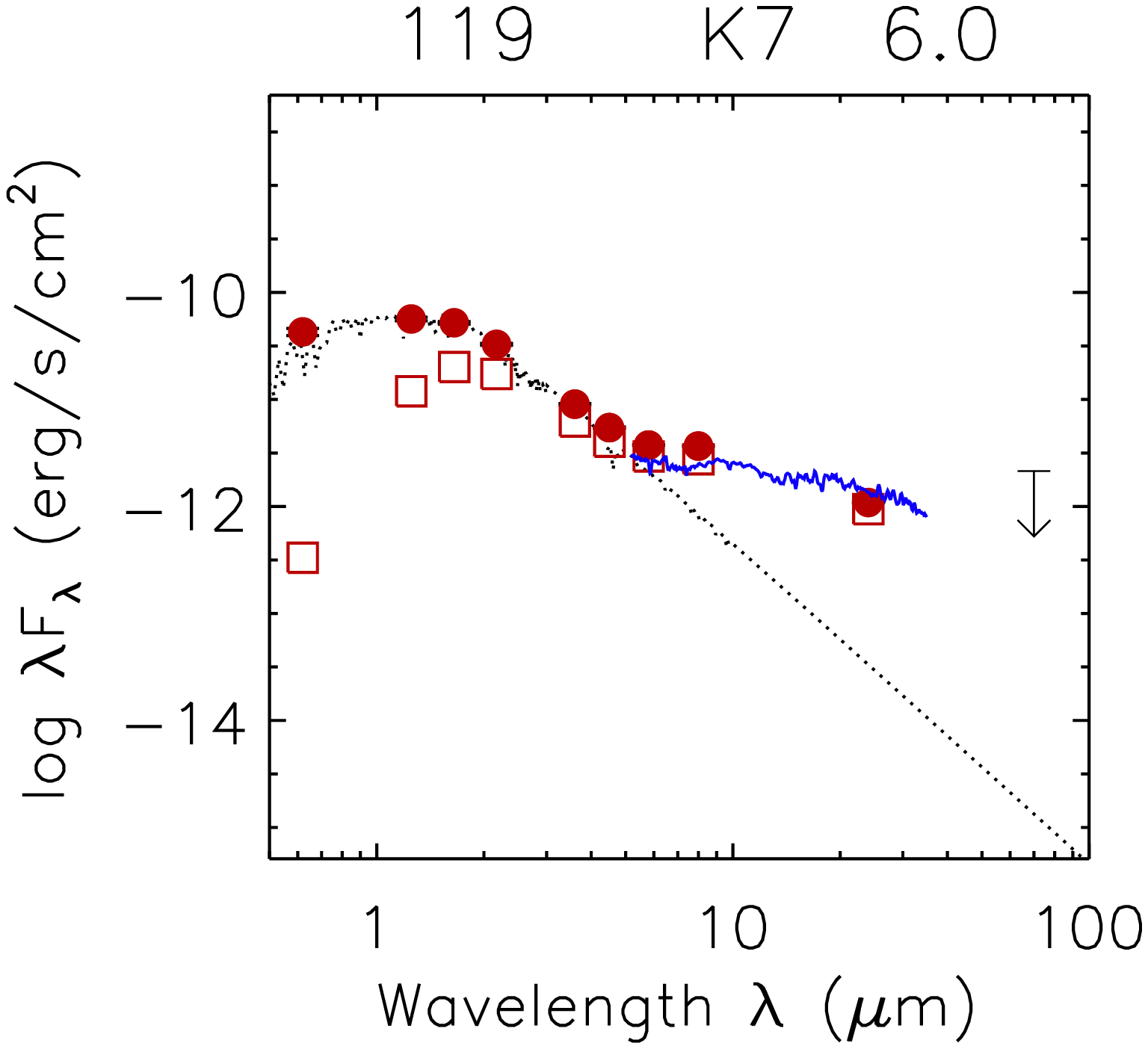}
\includegraphics[width=0.2\textwidth]{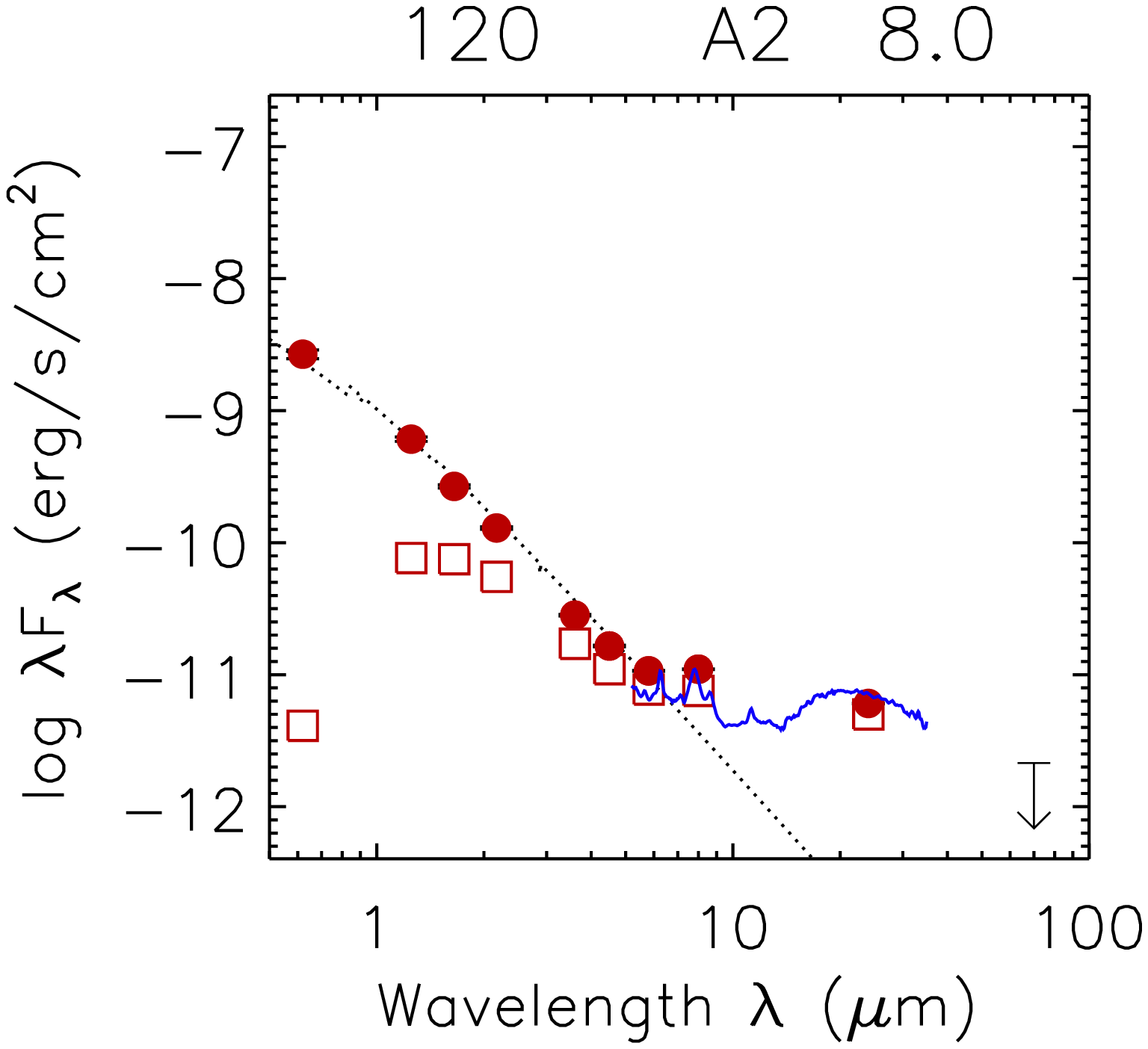}\\
\includegraphics[width=0.2\textwidth]{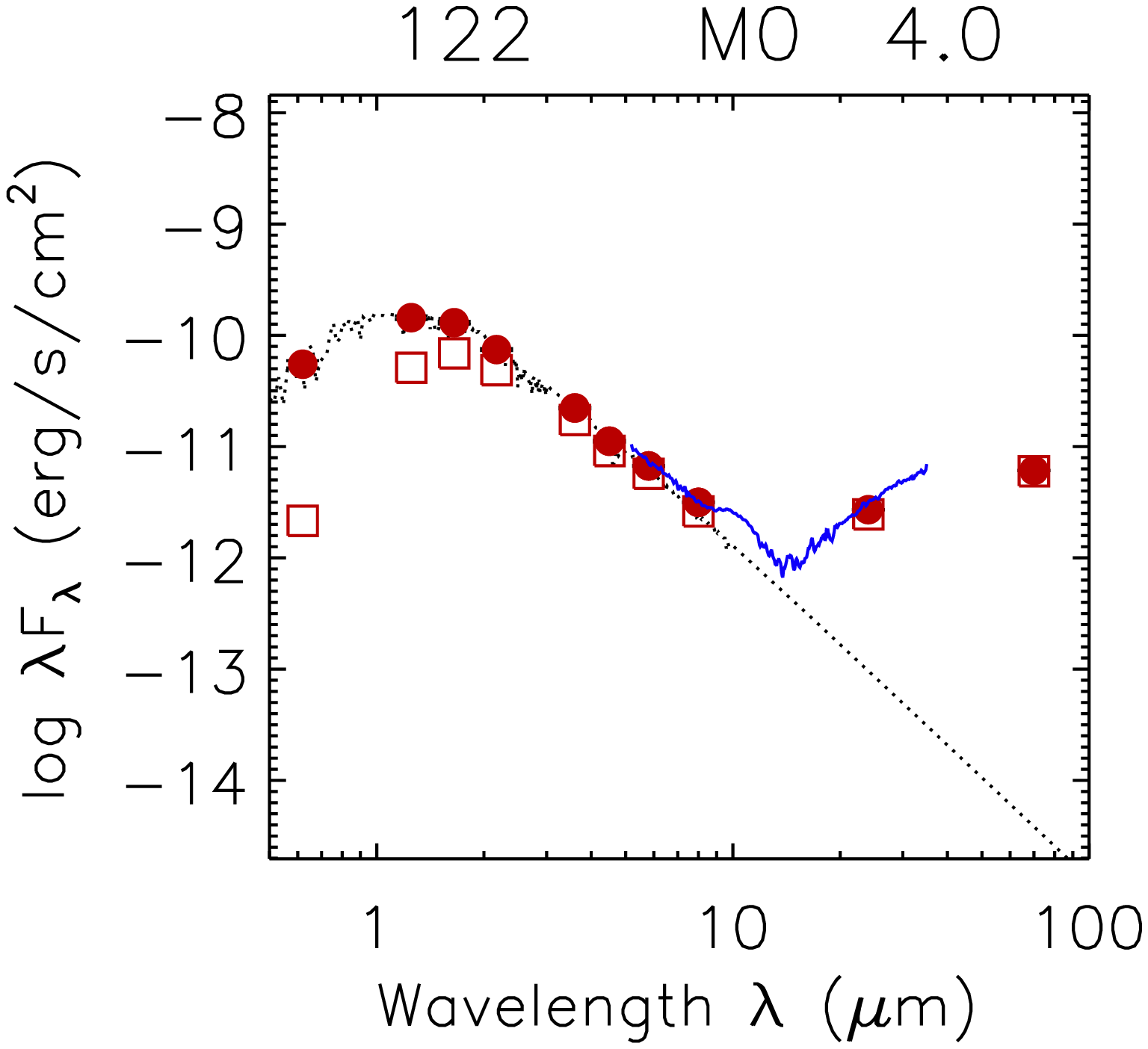}
\includegraphics[width=0.2\textwidth]{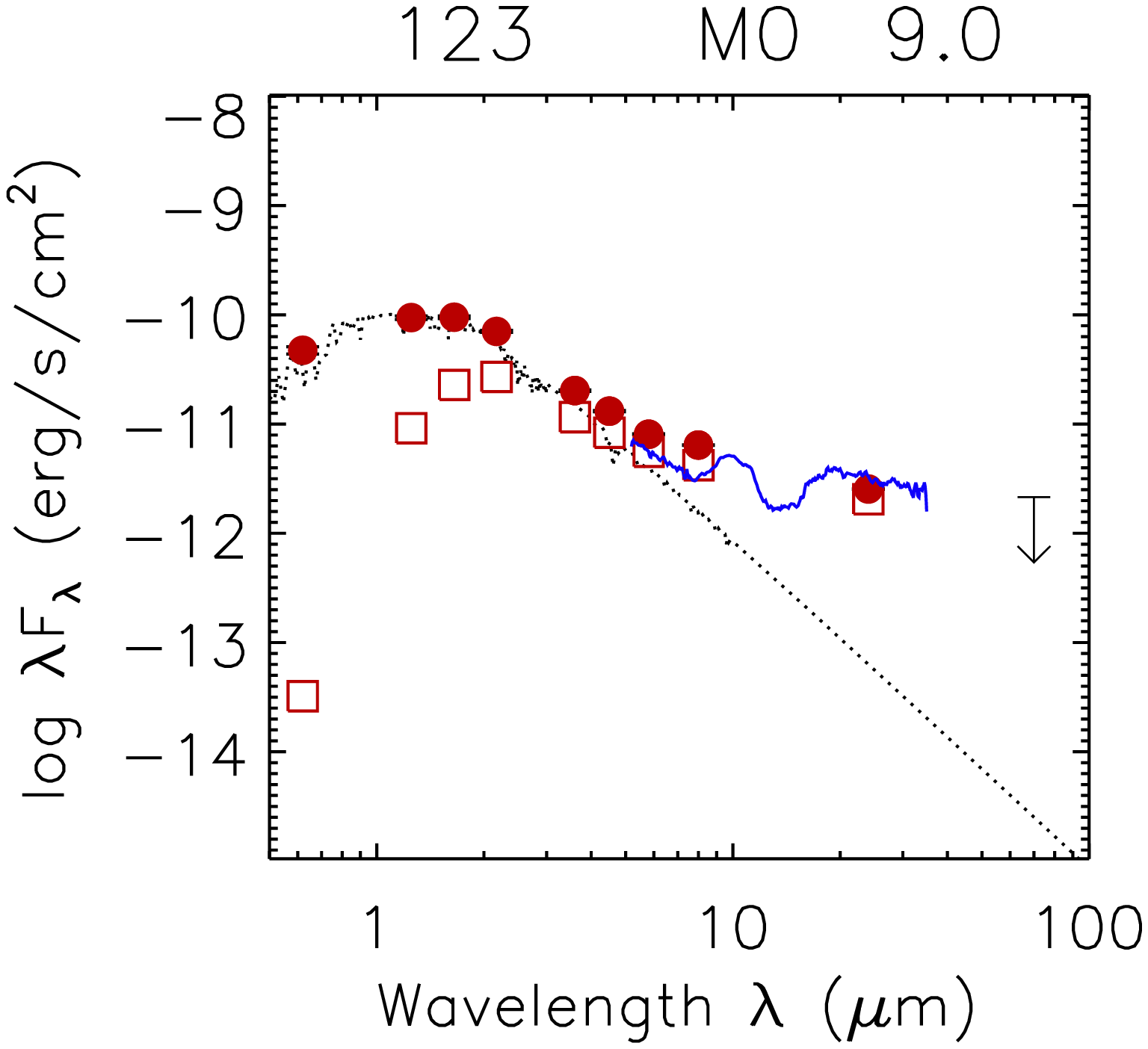}
\includegraphics[width=0.2\textwidth]{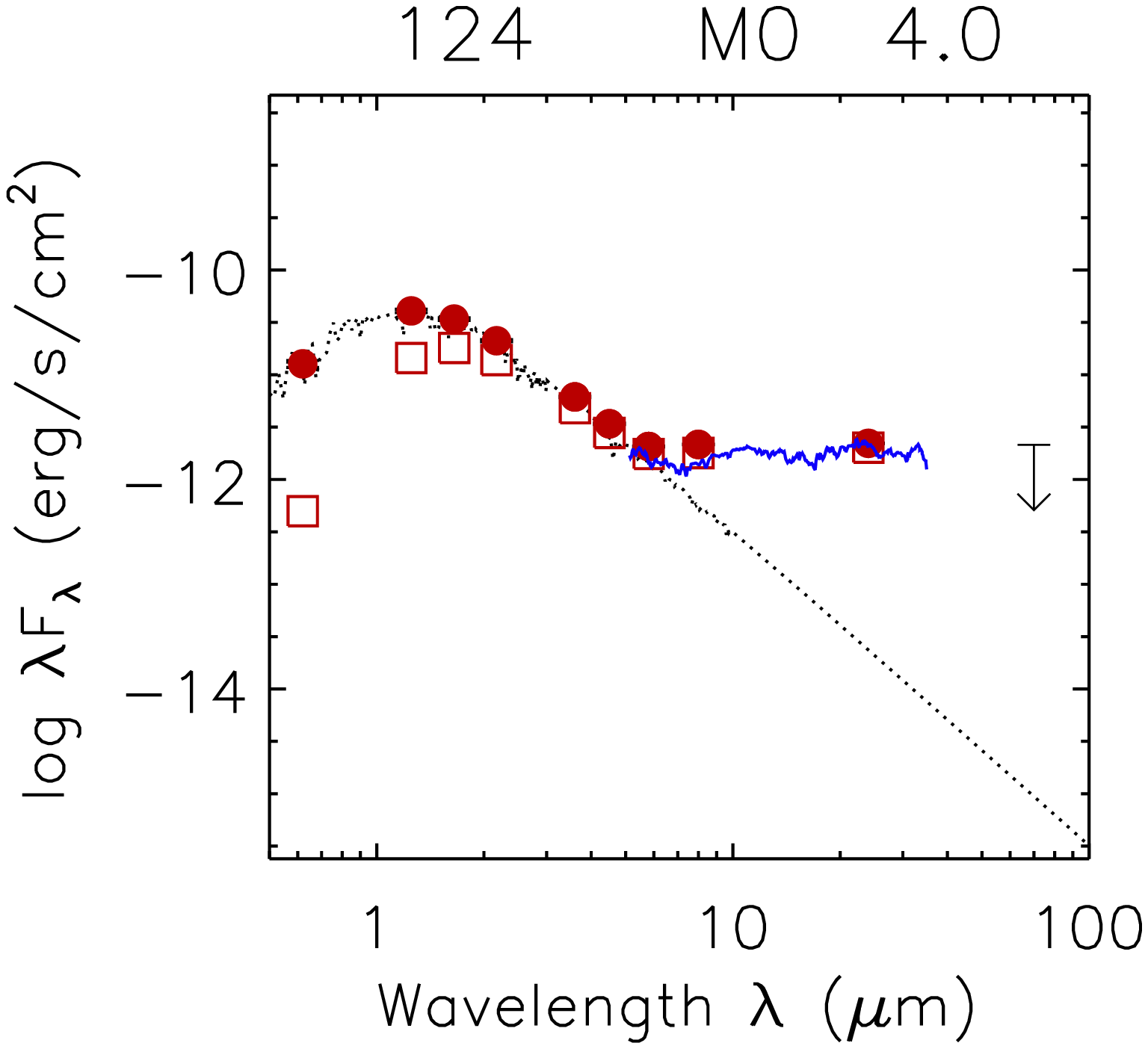}
\includegraphics[width=0.2\textwidth]{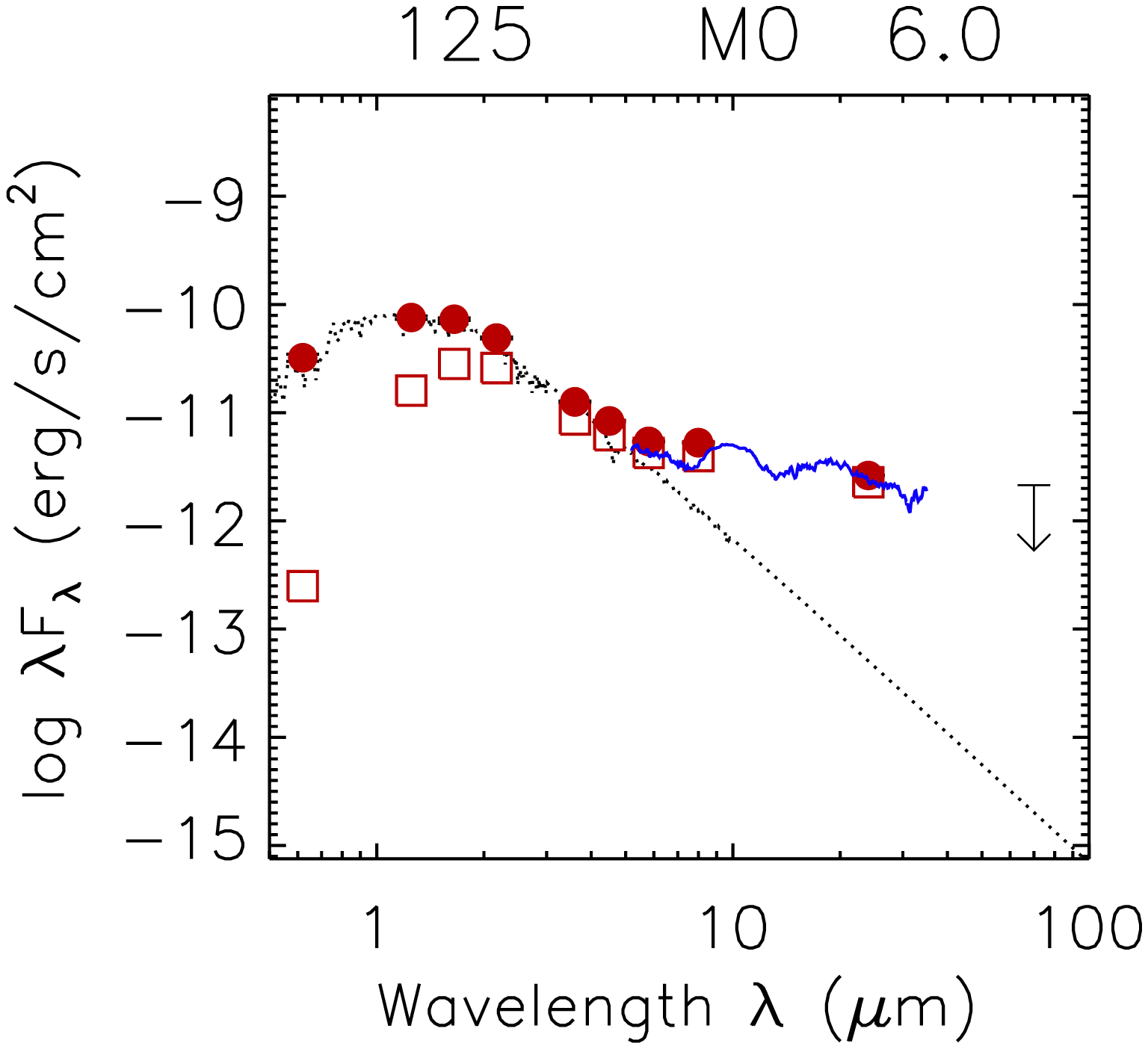}\\
\includegraphics[width=0.2\textwidth]{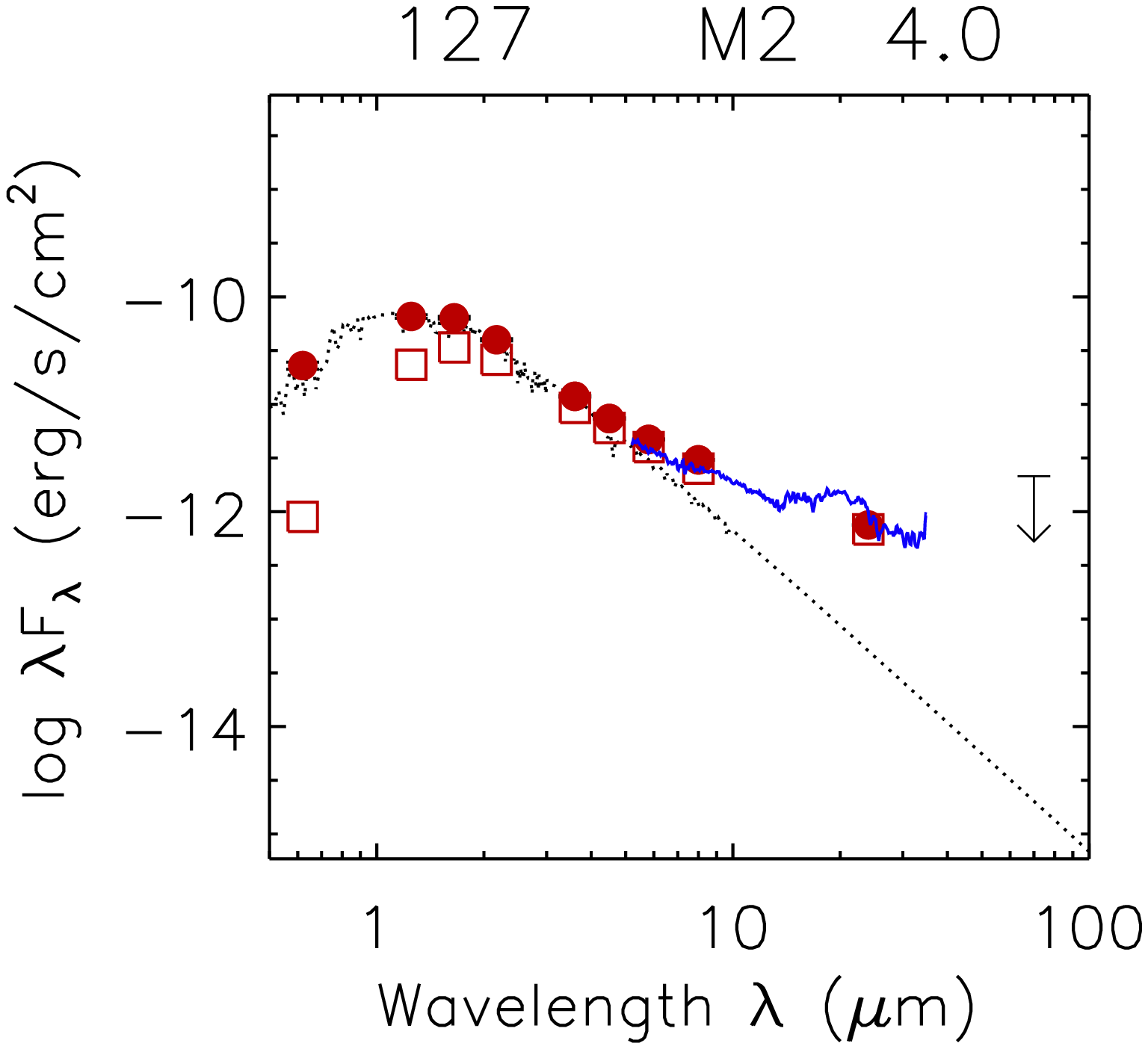}
\includegraphics[width=0.2\textwidth]{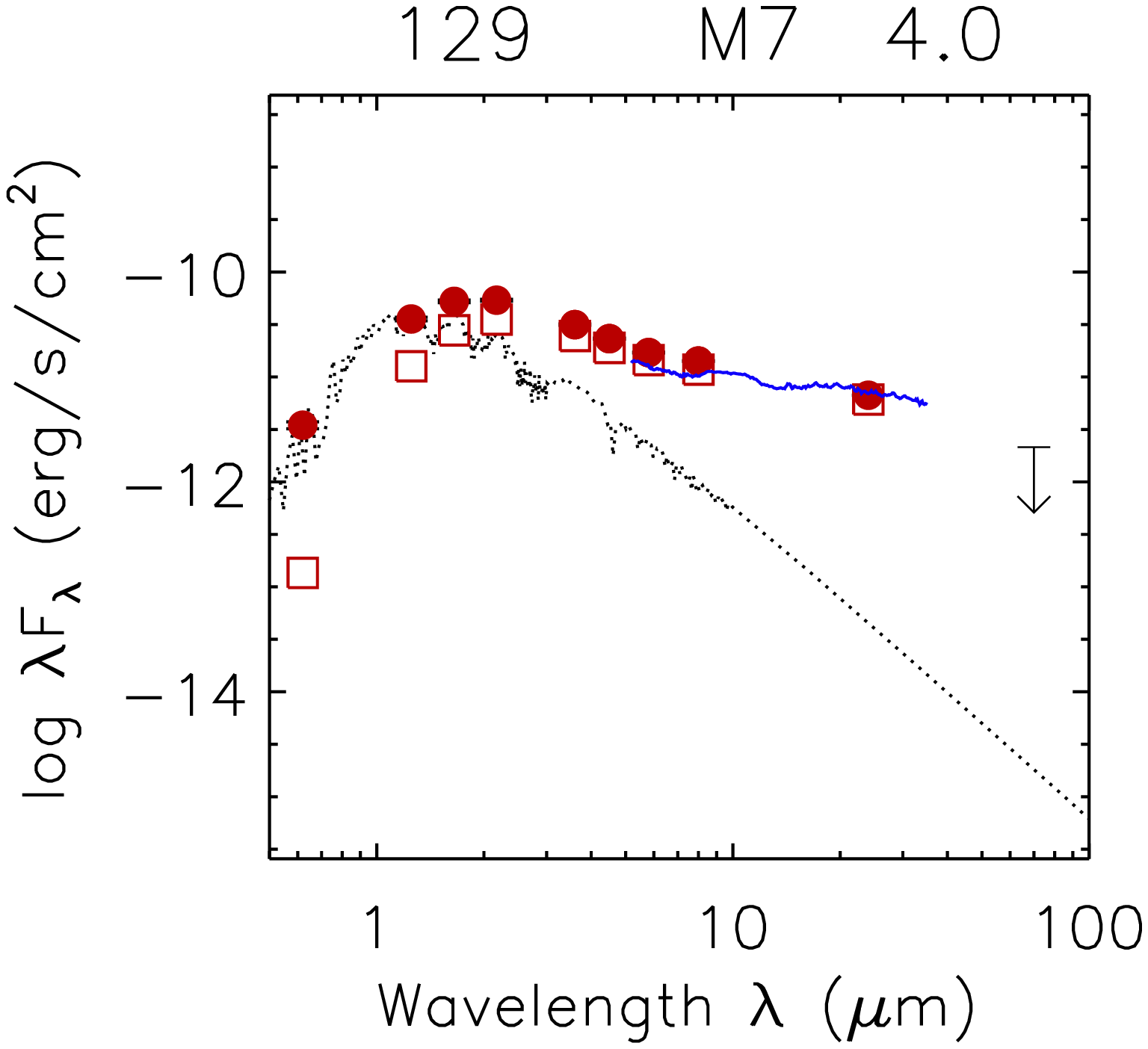}
\includegraphics[width=0.2\textwidth]{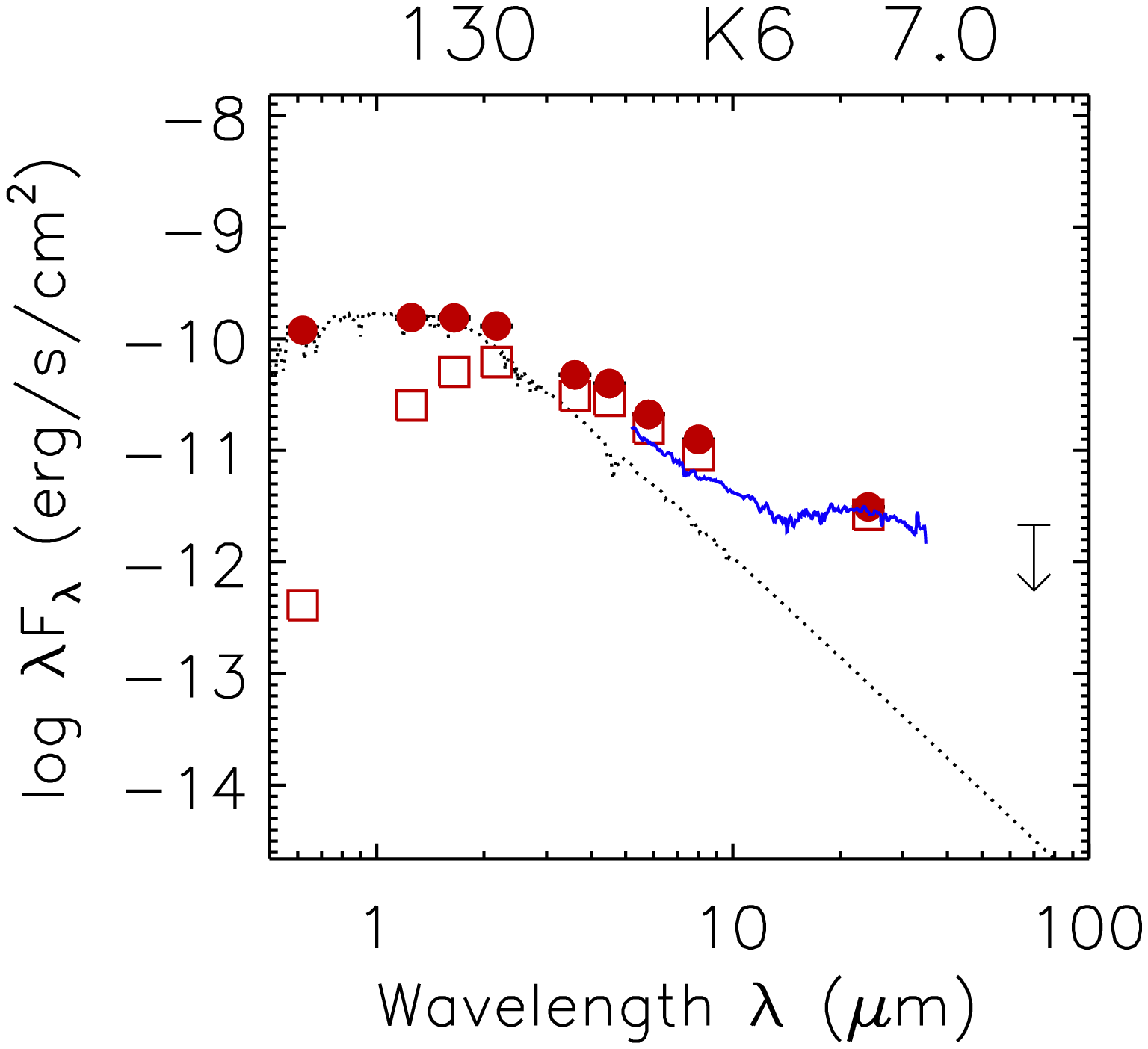}
\includegraphics[width=0.2\textwidth]{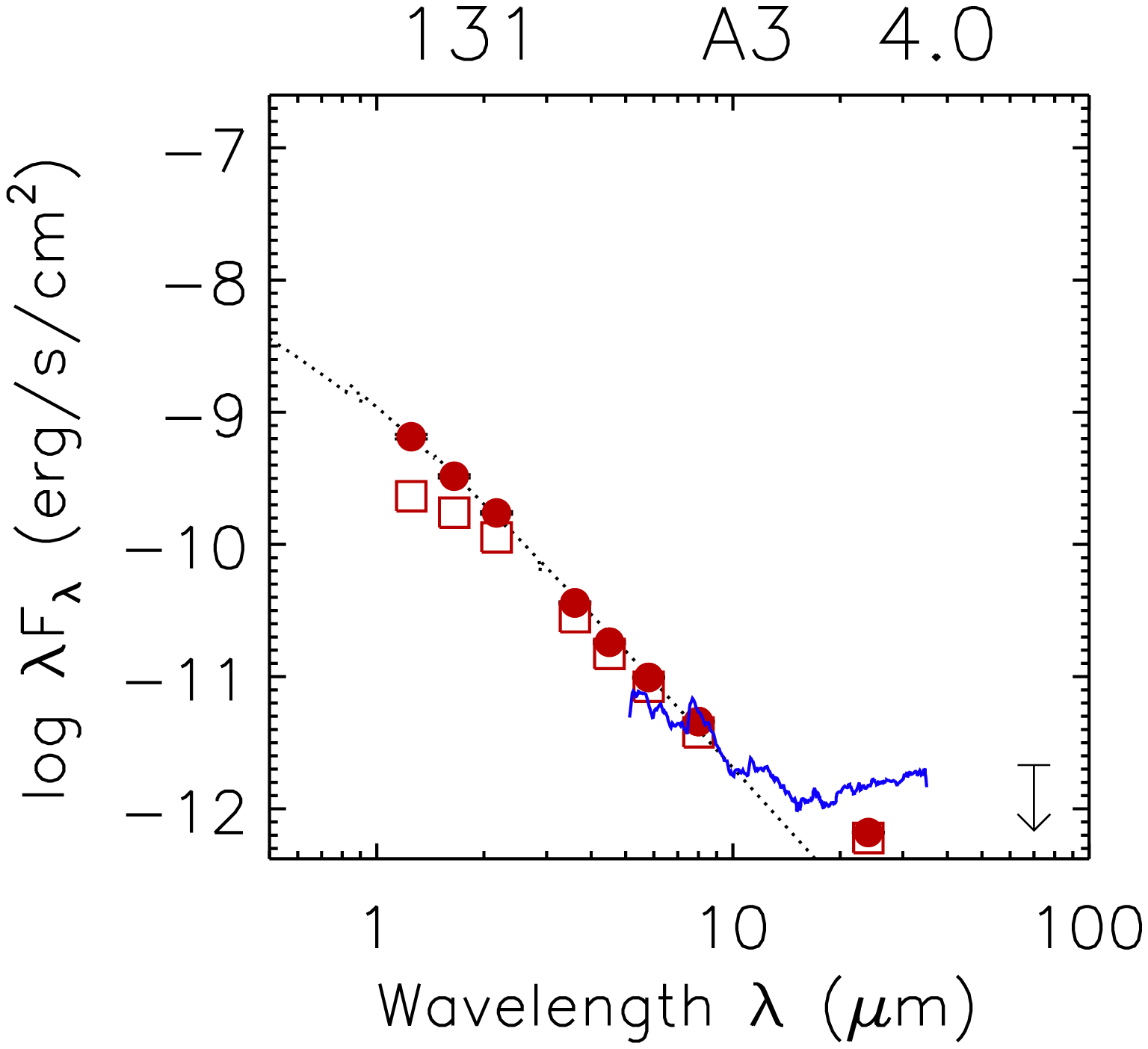}\\
\includegraphics[width=0.2\textwidth]{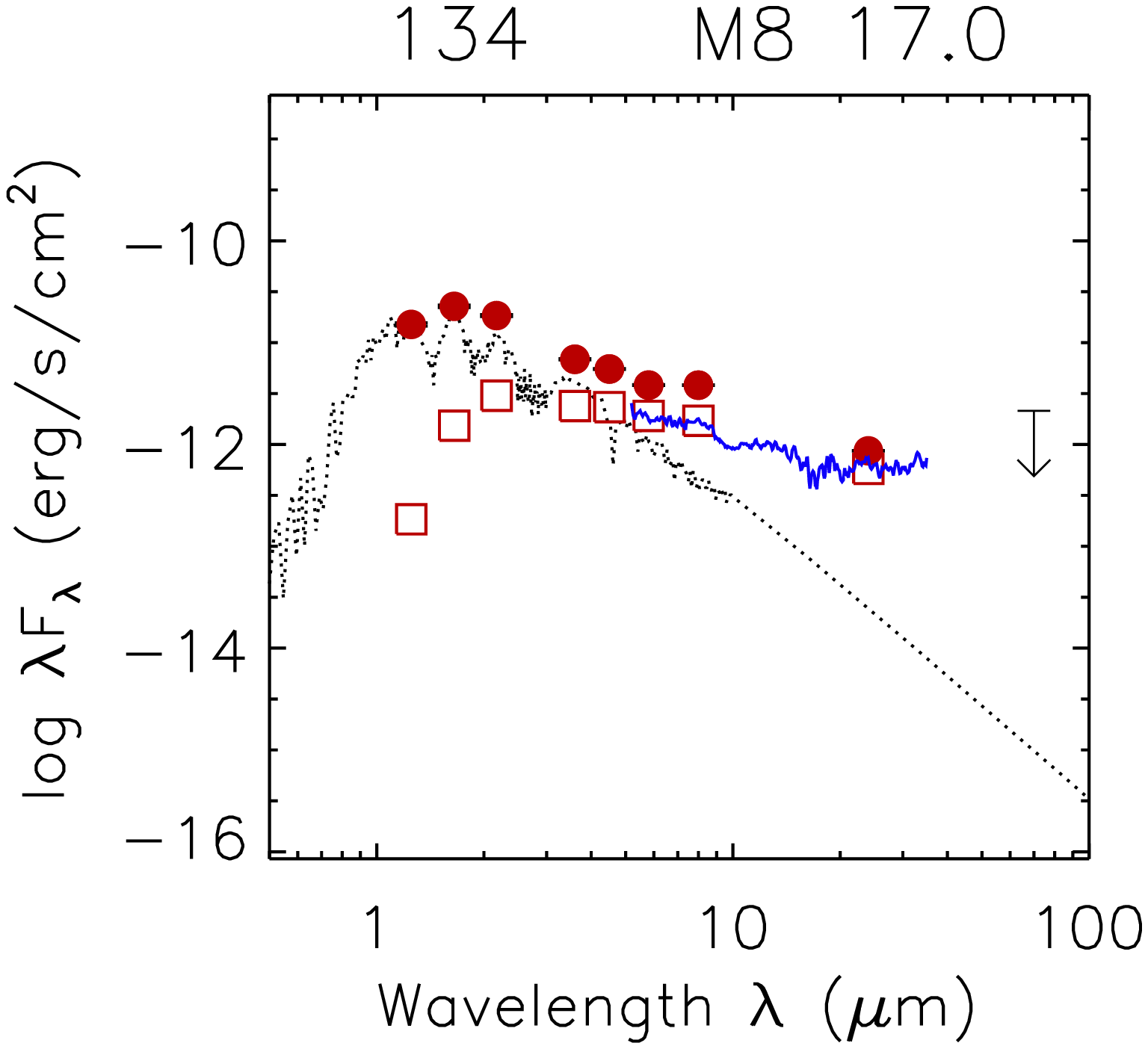}
\includegraphics[width=0.2\textwidth]{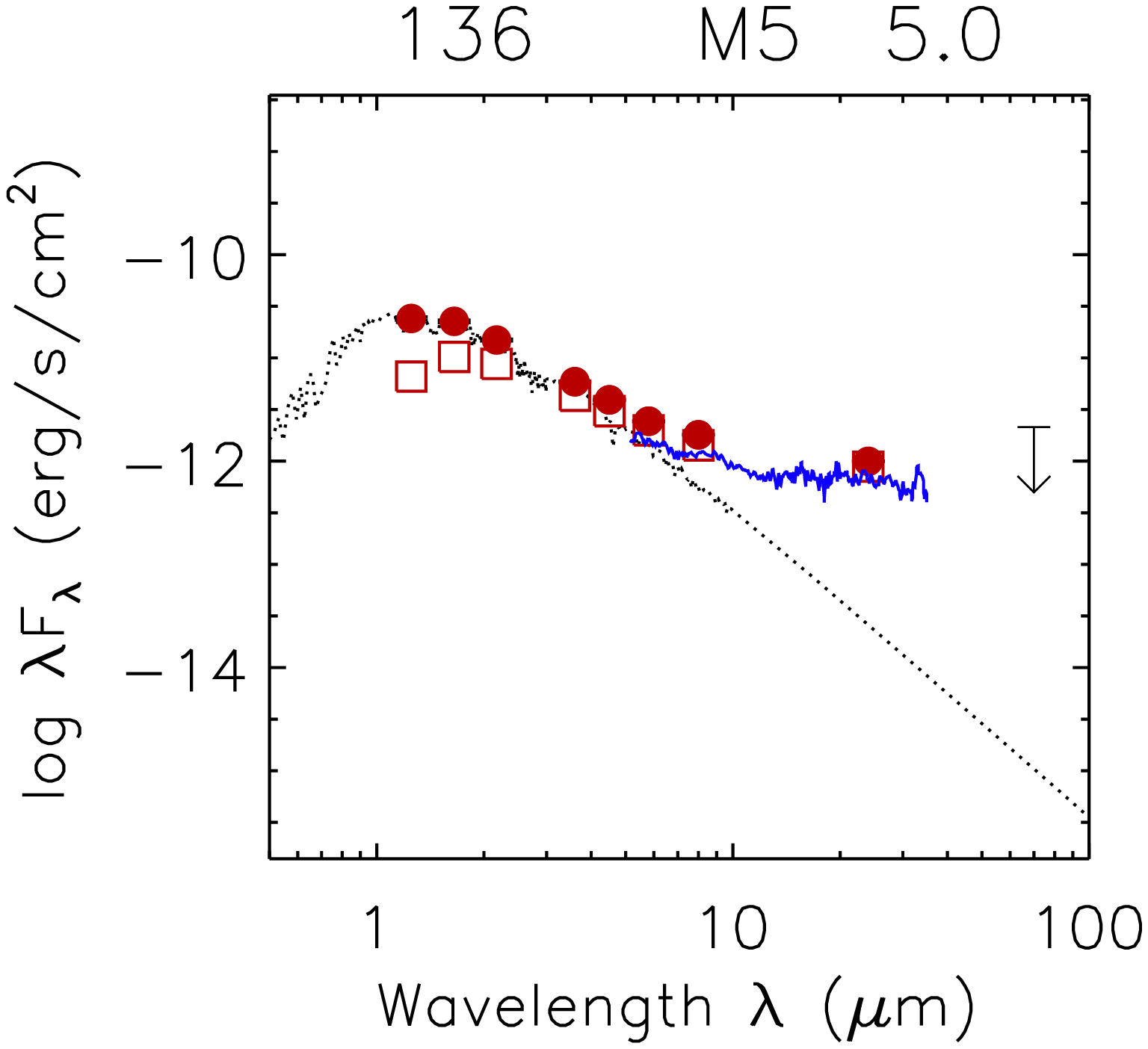}
\includegraphics[width=0.2\textwidth]{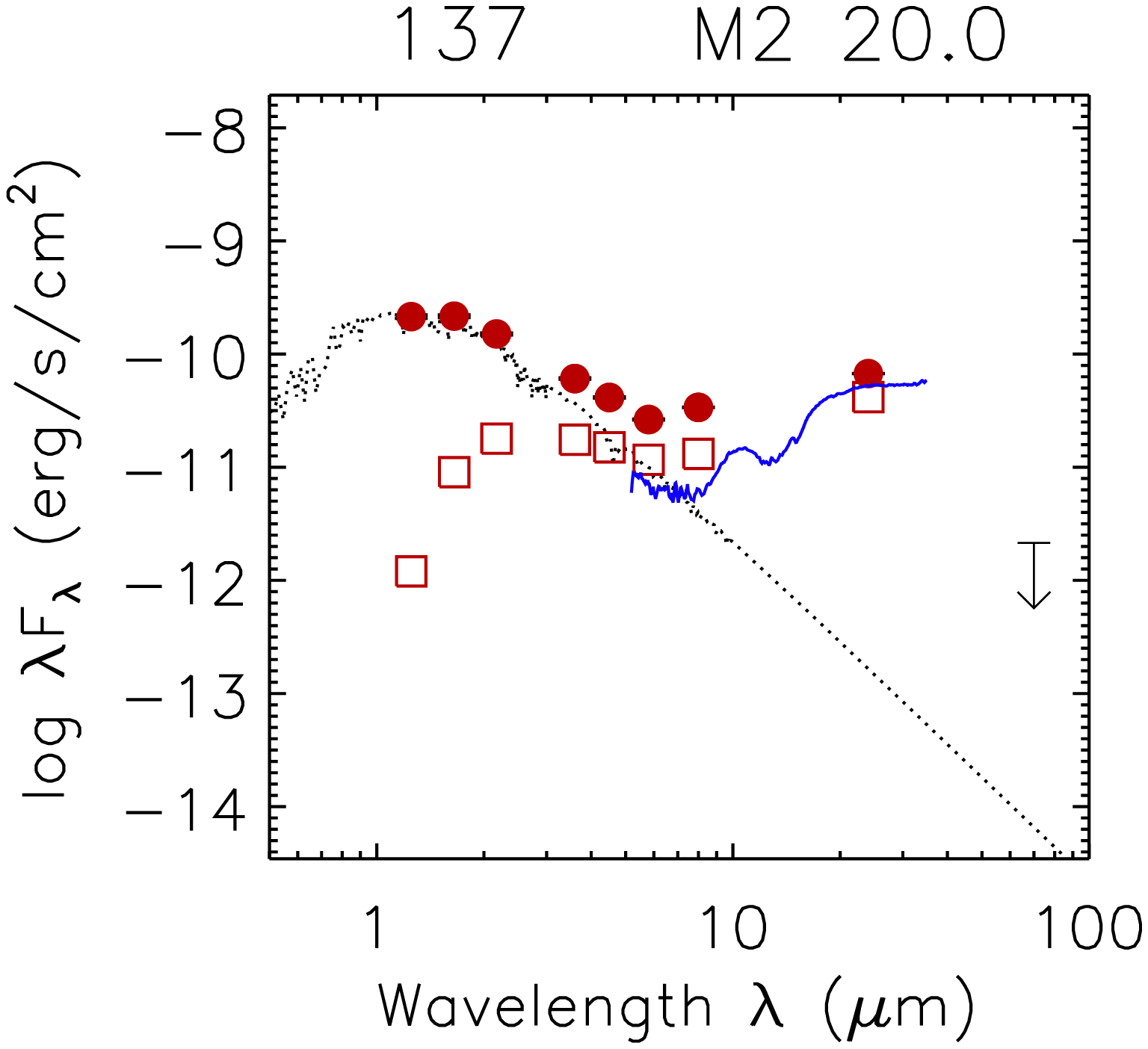}
\includegraphics[width=0.2\textwidth]{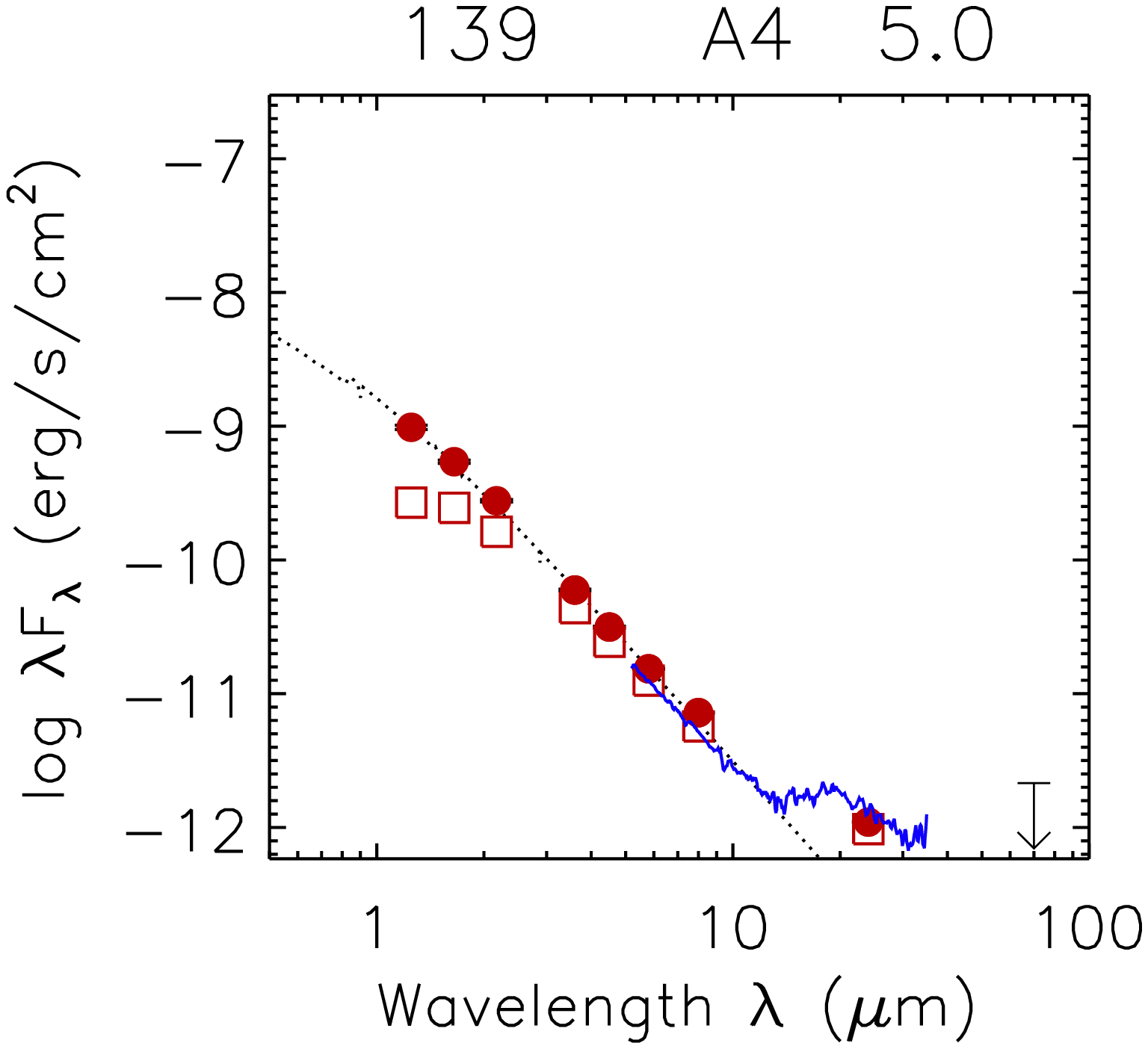}\\
\includegraphics[width=0.2\textwidth]{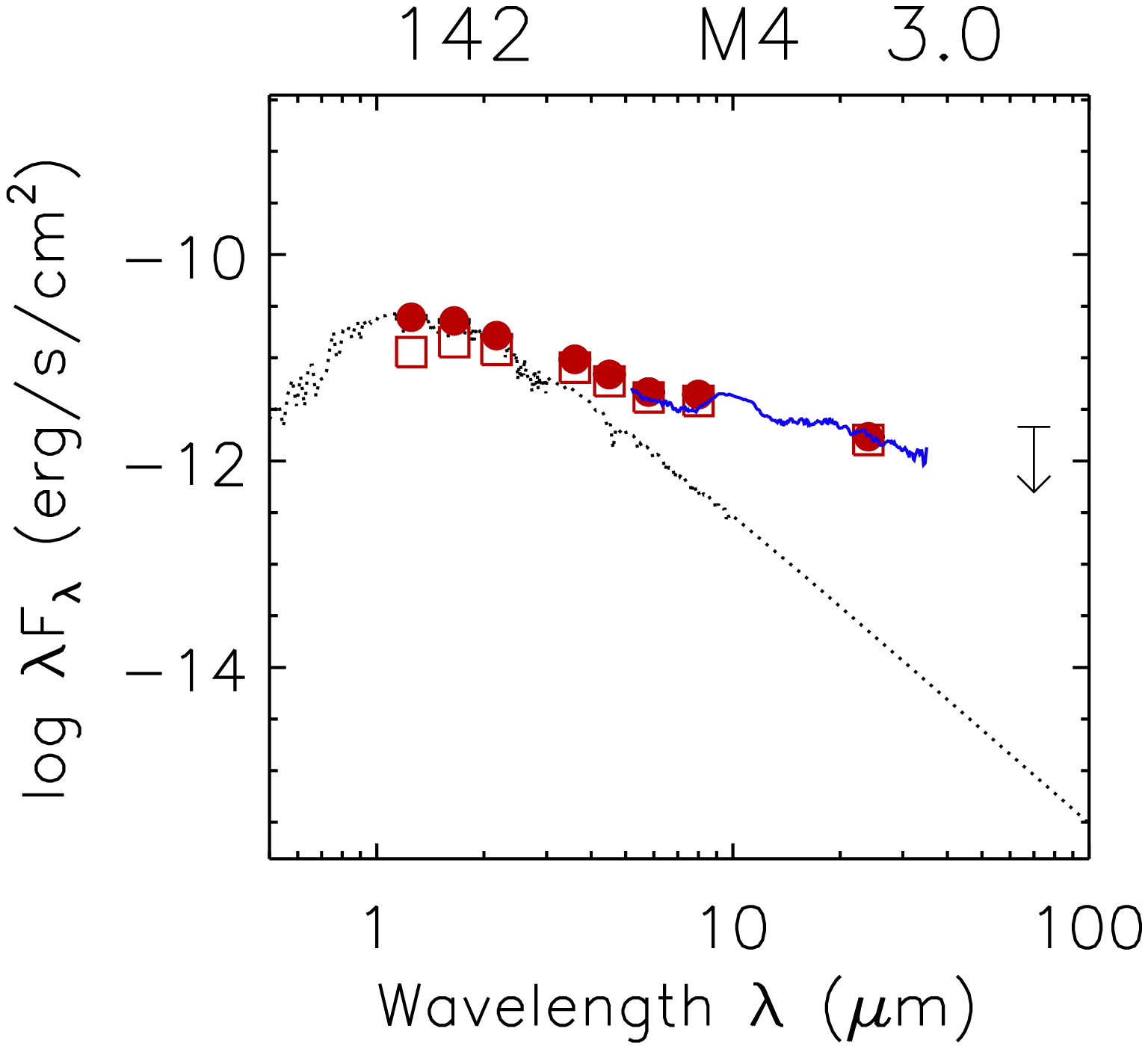}
\includegraphics[width=0.2\textwidth]{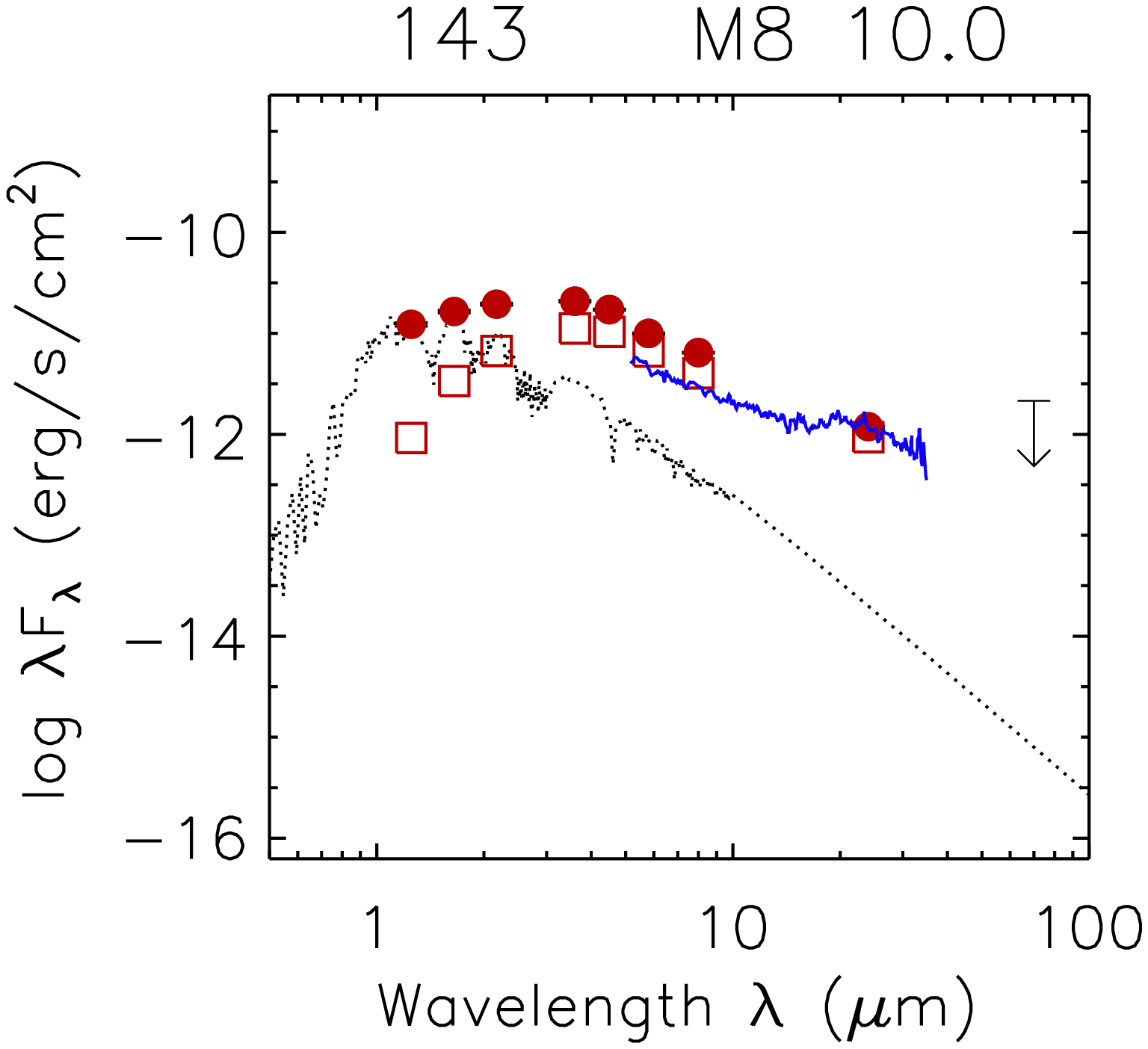}
\includegraphics[width=0.2\textwidth]{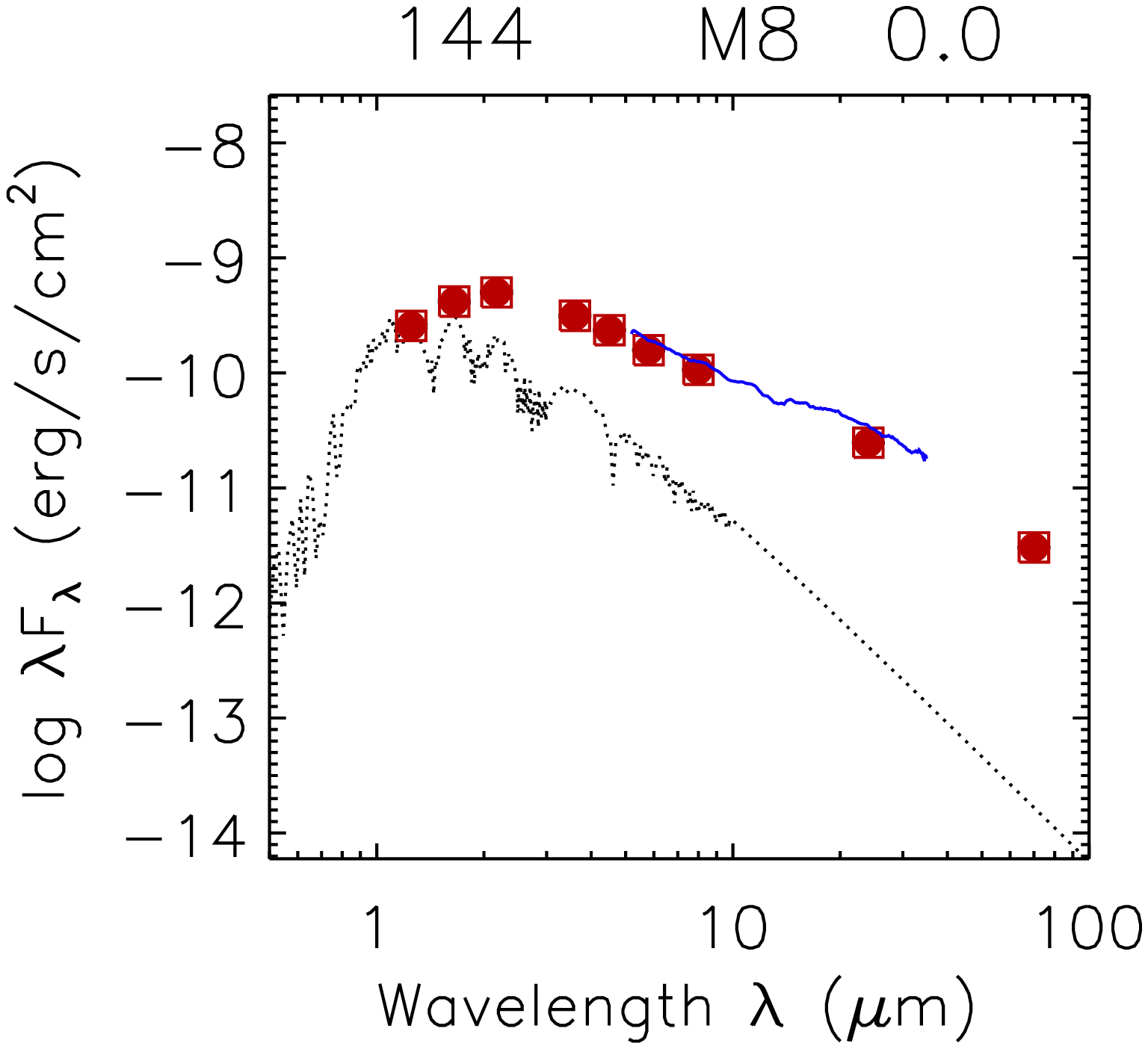}
\includegraphics[width=0.2\textwidth]{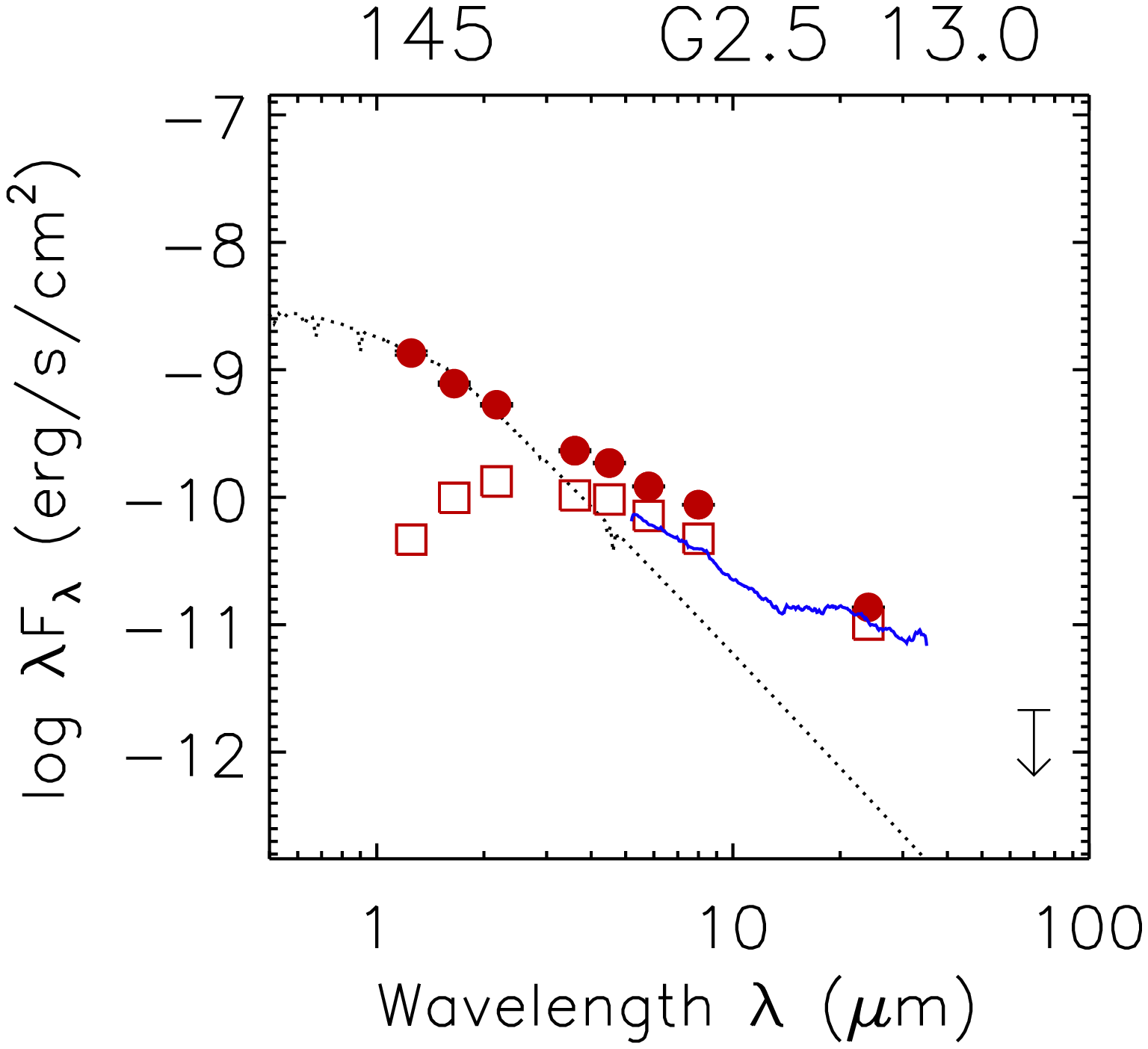}\\
\includegraphics[width=0.2\textwidth]{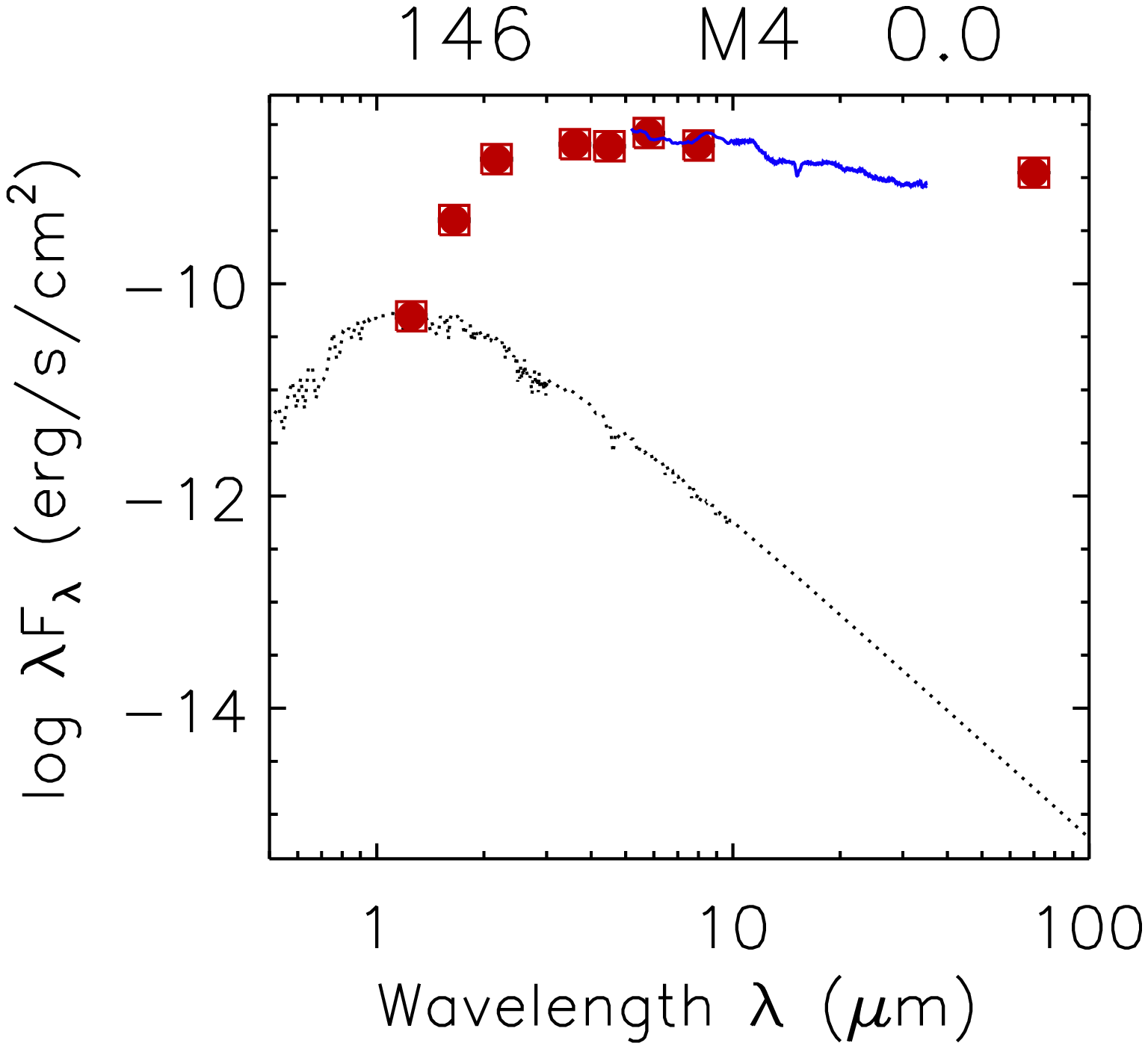}
\includegraphics[width=0.2\textwidth]{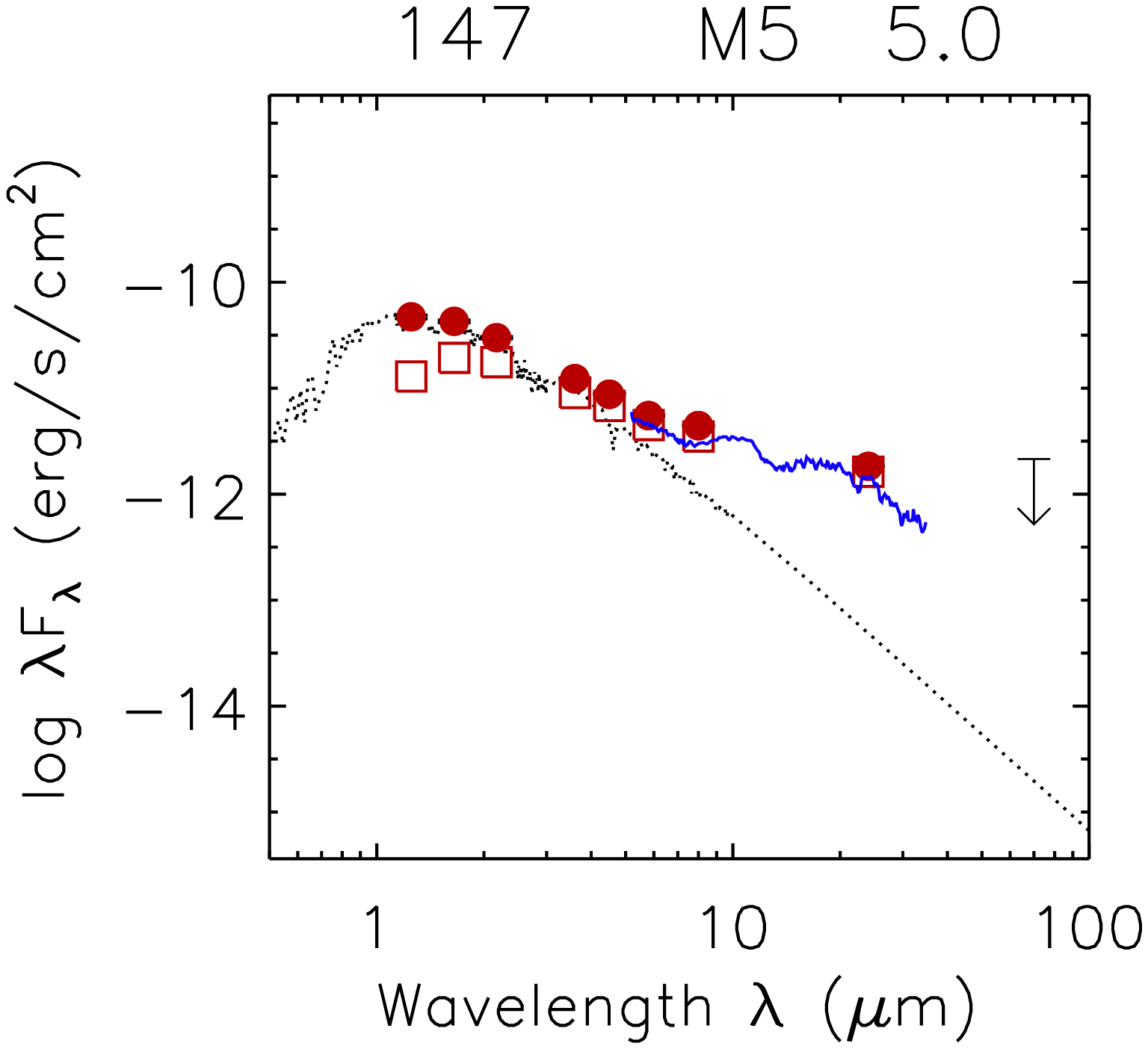}
\includegraphics[width=0.2\textwidth]{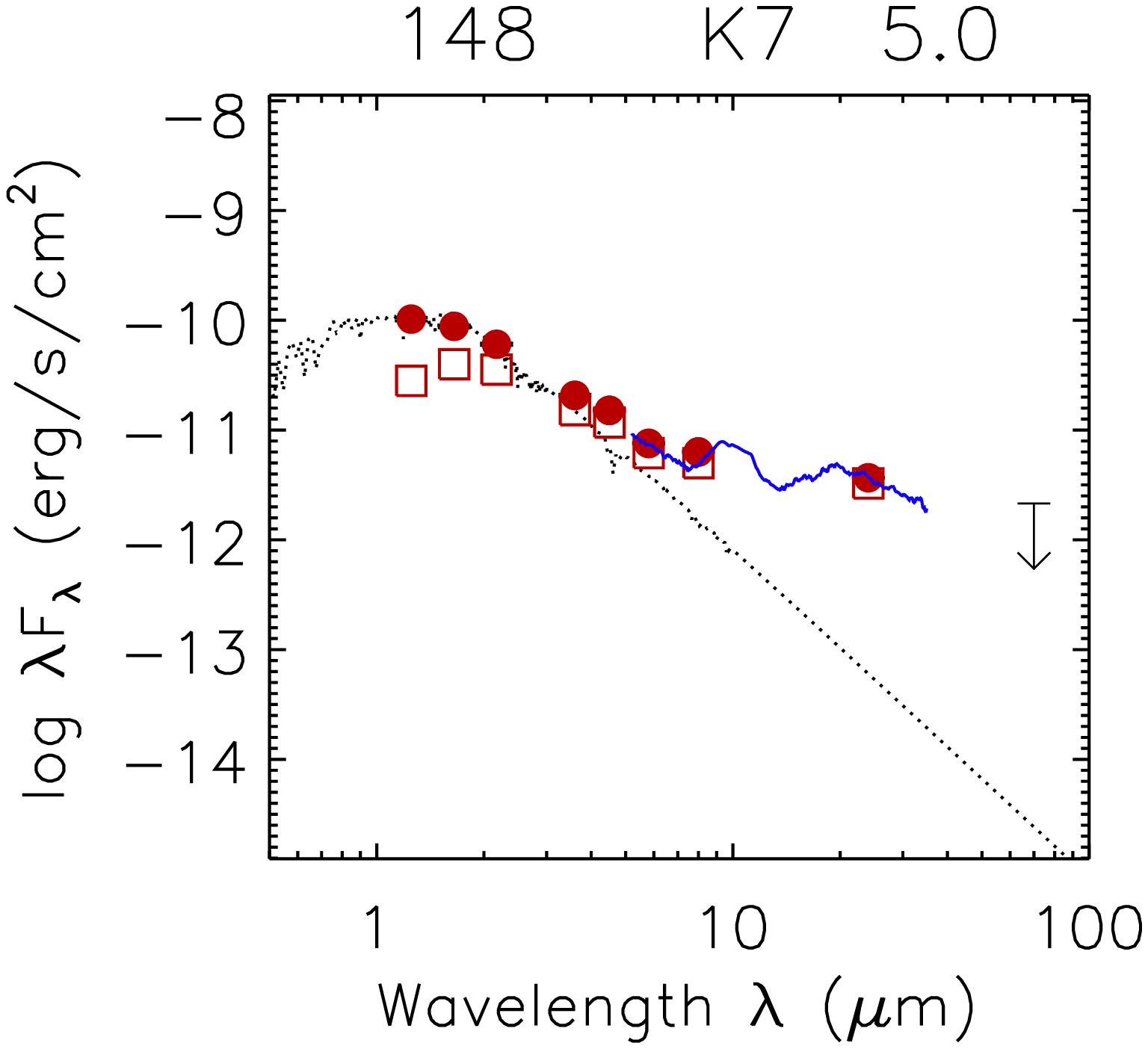}
\includegraphics[width=0.2\textwidth]{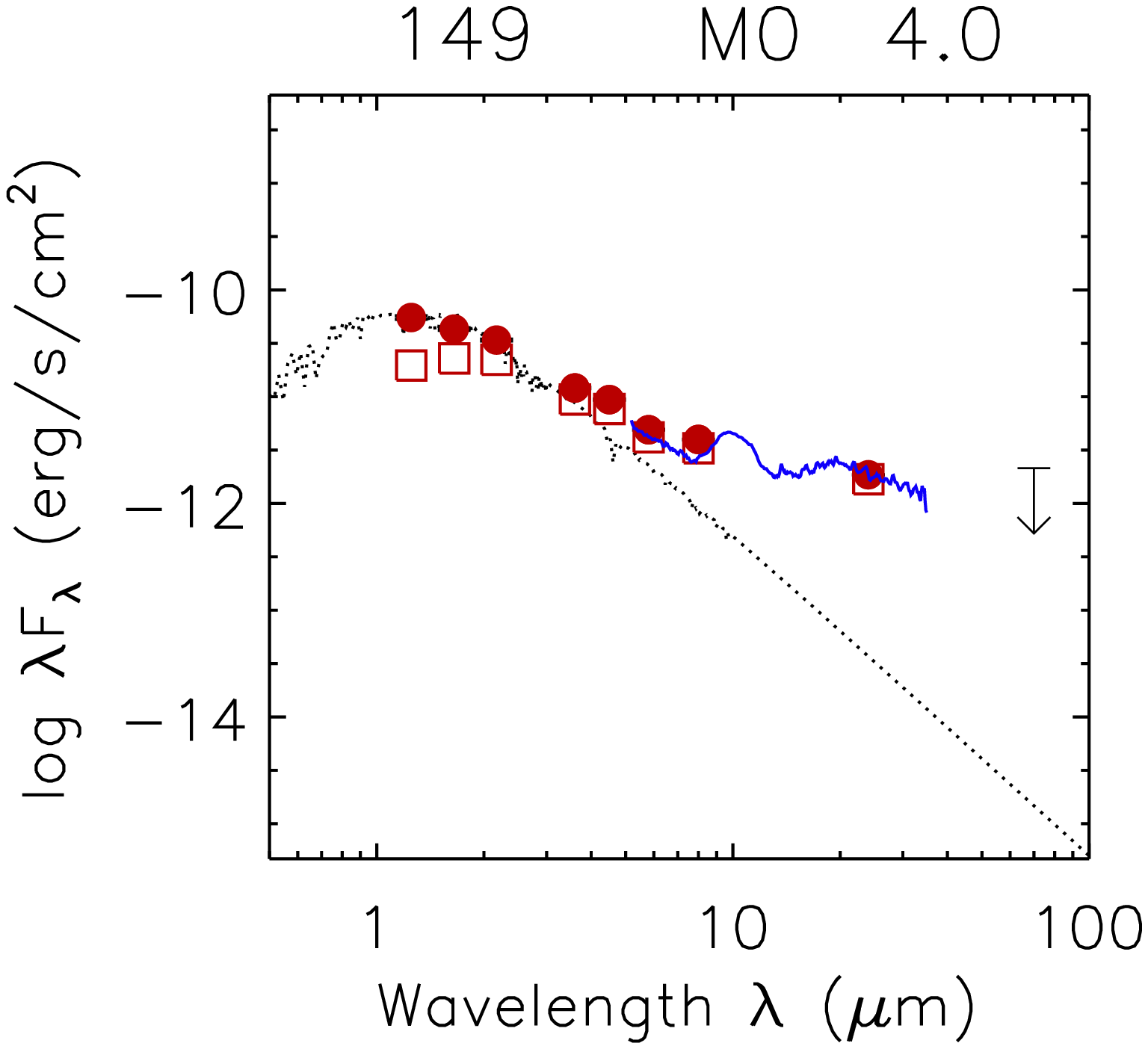}
\end{center}
\caption{\label{7f_ap_seds4} SEDs, continued.  }
\end{figure*}

\newpage

\begin{deluxetable}{r c c c c c c c c c }
\tabletypesize{\tiny}
\tablecolumns{9}
\tablewidth{0pt}
\tablecaption{Stellar and Disk Parameters in Serpens \label{7tab1}} 
\tablehead{\colhead{ID\tablenotemark{a}} & 
           \colhead{c2d ID (SSTc2dJ)}     &  
           \colhead{Spectral Type}       &  
           \colhead{T$_{{\rm eff}}$}        &
           \colhead{L$_{{\rm star}}$}       &
           \colhead{L$_{{\rm disk}}$}       &
           \colhead{A$_V$}               &
           \colhead{Age (Myr)}           &
           \colhead{Mass (M$_\odot$)}     &
           \colhead{Accreting?} 
}
\startdata
  1            & 18275383-0002335  &    K2  &  4900$^{+ 610}_{- 450}$  &    1.07$^{+  0.88}_{- 0.52}$   &   0.13$^{+ 0.14}_{- 0.14}$  &   2.6  &   10.86$^{+16.99}_{- 7.04}$  &    1.27$^{+ 0.31}_{-0.31}$  &  yes        \\
  3            & 18280845-0001064  &   M0  &  3850$^{+ 120}_{-  80}$  &    1.77$^{+  1.55}_{- 0.84}$   &   0.53$^{+ 0.94}_{- 0.94}$  &   3.0  &    0.50$^{+ 1.40}_{- 0.50}$  &    1.04$^{+ 0.18}_{-0.10}$  &  yes	\\
  6            & 18281350-0002491  &   K5  &  4350$^{+ 330}_{- 340}$  &    3.30$^{+  1.79}_{- 1.17}$   &   0.27$^{+ 0.89}_{- 0.89}$  &   3.0  &    2.37$^{+ 0.30}_{- 0.30}$  &    1.48$^{+ 0.27}_{-0.27}$  &  yes        \\
  7            & 18281501-0002588  &   M0  &  3850$^{+ 120}_{-  80}$  &    0.51$^{+  2.36}_{- 0.42}$   &   1.56$^{+ 0.80}_{- 0.80}$  &   6.0  &    4.36$^{+47.24}_{- 4.36}$  &    0.88$^{+ 0.28}_{-0.24}$  &  yes	\\
  8$^b$    & 18281519-0001407   &   M8  &  2640$^{+ 450}_{- 207}$  &    0.11$^{+  0.09}_{- 0.05}$   &   0.13$^{+ 0.01}_{- 0.01}$  &   2.0  &    0.78$^{+ 0.78}_{- 0.78}$  &    0.12$^{+ 0.12}_{-0.12}$  &  --	\\
  9$^b$    & 18281525-0002434   &   M0  &  3850$^{+ 190}_{- 220}$  &    3.23$^{+  2.82}_{- 1.53}$   &   0.38$^{+ 1.22}_{- 1.22}$  &   6.0  &    0.56$^{+ 0.56}_{- 0.56}$  &    1.03$^{+ 1.03}_{-1.03}$  &  --	\\
 10$^b$   & 18281629-0003164  &    M3  &  3470$^{+ 220}_{- 260}$  &    1.82$^{+  1.59}_{- 0.87}$   &   0.18$^{+ 0.32}_{- 0.32}$  &   6.0  &    0.44$^{+ 0.44}_{- 0.44}$  &    0.68$^{+ 0.68}_{-0.68}$  &  --	\\
 13$^b$   & 18281981-0001475  &    M7  &  2940$^{+ 340}_{- 390}$  &    0.12$^{+  0.10}_{- 0.06}$   &   0.15$^{+ 0.02}_{- 0.02}$  &   8.0  &    2.54$^{+ 2.54}_{- 2.54}$  &    0.24$^{+ 0.24}_{-0.24}$  &  --	\\
 14           & 18282143+0010411  &   M2  &  3580$^{+ 250}_{- 230}$  &    0.49$^{+  0.43}_{- 0.23}$   &   0.00$^{+ 0.00}_{- 0.00}$  &   3.0  &    2.27$^{+ 3.45}_{- 1.50}$  &    0.63$^{+ 0.28}_{-0.19}$  &  yes	\\
 15$^b$   & 18282159+0000162  &   M0  &  3850$^{+ 190}_{- 220}$  &    1.32$^{+  1.15}_{- 0.63}$   &   0.49$^{+ 0.64}_{- 0.64}$  &   4.0  &    1.01$^{+ 1.74}_{- 1.01}$  &    0.98$^{+ 0.16}_{-0.18}$  &  -- 	\\
 20$^b$   & 18282849+0026500  &   M0  &  3850$^{+ 190}_{- 220}$  &    0.29$^{+  0.25}_{- 0.14}$   &   0.05$^{+ 0.01}_{- 0.00}$  &  13.0  &    9.06$^{+14.45}_{- 5.16}$  &    0.81$^{+ 0.08}_{-0.25}$  &  --	\\
 21$^b$   & 18282905+0027560  &   M0  &  3850$^{+ 190}_{- 220}$  &    0.66$^{+  0.58}_{- 0.32}$   &   0.15$^{+ 0.10}_{- 0.10}$  &  12.0  &    2.89$^{+ 4.58}_{- 1.74}$  &    0.91$^{+ 0.14}_{-0.26}$  &  --	\\
 24$^b$   & 18284025+0016173  &   M7  &  2940$^{+ 340}_{- 390}$  &    0.15$^{+  0.13}_{- 0.07}$   &   0.14$^{+ 0.02}_{- 0.02}$  &  14.0  &    2.06$^{+ 2.06}_{- 2.06}$  &    0.24$^{+ 0.24}_{-0.24}$  &  --	\\
 29           & 18284481+0048085  &   M2  &  3580$^{+ 320}_{- 360}$  &    0.18$^{+  0.16}_{- 0.09}$   &   0.20$^{+ 0.04}_{- 0.04}$  &   4.0  &    8.13$^{+18.42}_{- 6.50}$  &    0.56$^{+ 0.23}_{-0.32}$  &  yes	\\
 30           & 18284497+0045239  &   M1  &  3720$^{+ 150}_{- 120}$  &    1.00$^{+  0.53}_{- 0.36}$   &   0.10$^{+ 0.10}_{- 0.10}$  &   2.0  &    1.18$^{+ 1.13}_{- 0.74}$  &    0.83$^{+ 0.18}_{-0.10}$  &  yes	\\
 31$^b$    & 18284559-0007132  &  M9  &  2510$^{+ 440}_{- 107}$  &   94.77$^{+ 75.01}_{-47.21}$   &   0.20$^{+18.94}_{-18.94}$  &  15.0  &                          &                         &  --      \\     
 32$^b$    & 18284614+0003016  &  M6  &  3050$^{+ 370}_{- 360}$  &    0.30$^{+  0.25}_{- 0.14}$   &   0.22$^{+ 0.06}_{- 0.06}$  &  13.0  &    1.73$^{+ 1.73}_{- 1.73}$  &    0.38$^{+ 0.38}_{-0.38}$  &  --	\\
 36           & 18285020+0009497  &   K5  &  4350$^{+ 680}_{- 480}$  &    2.88$^{+  1.56}_{- 1.02}$   &   0.50$^{+ 1.43}_{- 1.43}$  &  10.0  &    1.20$^{+ 4.56}_{- 1.20}$  &    1.21$^{+ 0.50}_{-0.66}$  &  yes       \\
 38$^b$    & 18285060+0007540  &   K7  &  4060$^{+ 350}_{-  80}$  &    0.18$^{+  0.16}_{- 0.09}$   &   0.02$^{+ 0.00}_{- 0.00}$  &   3.0  &   21.77$^{+39.11}_{-13.85}$  &    0.72$^{+ 0.02}_{-0.11}$  &  --       \\
 40           & 18285249+0020260  &   M7  &  2940$^{+ 570}_{- 507}$  &    0.36$^{+  0.30}_{- 0.17}$   &   4.46$^{+ 1.61}_{- 1.61}$  &  10.0  &    2.03$^{+ 2.03}_{- 2.03}$  &    0.48$^{+ 0.48}_{-0.48}$  &  no	\\
 41           & 18285276+0028466  &   K2  &  4900$^{+ 610}_{- 450}$  &    0.11$^{+  0.09}_{- 0.05}$   &   0.28$^{+ 0.03}_{- 0.03}$  &  15.0  &                          &                         &  no        \\     
 43           & 18285395+0045530  &  M0.5  &  3785$^{+ 225}_{- 380}$  &    0.18$^{+  0.16}_{- 0.09}$   &   0.56$^{+ 0.10}_{- 0.10}$  &   1.0  &   18.20$^{+25.92}_{-15.10}$  &    0.75$^{+ 0.08}_{-0.37}$  &  no	\\
 48           & 18285529+0020522  &  M5.5  &  3145$^{+ 425}_{- 525}$  &    0.34$^{+  0.29}_{- 0.16}$   &   1.32$^{+ 0.44}_{- 0.44}$  &  10.0  &    1.58$^{+ 1.07}_{- 1.07}$  &    0.33$^{+ 0.19}_{-0.19}$  &  yes	\\
 52           & 18285808+0017244  &   G3  &  5830$^{+ 400}_{- 230}$  &    8.14$^{+  6.40}_{- 4.07}$   &   0.05$^{+ 0.43}_{- 0.43}$  &   4.0  &    6.93$^{+ 6.33}_{- 2.63}$  &    1.82$^{+ 0.39}_{-0.38}$  &  no       \\
 53           & 18285860+0048594  &  M2.5  &  3525$^{+ 385}_{- 355}$  &    0.35$^{+  0.31}_{- 0.17}$   &   0.06$^{+ 0.02}_{- 0.02}$  &   6.0  &    2.34$^{+ 6.84}_{- 2.34}$  &    0.50$^{+ 0.41}_{-0.36}$  &  yes	\\
 54$^b$    & 18285946+0030029  &  M0  &  3850$^{+ 190}_{- 220}$  &    0.58$^{+  0.51}_{- 0.28}$   &   1.61$^{+ 0.94}_{- 0.94}$  &  12.0  &    3.44$^{+ 5.18}_{- 1.97}$  &    0.90$^{+ 0.14}_{-0.26}$  &  --	\\
 55           & 18290025+0016580  &   K2  &  4900$^{+ 450}_{- 210}$  &    2.44$^{+ 10.94}_{- 2.03}$   &   0.04$^{+ 0.09}_{- 0.09}$  &   7.0  &    4.02$^{+36.50}_{- 4.02}$  &    1.68$^{+ 1.13}_{-0.80}$  &  yes        \\
 56$^b$    & 18290057+0045079  &  M8  &  2640$^{+ 450}_{- 207}$  &    0.09$^{+  0.07}_{- 0.04}$   &   1.73$^{+ 0.15}_{- 0.15}$  &   3.0  &    1.27$^{+ 1.27}_{- 1.27}$  &    0.12$^{+ 0.12}_{-0.12}$  &  --	\\
 57$^b$    & 18290082+0027467  &  M8  &  2640$^{+ 450}_{- 207}$  &    0.04$^{+  0.03}_{- 0.02}$   &   2.54$^{+ 0.11}_{- 0.11}$  &   4.0  &    2.94$^{+ 2.94}_{- 2.94}$  &    0.11$^{+ 0.11}_{-0.11}$  &  --	\\
 58           & 18290088+0029315  &   K7  &  4060$^{+ 350}_{-  80}$  &    1.19$^{+  5.48}_{- 0.98}$   &   3.89$^{+ 4.62}_{- 4.62}$  &   5.0  &    2.31$^{+24.92}_{- 2.31}$  &    1.14$^{+ 0.25}_{-0.14}$  &  yes	\\
 59$^b$    & 18290107+0031451  &  M0  &  3850$^{+ 190}_{- 220}$  &    0.51$^{+  0.44}_{- 0.24}$   &   2.94$^{+ 1.49}_{- 1.49}$  &  14.0  &    4.42$^{+ 5.41}_{- 2.58}$  &    0.87$^{+ 0.14}_{-0.26}$  &  --	\\
 60           & 18290122+0029330  &  M0.5  &  3785$^{+  50}_{-  80}$  &    0.83$^{+  0.73}_{- 0.40}$   &   1.17$^{+ 0.97}_{- 0.97}$  &   6.0  &    2.20$^{+ 3.35}_{- 1.51}$  &    0.93$^{+ 0.08}_{-0.10}$  &  yes	\\
 61           & 18290175+0029465  &   M0  &  3850$^{+ 120}_{-  80}$  &    3.65$^{+  3.19}_{- 1.73}$   &   0.24$^{+ 0.87}_{- 0.87}$  &   5.0  &    0.40$^{+ 0.40}_{- 0.40}$  &    1.05$^{+ 1.05}_{-1.05}$  &  yes	\\
 62           & 18290184+0029546  &   K0  &  5250$^{+ 630}_{-1140}$  &   18.94$^{+ 15.34}_{- 9.34}$   &   0.33$^{+ 6.16}_{- 6.16}$  &   8.0  &                          &                         &  no        \\
 64$^b$   & 18290215+0029400  &   M5  &  3240$^{+ 270}_{- 260}$  &    0.13$^{+  0.11}_{- 0.06}$   &   0.58$^{+ 0.08}_{- 0.08}$  &   5.0  &    2.05$^{+ 5.19}_{- 2.05}$  &    0.21$^{+ 0.21}_{-0.11}$  &  --	\\
 65$^b$   & 18290286+0030089  &   M6  &  3050$^{+ 370}_{- 360}$  &    0.20$^{+  0.17}_{- 0.10}$   &   2.46$^{+ 0.49}_{- 0.49}$  &  10.0  &    1.48$^{+ 1.11}_{- 1.11}$  &    0.23$^{+ 0.12}_{-0.12}$  &  --	\\
 66           & 18290393+0020217  &   K5  &  4350$^{+ 330}_{- 340}$  &    5.11$^{+  4.35}_{- 2.46}$   &   0.39$^{+ 1.99}_{- 1.99}$  &   7.0  &    1.64$^{+ 0.17}_{- 0.17}$  &    1.56$^{+ 0.35}_{-0.35}$  &  yes       \\
 69$^b$   & 18290518+0038438  &   M5  &  3240$^{+ 270}_{- 260}$  &    0.23$^{+  0.20}_{- 0.11}$   &   0.00$^{+ 0.00}_{- 0.00}$  &  11.0  &    0.77$^{+ 2.22}_{- 0.77}$  &    0.21$^{+ 0.23}_{-0.07}$  &  --	\\
 70           & 18290575+0022325  &   A3  &  8720$^{+ 720}_{- 775}$  &   20.64$^{+ 15.11}_{-10.63}$   &   0.01$^{+ 0.22}_{- 0.22}$  &   3.0  &    8.18$^{+70.19}_{- 8.18}$  &    2.10$^{+ 0.38}_{-0.21}$  &  no       \\
 71           & 18290615+0019444  &   M3  &  3470$^{+  80}_{- 130}$  &    0.33$^{+  0.29}_{- 0.16}$   &   0.03$^{+ 0.01}_{- 0.01}$  &   4.0  &    2.48$^{+ 3.15}_{- 1.37}$  &    0.49$^{+ 0.09}_{-0.11}$  &  yes	\\
 74$^b$   & 18290699+0038377  &   M7  &  2940$^{+ 340}_{- 390}$  &    0.12$^{+  0.10}_{- 0.06}$   &   2.62$^{+ 0.31}_{- 0.31}$  &   7.0  &    2.66$^{+ 2.66}_{- 2.66}$  &    0.24$^{+ 0.24}_{-0.24}$  &  --	\\
 75$^b$   & 18290765+0052223  &   M5  &  3240$^{+ 270}_{- 260}$  &    0.11$^{+  0.10}_{- 0.05}$   &   0.23$^{+ 0.03}_{- 0.03}$  &   4.0  &    2.33$^{+ 5.95}_{- 2.33}$  &    0.21$^{+ 0.21}_{-0.11}$  &  --	\\
 76           & 18290775+0054037  &   M1  &  3720$^{+ 150}_{- 120}$  &    0.33$^{+  0.29}_{- 0.16}$   &   0.11$^{+ 0.04}_{- 0.04}$  &   4.0  &    5.66$^{+ 7.09}_{- 3.29}$  &    0.71$^{+ 0.16}_{-0.14}$  &  no	\\
 80$^b$   & 18290956+0037016  &    F0  &  7200$^{+ 380}_{- 310}$  &  370.99$^{+288.34}_{-86.29}$   &   0.17$^{+62.16}_{-62.16}$  &  24.0  &                          &                         &  --        \\     
 81           & 18290980+0034459  &   M5  &  3240$^{+ 520}_{- 690}$  &   60.77$^{+ 52.31}_{-29.09}$   &   0.12$^{+ 7.38}_{- 7.38}$  &  15.0  &                          &                         &  --        \\     
 82           & 18291148+0020387  &   M0  &  3850$^{+ 190}_{- 220}$  &    0.20$^{+  0.18}_{- 0.10}$   &   0.03$^{+ 0.01}_{- 0.01}$  &   4.0  &   15.80$^{+21.23}_{- 9.13}$  &    0.76$^{+ 0.08}_{-0.22}$  &  yes	\\
 83$^b$   & 18291249+0018152  &   M6  &  3050$^{+ 370}_{- 360}$  &    0.31$^{+  0.26}_{- 0.15}$   &   0.04$^{+ 0.01}_{- 0.01}$  &  14.0  &    1.64$^{+ 1.64}_{- 1.64}$  &    0.38$^{+ 0.38}_{-0.38}$  &  --	\\
 86           & 18291508+0052124  & M5.5  &  3145$^{+ 270}_{- 160}$  &   35.78$^{+ 30.57}_{-17.19}$   &   0.15$^{+ 5.50}_{- 5.50}$  &   6.0  &                          &                         &  --        \\     
 87          & 18291513+0039378  &    M4  &  3370$^{+ 320}_{- 460}$  &    0.82$^{+  0.72}_{- 0.39}$   &   0.34$^{+ 0.28}_{- 0.28}$  &   8.0  &    1.43$^{+ 0.24}_{- 0.24}$  &    0.64$^{+ 0.19}_{-0.19}$  &  no	\\
 88          & 18291539-0012519  &   M0.5  &  3785$^{+ 155}_{- 275}$  &    0.64$^{+  0.56}_{- 0.31}$   &   1.04$^{+ 0.67}_{- 0.67}$  &   0.0  &    2.98$^{+ 4.76}_{- 1.80}$  &    0.91$^{+ 0.12}_{-0.30}$  &  no	\\
 89           & 18291557+0039119  &  K5  &  4350$^{+ 850}_{- 550}$  &    0.95$^{+  0.81}_{- 0.46}$   &   0.21$^{+ 0.20}_{- 0.20}$  &  11.0  &    5.11$^{+21.66}_{- 3.76}$  &    1.18$^{+ 0.02}_{-0.62}$  &  yes        \\
 90$^b$   & 18291563+0039228  &   M7  &  2940$^{+ 680}_{- 507}$  &    0.19$^{+  0.16}_{- 0.09}$   &   0.40$^{+ 0.08}_{- 0.08}$  &   5.0  &    7.34$^{+ 7.34}_{- 7.34}$  &    0.54$^{+ 0.54}_{-0.54}$  &  --	\\
 92           & 18291969+0018031  &   M0  &  3850$^{+ 120}_{-  80}$  &    0.58$^{+  0.51}_{- 0.28}$   &   0.39$^{+ 0.23}_{- 0.23}$  &   6.0  &    3.40$^{+ 5.17}_{- 1.95}$  &    0.90$^{+ 0.10}_{-0.10}$  &  yes	\\
 96           & 18292184+0019386  &   M1  &  3720$^{+ 150}_{- 120}$  &    0.34$^{+  0.30}_{- 0.16}$   &   0.16$^{+ 0.06}_{- 0.06}$  &   5.0  &    5.51$^{+ 6.80}_{- 3.19}$  &    0.72$^{+ 0.16}_{-0.14}$  &  yes	\\
 97           & 18292250+0034118  &   M2  &  3580$^{+ 250}_{- 230}$  &    0.14$^{+  0.08}_{- 0.05}$   &   0.14$^{+ 0.02}_{- 0.02}$  &   5.0  &   10.11$^{+18.68}_{- 6.67}$  &    0.55$^{+ 0.19}_{-0.22}$  &  no	\\
 98           & 18292253+0034176  &   A3  &  8720$^{+3100}_{-1220}$  &   32.48$^{+ 23.78}_{-16.72}$   &   0.00$^{+ 0.12}_{- 0.12}$  &   8.0  &    4.99$^{+ 7.78}_{- 1.68}$  &    2.42$^{+ 0.45}_{-0.44}$  &  no       \\
100$^b$   & 18292640+0030042  &   M6  &  3050$^{+ 370}_{- 360}$  &    0.18$^{+  0.15}_{- 0.09}$   &   0.24$^{+ 0.04}_{- 0.04}$  &  12.0  &    1.61$^{+ 1.15}_{- 1.15}$  &    0.22$^{+ 0.12}_{-0.12}$  &  --	\\
\enddata
\tablenotetext{a}{As in \citet{OL10}}
\tablenotetext{b}{Spectral types from photometry}
\end{deluxetable}

\newpage

\begin{deluxetable}{r c c c c c c c c c }
\tabletypesize{\tiny}
\tablecolumns{9}
\tablewidth{0pt}
\tablecaption{Stellar and Disk Parameters in Serpens} 
\tablehead{\colhead{ID\tablenotemark{a}} & 
           \colhead{c2d ID (SSTc2dJ)}     &  
           \colhead{Spectral Type}       &  
           \colhead{T$_{{\rm eff}}$}        &
           \colhead{L$_{{\rm star}}$}       &
           \colhead{L$_{{\rm disk}}$}       &
           \colhead{A$_V$}               &
           \colhead{Age (Myr)}           &
           \colhead{Mass (M$_\odot$)}     &
           \colhead{Accreting?} 
}
\startdata
101$^b$   & 18292679+0039497  &   M8  &  2640$^{+ 450}_{- 207}$  &    0.06$^{+  0.05}_{- 0.03}$   &   3.86$^{+ 0.24}_{- 0.24}$  &  10.0  &    2.02$^{+ 2.02}_{- 2.02}$  &    0.12$^{+ 0.12}_{-0.12}$  &  --	\\
103$^b$   & 18292824-0022574  &   M6  &  3050$^{+ 370}_{- 360}$  &  166.68$^{+141.01}_{-80.49}$   &   0.06$^{+ 0.69}_{- 0.00}$  &  25.0  &                          &                         &  --        \\     
104$^b$   & 18292833+0049569  &   M5  &  3240$^{+ 270}_{- 260}$  &    0.12$^{+  0.11}_{- 0.06}$   &   0.10$^{+ 0.01}_{- 0.01}$  &   6.0  &    2.16$^{+ 5.49}_{- 2.16}$  &    0.21$^{+ 0.21}_{-0.11}$  &  --	\\
105$^b$   & 18292864+0042369  &   M5  &  3240$^{+ 270}_{- 260}$  &    0.09$^{+  0.05}_{- 0.03}$   &   0.00$^{+ 0.00}_{- 0.00}$  &   5.0  &    2.91$^{+ 7.96}_{- 2.91}$  &    0.21$^{+ 0.20}_{-0.12}$  &  --	\\
106           & 18292927+0018000  &  M3  &  3470$^{+  80}_{- 130}$  &    0.25$^{+  0.13}_{- 0.09}$   &   0.02$^{+ 0.01}_{- 0.01}$  &   5.0  &    3.10$^{+ 3.20}_{- 1.02}$  &    0.47$^{+ 0.08}_{-0.11}$  &  no	\\
107$^b$   & 18293056+0033377  &   M8  &  2640$^{+ 450}_{- 207}$  &    0.12$^{+  0.10}_{- 0.06}$   &   0.04$^{+ 0.01}_{- 0.01}$  &  13.0  &    0.63$^{+ 0.63}_{- 0.63}$  &    0.12$^{+ 0.12}_{-0.12}$  &  --	\\
109$^b$   & 18293300+0040087  &   M7  &  2940$^{+ 340}_{- 390}$  &    0.22$^{+  0.19}_{- 0.11}$   &   0.19$^{+ 0.04}_{- 0.04}$  &   7.0  &    1.16$^{+ 1.16}_{- 1.16}$  &    0.25$^{+ 0.25}_{-0.25}$  &  --	\\
111$^b$   & 18293337+0050136  &   M5  &  3240$^{+ 270}_{- 260}$  &    0.07$^{+  0.04}_{- 0.02}$   &   0.11$^{+ 0.01}_{- 0.01}$  &   7.0  &    4.26$^{+13.41}_{- 4.26}$  &    0.20$^{+ 0.20}_{-0.11}$  &  --	\\
113           & 18293561+0035038  &  K7  &  4060$^{+ 350}_{-  80}$  &    2.32$^{+  2.02}_{- 1.11}$   &   0.03$^{+ 0.08}_{- 0.00}$  &   6.0  &    0.68$^{+ 1.66}_{- 0.13}$  &    0.13$^{+ 0.91}_{-0.22}$  &  yes	\\
114           & 18293619+0042167  &  F9  &  6115$^{+ 390}_{- 400}$  &    3.68$^{+  2.87}_{- 1.84}$   &   0.07$^{+ 0.24}_{- 0.24}$  &   9.5  &   14.29$^{+17.38}_{-14.15}$  &  116.38$^{+ 1.39}_{-0.18}$  &  no	\\
115           & 18293672+0047579  &  M0.5  &  3785$^{+ 155}_{- 275}$  &    0.50$^{+  0.43}_{- 0.24}$   &   0.17$^{+ 0.09}_{- 0.09}$  &   7.0  &    4.58$^{+ 5.47}_{- 2.84}$  &    0.87$^{+ 0.12}_{-0.31}$  &  no	\\
116$^b$   & 18293882+0044380  &   M5  &  3240$^{+ 270}_{- 260}$  &    0.21$^{+  0.18}_{- 0.10}$   &   0.05$^{+ 0.01}_{- 0.01}$  &   9.0  &    0.93$^{+ 2.29}_{- 0.93}$  &    0.21$^{+ 0.22}_{-0.08}$  &  --	\\
117$^b$   & 18294020+0015131  &   M7  &  2940$^{+ 680}_{- 507}$  &    0.02$^{+  0.02}_{- 0.01}$   &   9.47$^{+ 0.20}_{- 0.20}$  &   1.0  &    3.14$^{+ 4.07}_{- 3.14}$  &    0.06$^{+ 0.73}_{-0.00}$  &  yes	\\
119           & 18294121+0049020  &  K7  &  4060$^{+ 350}_{-  80}$  &    0.46$^{+  0.40}_{- 0.22}$   &   0.02$^{+ 0.01}_{- 0.01}$  &   6.0  &    4.86$^{+ 9.08}_{- 2.33}$  &    0.73$^{+ 0.27}_{-0.09}$  &  yes        \\
120           & 18294168+0044270  &  A2  &  8970$^{+ 520}_{- 540}$  &   25.13$^{+ 18.67}_{-12.86}$   &   0.00$^{+ 0.06}_{- 0.00}$  &   8.0  &    6.69$^{+ 2.10}_{- 6.69}$  &    2.24$^{+ 0.35}_{-0.29}$  &  no        \\
122           & 18294410+0033561  &  M0  &  3850$^{+ 155}_{- 150}$  &    1.10$^{+  0.96}_{- 0.52}$   &   0.03$^{+ 0.03}_{- 0.00}$  &   4.0  &    1.43$^{+ 2.03}_{- 1.12}$  &    0.96$^{+ 0.15}_{-0.16}$  &  yes	\\
123           & 18294503+0035266  &  M0  &  3850$^{+ 120}_{-  80}$  &    0.72$^{+  0.63}_{- 0.34}$   &   0.12$^{+ 0.09}_{- 0.09}$  &   9.0  &    2.63$^{+ 4.10}_{- 1.66}$  &    0.92$^{+ 0.12}_{-0.10}$  &  no	\\
124           & 18294725+0039556  &  M0  &  3850$^{+ 155}_{- 150}$  &    0.27$^{+  0.15}_{- 0.10}$   &   0.08$^{+ 0.02}_{- 0.02}$  &   4.0  &    9.68$^{+10.12}_{- 3.71}$  &    0.81$^{+ 0.06}_{-0.17}$  &  no	\\
125           & 18294726+0032230  &  M0  &  3850$^{+ 120}_{-  80}$  &    0.58$^{+  0.51}_{- 0.28}$   &   0.07$^{+ 0.04}_{- 0.04}$  &   6.0  &    3.45$^{+ 5.18}_{- 1.98}$  &    0.90$^{+ 0.10}_{-0.10}$  &  yes	\\
127           & 18295001+0051015  &  M2  &  3580$^{+ 120}_{- 130}$  &    0.48$^{+  0.42}_{- 0.23}$   &   0.03$^{+ 0.02}_{- 0.02}$  &   4.0  &    2.35$^{+ 3.15}_{- 1.46}$  &    0.63$^{+ 0.15}_{-0.12}$  &  yes	\\
129$^b$   & 18295016+0056081  &   M7  &  2940$^{+ 340}_{- 390}$  &    0.22$^{+  0.18}_{- 0.11}$   &   1.32$^{+ 0.29}_{- 0.29}$  &   4.0  &    1.23$^{+ 1.23}_{- 1.23}$  &    0.25$^{+ 0.25}_{-0.25}$  &  --	\\
130           & 18295041+0043437  &  K6  &  4205$^{+ 150}_{- 140}$  &    1.33$^{+  1.15}_{- 0.64}$   &   0.22$^{+ 0.30}_{- 0.30}$  &   7.0  &    2.16$^{+ 2.43}_{- 2.16}$  &    0.91$^{+ 0.22}_{-0.16}$  &  yes        \\
131           & 18295130+0027479  &  A3  &  8720$^{+ 720}_{- 775}$  &   25.57$^{+ 18.72}_{-13.16}$   &   0.00$^{+ 0.02}_{- 0.02}$  &   4.0  &    6.49$^{+ 2.08}_{- 6.49}$  &    2.23$^{+ 0.37}_{-0.30}$  &  no        \\
134$^b$   & 18295244+0031496  &   M8  &  2640$^{+ 450}_{- 207}$  &    0.08$^{+  0.07}_{- 0.04}$   &   0.74$^{+ 0.06}_{- 0.06}$  &  17.0  &    1.30$^{+ 1.30}_{- 1.30}$  &    0.12$^{+ 0.12}_{-0.12}$  &  --	\\
136$^b$   & 18295304+0040105  &   M5  &  3240$^{+ 270}_{- 260}$  &    0.17$^{+  0.09}_{- 0.06}$   &   0.04$^{+ 0.01}_{- 0.01}$  &   5.0  &    1.44$^{+ 3.61}_{- 1.44}$  &    0.21$^{+ 0.22}_{-0.10}$  &  --	\\
137$^b$   & 18295305+0036065  &   M2  &  3580$^{+ 250}_{- 230}$  &    1.56$^{+  1.36}_{- 0.74}$   &   0.31$^{+ 0.48}_{- 0.48}$  &  20.0  &    1.01$^{+ 0.06}_{- 0.06}$  &    0.89$^{+ 0.18}_{-0.18}$  &  --	\\
139           & 18295422+0045076  &  A4  &  8460$^{+1120}_{- 820}$  &   33.71$^{+ 24.87}_{-17.30}$   &   0.00$^{+ 0.14}_{- 0.14}$  &   5.0  &    4.77$^{+ 4.60}_{- 1.79}$  &    2.43$^{+ 0.39}_{-0.45}$  &  no       \\
142           & 18295592+0040150  &  M4  &  3370$^{+ 180}_{- 350}$  &    0.17$^{+  0.09}_{- 0.06}$   &   0.26$^{+ 0.05}_{- 0.05}$  &   3.0  &    3.05$^{+ 4.84}_{- 3.05}$  &    0.36$^{+ 0.18}_{-0.23}$  &  yes	\\
143$^b$   & 18295620+0033391  &   M8  &  2640$^{+ 450}_{- 207}$  &    0.07$^{+  0.06}_{- 0.03}$   &   2.02$^{+ 0.14}_{- 0.14}$  &  10.0  &    1.76$^{+ 1.76}_{- 1.76}$  &    0.12$^{+ 0.12}_{-0.12}$  &  --	\\
144$^b$   & 18295701+0033001  &   M8  &  2640$^{+ 450}_{- 207}$  &    1.43$^{+  1.16}_{- 0.70}$   &   1.90$^{+ 2.71}_{- 2.71}$  &   0.0  &                          &                         &  --       \\     
145           & 18295714+0033185  &  G2.5  &  5845$^{+ 230}_{-  30}$  &   19.73$^{+ 15.51}_{- 9.86}$   &   0.05$^{+ 1.01}_{- 1.01}$  &  13.0  &    3.19$^{+ 2.66}_{- 0.89}$  &    2.47$^{+ 0.44}_{-0.52}$  &  no       \\
146           & 18295772+0114057  &  M4  &  3370$^{+ 180}_{- 350}$  &    0.34$^{+  0.30}_{- 0.16}$   &  89.13$^{+30.44}_{-30.44}$  &   0.0  &    1.65$^{+ 1.39}_{- 1.65}$  &    0.42$^{+ 0.16}_{-0.26}$  &  yes	\\
147$^b$   & 18295872+0036205  &   M5  &  3240$^{+ 270}_{- 260}$  &    0.31$^{+  0.27}_{- 0.15}$   &   0.10$^{+ 0.03}_{- 0.03}$  &   5.0  &    0.37$^{+ 1.98}_{- 0.37}$  &    0.20$^{+ 0.26}_{-0.05}$  &  --	\\
148           & 18300178+0032162  &   K7  &  4060$^{+ 350}_{-  80}$  &    0.83$^{+  0.72}_{- 0.40}$   &   0.04$^{+ 0.03}_{- 0.03}$  &   5.0  &    2.58$^{+ 2.87}_{- 1.48}$  &    0.70$^{+ 0.42}_{-0.08}$  &  yes        \\
149           & 18300350+0023450  &   M0  &  3850$^{+ 190}_{- 220}$  &    0.42$^{+  0.37}_{- 0.20}$   &   0.04$^{+ 0.02}_{- 0.02}$  &   4.0  &    5.82$^{+ 7.72}_{- 3.47}$  &    0.85$^{+ 0.14}_{-0.26}$  &  yes	\\
\enddata
\tablenotetext{a}{As in \citet{OL10}}
\tablenotetext{b}{Spectral types from photometry}
\end{deluxetable}


\begin{thebibliography}{}

\bibitem[{\'A}brah{\'a}m et al.(2009)]{AB09} {\'A}brah{\'a}m, P.,
  Juh{\'a}sz, A., Dullemond, C.~P., et al.\ 2009, \nat, 459, 224
\bibitem[Acke et al.(2004)]{BA04} Acke, B., van den Ancker, M.~E.,
  Dullemond, C.~P., van Boekel, R., \& Waters, L.~B.~F.~M.\ 2004,
  \aap, 422, 621
\bibitem[Alcal{\'a} et al.(2008)]{JA08} Alcal{\'a}, J.~M., Spezzi, L.,
  Chapman, N., et al.\ 2008, \apj, 676, 427
\bibitem[Alves \& Bouy(2012)]{AB12} Alves, J., \& Bouy, H.\ 2012,
  arXiv:1209.3787
\bibitem[Andrews \& Williams(2005)]{AW05} Andrews, S.~M., \& Williams,
  J.~P.\ 2005, \apj, 631, 1134
\bibitem[Baraffe et al.(1998)]{BA98} Baraffe, I., Chabrier, G.,
  Allard, F., \& Hauschildt, P.~H.\ 1998, \aap, 337, 403
\bibitem[Blaauw(1978)]{BL78} Blaauw, A.\ 1978, Problems of Physics and
  Evolution of the Universe, 101
\bibitem[Blum \& Wurm(2008)]{BW08} Blum, J., \& Wurm, G.\ 2008, \araa,
  46, 21 
\bibitem[Bouvier et al.(2007)]{BO07} Bouvier, J., Alencar, S.~H.~P.,
  Boutelier, T., et al.\ 2007, \aap, 463, 1017
\bibitem[Bouwman et al.(2008)]{BO08} Bouwman, J., et al.\ 2008, \apj,
  683, 479
\bibitem[Brown et al.(2007)]{JB07} Brown, J.~M., et al.\ 2007, \apjl,
  664, L107
\bibitem[Carpenter et al.(2001)]{CA01} Carpenter, J.~M., Hillenbrand,
  L.~A., \& Skrutskie, M.~F.\ 2001, \aj, 121, 3160
\bibitem[Carpenter et al.(2006)]{CA06} Carpenter, J.~M., Mamajek,
  E.~E., Hillenbrand, L.~A., \& Meyer, M.~R.\ 2006, \apjl, 651, L49
\bibitem[Chavarria(1981)]{CH81} Chavarria, C.\ 1981, \aap, 101, 105
\bibitem[Chen et al.(2006)]{CC05} Chen, C.~H., Sargent, B.~A., Bohac,
  C., et al.\ 2006, \apjs, 166, 351
\bibitem[Chiang \& Goldreich(1997)]{CG97} Chiang, E.~I., \& Goldreich,
  P.\ 1997, \apj, 490, 368
\bibitem[Ciesla(2007)]{CI07} Ciesla, F.~J.\ 2007, Science, 318, 613
\bibitem[Cieza et al.(2007)]{LC07} Cieza, L., et al.\ 2007, \apj, 667,
  308 
\bibitem[Clarke et al.(2001)]{CC01} Clarke, C.~J., Gendrin, A., \&
  Sotomayor, M.\ 2001, \mnras, 328, 485
\bibitem[Comer{\'o}n(2008)]{FC08} Comer{\'o}n, F.\ 2008, Handbook of
  Star Forming Regions, Volume II, 295
\bibitem[Currie \& Kenyon(2009)]{CU09} Currie, T., \& Kenyon, S.~J.\
  2009, \aj, 138, 703
\bibitem[Dahm \& Carpenter(2009)]{DC09} Dahm, S.~E., \& Carpenter,
  J.~M.\ 2009, \aj, 137, 4024
\bibitem[de Zeeuw et al.(1999)]{DZ99} de Zeeuw, P.~T., Hoogerwerf, R.,
  de Bruijne, J.~H.~J., Brown, A.~G.~A., \& Blaauw, A.\ 1999, \aj,
  117, 354
\bibitem[Dominik \& Tielens(1997)]{DT97} Dominik, C., \& Tielens,
  A.~G.~G.~M.\ 1997, \apj, 480, 647
\bibitem[Dullemond \& Dominik(2004)]{DD04b} Dullemond, C.~P., \&
  Dominik, C.\ 2004, \aap, 421, 1075
\bibitem[Dzib et al.(2010)]{DZ10} Dzib, S., Loinard, L., Mioduszewski,
  A.~J., Boden, A.~F., Rodr{\'{\i}}guez, L.~F., \& Torres,
  R.~M.\ 2010, \apj, 718, 610
\bibitem[Eiroa et al.(2002)]{CE02} Eiroa, C., Oudmaijer, R.~D.,
  Davies, J.~K., et al.\ 2002, \aap, 384, 103
\bibitem[Eiroa et al.(2008)]{CE08} Eiroa, C., Djupvik, A.~A., \&
  Casali, M.~M.\ 2008, Handbook of Star Forming Regions, Volume II,
  693
\bibitem[Ercolano et al.(2009)]{ER09} Ercolano, B., Clarke, C.~J., \&
  Drake, J.~J.\ 2009, \apj, 699, 1639
\bibitem[Evans et al.(2003)]{EV03} Evans, N.~J., II, et al.\ 2003,
  \pasp, 115, 965
\bibitem[Evans et al.(2009)]{EV09} Evans, N.~J., II, Dunham, M.~M.,
  J{\o}rgensen, J.~K., et al.\ 2009, \apjs, 181, 321
\bibitem[Furlan et al.(2006)]{FU06} Furlan, E., et al.\ 2006, \apjs,
  165, 568
\bibitem[Glauser et al.(2009)]{GL09} Glauser, A.~M., G{\"u}del, M.,
  Watson, D.~M., et al.\ 2009, \aap, 508, 247
\bibitem[Gorti \& Hollenbach(2009)]{GO09a} Gorti, U., \& Hollenbach,
  D.\ 2009, \apj, 690, 1539
\bibitem[Gorti et al.(2009)]{GO09b} Gorti, U., Dullemond, C.~P., \&
  Hollenbach, D.\ 2009, \apj, 705, 1237
\bibitem[G{\"u}del et al.(2007)]{MG07} G{\"u}del, M., et al.\ 2007,
  \aap, 468, 353
\bibitem[Gutermuth et al.(2008)]{GU08} Gutermuth, R.~A., Myers,
  P.~C., Megeath, S.~T., et al.\ 2008, \apj, 674, 336
\bibitem[Greaves \& Rice(2010)]{GR10} Greaves, J.~S., \& Rice,
  W.~K.~M.\ 2010, \mnras, 407, 1981
\bibitem[Haisch et al.(2001)]{HA01} Haisch, K.~E., Jr., Lada, E.~A.,
  \& Lada, C.~J.\ 2001, \apjl, 553, L153
\bibitem[Haisch et al.(2005)]{HA05} Haisch, K.~E., Jr., Jayawardhana,
  R., \& Alves, J.\ 2005, \apjl, 627, L57
\bibitem[Hartmann et al.(2001)]{LH01} Hartmann, L.,
  Ballesteros-Paredes, J., \& Bergin, E.~A.\ 2001, \apj, 562, 852
\bibitem[Harvey et al.(2006)]{HA06} Harvey, P.~M., et al.\ 2006, \apj,
  644, 307
\bibitem[Harvey et al.(2007a)]{HB07} Harvey, P.~M., et al.\ 2007,
  \apj, 663, 1139
\bibitem[Harvey et al.(2007b)]{HA07} Harvey, P., Mer{\'{\i}}n, B.,
  Huard, T.~L., Rebull, L.~M., Chapman, N., Evans, N.~J., II, \&
  Myers, P.~C.\ 2007, \apj, 663, 1149
\bibitem[Hauschildt et al.(1999)]{HA99} Hauschildt, P.~H., Allard, F.,
  Ferguson, J., Baron, E., \& Alexander, D.~R.\ 1999, \apj, 525, 871
\bibitem[Henning(2010)]{HE10} Henning, T.\ 2010, \araa, 48, 21
\bibitem[Hern{\'a}ndez et al.(2007)]{HE07} Hern{\'a}ndez, J., Calvet,
  N., Brice{\~n}o, C., et al.\ 2007, \apj, 671, 1784
\bibitem[Hern{\'a}ndez et al.(2008)]{HE08} Hern{\'a}ndez, J.,
  Hartmann, L., Calvet, N., Jeffries, R.~D., Gutermuth, R., Muzerolle,
  J., \& Stauffer, J.\ 2008, \apj, 686, 1195
\bibitem[Juh{\'a}sz et al.(2012)]{JU12} Juh{\'a}sz, A., Dullemond,
  C.~P., van Boekel, R., et al.\ 2012, \apj, 744, 118
\bibitem[Kennedy \& Kenyon(2009)]{KK09} Kennedy, G.~M., \& Kenyon, S.~J.\ 
2009, \apj, 695, 1210 
\bibitem[Kenyon \& Hartmann(1987)]{KH87} Kenyon, S.~J., \& Hartmann,
  L.\ 1987, \apj, 323, 714
\bibitem[Kenyon \& Hartmann(1995)]{KH95} Kenyon, S.~J., \& Hartmann,
  L.\ 1995, \apjs, 101, 117
\bibitem[Kessler-Silacci et al.(2007)]{KE07} Kessler-Silacci, J.~E.,
  et al.\ 2007, \apj, 659, 680
\bibitem[Knude(2010)]{JK10} Knude, J.\ 2010, arXiv:1006.3676
\bibitem[Lada et al.(2006)]{LA06} Lada, C.~J., Muench, A.~A., Luhman,
  K.~L., et al.\ 2006, \aj, 131, 1574
\bibitem[Lommen et al.(2010)]{LO10} Lommen, D.~J.~P., van Dishoeck,
  E.~F., Wright, C.~M., et al.\ 2010, \aap, 515, A77
\bibitem[Luhman et al.(2003)]{LU03} Luhman, K.~L., Stauffer, J.~R.,
  Muench, A.~A., Rieke, G.~H., Lada, E.~A., Bouvier, J., \& Lada,
  C.~J.\ 2003, \apj, 593, 1093
\bibitem[Luhman(2004)]{KL04} Luhman, K.~L.\ 2004, \apj, 602, 816
\bibitem[Luhman \& Steeghs(2004)]{LS04} Luhman, K.~L., \& Steeghs,
  D.\ 2004, \apj, 609, 917
\bibitem[Luhman(2008)]{KL08} Luhman, K.~L.\ 2008, Handbook of Star
  Forming Regions, Volume II, 169
\bibitem[Luhman et al.(2010)]{LU10} Luhman, K.~L., Allen, P.~R.,
  Espaillat, C., Hartmann, L., \& Calvet, N.\ 2010, \apjs, 186, 111
\bibitem[Mamajek et al.(2002)]{MA02} Mamajek, E.~E., Meyer, M.~R., \&
  Liebert, J.\ 2002, \aj, 124, 1670
\bibitem[Mamajek(2009)]{MA09} Mamajek, E.~E.\ 2009, American Institute
  of Physics Conference Series, 1158, 3
\bibitem[Meeus et al.(2001)]{ME01} Meeus, G., Waters, L.~B.~F.~M.,
  Bouwman, J., van den Ancker, M.~E., Waelkens, C., \& Malfait,
  K.\ 2001, \aap, 365, 476
\bibitem[Megeath et al.(2005)]{ME05} Megeath, S.~T., Hartmann, L.,
  Luhman, K.~L., \& Fazio, G.~G.\ 2005, \apjl, 634, L113
\bibitem[Mer{\'{\i}}n et al.(2010)]{BM10} Mer{\'{\i}}n, B., et al.\
  2010, \apj, 718, 1200
\bibitem[Mortier et al.(2011)]{AM11} Mortier, A., Oliveira, I., \& van
  Dishoeck, E.~F.\ 2011, \mnras, 418, 1194
\bibitem[Muzerolle et al.(2009)]{MU09} Muzerolle, J., Flaherty, K.,
  Balog, Z., et al.\ 2009, \apjl, 704, L15
\bibitem[Muzerolle et al.(2010)]{MU10} Muzerolle, J., Allen, L.~E.,
  Megeath, S.~T., Hern{\'a}ndez, J., \& Gutermuth, R.~A.\ 2010, \apj,
  708, 1107
\bibitem[Myers et al.(2000)]{MY00} Myers, P. C.; Evans, N. J., II; Ohashi, N.\ 
2000, Protostars and Planets IV, 4
\bibitem[Natta et al.(2004)]{NA04} Natta, A., Testi, L., Muzerolle,
  J., et al.\ 2004, \aap, 424, 603
\bibitem[Oliveira et al.(2009)]{OL09} Oliveira, I., Mer{\'{\i}}n, B.,
  Pontoppidan, K.~M., et al.\ 2009, \apj, 691, 672
\bibitem[Oliveira et al.(2010)]{OL10} Oliveira, I., Pontoppidan,
  K.~M., Mer{\'{\i}}n, B., et al.\ 2010, \apj, 714, 778
\bibitem[Oliveira et al.(2011)]{OL11} Oliveira, I., Olofsson, J.,
  Pontoppidan, K.~M., et al.\ 2011, \apj, 734, 51
\bibitem[Olofsson et al.(2010)]{OF10} Olofsson, J., Augereau, J.-C.,
  van Dishoeck, E.~F., Mer{\'{\i}}n, B., Grosso, N., M{\'e}nard, F.,
  Blake, G.~A., \& Monin, J.-L.\ 2010, \aap, 520, A39
\bibitem[Padgett et al.(1999)]{DP99} Padgett, D.~L., Brandner, W.,
  Stapelfeldt, K.~R., Strom, S.~E., Terebey, S., \& Koerner, D.\ 1999,
  \aj, 117, 1490
\bibitem[Pecaut et al.(2012)]{PE12} Pecaut, M.~J., Mamajek, E.~E., \&
  Bubar, E.~J.\ 2012, \apj, 746, 154
\bibitem[Pontoppidan \& Brearley(2010)]{PB10} Pontoppidan, K.~M., \&
  Brearley, A.~J.\ 2010, Protoplanetary Dust: Astrophysical and
  Cosmochemical Perspectives, 191
\bibitem[Preibisch et al.(2002)]{PR02} Preibisch, T., Brown, A.~G.~A.,
  Bridges, T., Guenther, E., \& Zinnecker, H.\ 2002, \aj, 124, 404
\bibitem[Preibisch \& Zinnecker(1999)]{PZ99} Preibisch, T., \&
  Zinnecker, H.\ 1999, \aj, 117, 2381
\bibitem[Shu et al.(1993)]{SH93} Shu, F.~H., Johnstone, D., \&
  Hollenbach, D.\ 1993, Icarus, 106, 92
\bibitem[Sicilia-Aguilar et al.(2006)]{SI06} Sicilia-Aguilar, A.,
  Hartmann, L., Calvet, N., et al.\ 2006, \apj, 638, 897
\bibitem[Sicilia-Aguilar et al.(2007)]{SI07} Sicilia-Aguilar, A.,
  Hartmann, L.~W., Watson, D., et al.\ 2007, \apj, 659, 1637
\bibitem[Sicilia-Aguilar et al.(2009)]{SI09} Sicilia-Aguilar, A., et
  al.\ 2009, \apj, 701, 1188
\bibitem[Sicilia-Aguilar et al.(2011)]{SI11} Sicilia-Aguilar, A.,
  Henning, T., Dullemond, C.~P., et al.\ 2011, \apj, 742, 39
\bibitem[Siess et al.(2000)]{SI00} Siess, L., Dufour, E., \&
  Forestini, M.\ 2000, \aap, 358, 593
\bibitem[Spezzi et al.(2008)]{SP08} Spezzi, L., Alcal{\'a}, J.~M.,
  Covino, E., et al.\ 2008, \apj, 680, 1295
\bibitem[Spezzi et al.(2010)]{SP10} Spezzi, L., Merin, B., Oliveira,
  I., van Dishoeck, E.~F., \& Brown, J.~M.\ 2010, \aap, 513, A38
\bibitem[Straizys et al.(1996)]{ST96} Straizys, V., Cernis, K. \&
  Bartasiute, S.\ 1996, Baltic Astron., 5, 125
\bibitem[Strom et al.(1989)]{ST89} Strom, K.~M., Strom, 
S.~E., Edwards, S., Cabrit, S., \& Skrutskie, M.~F.\ 1989, \aj, 97, 1451
\bibitem[Udry \& Santos(2007)]{US07} Udry, S., \& Santos, N.~C.\ 2007,
  \araa, 45, 397
\bibitem[Visser \& Dullemond(2010)]{VD10} Visser, R., \& Dullemond,
  C.~P.\ 2010, \aap, 519, A28
\bibitem[Wahhaj et al.(2010)]{ZW10} Wahhaj, Z., Cieza, L., Koerner,
  D.~W., et al.\ 2010, \apj, 724, 835
\bibitem[Watson et al.(2009)]{WA09} Watson, D.~M., Leisenring, J.~M.,
  Furlan, E., et al.\ 2009, \apjs, 180, 84
\bibitem[Weidenschilling(1980)]{WE80} Weidenschilling, S.~J.\ 1980, Icarus, 
44, 172 
\bibitem[Weingartner \& Draine(2001)]{WD01} Weingartner, J.~C., \&
  Draine, B.~T.\ 2001, \apj, 548, 296
\bibitem[White \& Basri(2003)]{WB03} White, R.~J., \& Basri, G.\ 2003,
  \apj, 582, 1109
\bibitem[Winston et al.(2010)]{WI10} Winston, E., Megeath, S.~T.,
  Wolk, S.~J., et al.\ 2010, \aj, 140, 266
\bibitem[Wooden et al.(2007)]{WO07} Wooden, D., Desch, S., Harker, D.,
  Gail, H.-P., \& Keller, L.\ 2007, Protostars and Planets V, 815

\end{thebibliography}
\end{document}